\documentclass[journal]{IEEEtran}
\usepackage{times}
\usepackage{epsfig}
\usepackage{graphicx}
\usepackage{amsmath}
\usepackage{amssymb}
\usepackage{booktabs}
\usepackage{amsmath}
\usepackage{tabularx}
\usepackage{color}
\usepackage{bm}
\usepackage{algorithm}
\usepackage{algorithmic}
\usepackage{multirow}
\usepackage{subfigure}
\usepackage{booktabs}
\usepackage{graphicx}
\usepackage{subeqnarray}
\usepackage{cases}
\usepackage{mathrsfs}

\usepackage{cuted}
\usepackage{capt-of}

\usepackage{hyperref}

\ifCLASSINFOpdf
\else
\fi
\begin{document}
%
\title{Learning Deep Context-Sensitive Decomposition for Low-Light Image Enhancement}
%
%
%

\author{Long~Ma,
	Risheng~Liu,~\IEEEmembership{Member,~IEEE,}
	Jiaao~Zhang,
	Xin~Fan,~\IEEEmembership{Senior Member,~IEEE,}
	and Zhongxuan~Luo
	\thanks{This work is partially supported by the National Natural Science Foundation of China (Nos.	61922019, 61733002, 62027826, 61722105, and 61672125), LiaoNing Revitalization Talents Program (XLYC1807088), and the Fundamental Research Funds for the Central Universities.}
	\thanks{L. Ma and J. Zhang are with the School of Software Technology, Dalian University of Technology, Dalian, 116024, China. (email: malone94319@gmail.com, jiaaozhang@mail.dlut.edu.cn).}
	\thanks{R. Liu is with the DUT-RU International School of Information Science \& Engineering, Dalian University of Technology, Dalian, 116024, China. 	He is also with the Peng Cheng Laboratory, Shenzhen, 518052, China. (Corresponding author, email: rsliu@dlut.edu.cn).}
	\thanks{X. Fan is with the DUT-RU International School of Information Science \& Engineering, Dalian University of Technology, Dalian, 116024, China. He is also with the Peng Cheng Laboratory, Shenzhen, 518052, China. (email: xin.fan@dlut.edu.cn).}
	\thanks{Z. Luo is with the School of Software, Dalian University of Technology, Dalian, 116024, China. He is also with the Peng Cheng Laboratory, Shenzhen, 518052, China. (email: zxluo@dlut.edu.cn).}
	\thanks{Manuscript received April 19, 2005; revised August 26, 2015.}
}
%
%

\markboth{Journal of \LaTeX\ Class Files,~Vol.~14, No.~8, August~2015}%
{Shell \MakeLowercase{\textit{et al.}}: Bare Demo of IEEEtran.cls for IEEE Journals}
%



\maketitle

{
\begin{abstract}
	Enhancing the quality of low-light images plays a very important role in many image processing and multimedia applications.
	In recent years, a variety of deep learning techniques have been developed to address this challenging task. A typical framework is to simultaneously estimate the illumination and reflectance, but they disregard the scene-level contextual information encapsulated in feature spaces, causing many unfavorable outcomes, e.g., details loss, color unsaturation, artifacts, and so on. To address these issues, we develop a new context-sensitive decomposition network architecture to exploit the scene-level contextual dependencies on spatial scales. More concretely, we build a two-stream estimation mechanism including reflectance and illumination estimation network. We design a novel context-sensitive decomposition connection to bridge the two-stream mechanism by incorporating the physical principle. The spatially-varying illumination guidance is further constructed for achieving the edge-aware smoothness property of the illumination component. According to different training patterns, we construct CSDNet (paired supervision) and CSDGAN (unpaired supervision) to fully evaluate our designed architecture. 
	We test our method on seven testing benchmarks (including MIT-Adobe FiveK, LOL, ExDark, NPE, etc.) to conduct plenty of analytical and evaluated experiments. Thanks to our designed context-sensitive decomposition connection, we successfully realized excellent enhanced results (with sufficient details, vivid colors, and few noises), which fully indicates our superiority against existing state-of-the-art approaches. Finally, considering the practical needs for high-efficiency, we develop a lightweight CSDNet (named LiteCSDNet) by reducing the number of channels. Further, by sharing an encoder for these two components, we obtain a more lightweight version (SLiteCSDNet for short). SLiteCSDNet just contains 0.0301M parameters but achieves the almost same performance as CSDNet.   
	\textit{Code is available at \url{https://github.com/KarelZhang/CSDNet-CSDGAN}.}
\end{abstract}
	}

%
\IEEEpeerreviewmaketitle

\begin{figure}[t]
	\centering
	\begin{tabular}{c} 
		\includegraphics[width=0.47\textwidth]{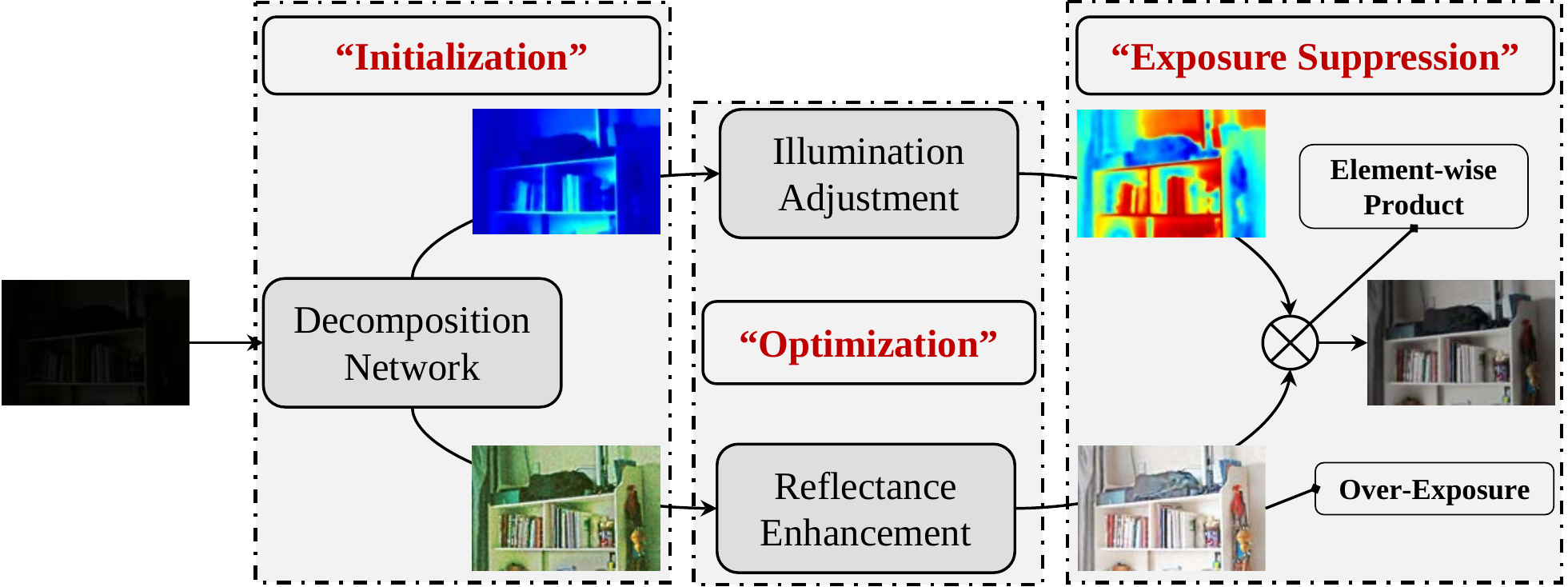}\\
		\footnotesize (a) The architecture of existing representative networks, e.g., KinD~\cite{zhang2019kindling}.\\
		\includegraphics[width=0.47\textwidth]{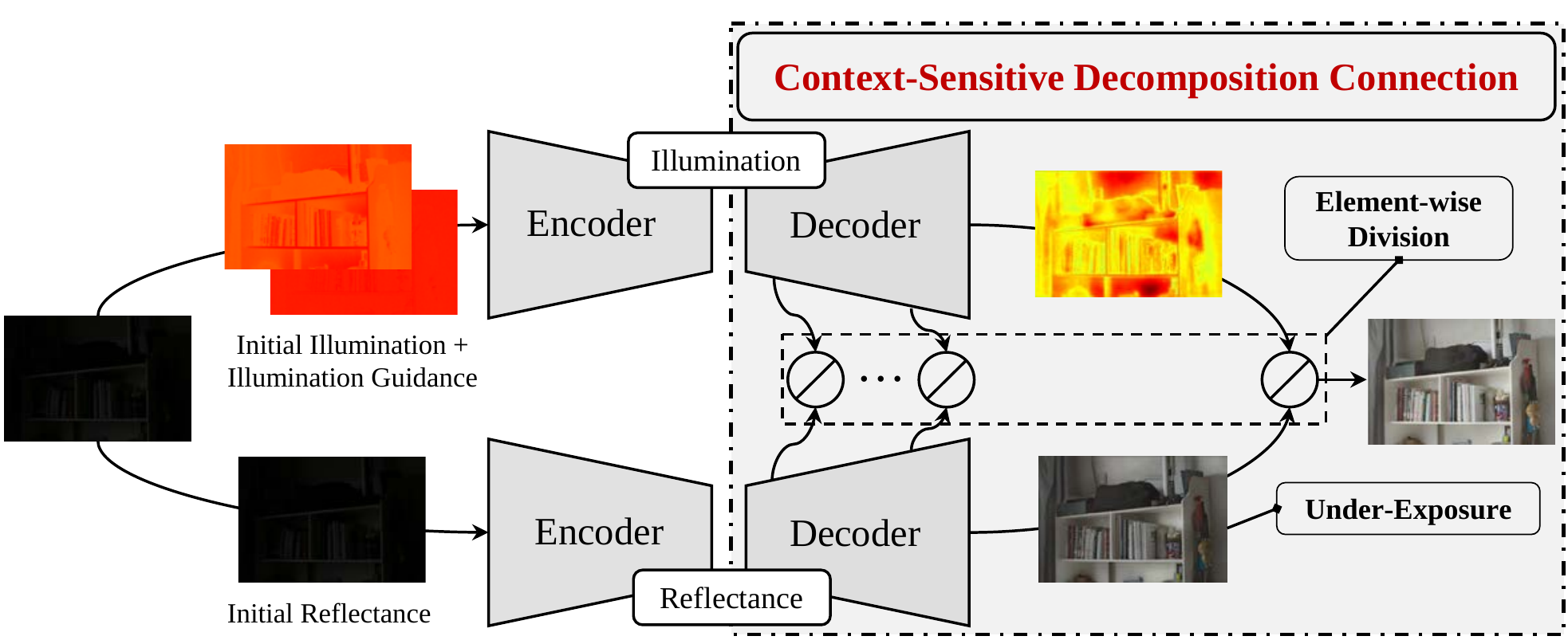}\\
		\footnotesize (b) Our proposed context-sensitive decomposition network architecture.\\
	\end{tabular}
	\caption{Flowchart comparison among existing representative deep networks (e.g., KinD~\cite{zhang2019kindling}, RetinexNet~\cite{Chen2018Retinex}) and our method. (a) and (b) have the same goal, i.e., simultaneously optimizing the illumination and reflectance. As for (a), its entire architecture is redundant, it can be roughly partitioned into three parts including ``Initialization'', ``Optimization'', and ``Exposure Suppression". In which, ``Initialization" part generates two inaccurate initial estimations according to the Retinex-based fidelity term. In the phase of ``Optimization", the illumination is further optimized to ensure spatially-varying smoothness, and the reflectance is further enhanced to eliminate noises and improve brightness. `Exposure Suppression" stage suppresses the over-exposure by utilizing the final illumination. Actually, the pipeline of ``{under-exposure$\Rightarrow$over-exposure$\Rightarrow$normal-exposure}" is superfluous and unnecessary. In other words, this pipeline takes a detour route to address low-light image enhancement and ignores the exploitation of the contextual information. Different from it, our designed architecture considers a straightforward route to progressively improve visual quality. On the one hand, we just need to obtain the initial values by operating the low-light input, not a network. We also define the illumination guidance to accurately estimate the illumination. On the other hand, we construct the context-sensitive decomposition connection inspired by physical knowledge. This connection acts on the feature and image levels of two components, to fully exploit the contextual dependencies on spatial scales. In a word, our designed architecture is more reasonable and meaningful.}
	\label{fig:FirstFig}
\end{figure}

\begin{figure*}[t]
	\centering
	\begin{tabular}{c@{\extracolsep{0.3em}}c@{\extracolsep{0.3em}}c@{\extracolsep{0.3em}}c@{\extracolsep{0.3em}}c} 
		\includegraphics[width=0.19\textwidth]{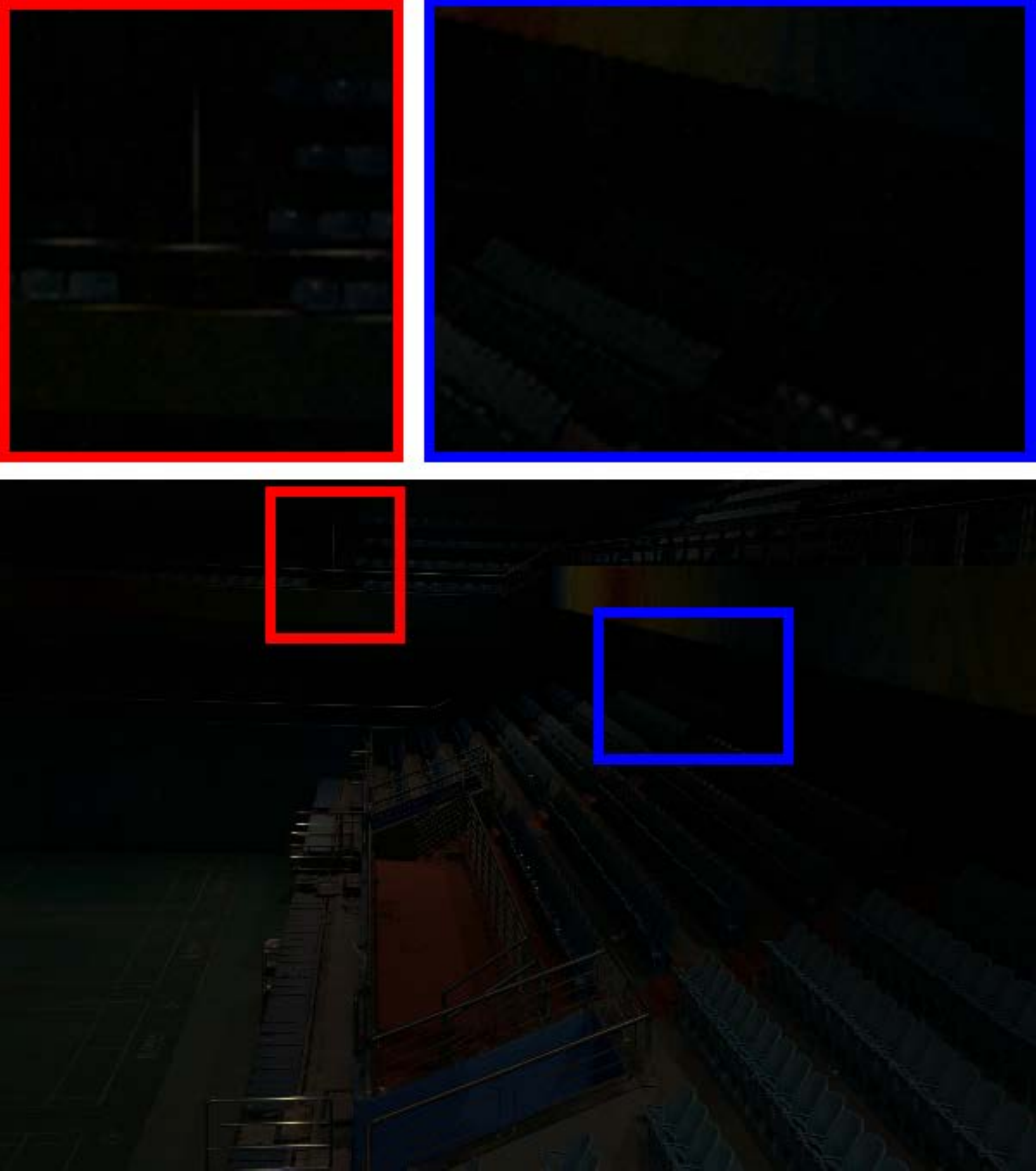}&
		\includegraphics[width=0.19\textwidth]{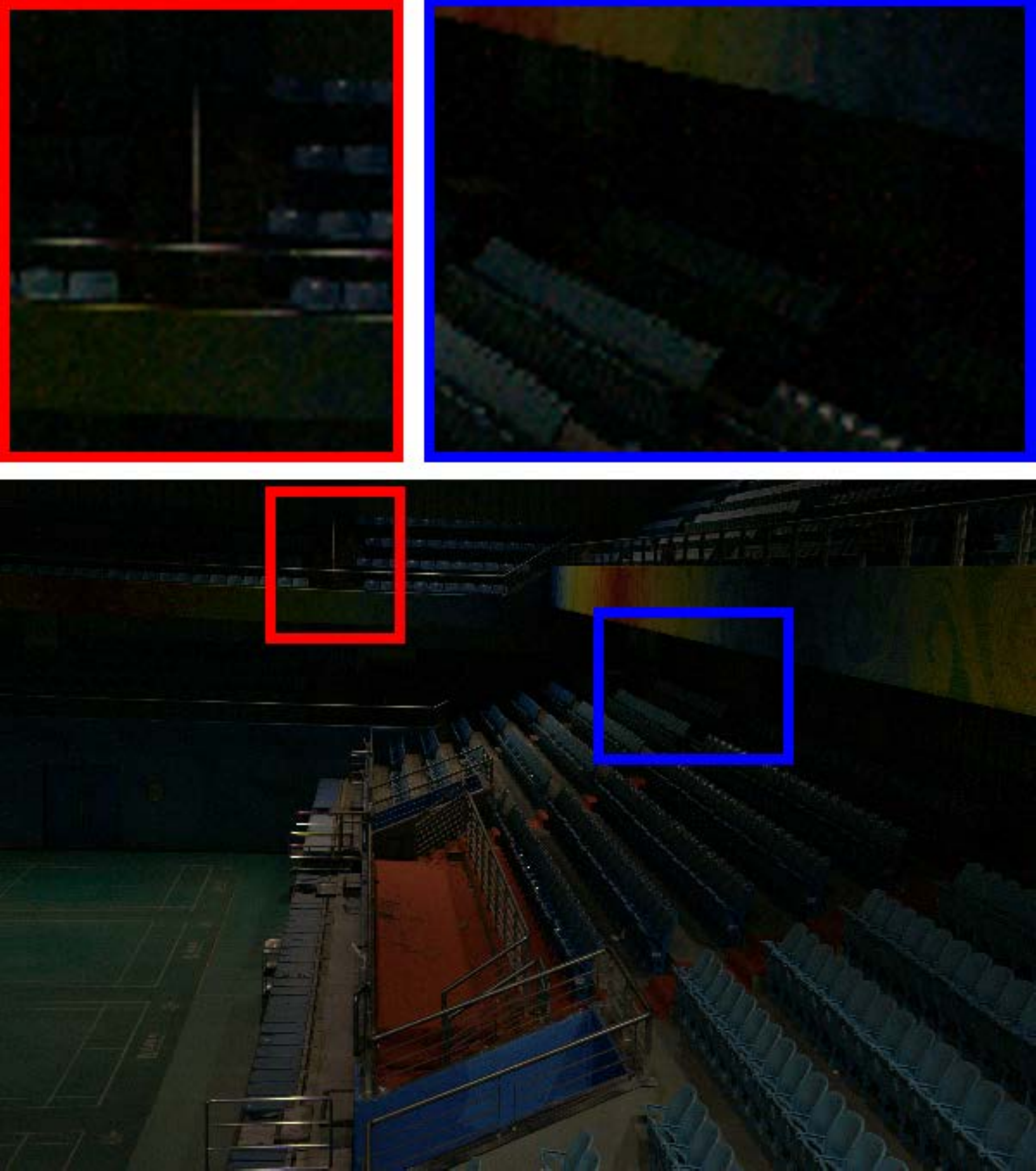}&
		\includegraphics[width=0.19\textwidth]{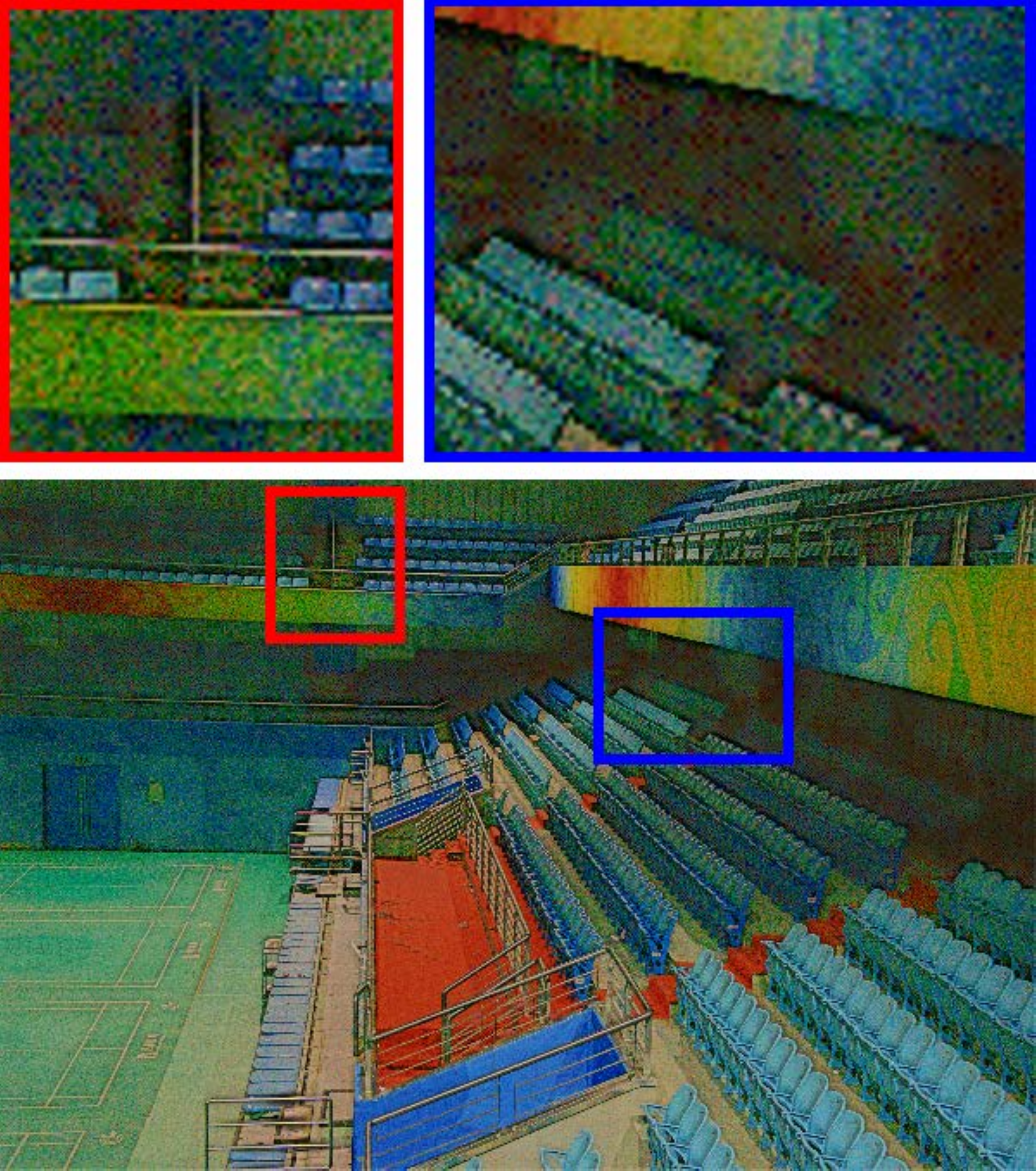}&
		\includegraphics[width=0.19\textwidth]{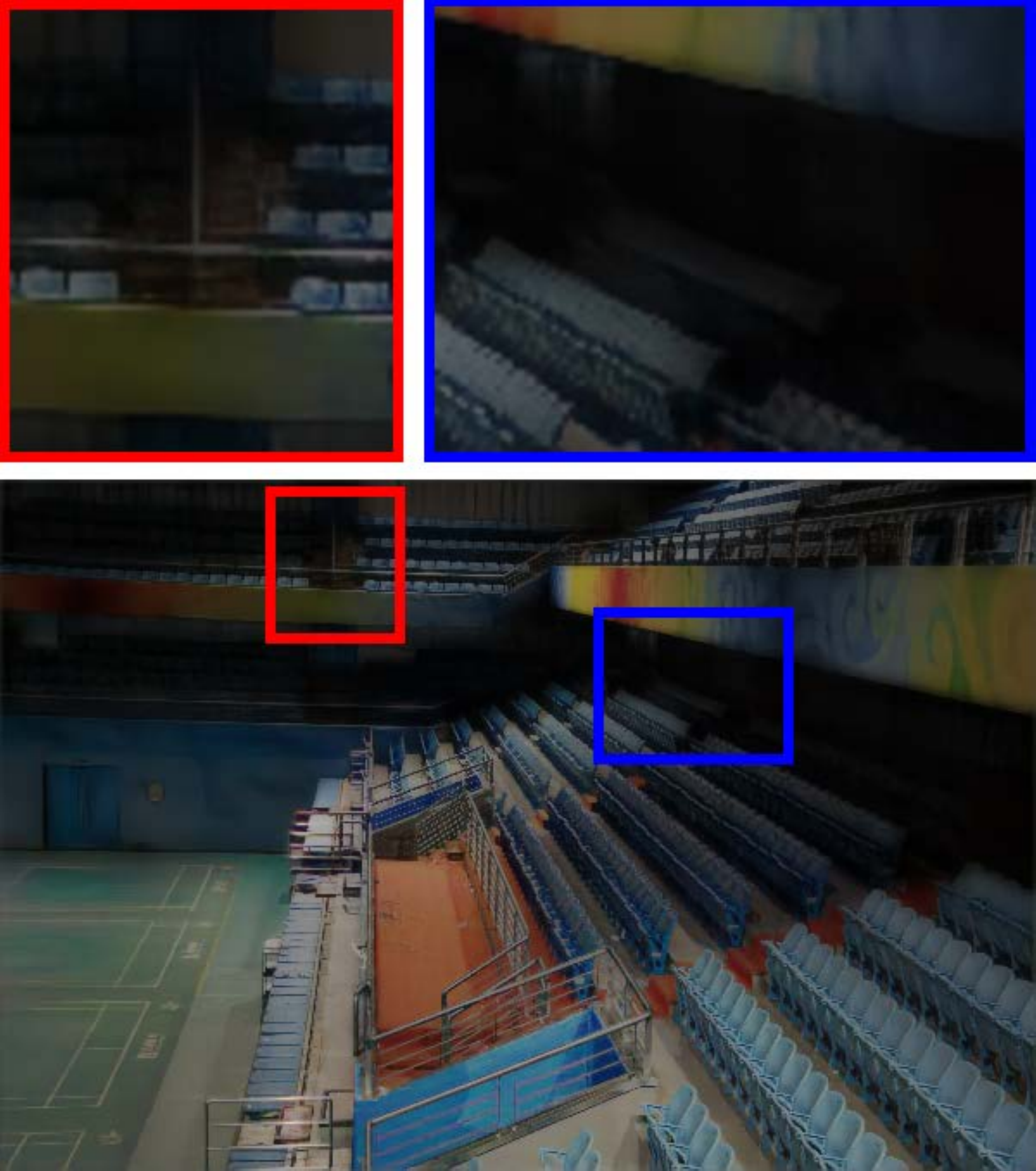}&
		\includegraphics[width=0.19\textwidth]{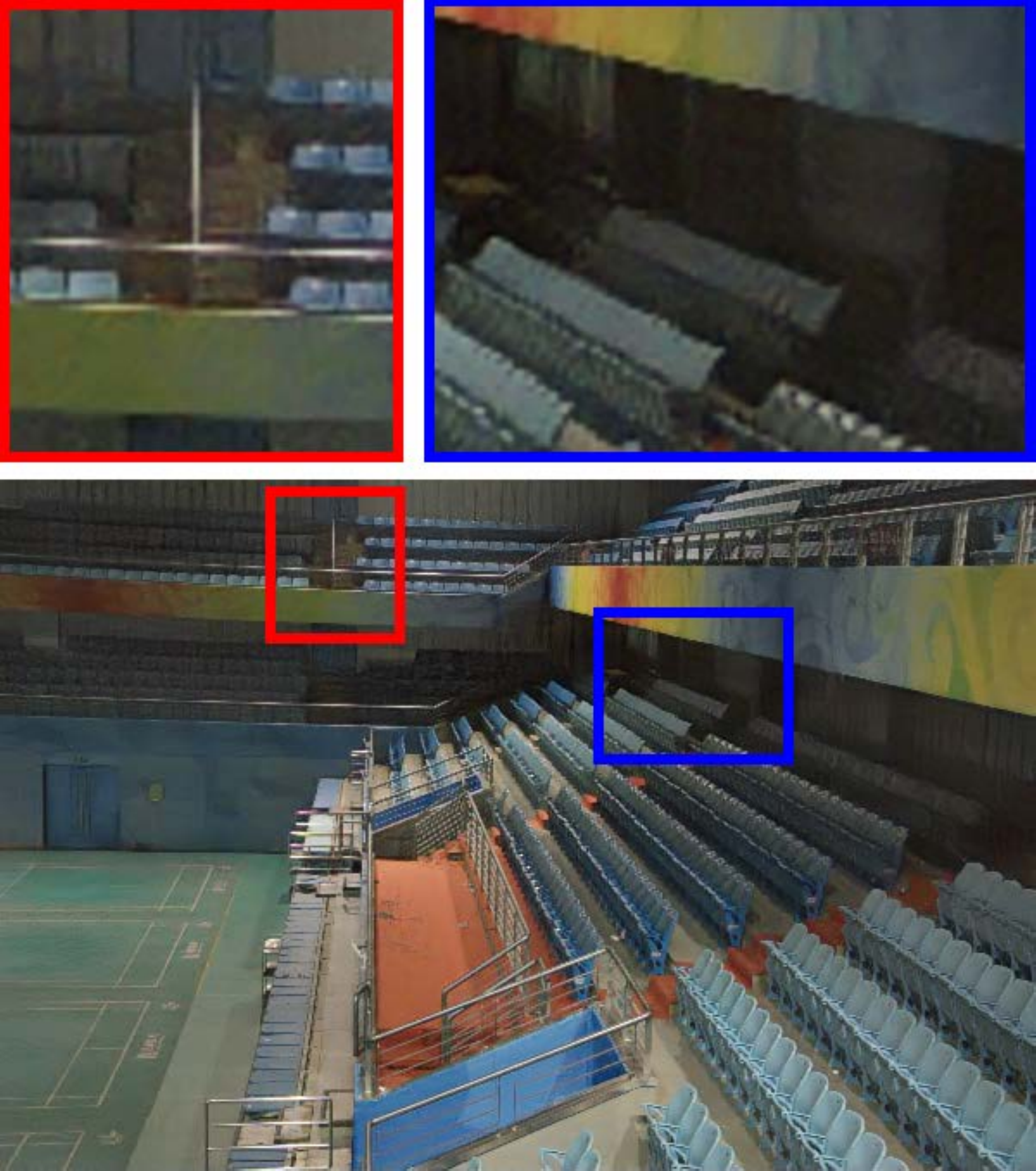}\\
		\footnotesize Input&\footnotesize LightenNet~\cite{li2018lightennet}&\footnotesize RetinexNet~\cite{Chen2018Retinex}&\footnotesize KinD~\cite{zhang2019kindling}&\footnotesize Ours\\
	\end{tabular}
	\caption{Visual comparison on an extremely dark example from the LOL dataset~\cite{Chen2018Retinex}. Obviously, LightenNet~\cite{li2018lightennet} fails to enhance the exposure, RetinexNet~\cite{Chen2018Retinex} generates the visible noises and color distortion, KinD suppresses the noises but some details are still invisible, especially in the zoomed-in regions. Thanks to the exploitation of contextual dependencies on spatial scales, our proposed CSDNet achieves the best visual performance with normal exposure, vivid color and fewer noises with no need for the additional denoising operators. }
	\label{fig:SecondFig}
\end{figure*}

\section{Introduction}
\IEEEPARstart{L}{ow-light} is one of the most common factors to degrade image quality in computer vision and multimedia communities~\cite{zhang2018high,li2018lightennet,wang2019underexposed}. The cause of low-light is diverse and unpredictable, e.g., scene luminance, shutter speed and so on. A lot of information contains the shape and chrominance of the object that are hidden in the dark, which is adverse to the further operation~\cite{chen2018learning}. Indeed, with the development and popularity of portable devices (e.g., mobile), the demands for clear and bright high-quality images under some challenging light conditions (e.g., night)~\cite{loh2019getting} have become significantly urgent. Therefore, it is extremely necessary to construct an effective and practical algorithmic framework to tackle the so-called Low-Light Image Enhancement (LLIE) task in diversified real-world scenarios.
\thispagestyle{empty}

Value-based mapping method (e.g., HE~\cite{cheng2004simple}, Gamma Correction) is a kind of direct and easy way to enhance the exposure of low-light images. Although the exposure levels indeed get a promotion, the non-uniform illumination and unrealistic color cause the unnaturalness performance. Retinex theory makes an assumption between low and normal light data to provide a physical explanation for low-light enhancement. According to this theory, many researchers strive to build model-based optimization algorithms~\cite{fu2016weighted,cai2017joint,guo2017lime}. These methods built their Retinex-based data fidelity and designed prior regularizations to present the intrinsic property of the desired components. Unfortunately, limited to the designed priors, these traditional algorithms tend to produce the result with under/over-exposure, insufficient details, unsaturated colors, prominent artifacts (or noises). 

To ameliorate the drawbacks of the above traditional works, it has become a predominant pattern that cascading the heuristic network modules to build an explicit mapping between low-light inputs and desired enhanced outputs~\cite{wang2019underexposed,Chen2018Retinex,zhang2019kindling}, especially in recent few years. As is known to all, acquiring training data is crucial, it is the cornerstone for deep learning techniques to go on wheels. Utilizing the physical degradation model to synthesize data is the most familiar way to provide effective training pairs in many conventional image processing problems~\cite{guo2019toward}. 
However, in the task of LLIE, illumination is the key degraded factor but difficult to obtain by manual operation. Some works~\cite{Chen2018Retinex, chen2018learning} changed the exposure time to acquire low and high exposure image pairs in real-world scenarios. Some works~\cite{fivek,wang2019underexposed} generated desired outputs with normal exposure based on expert-retouched. There also exist some datasets without labels~\cite{jiang2019enlightengan,loh2019getting}.

Based on these mentioned datesets, many end-to-end deep learning techniques (e.g., DeepUPE~\cite{wang2019underexposed}, RetinexNet~\cite{Chen2018Retinex}, KinD~\cite{zhang2019kindling}, EnlightenGAN~\cite{jiang2019enlightengan} and so on) have been developed to address LLIE. 
Introducing the task-specific physical principle and prior regularization become the mainstream designed idea. As is shown in Fig.~\ref{fig:FirstFig}, the representative data-driven deep network (e.g., RetinexNet~\cite{Chen2018Retinex} and KinD~\cite{zhang2019kindling}) built a ``three stages" architecture. They roughly estimated the initial illumination and reflectance by a decomposition network. Then further optimized these two components by adopting the designed architectures, respectively. However, the generated reflectance became over-exposure, thus they consider executing the Retinex theory to obtain the final normal-exposure result.
Actually, this pipeline is too redundant, especially the part of ``Initialization" and ``Exposure Suppression". More importantly, it ignored the exploitation of scene-level contextual information on spatial scales~\cite{wang2019erl}.

\begin{figure*}[t]
	\centering
	\begin{tabular}{c@{\extracolsep{0.3em}}c@{\extracolsep{0.3em}}c@{\extracolsep{0.35em}}c} 
		\includegraphics[width=0.24\textwidth]{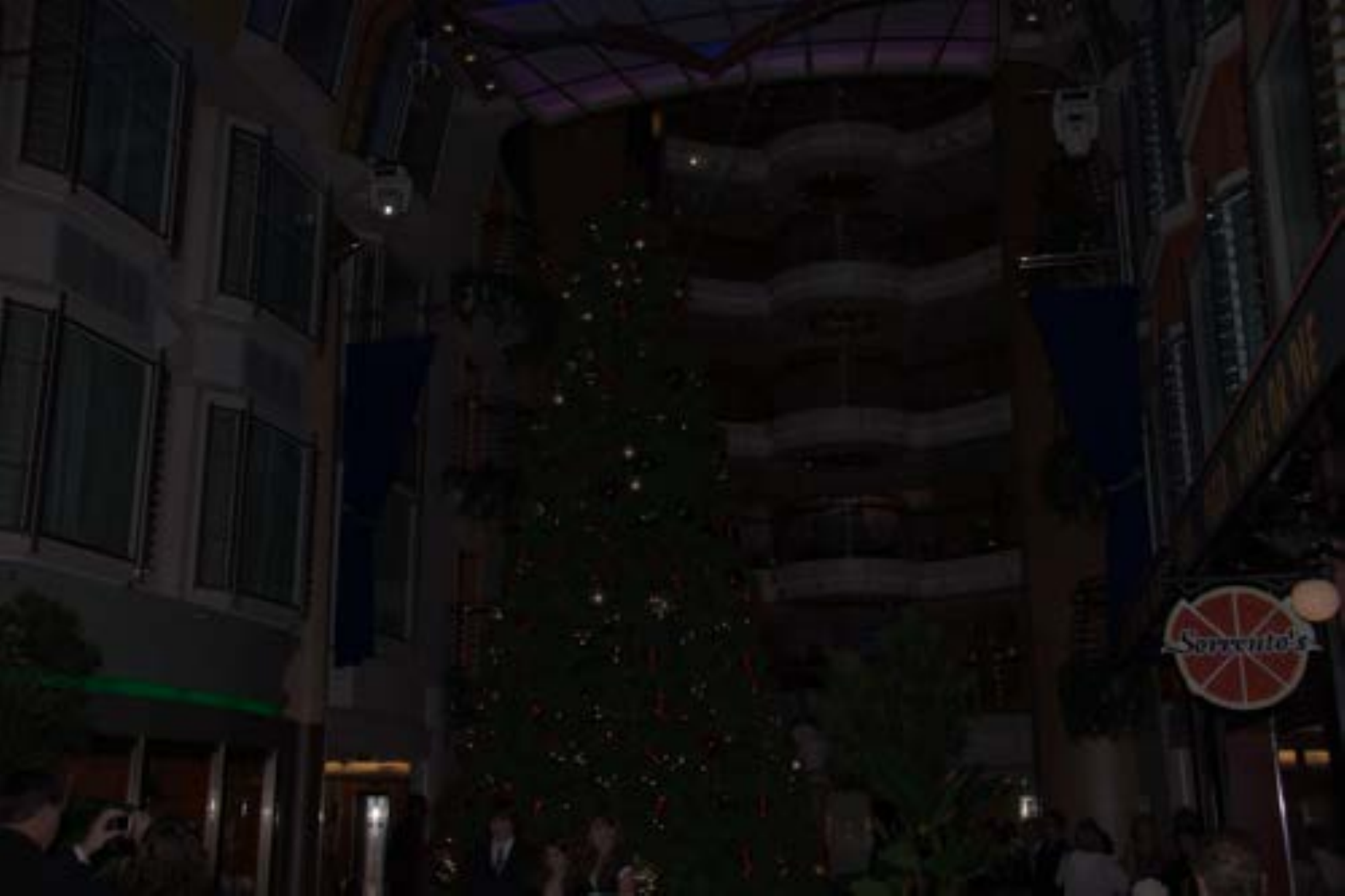}&
		\includegraphics[width=0.24\textwidth]{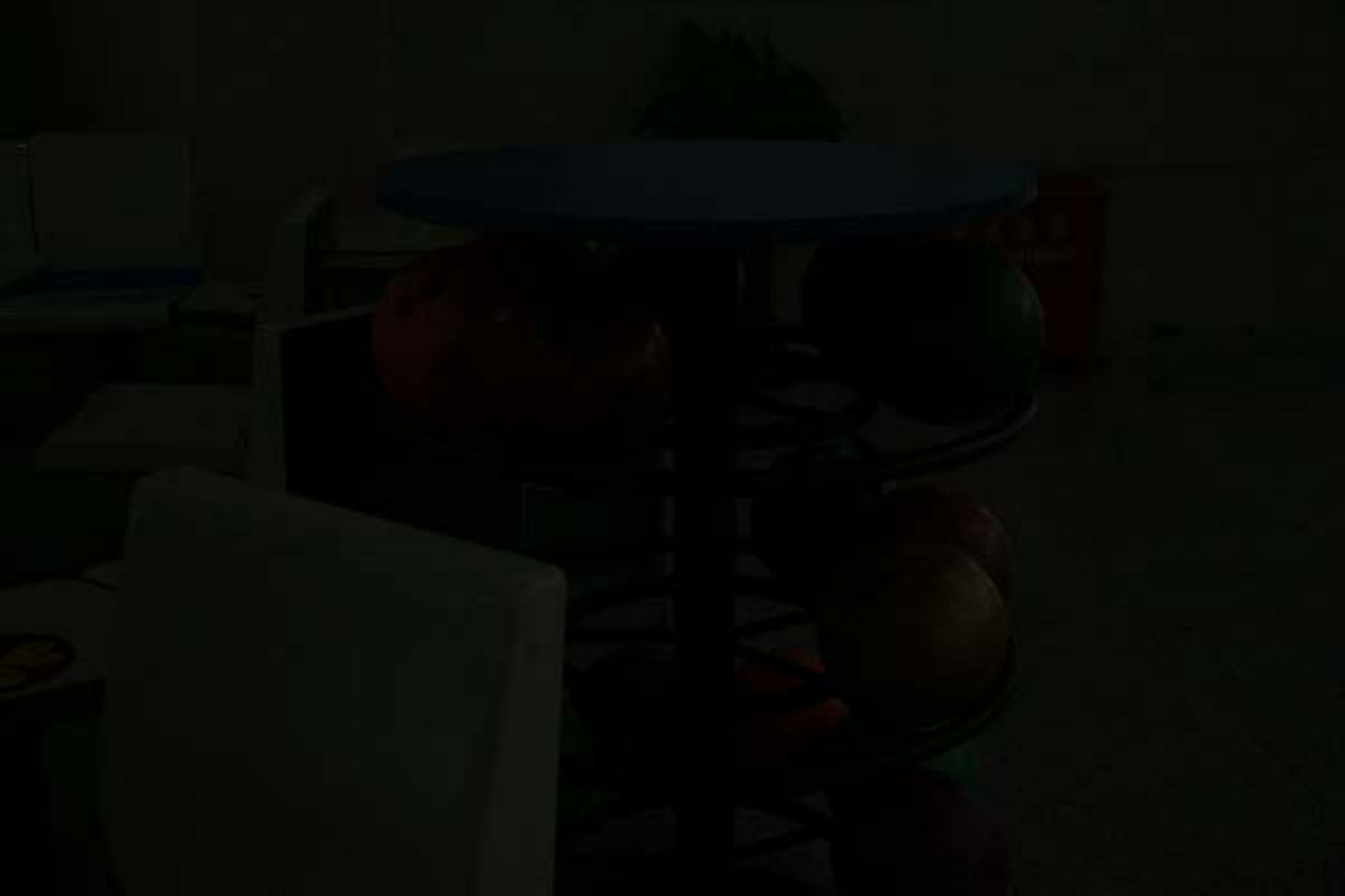}&
		\includegraphics[width=0.24\textwidth]{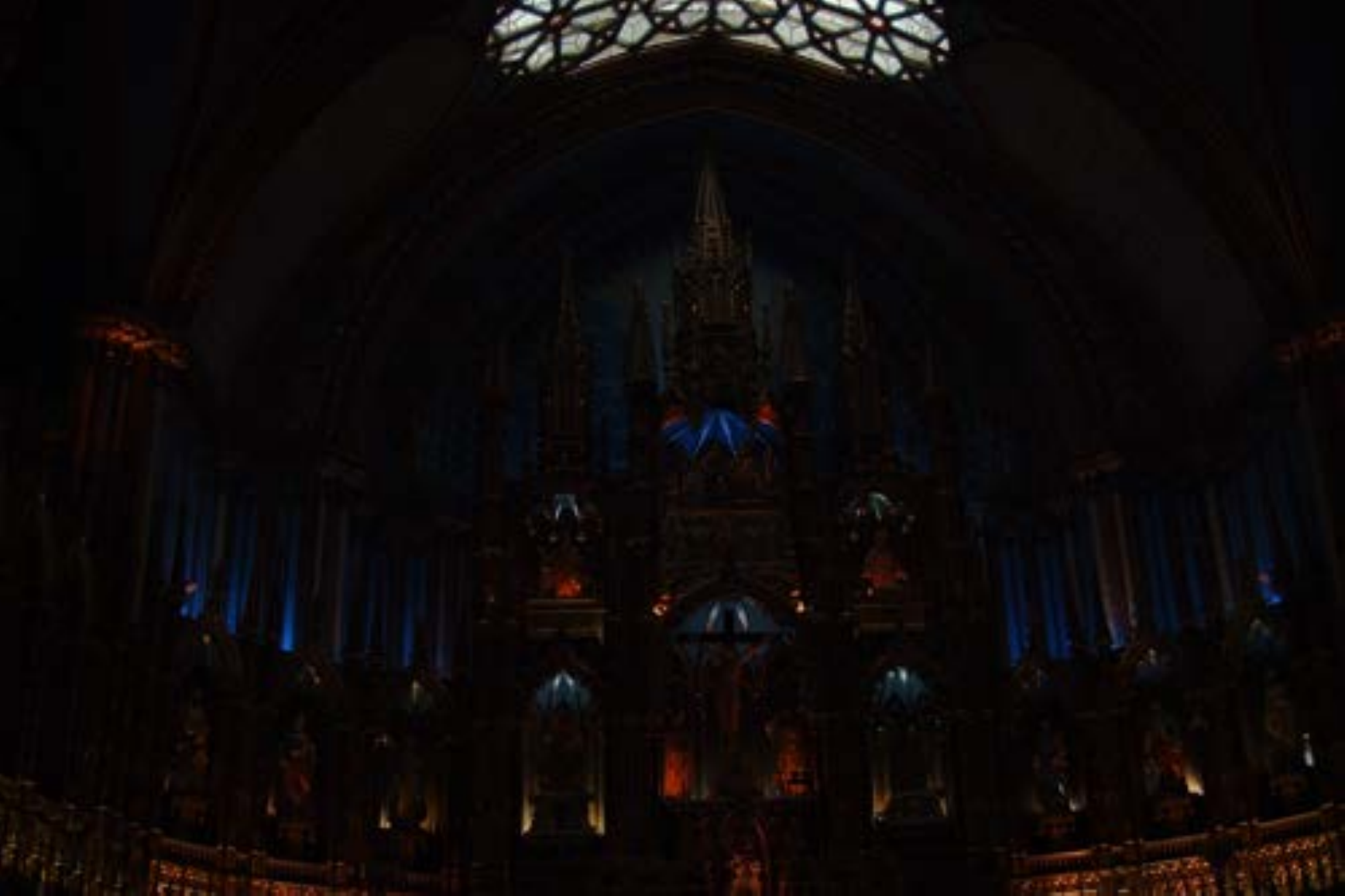}&
		\includegraphics[width=0.24\textwidth]{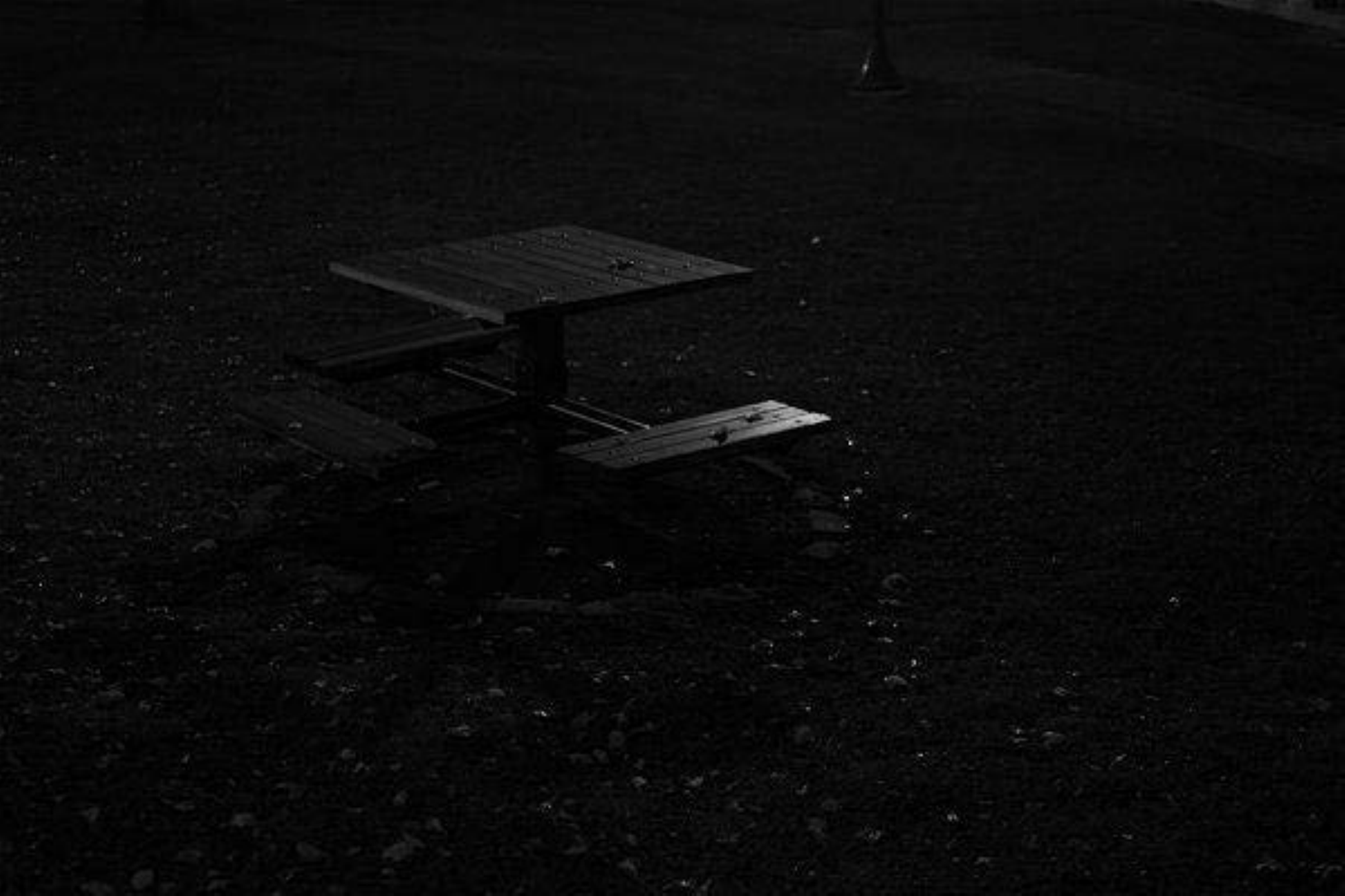}\\
		\includegraphics[width=0.24\textwidth]{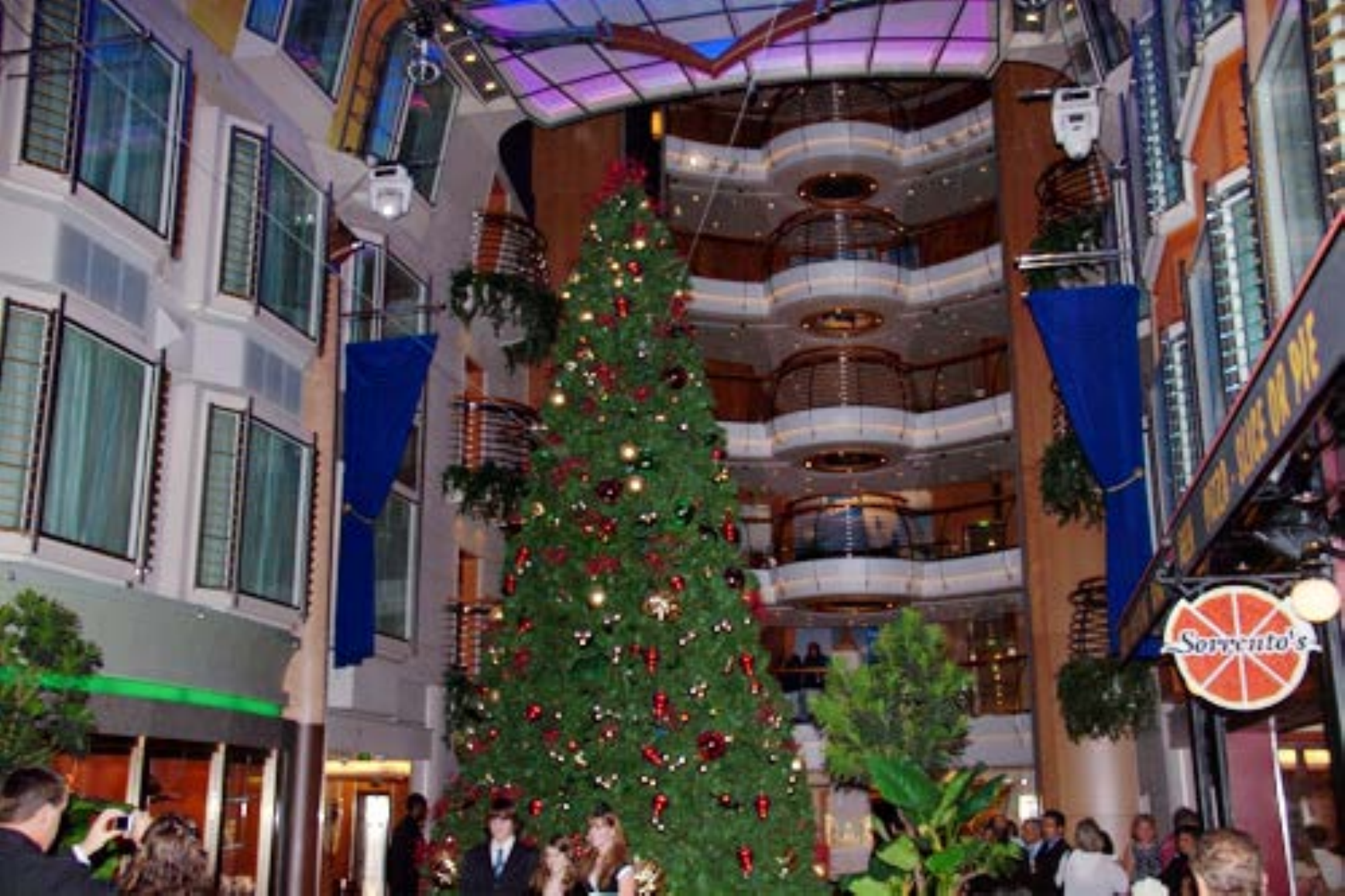}&
		\includegraphics[width=0.24\textwidth]{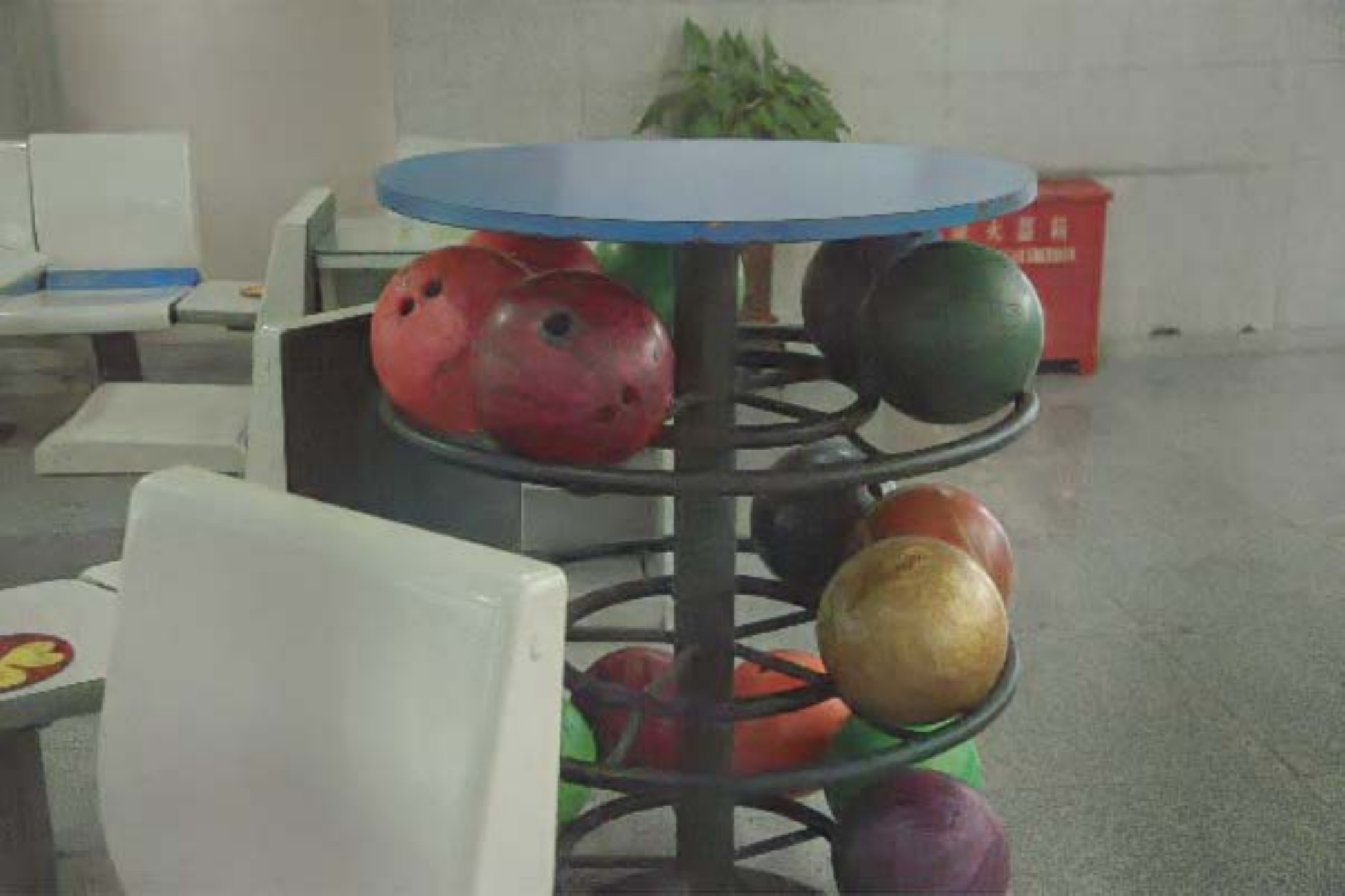}&
		\includegraphics[width=0.24\textwidth]{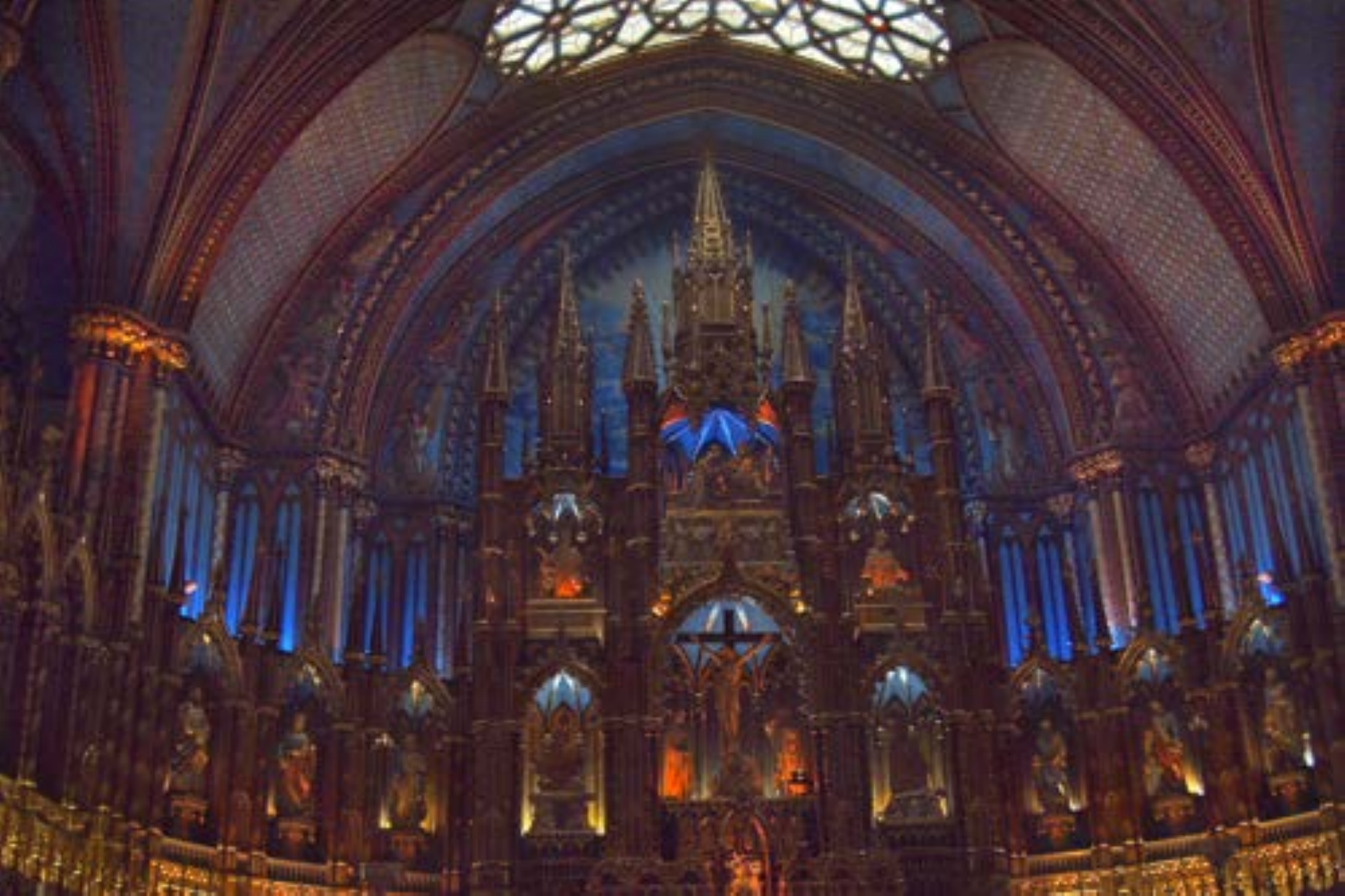}&
		\includegraphics[width=0.24\textwidth]{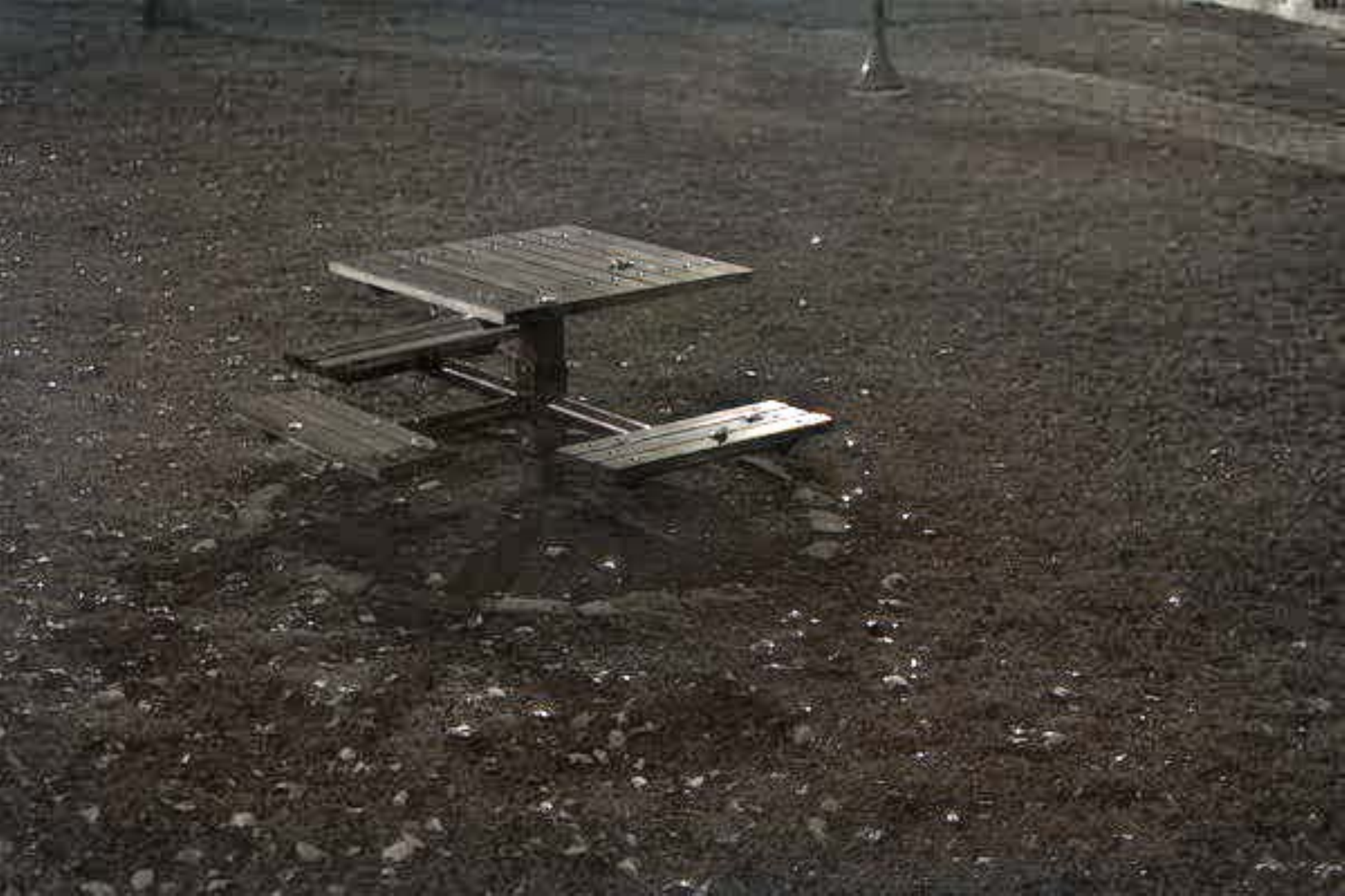}\\
		\footnotesize (a)&\footnotesize (b) &\footnotesize (c) &\footnotesize (d)\\
	\end{tabular}
	\caption{Visual comparison on some extremely low-light examples. \textbf{Top Row:} Low-light inputs. \textbf{Bottom Row:} The first two columns are the results of CSDNet, and the last two columns are the results of CSDGAN. (a)-(d) come from MIT-Adobe FiveK~\cite{fivek}, LOL~\cite{Chen2018Retinex}, NPE~\cite{wang2013naturalness}, ExDark~\cite{loh2019getting} benchmarks, respectively. Our proposed method (CSDNet and CSDGAN) can handle various complex and difficult low-light scenarios.}
	\label{fig:ThirdFig}
\end{figure*}

\subsection{Our Contributions}
In this work, to address LLIE, we develop a novel network architecture to exploit the scene-level contextual dependencies on spatial scales. Different from existing works to simultaneously estimate the illumination and reflectance using redundant architectures (as is shown in Fig.~\ref{fig:FirstFig} (a)), our designed architecture (as is shown in Fig.~\ref{fig:FirstFig} (b)) takes on a straight manner to progressively enhance the low-light input on spatial-scales. 
As is shown in Fig.~\ref{fig:SecondFig}, our proposed method successfully overcomes drawbacks (inappropriate exposure, color distortion, and visible noises) in these three deep learning works and achieves the outstanding performance\footnote{More visual comparison can be found in Sec.~\ref{sec:experiment}.}. In Fig.~\ref{fig:ThirdFig}, we also present the enhanced results of our designed CSDNet and CSDGAN in different challenging real-world scenarios. 

To be specific, we define illumination and reflectance estimation networks to obtain a two-stream estimation mechanism, for simultaneously estimating the illumination and reflectance. To bridge these two estimation sub-networks, we design the context-sensitive decomposition module. We also define the spatially-varying illumination guidance to ensure accurately estimating the illumination. 
	Extensive qualitative and quantitative evaluations demonstrate that our proposed method outperforms existing state-of-the-art approaches under different metrics and visual quality. In order to satisfy the actual demand for execution efficiency, LiteCSDNet is built by reducing the channel numbers, then a more lightweight version SLiteCSDNet is constructed by sharing an encoder for two-stream sub-networks.

In summary, our main contributions can be described as the following four-folds:
\begin{enumerate}
	\item
	 We build a two-stream estimation mechanism to simultaneously optimize the illumination and reflectance. To bridge these two sub-networks, a new context-sensitive decomposition connection is designed. This connection successfully aggregates the scene-level contextual dependencies of the reflectance and illumination on spatial scales by incorporating the physical principle of LLIE. In this way, we directly establish an explicit mapping from low-light to normal-exposure, instead of the detour route (see Fig.~\ref{fig:FirstFig}).
	 
	\item 
	 We construct the spatially-varying illumination guidance as an additional input channel of the illumination estimation network. This guidance guarantees the desired edge-aware smoothness property of the illumination component, to further realize the stable implementation when performing the context-sensitive decomposition connection on feature levels. Experiments about it also demonstrate that illumination guidance has the ability to significantly improve the brightness.
	
	\item
	 Sufficient analyses and evaluations in terms of our CSDNet (paired supervision) and CSDGAN (unpaired supervision) are conducted on seven commonly-used benchmarks. These results fully indicate that our superiority in the desirable visual quality (with sufficient details, vivid colors, and fewer noises) and the best quantitative scores against state-of-the-art approaches. 
	 \item 
	 Considering the need for fast inference time in practical applications, by reducing the number of channels in CSDNet, we construct a lightweight CSDNet, named LiteCSDNet. Further, we share an encoder for two components to obtain a more lightweight version, SLiteCSDNet for short. Surprisingly, it just needs 0.0301M parameters and achieves the parameter reduction rate of more than 98.5\% against existing state-of-the-art methods. More importantly, LiteCSDNet and SLiteCSDNet all achieve almost the same performance as CSDNet.
\end{enumerate}

\section{Related Works}

Enhancing the low-light images has attracted much attention in the past few decades. There emerge plentiful techniques that roughly involve value-based mapping, model-based optimization, and data-driven deep learning, where the last one has become the current mainstream in the recent few years. We will make brief but comprehensive reviews in the following.
\begin{figure*}[t]
	\centering
	\begin{tabular}{c}
		\includegraphics[width=0.98\textwidth]{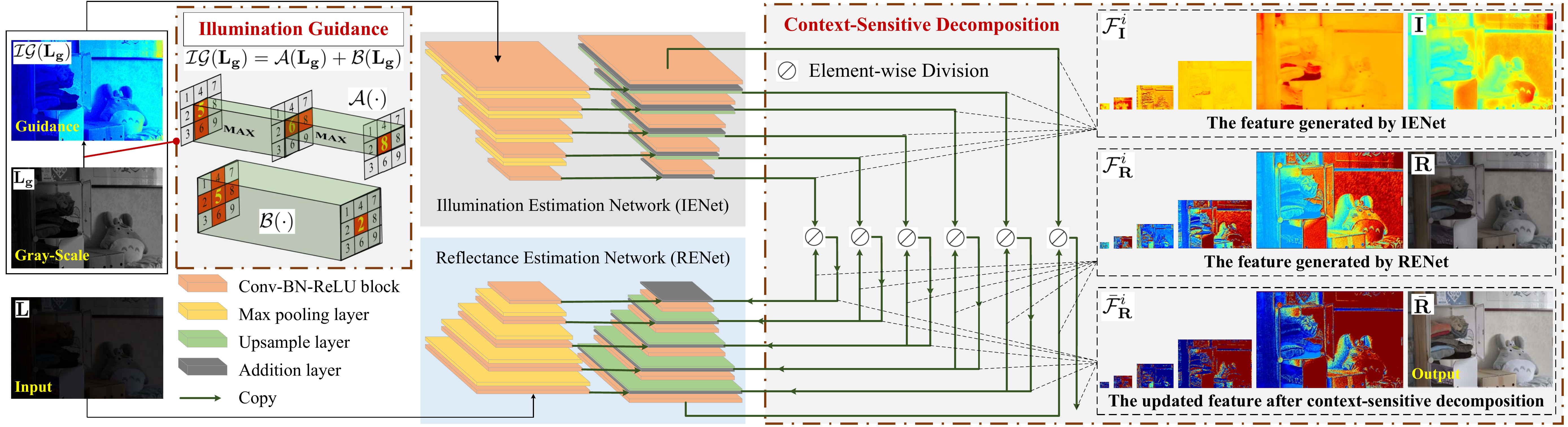}\\
	\end{tabular}
	\caption{The pipeline of our proposed context-sensitive decomposition network architecture, whose basic architecture is a U-Net based two-stream estimation mechanism consists of Illumination Estimation Network (IENet) and Reflectance Estimation Network (RENet). The context-sensitive decomposition connection is applied to each Conv-BN-ReLU block in the Decoder module. In the right corner of this figure, we demonstrate the immediate outputs of the feature maps before and after context-sensitive decomposition connection. It is easy to see that context-sensitive decomposition connection indeed aggregates the contextual dependencies of illumination and reflectance, to present more conspicuous and valuable structures without unwanted artifacts and noises. Additionally, we present all the single-channel maps in jet colormap to enlarge the difference and the same below. The spatially-varying illumination guidance is also normalized for a better visual presentation.}
	\label{fig:Flow}
\end{figure*}

\subsection{Value-based Mapping}
Mapping the value distribution of the low-light inputs to amplify small values (showed as dark) is an intuitive idea to be easily aware of and achieved. Histogram Equalization (HE) and S-curve based approaches are two representative works. 
The naive HE method~\cite{cheng2004simple} tends to generate enhanced results with improper exposure, inadequate details, distorted color, and unknown artifacts. A series of variants of HE are proposed to ameliorate the above drawbacks. For instance, Brightness Bi-Histogram Equalization (BBHE)~\cite{kim1997contrast} and Dualistic Sub-Image Histogram Equalization (DSIHE)~\cite{wang1999image} methods are designed to achieve natural exposure. To handle the loss of details, Adaptive Histogram Equalization (AHE)~\cite{pizer1987adaptive} and Contrast Limited Adaptive Histogram Equalization (CLAHE)~\cite{pisano1998contrast} are developed. However, the unnatural enhanced performance is still fatal flaws of these works.

Gamma Correction (GC) is one of the most well-known S-curve based techniques, it is a function to map luminance levels to compensate for the non-linear luminance effect of display devices. But for low-light enhancement, the enhanced results of GC are unnatural and unrealistic, especially in exposure level and details. Some improved versions are produced to overcome these issues. In~\cite{BennettVideo}, the low-light input is decomposed as two layers using a bilateral filter, and then recombined these processed two layers after different S-curve methods. The method proposed in~\cite{yuan2012automatic} tried to perform S-curve function for each subregion generated by segmenting the input. Unfortunately, for these S-curve based methods, the performance of uneven exposure is still hard to be addressed.

\subsection{Model-based Optimization}

Retinex theory provides an intuitive physical description for the procedure of enhancing the low-light images. This theory assumes that desired normal image (i.e., reflectance) can be obtained by removing the illumination of low-light inputs. Retinex theory~\cite{land1971lightness} is formulated as 
$	\mathbf{L} = \mathbf{R}\odot\mathbf{I},$
where $\mathbf{L}\in\mathbb{R}^{m\times n}$ is the low-light input, $\mathbf{R}\in\mathbb{R}^{m\times n}$ and $\mathbf{I}\in\mathbb{R}^{m\times n}$ are the reflectance and illumination, respectively. $\odot$ is the pixel-wise multiplication operation. 

Jobson \emph{et al.}~\cite{jobson1997multiscale} made some basic attempts but obtained an unrealistic appearance. Wang \emph{et al.}~\cite{wang2013naturalness} proposed a naturalness preservation algorithm and an evaluated metric. Insufficient details and inappropriate exposure are main drawbacks of this method. The work in~\cite{fu2015probabilistic} built a MAP-based energy model with different prior constraints for different layers. A weighted variational model in log-domain was also proposed for simultaneously estimating illumination and reflectance in~\cite{fu2016weighted}. The issues of these works lies in the production of ghosting artifacts because of the simplicity of designed priors.~\cite{cai2017joint} developed a jointly intrinsic-extrinsic prior model for LLIE, but it tended to generate the under-exposure results with insufficient structural information. 

The work in~\cite{guo2017lime} firstly focus on the illumination estimation, which optimized the initial illumination using edge-preserving smoothing. This work achieves nice performance but is over-exposed in most cases.~\cite{zhang2018high} further added extra constraints on~\cite{guo2017lime} to overcome the over-exposure, but it brings a high computational burden and unsatisfying exposure level. Additionally, because the noises and artifacts are always along with the procedure of enhancement, Li~\emph{et al.}~\cite{li2018structure} developed a denoising-type Retinex-based model and further built the optimization goal by defining the prior regularization of different components. However, the enhanced results using this work is occasionally over-smoothed and under-exposed because of inadequate modeling procedure. 

On the whole, limited to the restricted model capacity, these model-based works cannot come true to the consistent performance in some challenging scenarios.

\subsection{Data-driven Deep Learning}
 Different from other conventional vision tasks (e.g., image denoising, super-resolution), utilizing deep networks to address LLIE is still not mature enough. The key obstacle lies in the lack of effective training pairs. Until the recent three years, some paired datasets with low-light inputs and normal exposure labels are developed, push forward the development of the deep network. 

Concretely, Chen \emph{et al.}~\cite{Chen2018Retinex} built a new dataset (i.e., LOL dataset) obtained by changing the exposure time, and developed a RetinexNet based on Retinex theory to simultaneously estimate the illumination and reflectance. KinD~\cite{zhang2019kindling} is then designed based on the architecture of RetinexNet by adding some effective loss functions. Yang \emph{et al.}~\cite{yang2020fidelity} proposed a semi-supervised learning framework based on a deep recursive band network. This work trained the designed network in LOL dataset. 
In~\cite{li2018lightennet}, a convolutional neural network (LightenNet) for weakly illuminated image enhancement is proposed. They trained the proposed network by utilizing the synthetic training pairs created based on Retinex theory. {Ren~\emph{et al.}~\cite{ren2019low} designed a hybrid network with content and edge streams to recover more accurate scene content for low-light enhancement. }

Wang~\emph{et al.}~\cite{wang2019underexposed} adopted a similar strategy with MIT-Adobe FiveK~\cite{fivek} to produce the training pairs. This work used the architecture of HDRNet~\cite{hasinoff2016burst} and defined three different loss functions to address this task. 
LLNet~\cite{lore2017llnet} is proposed to train an autoencoder-based network by using the synthetic dataset. This method leads to the unrealistic and still under-exposure results. The work in~\cite{chen2018learning} built a paired dataset of raw by adjusting different exposure time, and trained an end-to-end fully convolutional network for handling the LLIE. The paper in~\cite{xu2020learning} prepared a low-light image dataset with real noise
and corresponding ground truth images. Based on this dataset, they proposed a frequency-based decomposition-and-enhancement model for noise suppression and detail enhancement. 

Generative Adversarial Network (GAN) has realized successes in a series of image-to-image translation~\cite{liu2017unsupervised}. Thus it is intuitive that GAN can gain a worth expecting performance in LLIE.
Inspired by the adversarial mechanism, some recent works proposed to handle LLIE by using GAN. 
RetinexGAN~\cite{shi2019low} proposed a generator and utilized the converted dataset derived from~\cite{chen2018learning} to execute the paired training manner. Unfortunately, in real-world scenarios, the enhanced results are frequently appeared to be unnatural. Besides, inspired by attention mechanism, EnlightenGAN~\cite{jiang2019enlightengan} established a generative adversarial network with the self-regularization in an unpaired supervision manner. However, it always produce some unknown artifacts since the neglect of physical principle. 

In brief, existing data-driven deep networks for settling LLIE all ignore the exploitation of the contextual information on spatial scales. As is shown in Fig.~\ref{fig:FirstFig} (a), the representative data-driven methods integrate the physical principle to design the architecture. 
However, they followed the detour route of ``{under-exposure$\Rightarrow$over-exposure$\Rightarrow$normal-exposure}" which brought about some superfluous process. In addition, they only perform the physical principle on image level, i.e., the output terminal, so that neglecting to exploit the feature-level contextual information by using the physical principle.

\section{The Proposed Approach}
In this section, we first present a two-stream estimation mechanism as our basic network architecture. The context-sensitive decomposition connection is then designed to bridge these two sub-networks. We construct the spatially-varying illumination guidance to assist in optimizing the illumination. Based on our designed architecture, we defined CSDNet and CSDGAN according to different training manners.

\subsection{Two-Stream Estimation Mechanism}
Retinex theory has been widely used and shows its effectiveness in addressing LLIE. Taking this into account, we define a two-stream estimation mechanism consists of Illumination Estimation Network (IENet) and Reflectance Estimation Network (RENet). The procedure of two-stream simultaneously optimizes the illumination and reflectance. These two estimation networks can be formulated as 
\begin{equation}
	\mathbf{I} = \mathcal{N}_{\mathtt{IENet}}(\mathbf{L_g}),\;\;\;\;
	\mathbf{R} = \mathcal{N}_{\mathtt{RENet}}(\mathbf{L}),
\end{equation}
where $\mathcal{N}_{\mathtt{IENet}}$ and $\mathcal{N}_{\mathtt{RENet}}$ represent the IENet and RENet, respectively. $\mathbf{I}$ and $\mathbf{R}$ are the estimated illumination and reflectance, respectively. $\mathbf{L_g}$ is the gray-scale image of the low-light input $\mathbf{L}$. 

Our built two-stream estimation mechanism is different from existing works e.g., RetinexNet~\cite{Chen2018Retinex}, KinD~\cite{zhang2019kindling}. These existing methods estimate the illumination and reflectance together by defining a circuitous pipeline (as is shown in Fig.~\ref{fig:FirstFig} (a)). Their two-stream estimation part generates the over-exposure reflectance and the adjusted illumination. Then by performing the Retinex theory, the adjusted illumination is used to overcome the over-exposure of the reflectance. But actually, the normal-exposure result can be generated in the procedure of generating the over-exposure intermediate result. In other words, their pipeline may be redundant, even a little irrational. 
In contrast, our two-stream estimation mechanism keeps an explicit mapping to improve image quality. Additionally, we do not need to consider a decomposition network (see Fig.~\ref{fig:FirstFig} (a)) to obtain a better initial estimation.

Actually, any fashionable network architectures can be utilized for easily catering to our requirements~\cite{ronneberger2015u, huang2017densely}. 
Here, we provide a choice, i.e., U-Net~\cite{ronneberger2015u}, a widely-used network architecture. We instantiate our IENet and RENet using the identical standard U-Net. 
The network architecture of U-Net contains 10 basic convolutional blocks. Each block consists of a convolutional layer, a batch normalization layer, and a ReLU layer. The first 4 blocks and last 4 blocks are followed by the Max polling (Encoder)/upsampling (Decoder) layer. As for the channels of each block, they are 32, 64, 128, 256, 512, 256, 128, 64, 32 in order. When it comes to the {lightweight CSDNet}\footnote{We define two lightweight versions of CSDNet to cater to the practical needs. Please refer to Sec.~\ref{sec: LiteCSDNet} for more details.}, the channel of each block is set to 12.

Now, our focus converts to how to utilize the physical principle to bridge these two sub-networks. 

\subsection{Context-Sensitive Decomposition Connection}
Generally, most of two-stream type works (\cite{yan2018two,zhang2019two,liu2017two}) learn to fuse two outputs of this network. Indeed, we can directly add the connection between outputs of IENet and RENet. But doing this ignores the exploitation of scene-level contextual dependencies on feature spaces. 
We know that the utilization of contextual information is extremely essential for deep network methods. The verification for this point has been demonstrated in many high-level vision tasks, such as object detection~\cite{liu2020deep}, visual tracking~\cite{han2019state}, semantic segmentation~\cite{yu2020context} and so on.

Inspired by the Retinex theory, we design the following context-sensitive decomposition connection to bridge our designed two-stream sub-networks
\begin{equation}\label{eq:RetinexConnection}
\bar{\mathcal{F}}^{i}_{\mathbf{R}}= {\mathcal{F}^{i}_{\mathbf{R}}}\oslash({\mathcal{F}^{i}_{\mathbf{I}}}),\;\;\;\;
{\mathcal{F}^{i}_{\mathbf{R}}} = \bar{\mathcal{F}}^{i}_{\mathbf{R}},
\end{equation}
where $\oslash$ is the element-wise division. ${\mathcal{F}}^{i}_{\mathbf{R}}$ and $\mathcal{F}^{i}_{\mathbf{I}}$ are the feature maps generated by $i$-th connected layer (except the last layer) of RENet and IENet, respectively. $\bar{\mathcal{F}}_{\mathbf{R}}$ represents the updated feature after context-sensitive decomposition connection. 
Notice that when meeting the last layer, that is 
\begin{equation}
	\bar{\mathbf{R}}=\mathbf{R}\oslash\mathbf{I},
\end{equation}
where $\bar{\mathbf{R}}$ is exactly the final enhanced result.

As is known to all, the feature maps generated by the layers of the Encoder are attached to the corresponding layers of the Decoder. According to this point, we execute the proposed context-sensitive decomposition connection on the feature maps of the Decoder. Notice that operated feature maps are composed of the copied features from the Encoder and the generated features by the Decoder. We will make a meticulous analysis about this in Sec.~\ref{sec:CSDEffects}.

Actually, we aggregate the contextual information of the reflectance and illumination generated by networks, to inference more valuable and helpful context for accurately estimating reflectance. As is shown in Fig.~\ref{fig:Flow}, executing the context-sensitive decomposition connection indeed improves the brightness, suppresses artifacts and highlights structural information, especially the output terminal.

From the perspective of incorporating the physical principle in the network.
Most of the existing approaches integrating the physical principle in the networks can be roughly divided into two types. One is designing optimization-based deep unrolling strategies (iterative scheme or end-to-end learning)~\cite{liu2019learning,dong2018denoising,brifman2019unified}. The other is constructing principle-based architecture and training costs~\cite{Chen2018Retinex,yang2019scale,li2019semi}. Actually, these works have a commonality, i.e., only focusing on the image domain. It leads to ignoring the exploitation of larger solution space with multiple spatial scales. By comparison, our context-sensitive decomposition connection endows the powerful characterization for the network by focusing on the inner structure.

\begin{figure}[t]
	\centering
	\begin{tabular}{c@{\extracolsep{0.35em}}c@{\extracolsep{0.35em}}c}
		\includegraphics[width=0.15\textwidth]{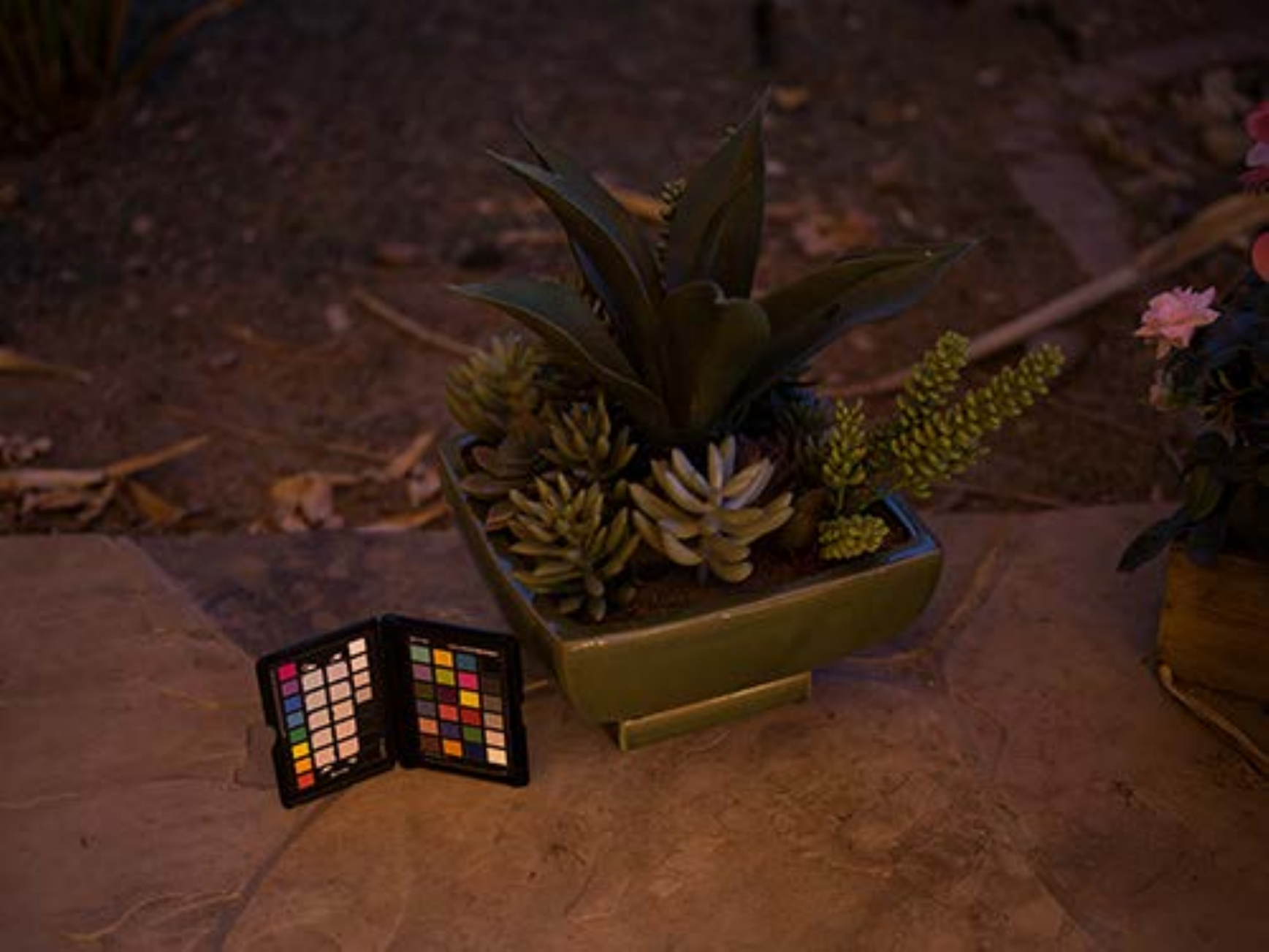}&
		\includegraphics[width=0.15\textwidth]{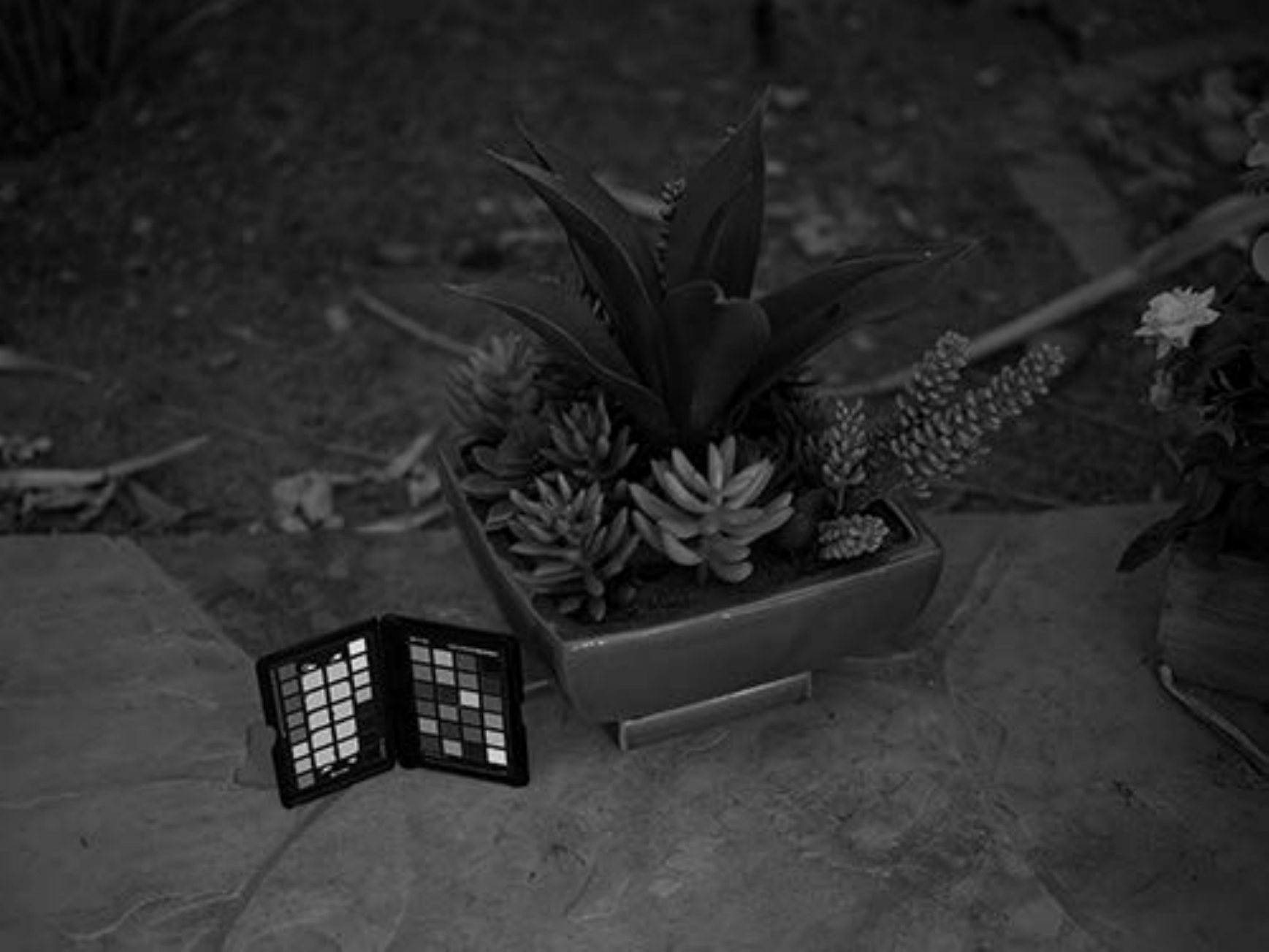}&
		\includegraphics[width=0.15\textwidth]{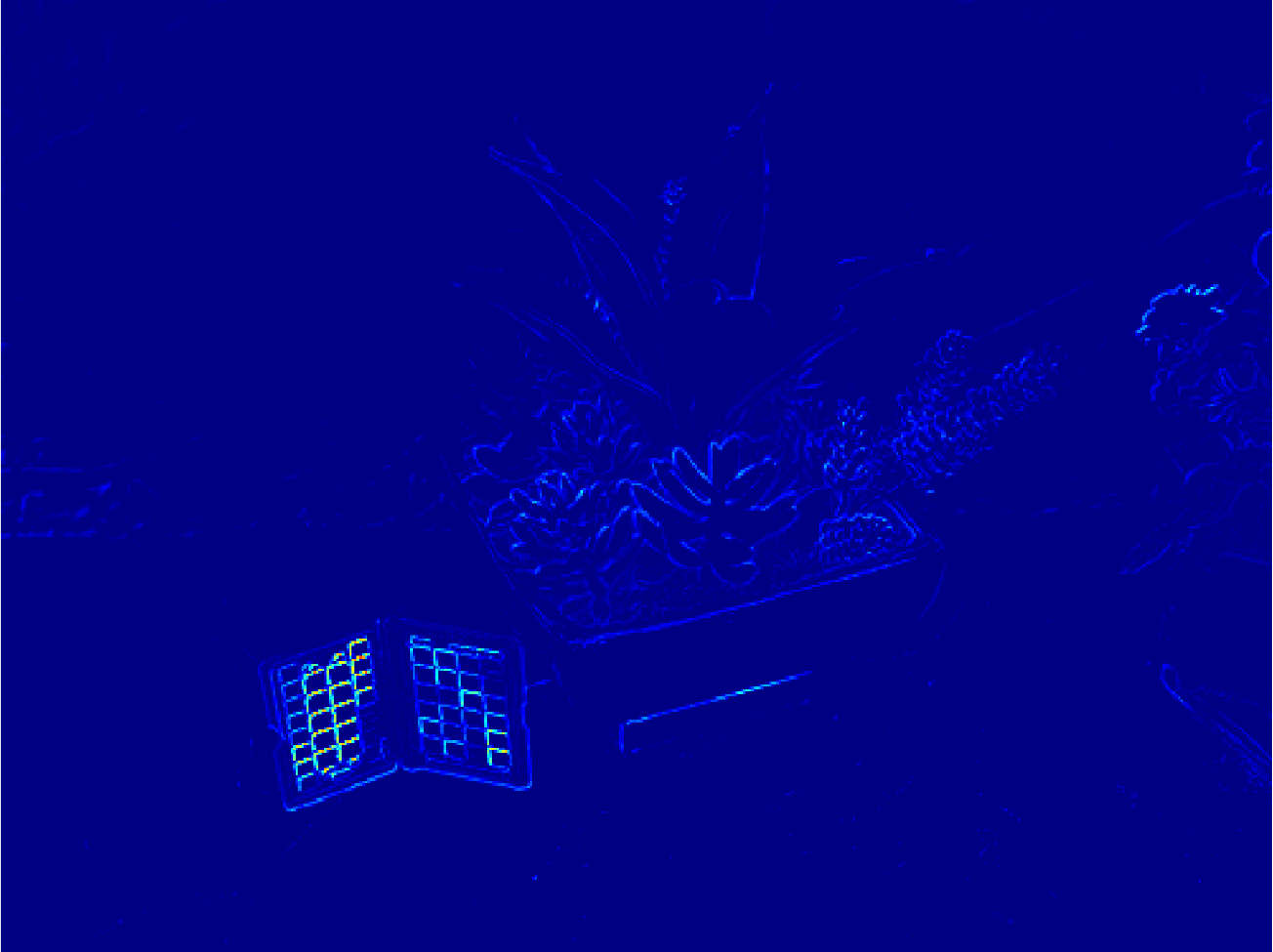}\\	
		\includegraphics[width=0.15\textwidth]{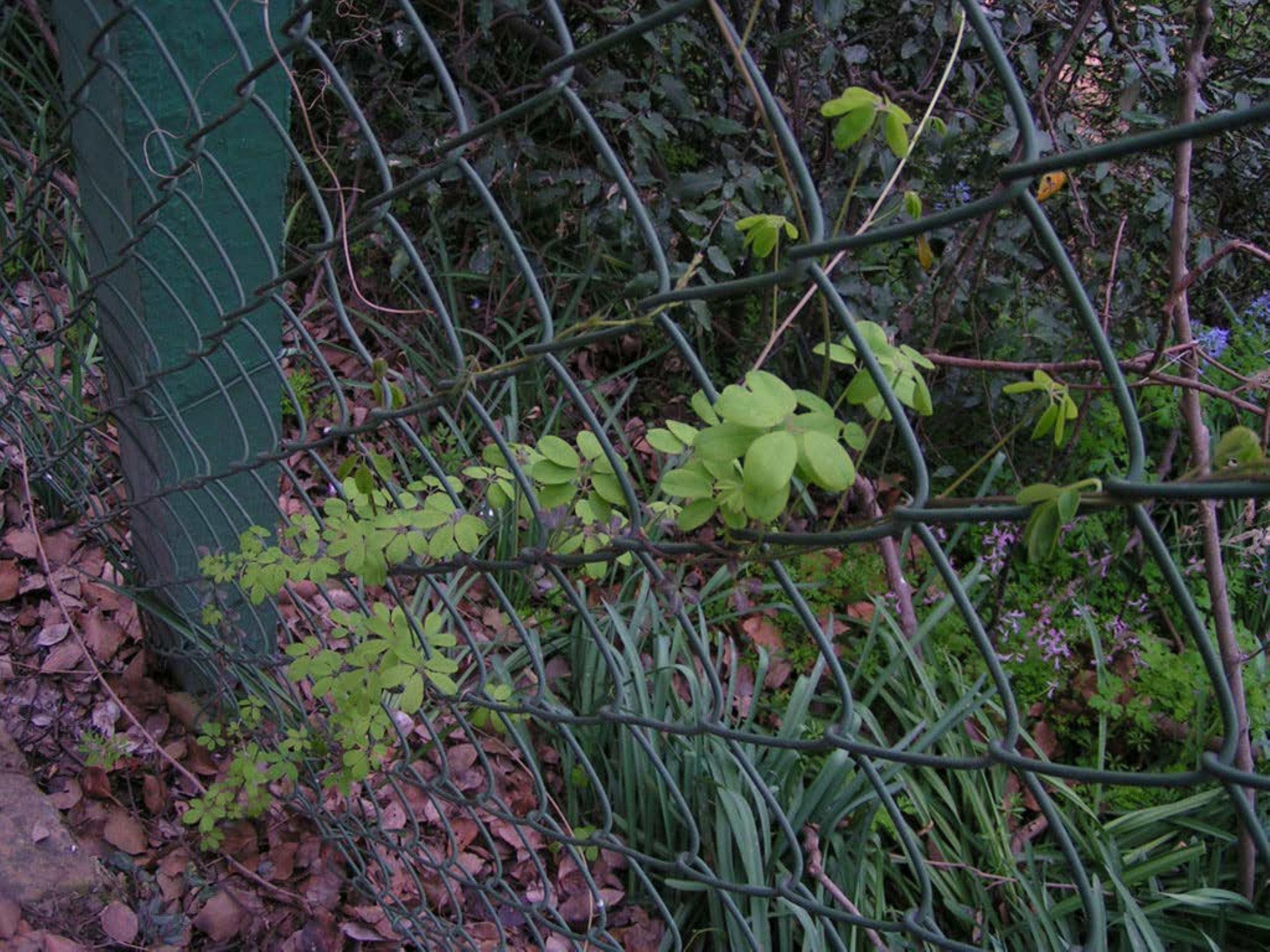}&
		\includegraphics[width=0.15\textwidth]{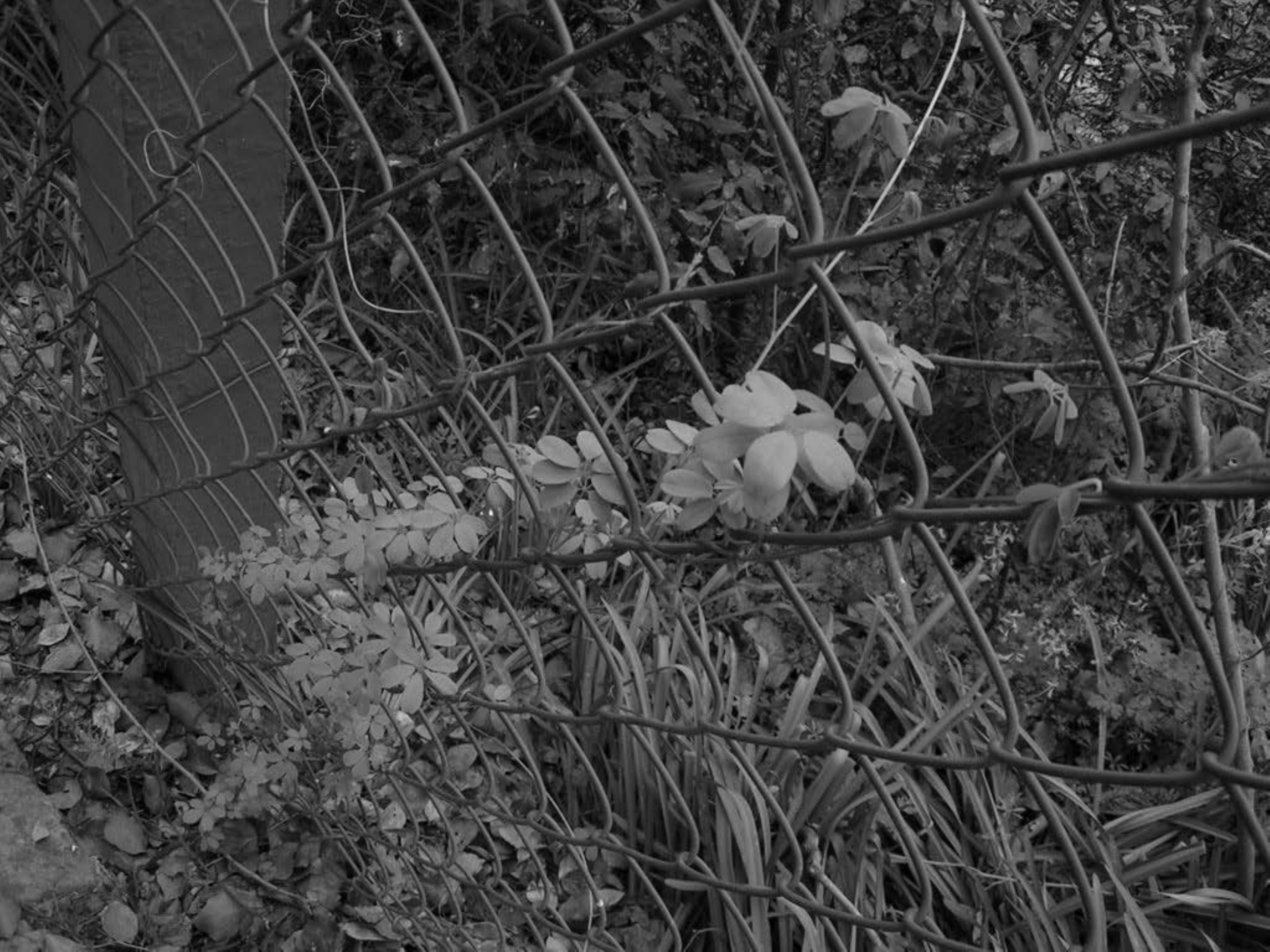}&
		\includegraphics[width=0.15\textwidth]{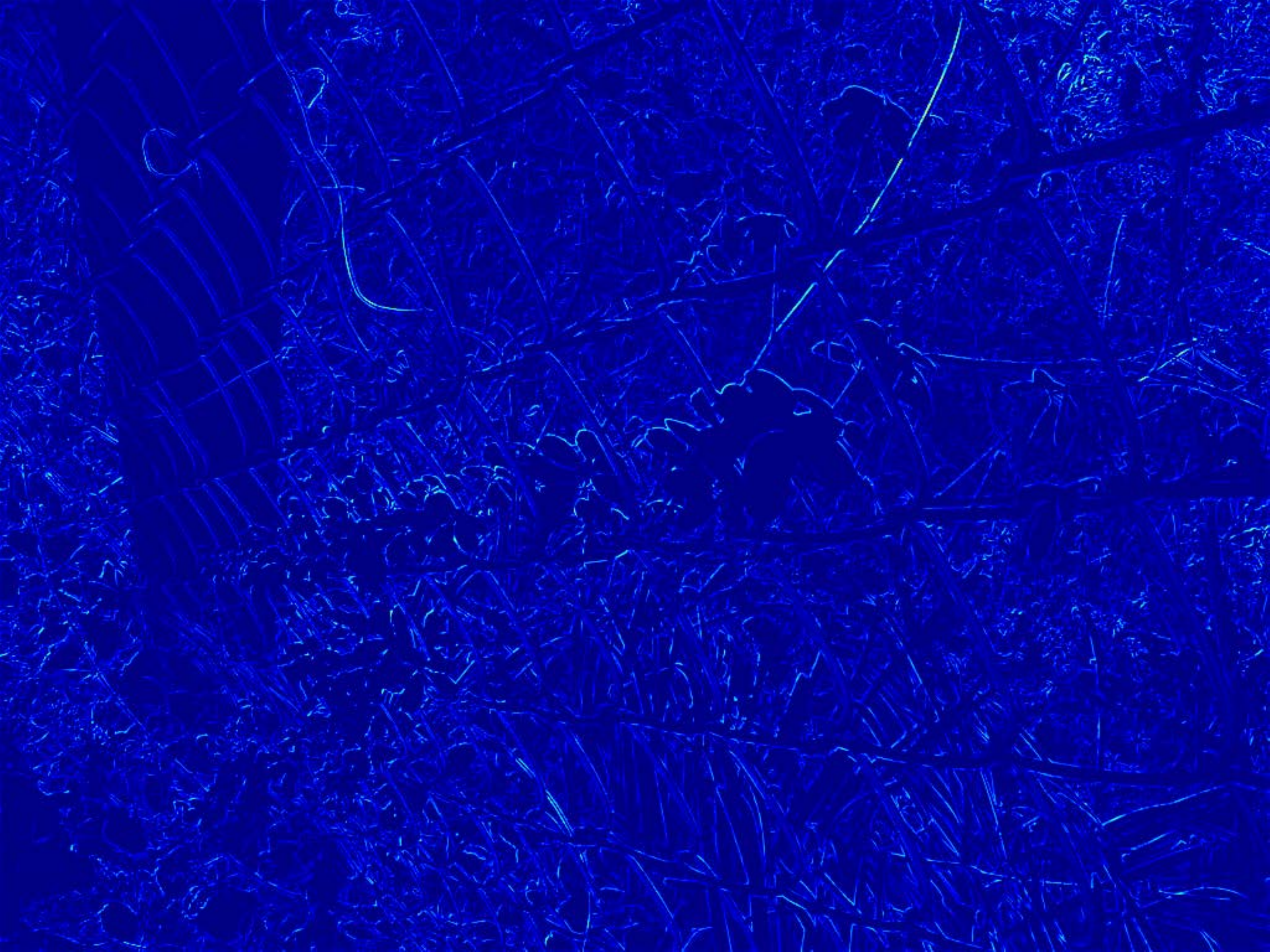}\\		
		\footnotesize Input&\footnotesize $\mathbf{L_g}$&\footnotesize $\mathcal{IG}(\mathbf{L_g})-\mathbf{L_g}$\\
	\end{tabular}
	\caption{Visualization of our proposed spatially-varying illumination guidance.}
	\label{fig:Edge}
\end{figure}	

\subsection{Spatially-Varying Illumination Guidance}
We find that the estimated illumination by directly performing the above-proposed architecture is unsatisfactory, especially in the structural and edge presentation. It causes the appearance of artifacts and inexact exposure in the enhanced results\footnote{Please see the ablation study in Sec.~\ref{sec:IGEffects}.}. To tackle this issue, previous end-to-end learning techniques~\cite{wang2019underexposed,zhang2019kindling} try to add complex prior regularizations as the loss function to constrain the illumination. Although it brings improvement, it needs a high computational cost. Instead of designing the complex regularization constraints, we introduce a spatially-varying illumination guidance operator to attain the edge-aware structural illumination to ensure the stable implementation of context-sensitive decomposition connection.
The illumination guidance operator is defined as $\mathcal{IG}(\cdot)$, formulated as

\begin{equation}\label{eq:IlluminationGuidance}
\mathcal{IG}(\mathbf{L}_\mathbf{g})=\mathcal{A}(\mathbf{L}_\mathbf{g})+\mathcal{B}(\mathbf{L}_\mathbf{g}),\\
\end{equation}
where 
\begin{equation}
\left\{
\begin{aligned}
\mathbf{U}&=\max_{i=1,...,m-1}\left(\mathbf{L}_\mathbf{g}(i,:),\mathbf{L}_\mathbf{g}(i+1,:)\right),\\
\mathcal{A}(\mathbf{L}_\mathbf{g})&=\max_{j=1,...,n-1}\left(\mathbf{U}(:,j),\mathbf{U}(:,j+1)\right),\\
\end{aligned}
\right.
\end{equation}
\begin{equation}
\left\{
\begin{aligned}
f(a,b)&=|\mathbf{L_g}(i,j)-\mathbf{L_g}(i+a,j+b)|,\\
\mathcal{B}(\mathbf{L}_\mathbf{g}) &= \frac{1}{4}\sum_{k=\{-1,1\}}\left(f(0,k)+f(k,0)\right),\\
\end{aligned}
\right.
\end{equation}
where $i$ and $j$ represent the pixel index of horizontal and vertical directions of the image, respectively. 

The visualization of the illumination guidance is shown in Fig.~\ref{fig:Edge}. It indicates that this guidance presents effective edge and texture information.

Then the estimation procedure of IENet is reformulated as 
\begin{equation}
	\mathbf{I} 
	= \mathcal{N}_{\mathtt{IENet}}(\mathbf{L_g};\mathcal{IG}(\mathbf{L}_\mathbf{g})).
\end{equation}

From Eq.~\eqref{eq:IlluminationGuidance}, we can find that this operator actually first enlarges the pixel distribution of the original input image by pixel values of neighborhood regions, then the illumination guidance map is defined as the difference between the enlarged input and the original input. In this way, we successfully obtain the desired information presenting edges and textures. 
In fact, compared with those defined similar maps in existing works~\cite{Fan2018ECCV,fan2018revisiting}, the computational procedure of our illumination guidance is more efficient. In a word, our illumination guidance is powerful and practical which can also be applied to other related tasks such as image deblurring, image dehazing and so on. 

Above all, the entire pipeline of our designed network architecture can be seen in Fig.~\ref{fig:Flow}.

\subsection{CSDNet}
	Based on our designed network architecture, we define the CSDNet which utilized the paired datasets for training. In the following, we make a detailed description of training loss for the CSDNet. 

\textbf{MSE Loss.}$\;$ We utilize the Mean Square Error (MSE) loss to constrain the final enhanced output, described as 
$\mathcal{L}_{MSE}=\|\bar{\mathbf{R}}-\hat{\mathbf{R}}\|^2,$
where $\hat{\mathbf{R}}$ represents the ground truth. 

\textbf{Perceptual Loss.}$\;$ We use the following perceptual loss presented in the work~\cite{johnson2016perceptual} to ensure the perceptual similarity.
{
\begin{equation}\label{eq:VGG}
\mathcal{L}_{P_{1}}=\frac{1}{{W}_{i,j}{H}_{i,j}}\sum_{x=1}^{{W}_{i,j}}\sum_{y=1}^{{H}_{i,j}}\left({{\phi}_{i,j}\left(\bar{\mathbf{R}}\right)}_{x,y}-{{\phi}_{i,j}\left(\hat{\mathbf{R}}\right)}_{x,y}\right),
\end{equation}
}
where we set $i=5$, $j=1$. ${W}_{i,j}$ and ${H}_{i,j}$ are the dimensions of the extracted feature maps. 

\textbf{Smooth $\mathcal{L}_1$ Loss.}$\;$ We utilize a simple smooth regularization to constrain the output of IENet, expressed as
\begin{equation}
\mathcal{L}_{S}=\frac{1}{mn}\sum_{i=1}^{m}\sum_{j=1}^{n}{smooth}_{\mathcal{L}_{1}}\left({\mathbf{I}(i,j)-\mathbf{L}_g(i,j)}\right). 
\end{equation}
The definition of ${smooth}_{\mathcal{L}_{1}}(u)$ is described as, if $|u|
\leq1$, then ${smooth}_{\mathcal{L}_{1}}(u)=0.5{u}^2$, otherwise ${smooth}_{\mathcal{L}_{1}}(u)=|u|-0.5$, $N$ is the number of pixel. 

In a word, we train our CSDNet using the following loss
{
\begin{equation}
	\mathcal{L}_\mathtt{CSDNet}=\mathcal{L}_{MSE}+\mathcal{L}_{P_1}+\mathcal{L}_{S},
\end{equation}
}

\begin{table*}[t]
	\centering
	\caption{Benchmarks Description.}
	\begin{tabular}{ccccccccc}
		\toprule
		Dataset&MIT-Adobe FiveK&LOL
		&EnlightenGAN&NPE&NASA&MEF&LIME&ExDark (Part)\\
		\midrule 
		Paired/Unpaired&Paired&Paired&Unpaired&Unpaired&Unpaired&Unpaired&Unpaired&Unpaired\\
		\midrule 
		Numbers&5000&500&914&130&23&17&10&2\\
		\midrule 
		Training Set&$\surd$&$\surd$&$\surd$&$\times$&$\times$&$\times$&$\times$&$\times$\\
		\midrule
		Testing Set&$\surd$&$\surd$&$\times$&$\surd$&$\surd$&$\surd$&$\surd$&$\surd$\\
		\bottomrule
	\end{tabular}
	\label{tab:datasets}
\end{table*}

\begin{figure*}[t]
	\centering
	\begin{tabular}{c@{\extracolsep{0.3em}}c@{\extracolsep{0.3em}}c@{\extracolsep{0.3em}}c@{\extracolsep{0.3em}}c}
		\includegraphics[width=0.19\textwidth]{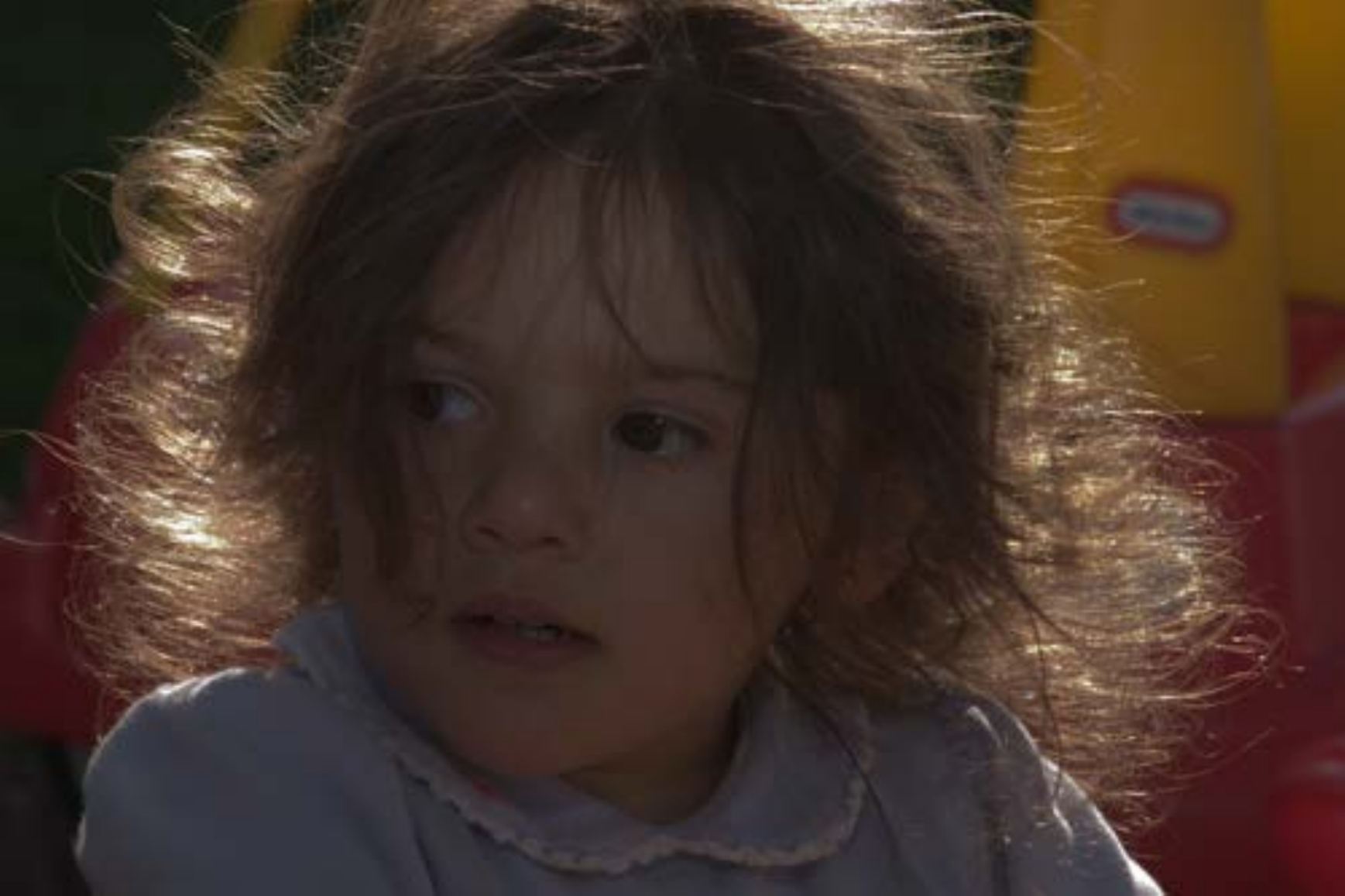}&
		\includegraphics[width=0.19\textwidth]{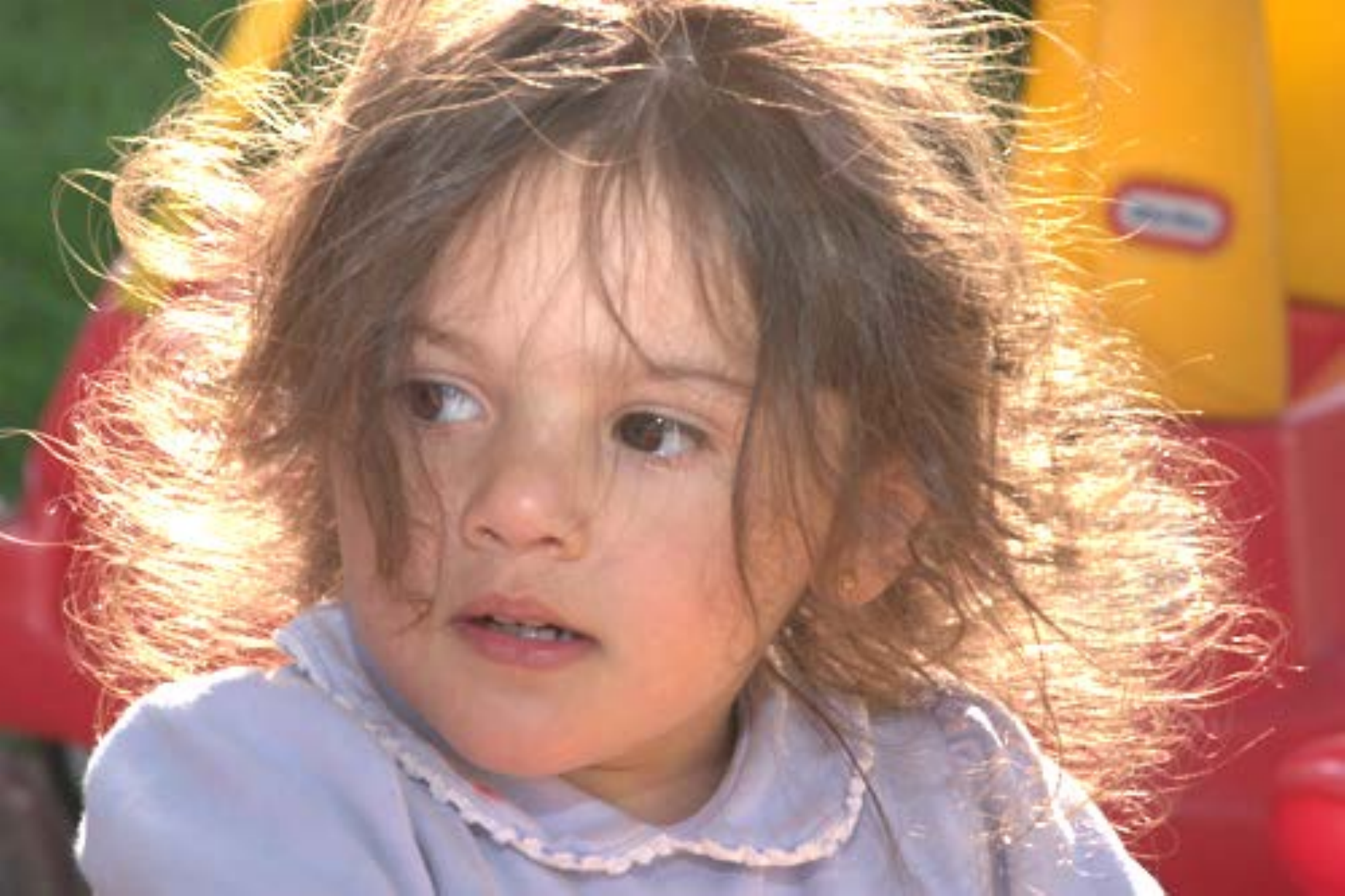}&
		\includegraphics[width=0.19\textwidth]{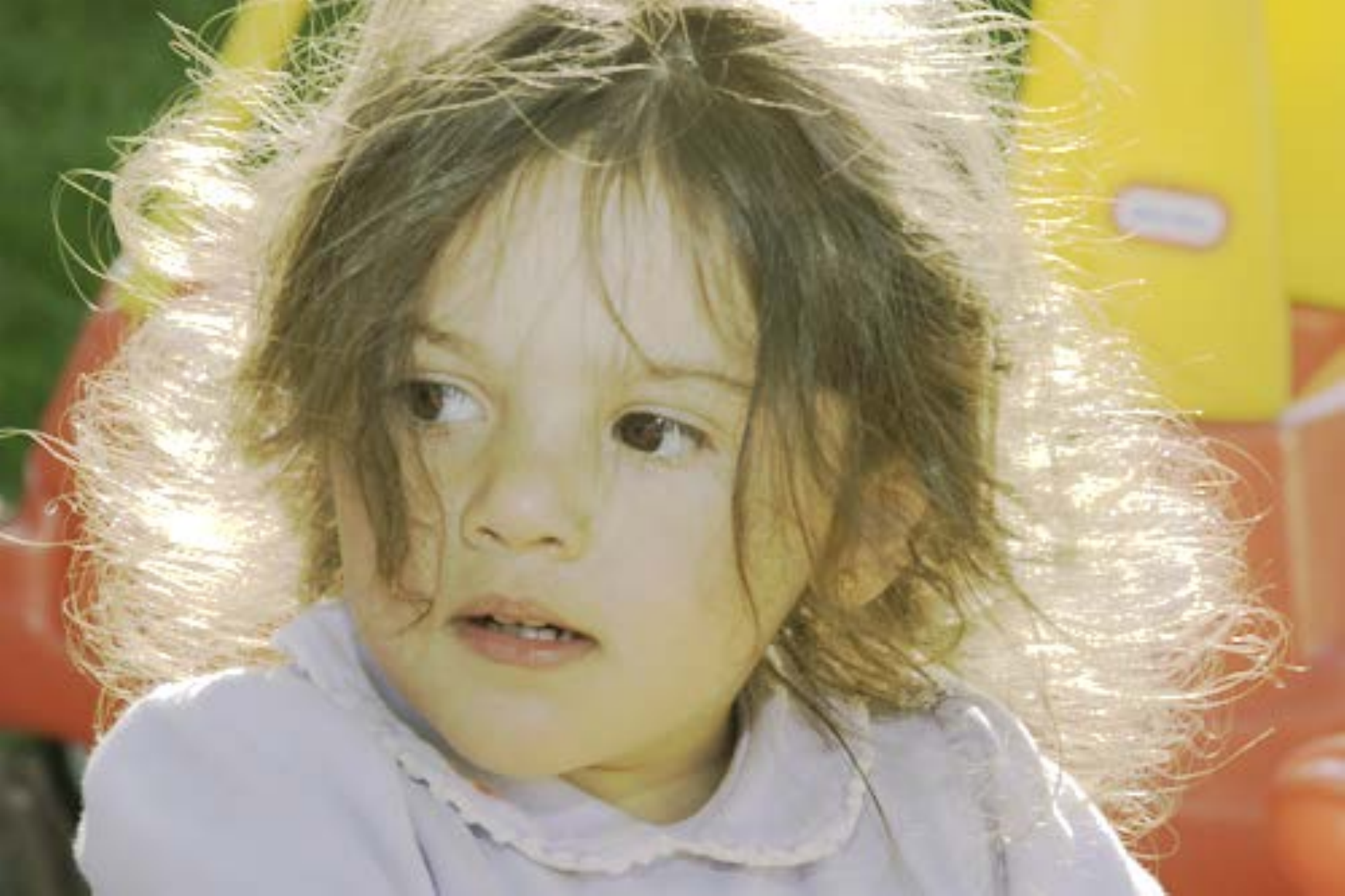}&
		\includegraphics[width=0.19\textwidth]{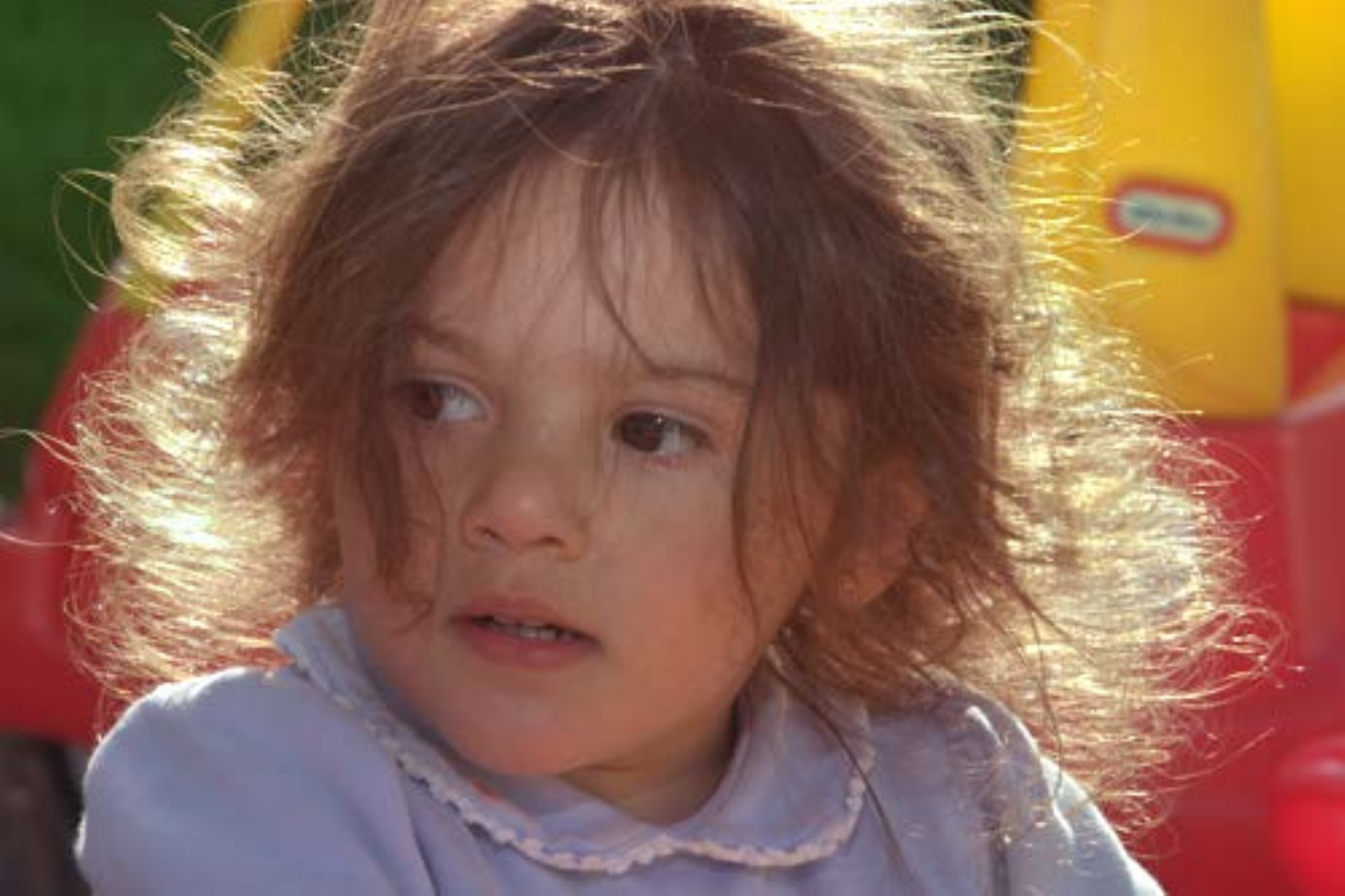}&
		\includegraphics[width=0.19\textwidth]{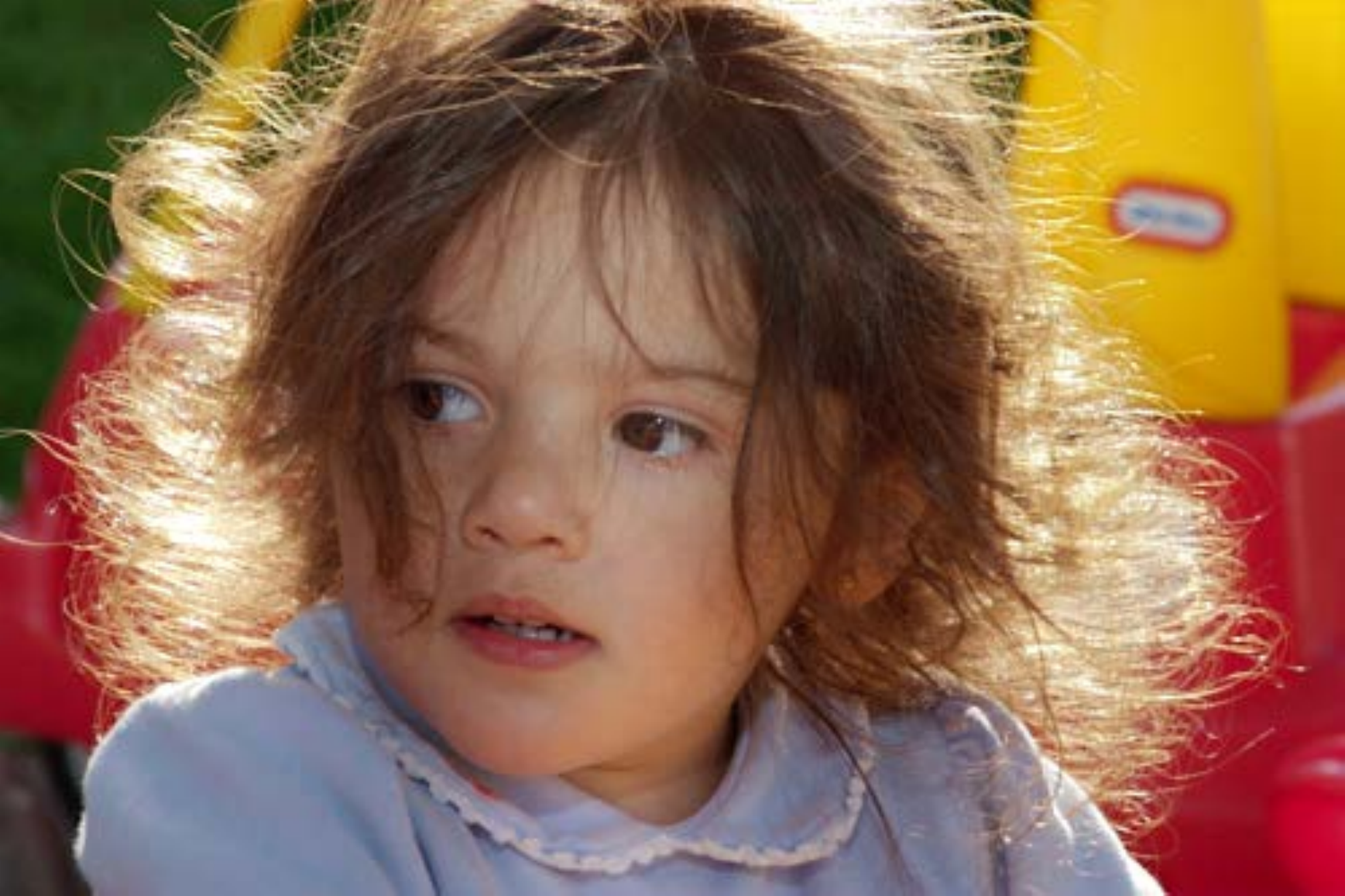}\\	
		\includegraphics[width=0.19\textwidth]{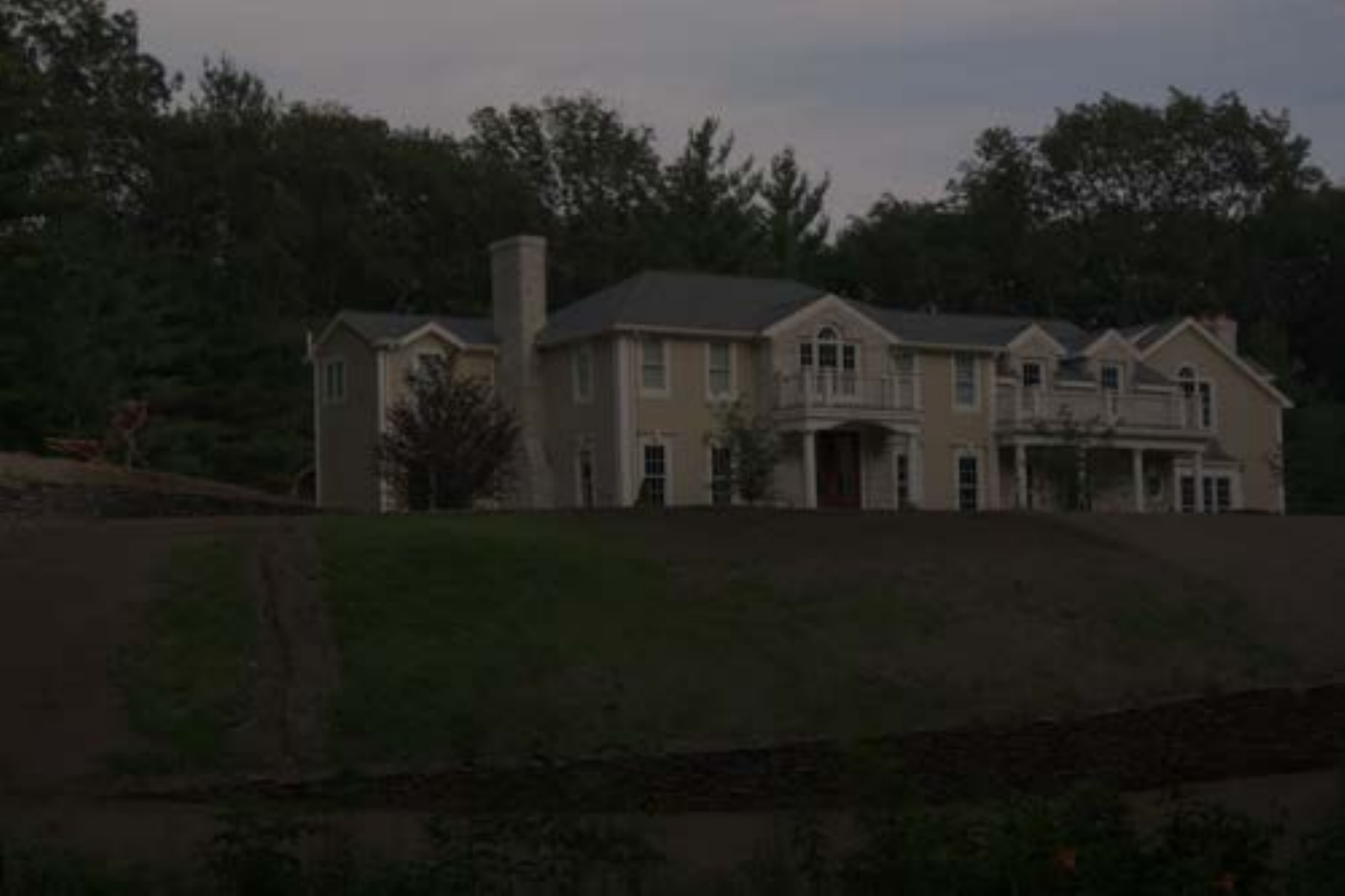}&	
		\includegraphics[width=0.19\textwidth]{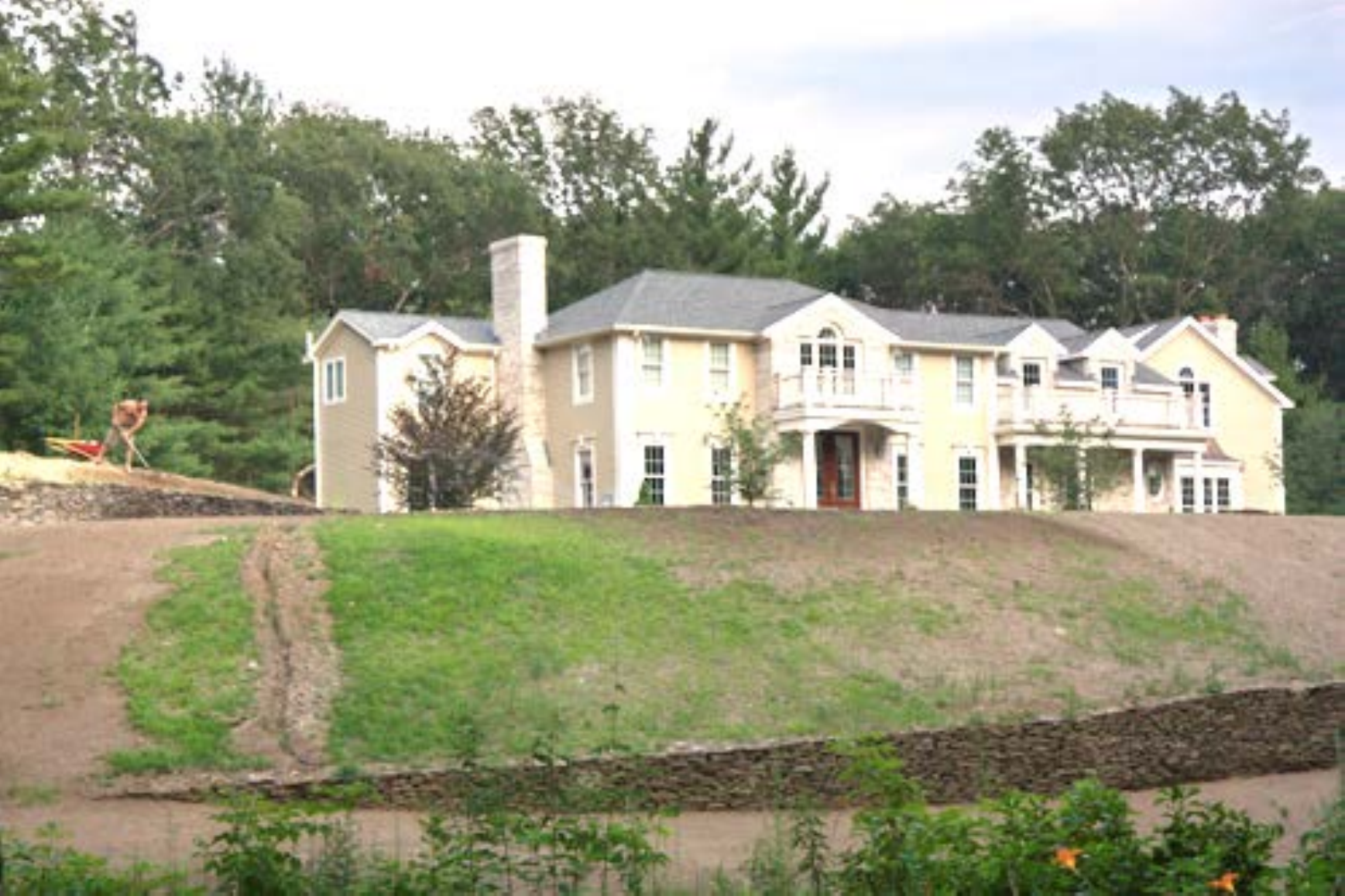}&
		\includegraphics[width=0.19\textwidth]{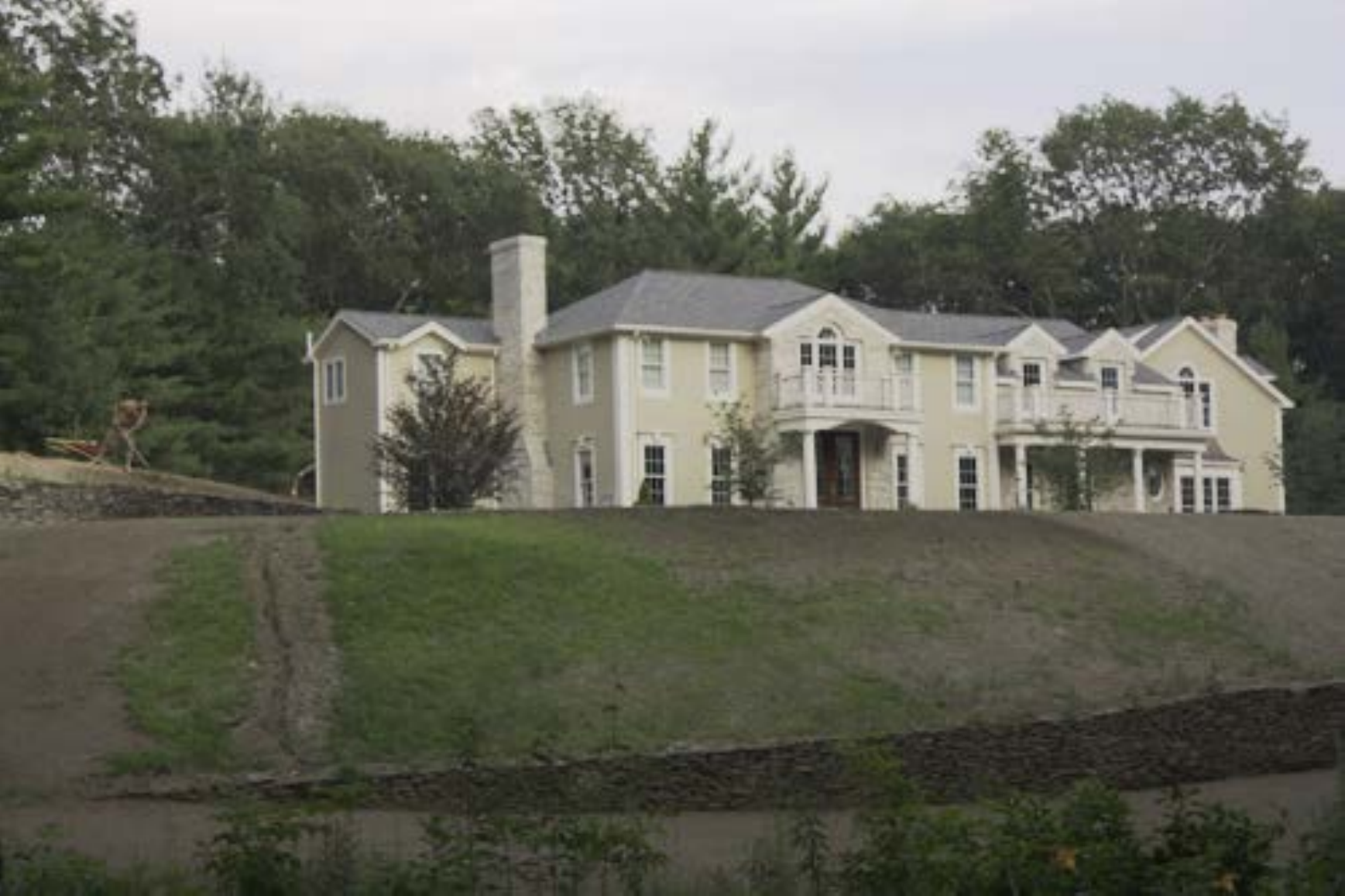}&
		\includegraphics[width=0.19\textwidth]{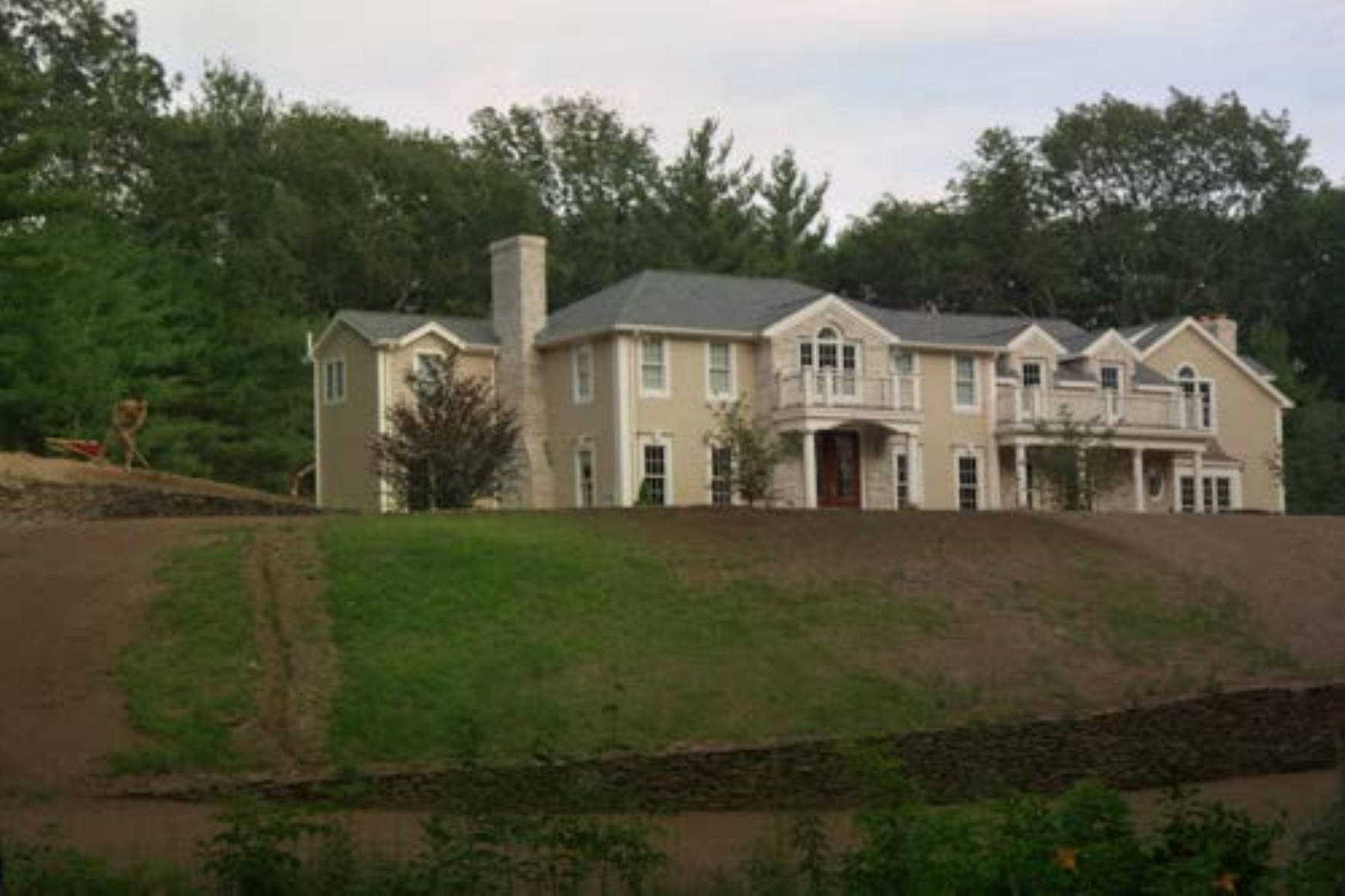}&
		\includegraphics[width=0.19\textwidth]{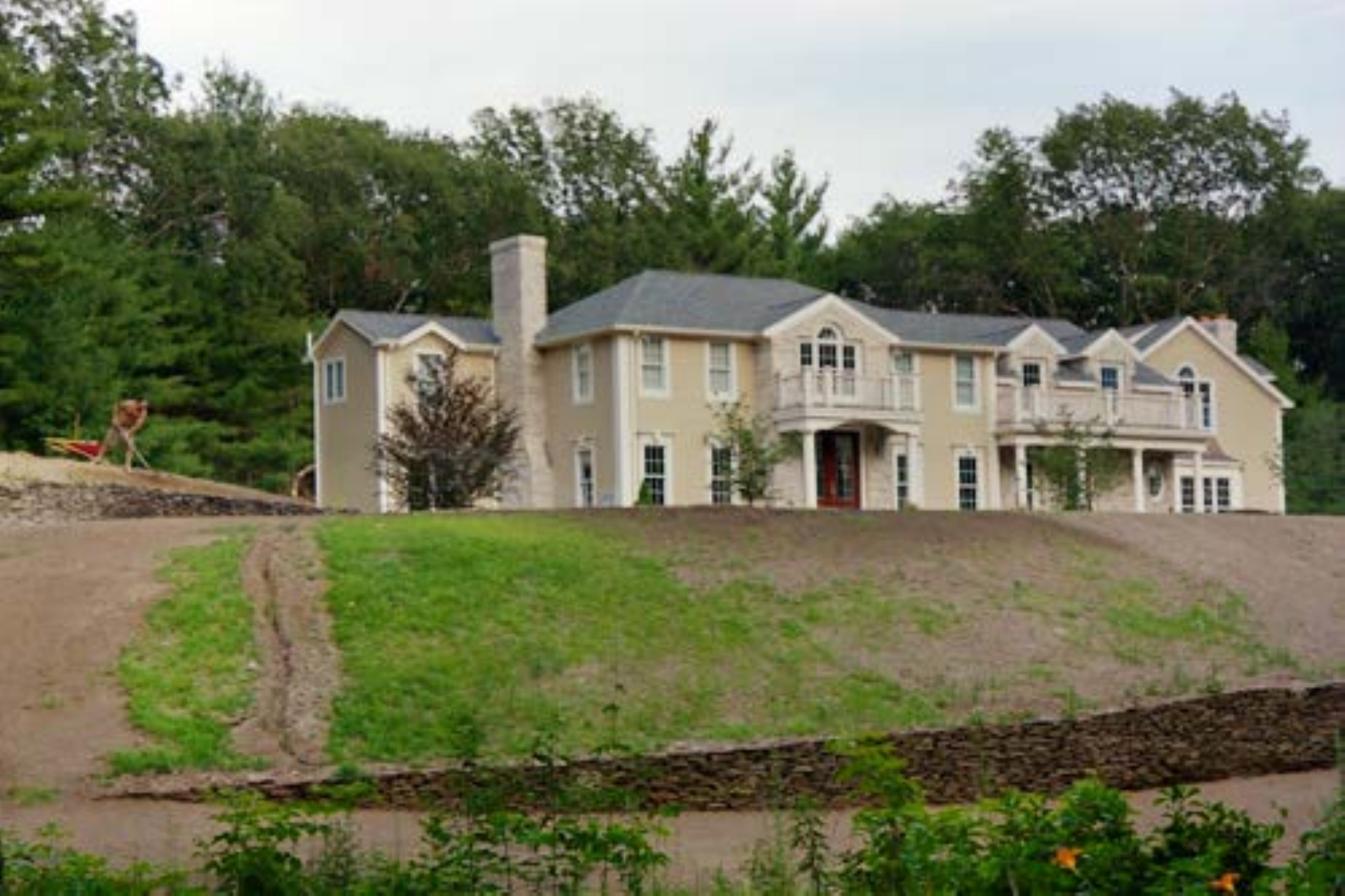}\\			
		\includegraphics[width=0.19\textwidth]{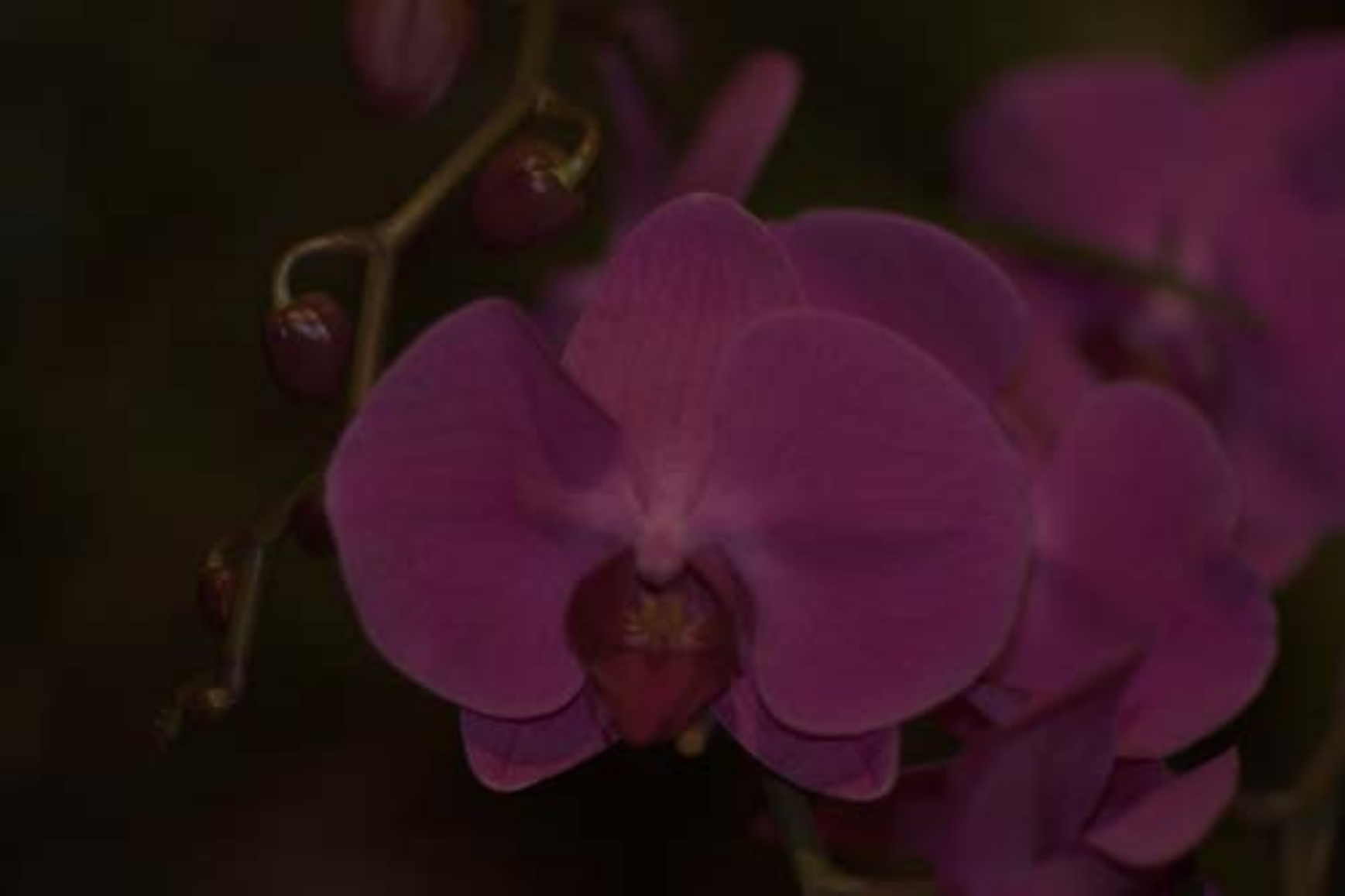}&		
		\includegraphics[width=0.19\textwidth]{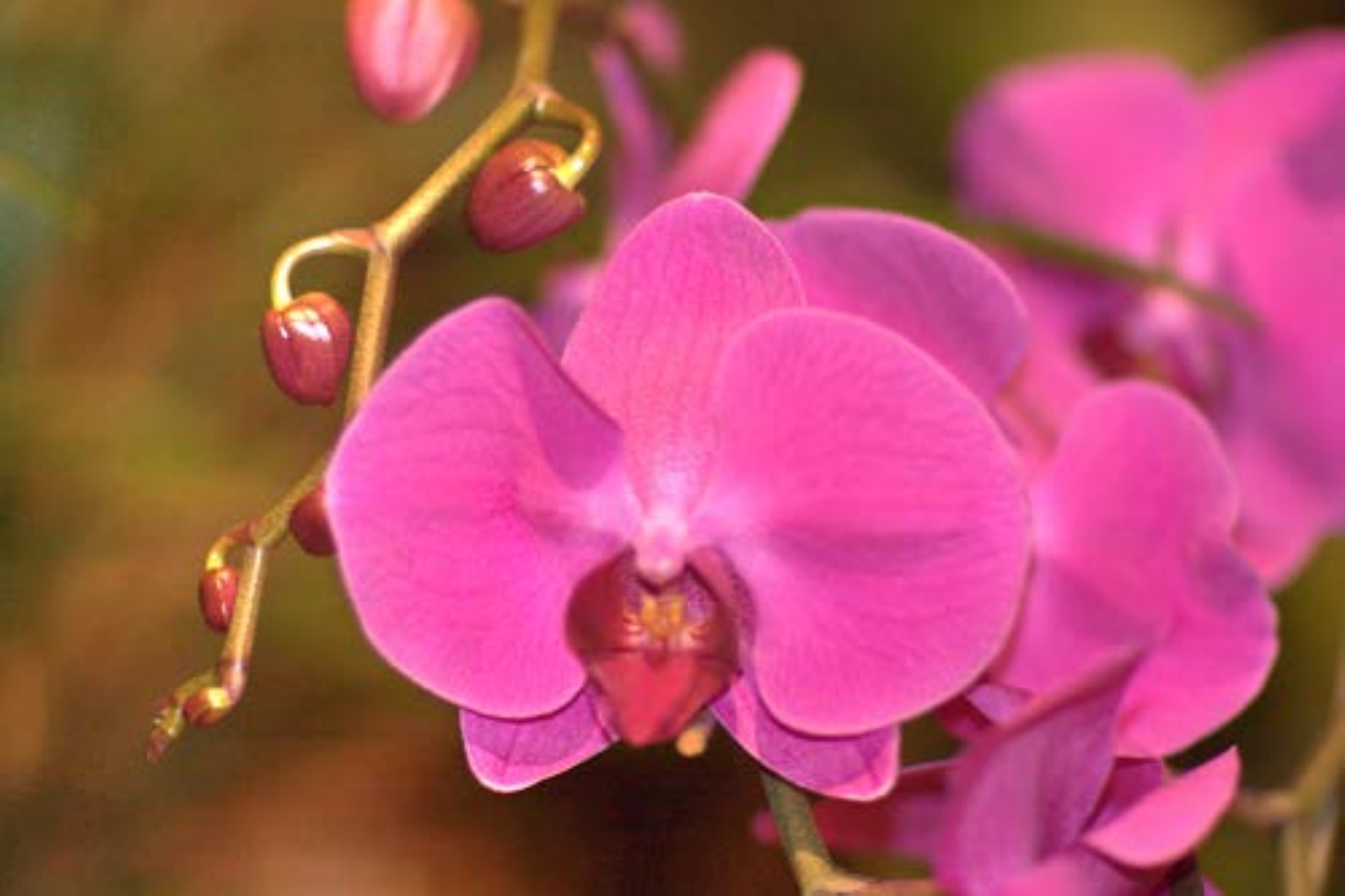}&
		\includegraphics[width=0.19\textwidth]{Figures/Comparison/FiveK/4/Input}&
		\includegraphics[width=0.19\textwidth]{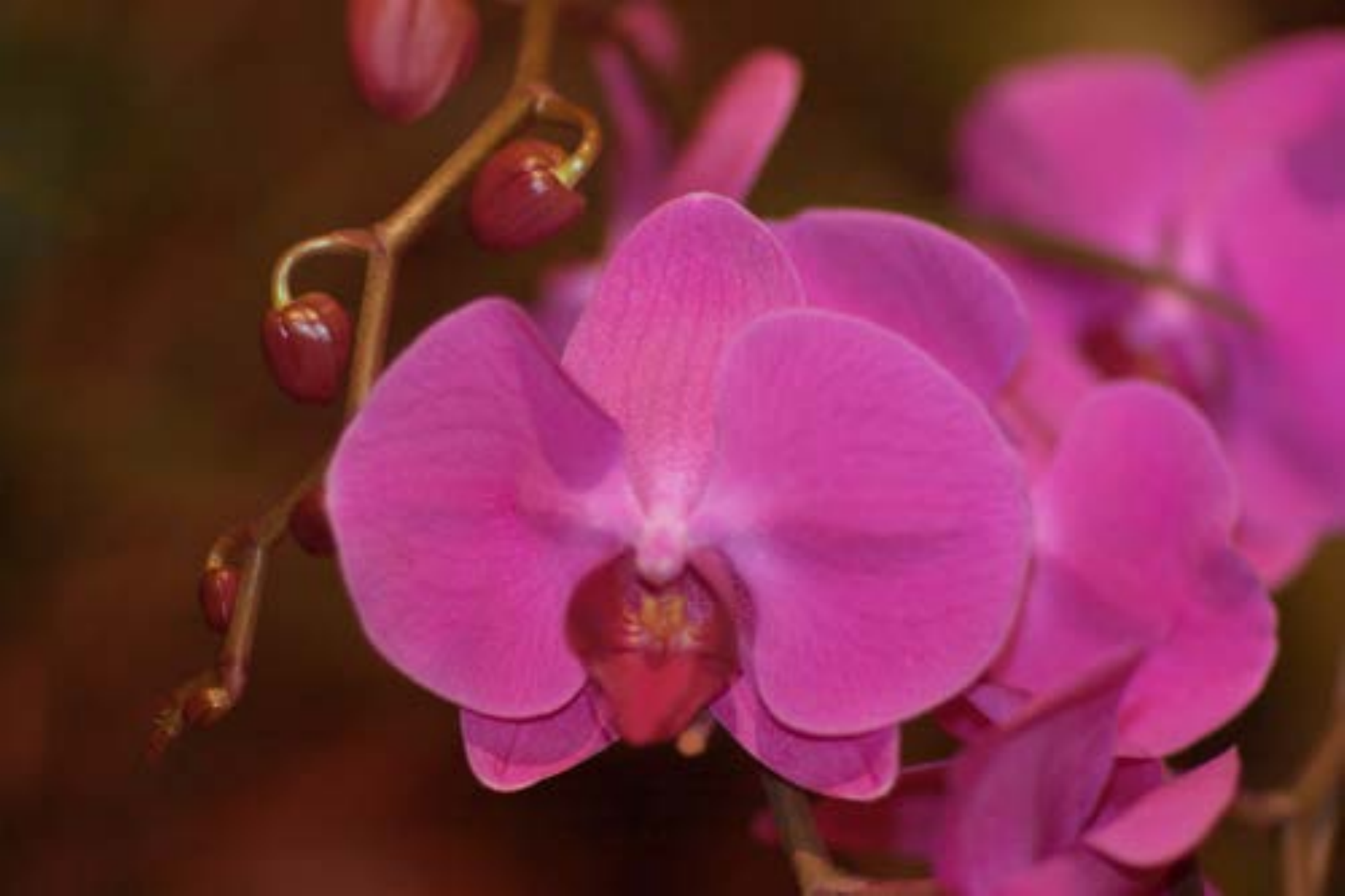}&
		\includegraphics[width=0.19\textwidth]{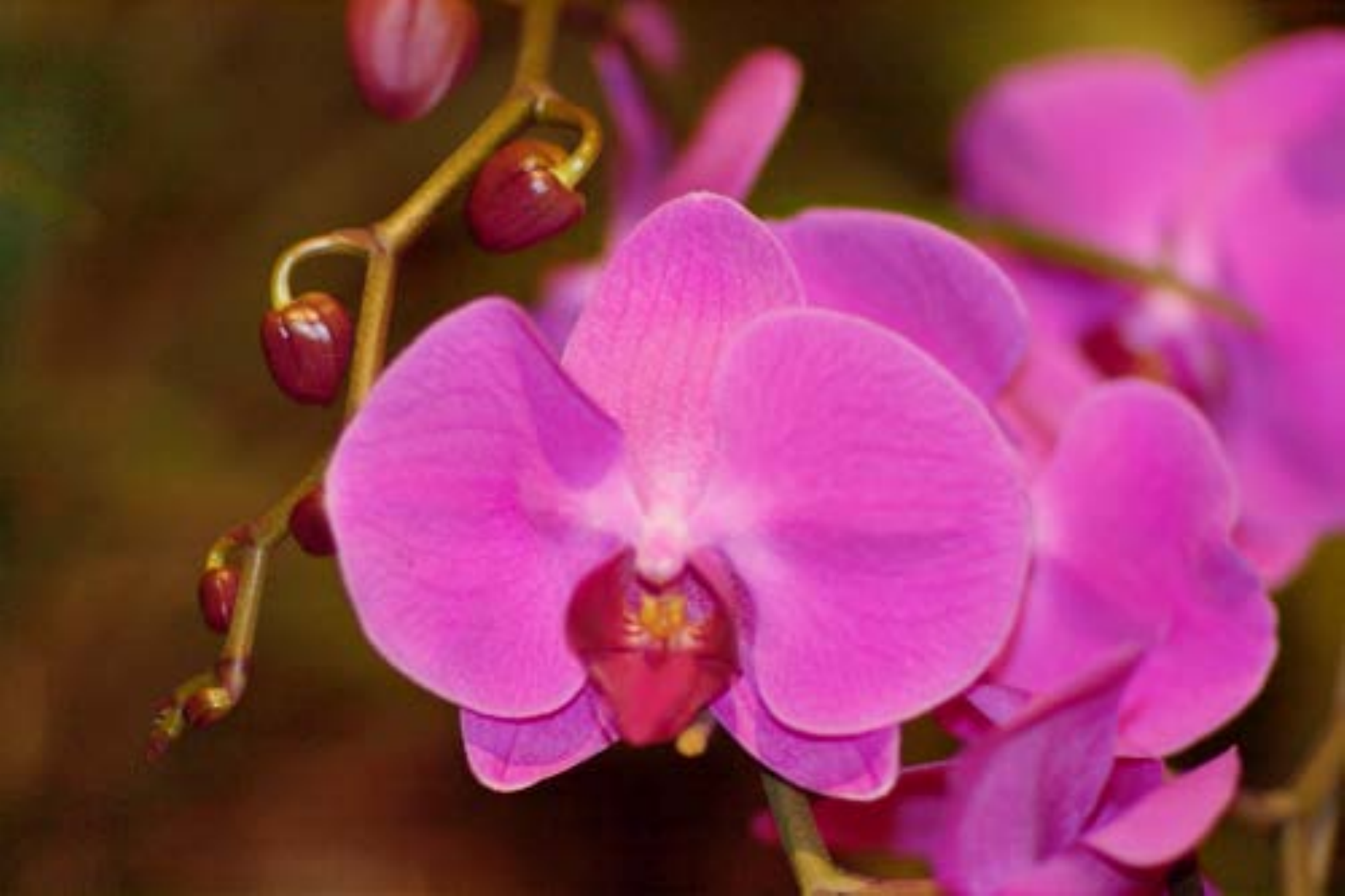}\\	
		\footnotesize Input&\footnotesize LIME~\cite{guo2017lime}&\footnotesize WhiteBox~\cite{hu2018exposure}&\footnotesize DeepUPE~\cite{wang2019underexposed}&\footnotesize CSDNet\\
	\end{tabular}
	\caption{Visual comparison on some example images in MIT-Adobe FiveK dataset~\cite{fivek}.}
	\label{fig:MITComparison}
\end{figure*}
\subsection{CSDGAN}
Considering the paired datasets exist the limitation in real-world scenarios, we also define the CSDGAN to ensure the stronger practical values. To be concrete, we adopt the following losses to train it by using the unpaired dataset. 

\textbf{Adversarial Loss.}$\;$ We adopt the following commonly-used adversarial loss to achieve the supervision
\begin{equation}\label{eq:adverloss}
\begin{split}
\mathcal{L}_{ADV}={E}_{{z}_{f}\sim{\mathbb{Q}}}\left[{g}_{1}\left({D}\left({z}_{f}\right)-{E}_{{z}_{r}\sim{\mathbb{P}}}\left[{D}\left({z}_{r}\right)\right]\right)\right]+ \\
{E}_{{z}_{r}\sim{\mathbb{P}}}\left[{g}_{2}\left({D}\left({z}_{r}\right)-{E}_{{z}_{f}\sim{\mathbb{Q}}}\left[{D}\left({z}_{f}\right)\right]\right)\right] +\\
{E}_{{z}_{f}\sim{\mathbb{Q}}}\left[{f}_{1}\left({D}\left({z}_{f}\right)-{E}_{{z}_{r}\sim{\mathbb{P}}}\left[{D}\left({z}_{r}\right)\right]\right)\right]+ \\
{E}_{{z}_{r}\sim{\mathbb{P}}}\left[{f}_{2}\left({D}\left({z}_{r}\right)-{E}_{{z}_{f}\sim{\mathbb{Q}}}\left[{D}\left({z}_{f}\right)\right]\right)\right],
\end{split}
\end{equation}
where ${z}_{f}$ and ${z}_{r}$ represent the samples randomly sampled (including global image and local patch) from the fake ($\mathbb{Q}$) and real ($\mathbb{{P}}$) distribution, respectively. ${g}_{1}\left({u}\right)={f}_{1}\left({u}\right)={u}^{2},{g}_{2}\left({u}\right)={f}_{2}\left({u}\right)=\left({u}-{1}\right)^{2}$.

\textbf{Perceptual Loss.}$\;$ Different from CSDNet, because of without the ground truth, we use the following perceptual loss to ensure the perceptual similarity between the low-light input and the final output.
{
\begin{equation}
\mathcal{L}_{P_2}=\frac{1}{{W}_{i,j}{H}_{i,j}}\sum_{x=1}^{{W}_{i,j}}\sum_{y=1}^{{H}_{i,j}}\left({{\phi}_{i,j}\left({\mathbf{L}}\right)}_{x,y}-{{\phi}_{i,j}\left(\hat{\mathbf{R}}\right)}_{x,y}\right),
\end{equation}
}
where we set $i=5$, $j=1$. ${W}_{i,j}$ and ${H}_{i,j}$ are the dimensions of the extracted feature maps. 

{
	In a word, we train our CSDGAN using the following loss
\begin{equation}
\mathcal{L}_\mathtt{CSDGAN}=\mathcal{L}_{ADV}+\mathcal{L}_{P_2}+\mathcal{L}_{S}.
\end{equation}
}

\begin{figure}[t]
	\centering
	\begin{tabular}{c@{\extracolsep{0.2em}}c@{\extracolsep{0.2em}}c}
		\footnotesize Input&\footnotesize DeepUPE~\cite{wang2019underexposed}&\footnotesize CSDNet\\
		\includegraphics[width=0.15\textwidth]{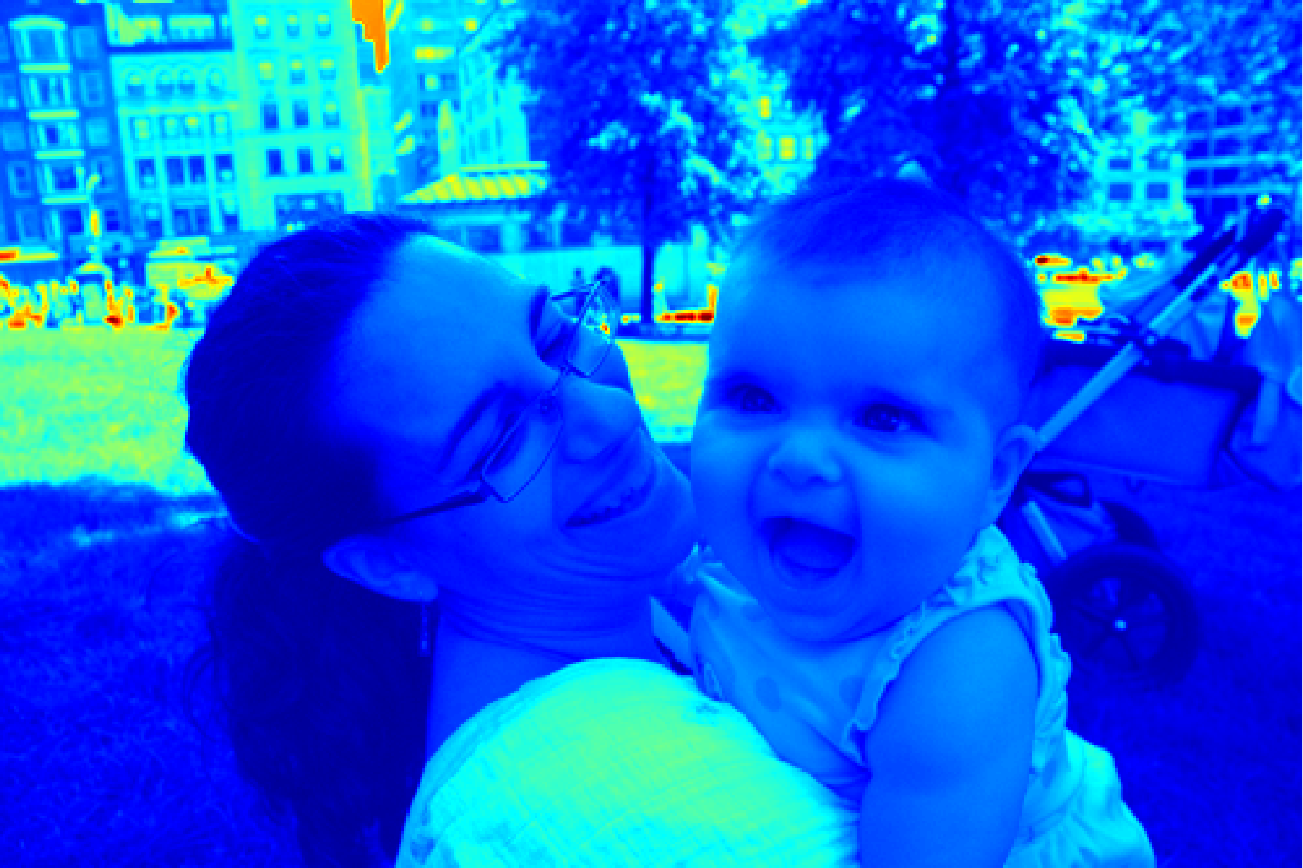}&
		\includegraphics[width=0.15\textwidth]{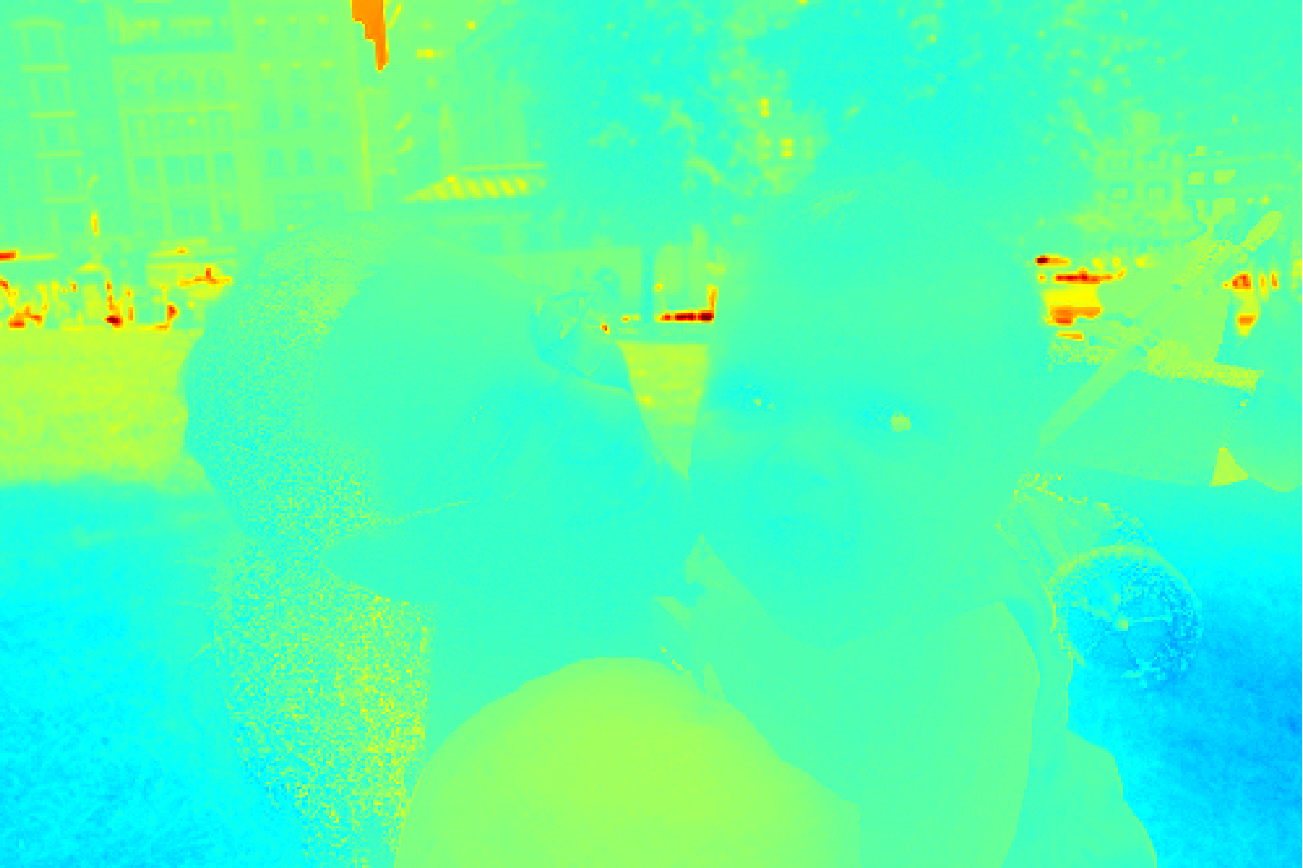}&
		\includegraphics[width=0.15\textwidth]{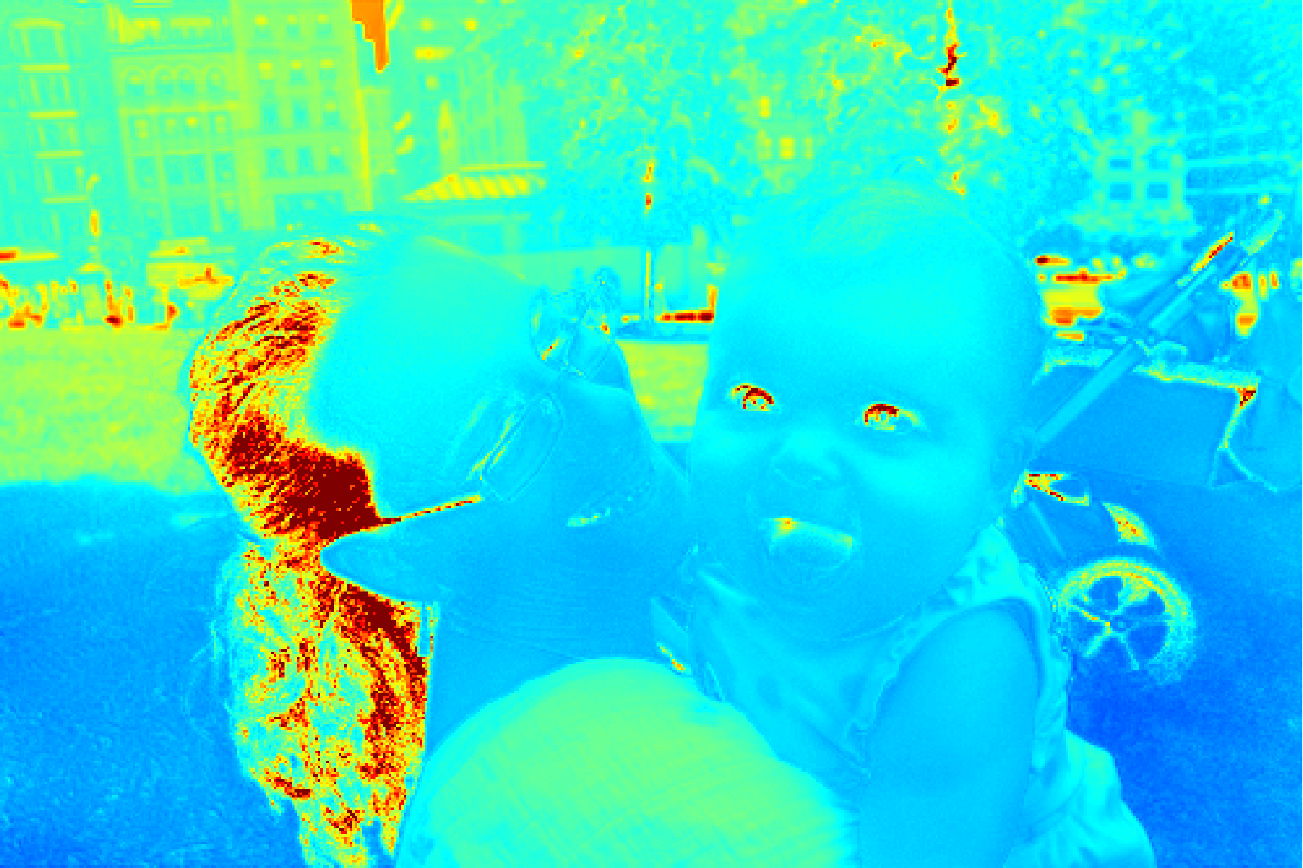}\\
		\multicolumn{3}{c}{\footnotesize The estimated illumination}\\ 	
		\includegraphics[width=0.15\textwidth]{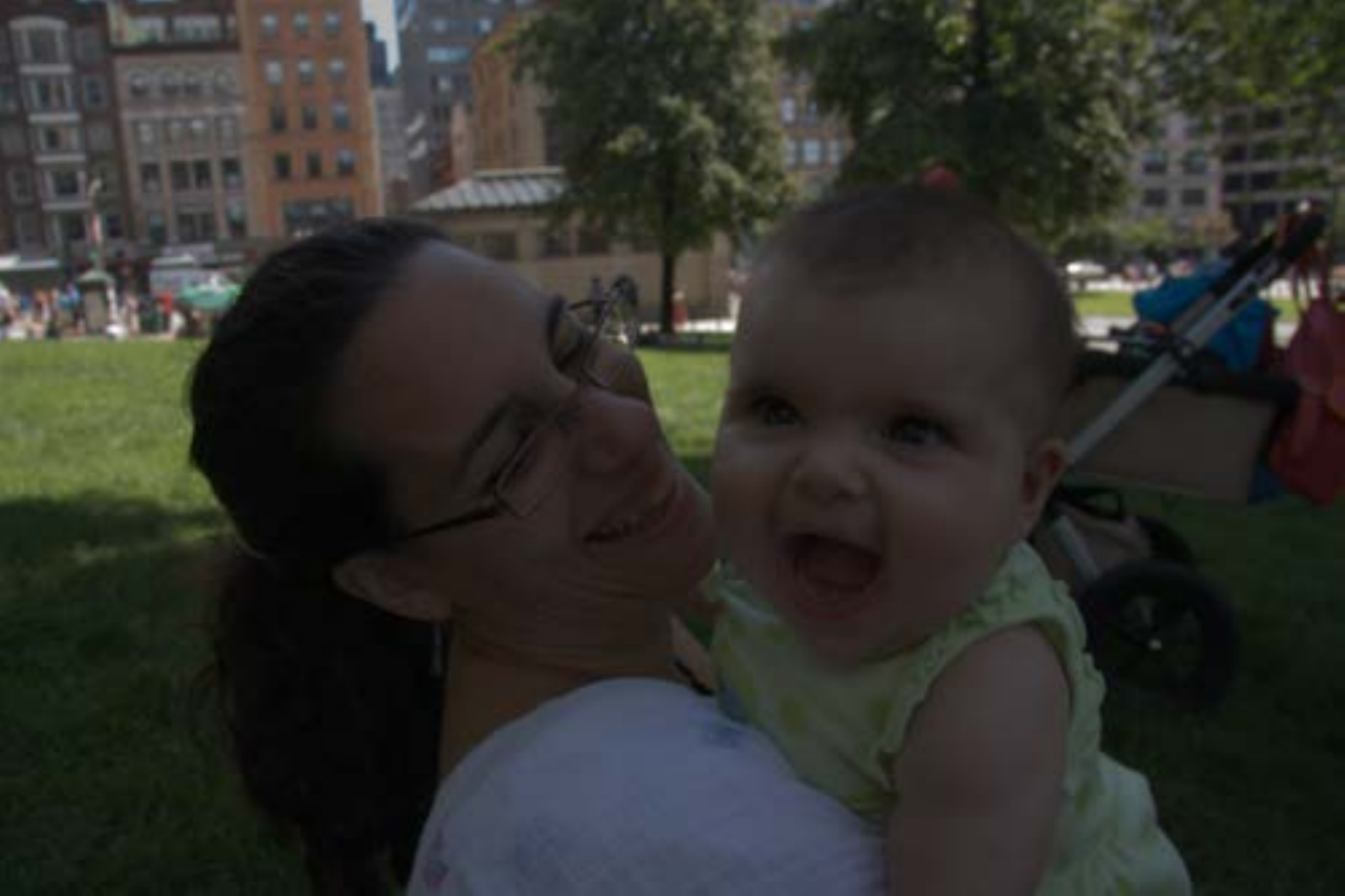}&
		\includegraphics[width=0.15\textwidth]{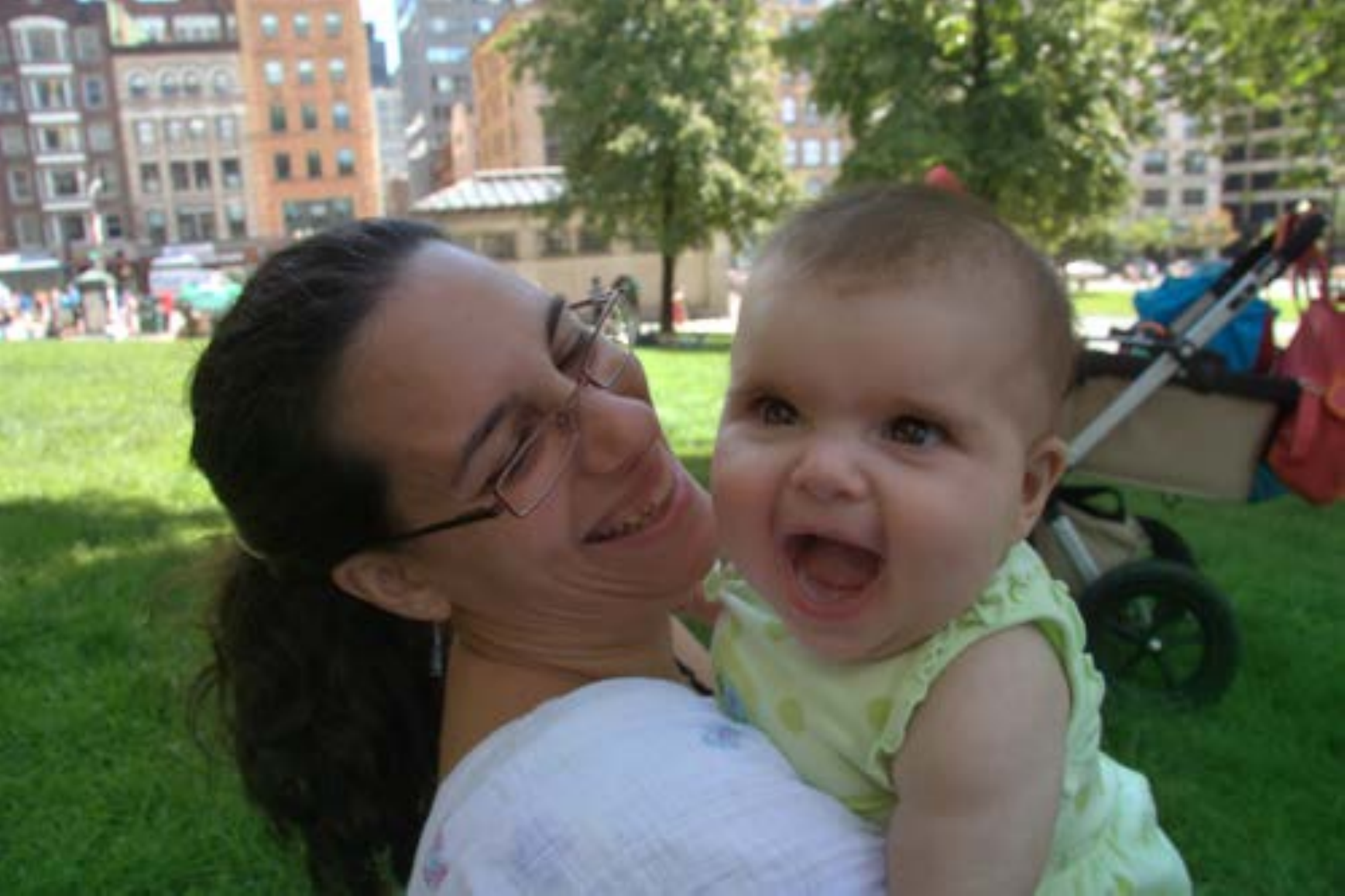}&
		\includegraphics[width=0.15\textwidth]{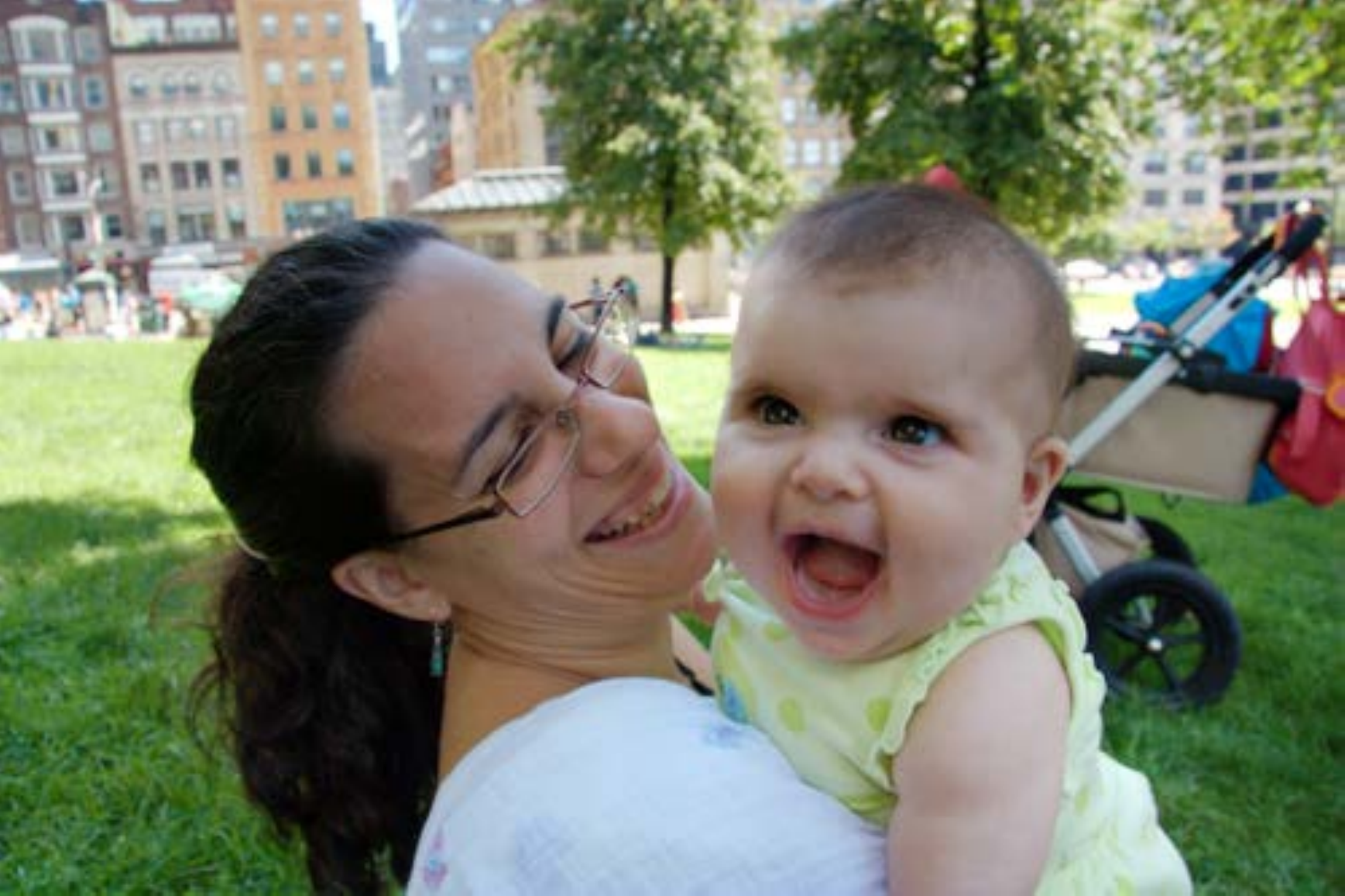}\\	
		\multicolumn{3}{c}{\footnotesize The enhanced output}\\ 	
	\end{tabular}
	\caption{Visual comparison of different components among two networks trained in MIT-Adobe FiveK dataset. }
	\label{fig:CompDeepUPE}
\end{figure}

\begin{table*}[t]
	\centering
	\caption{Quantitative comparison on LOL testing dataset in terms of PSNR and SSIM.}
	\begin{tabular}{cccccccccc}
		\toprule
		Benchmark&SRIE~\cite{fu2015probabilistic}&WVM~\cite{fu2016weighted}&LIME~\cite{guo2017lime}&JIEP~\cite{cai2017joint}&RRM~\cite{li2018structure}&LightenNet~\cite{li2018lightennet}&RetinexNet~\cite{Chen2018Retinex}&KinD~\cite{jiang2019enlightengan}&CSDNet\\
		\midrule 
		PSNR&12.4691&12.0367&16.4435&12.1805&13.8770&10.4461&16.4823&18.0891&\textbf{21.6370}\\
		SSIM&0.5388&0.5247&0.5029&0.5359 &0.6803&0.4008&0.4898&0.8122&\textbf{{0.8526}}\\
		\bottomrule
	\end{tabular}
	\label{tab:LOLTab}
\end{table*}

\begin{figure*}[t]
	\centering
	\begin{tabular}{c@{\extracolsep{0.3em}}c@{\extracolsep{0.3em}}c@{\extracolsep{0.3em}}c@{\extracolsep{0.3em}}c@{\extracolsep{0.3em}}c} 
		\includegraphics[width=0.158\textwidth]{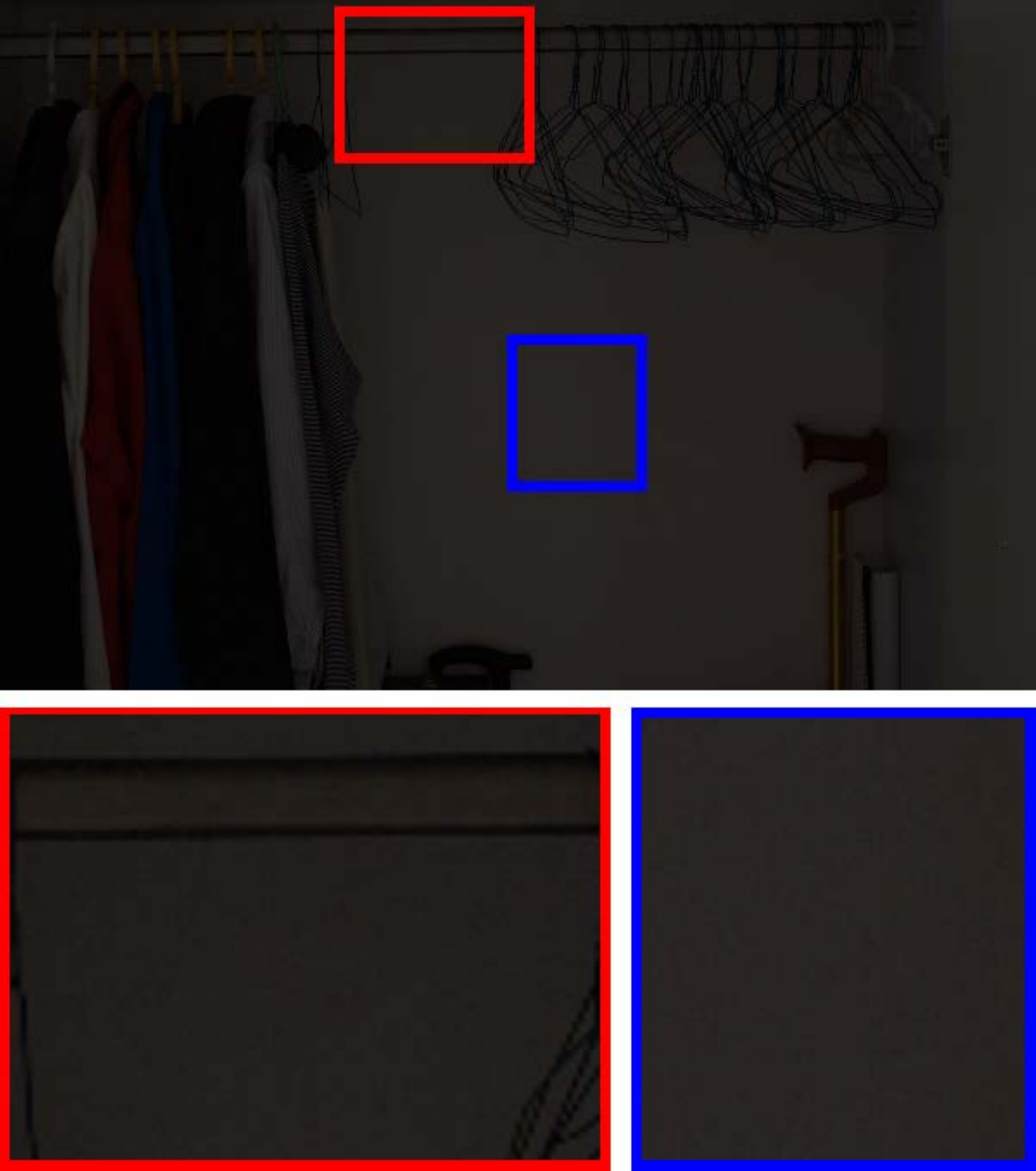}&
		\includegraphics[width=0.158\textwidth]{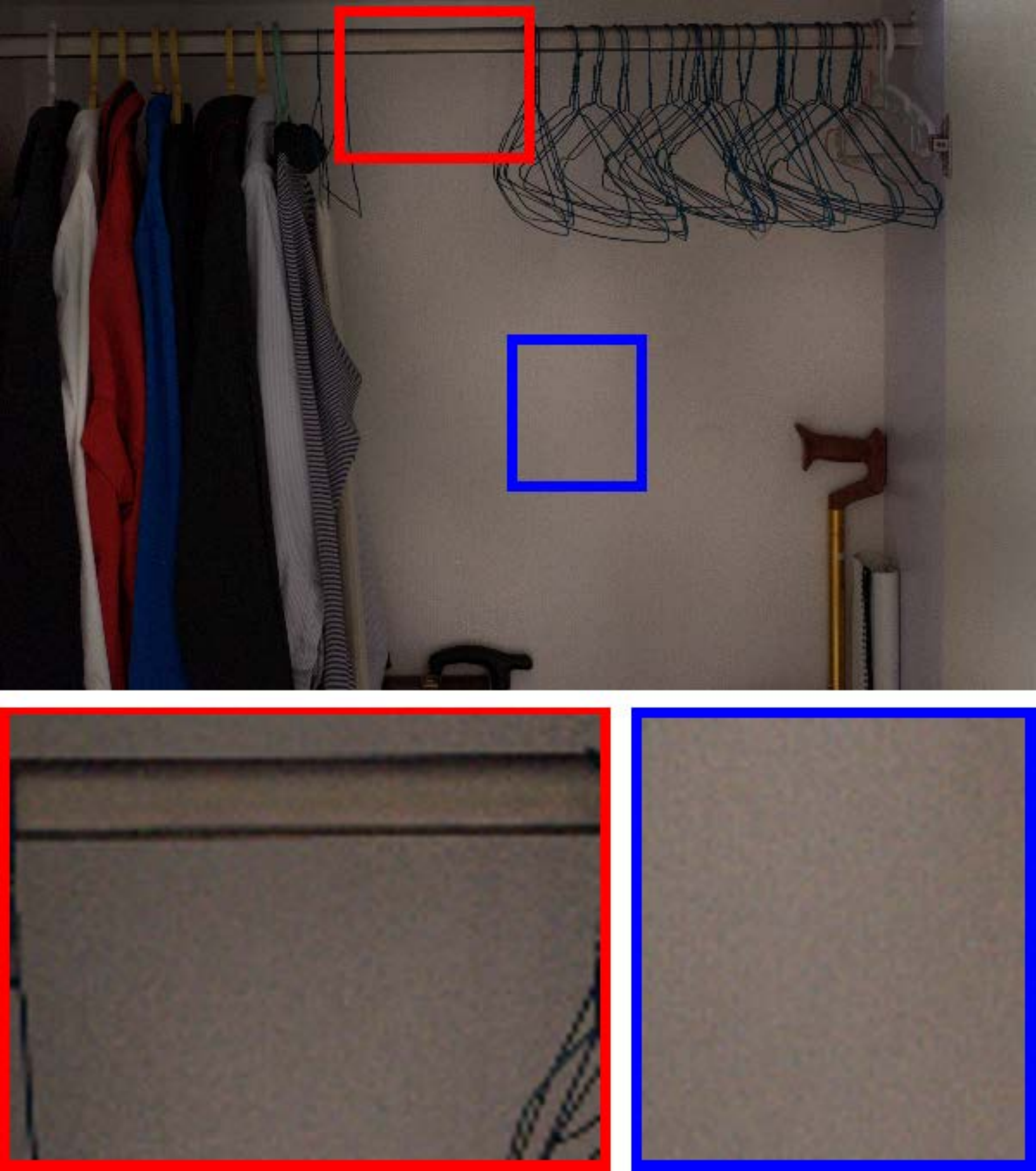}&
		\includegraphics[width=0.158\textwidth]{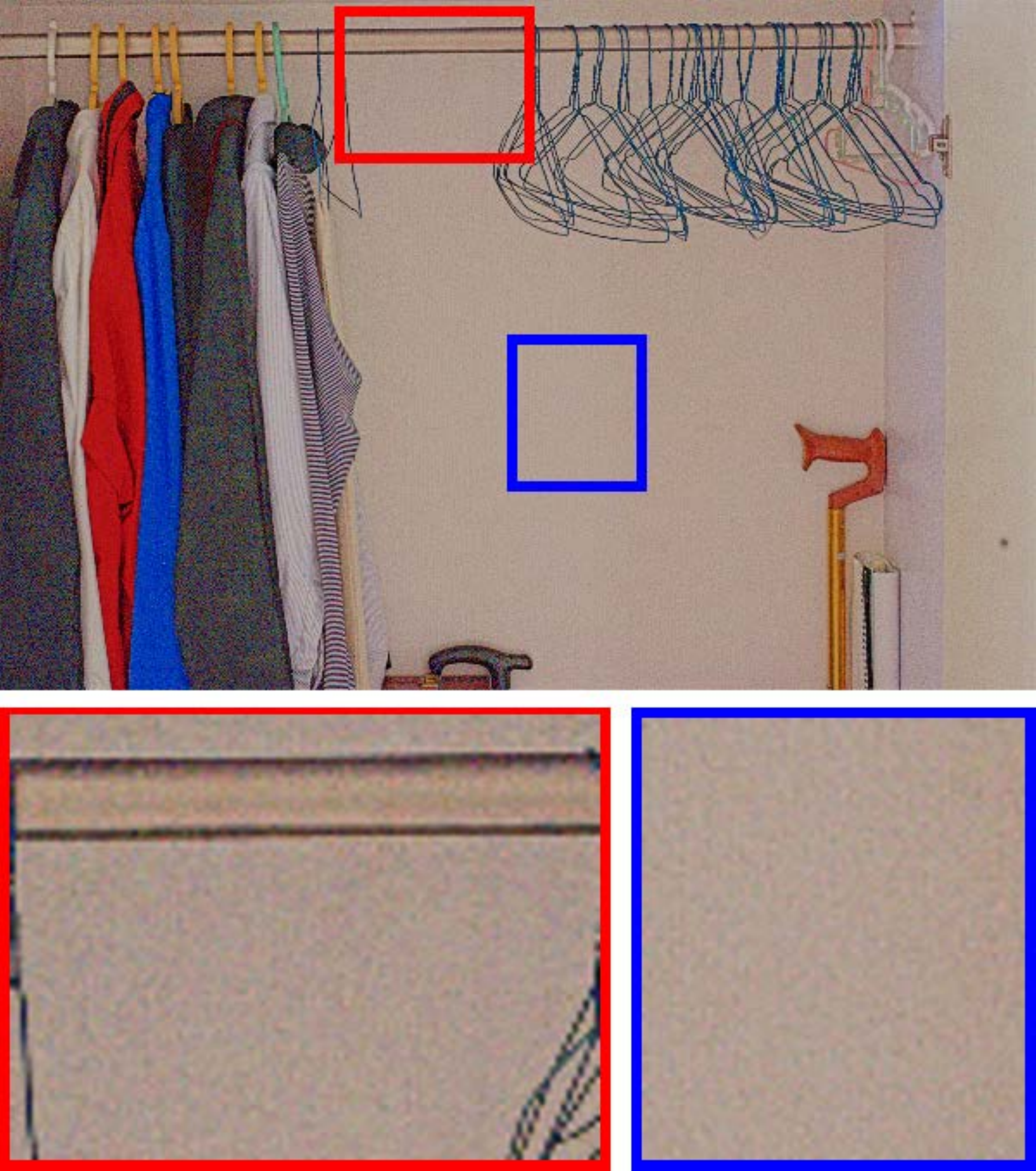}&
		\includegraphics[width=0.158\textwidth]{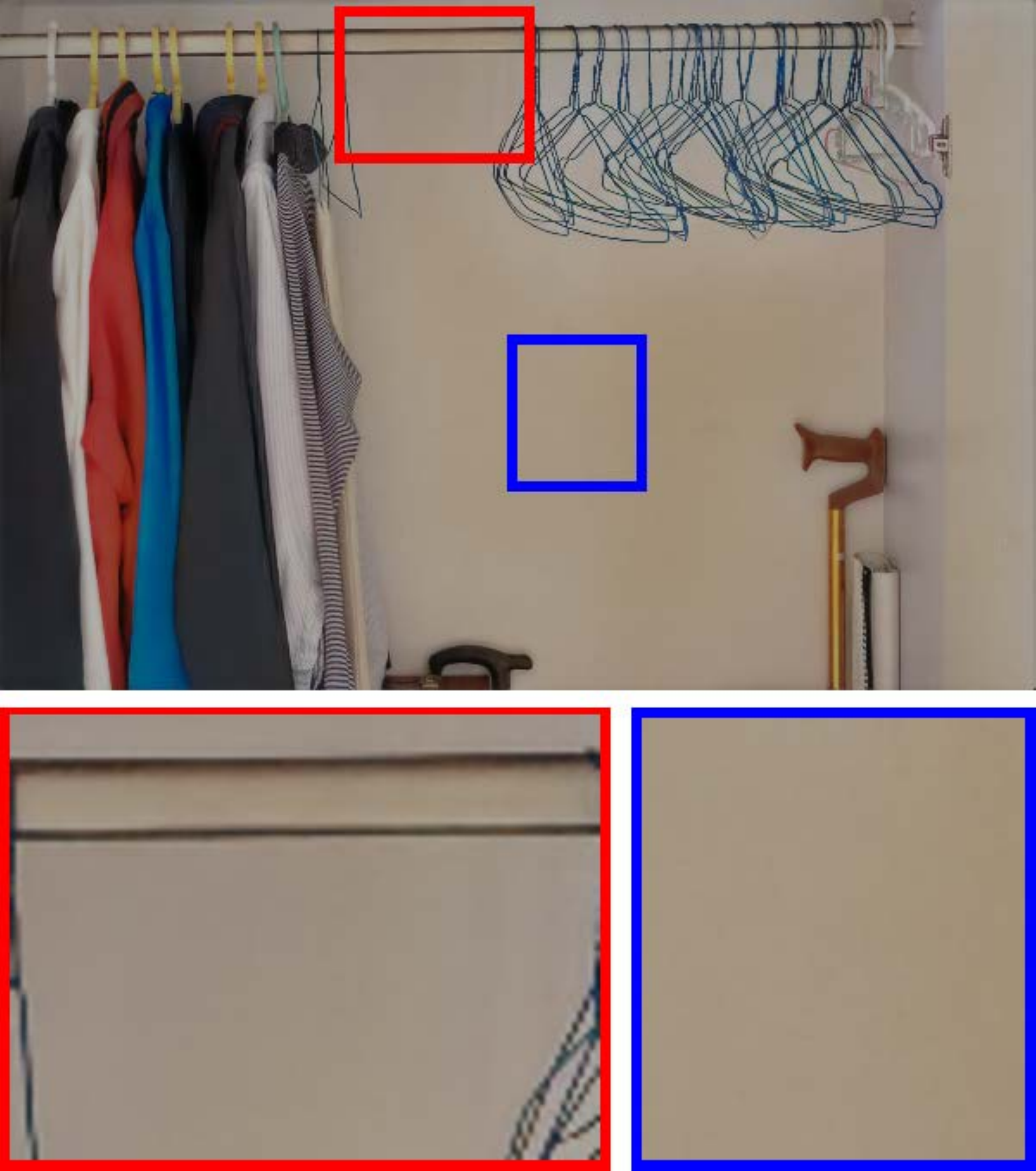}&
		\includegraphics[width=0.158\textwidth]{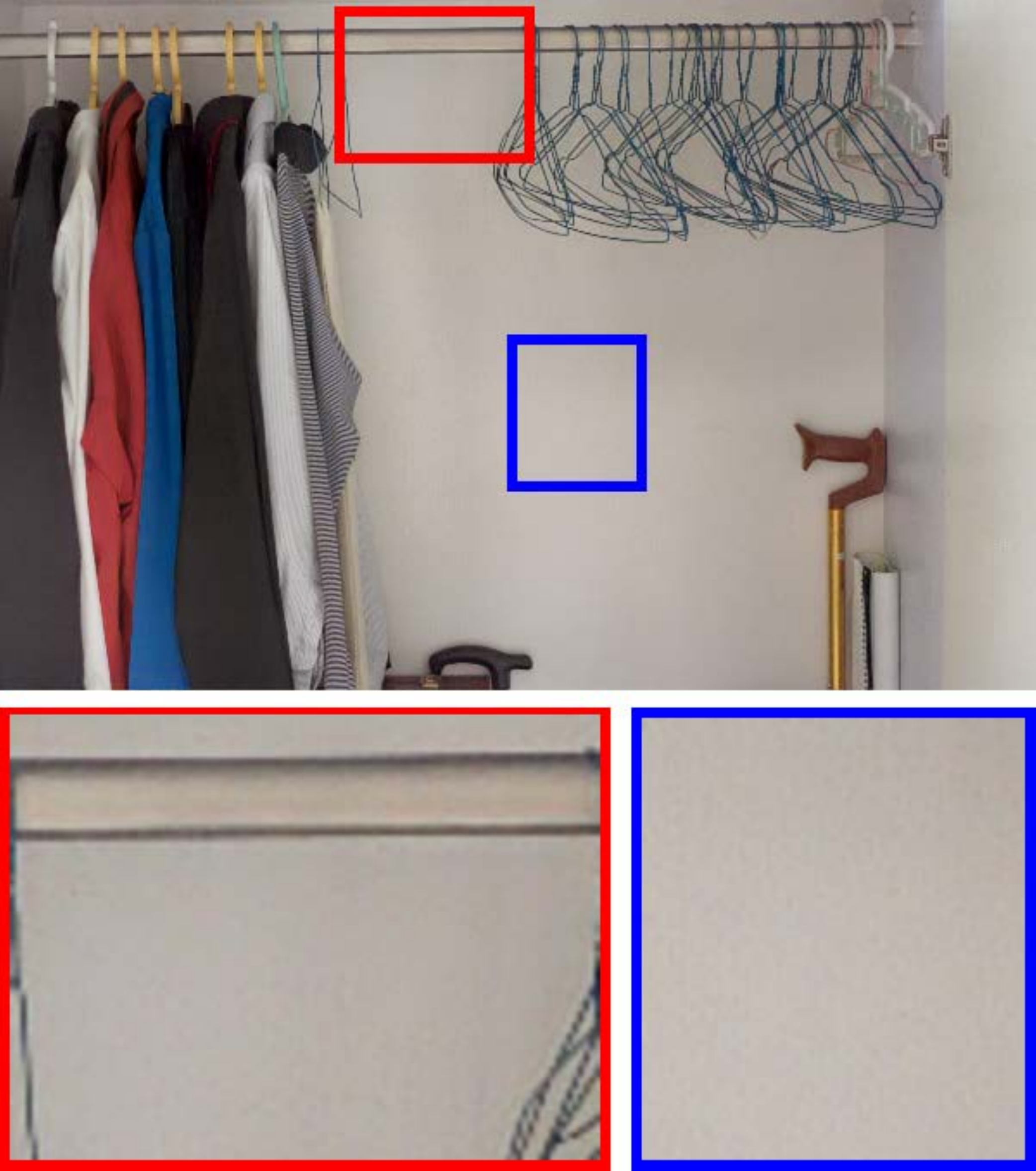}&
		\includegraphics[width=0.158\textwidth]{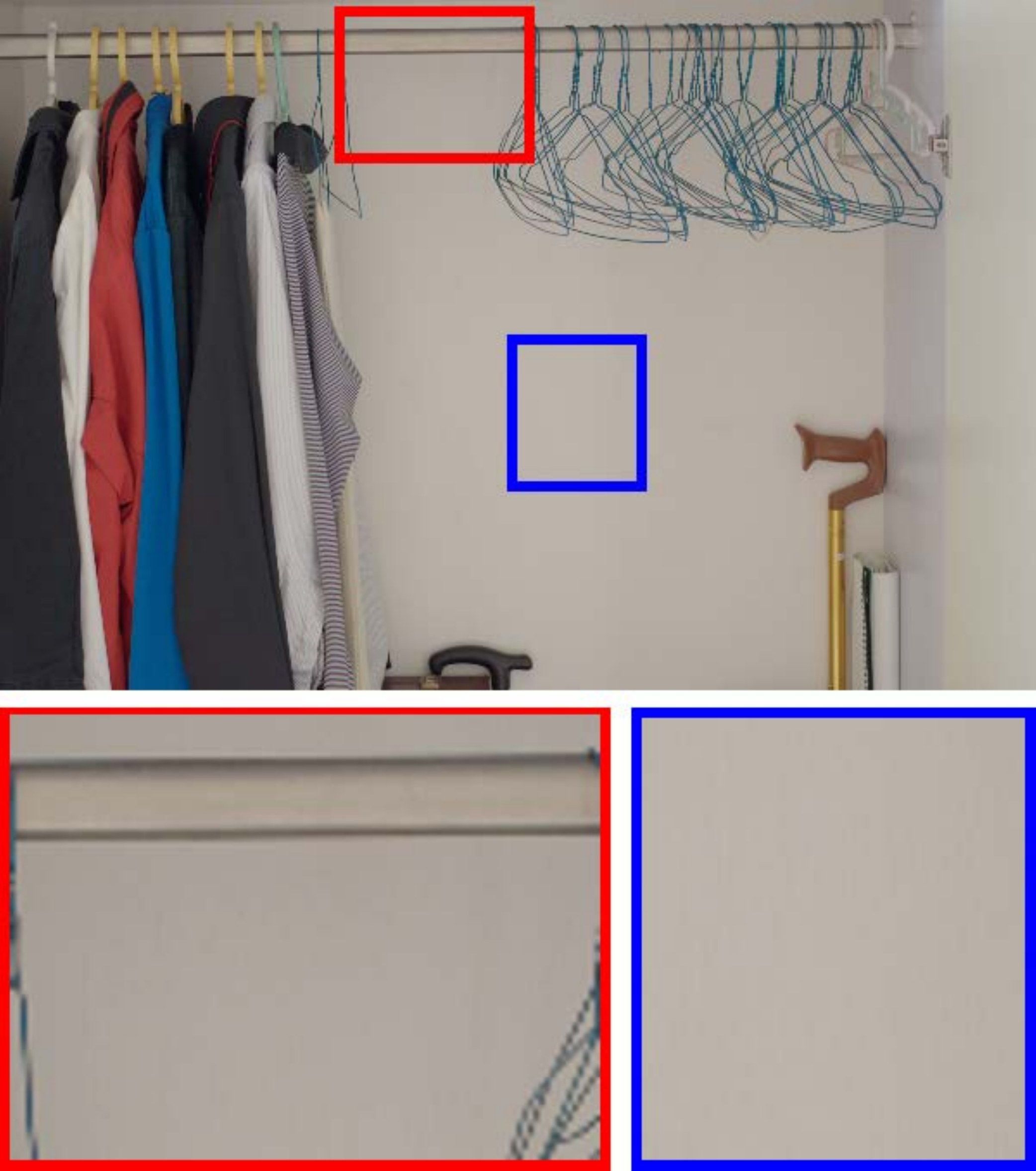}\\	
		\footnotesize 6.6742/0.2274&\footnotesize 11.2311/0.6239&\footnotesize 21.3033/0.6779& \footnotesize 21.1558/0.8750&\footnotesize \textbf{26.2065/0.9144}&\textendash\\
		\footnotesize Input&\footnotesize LightenNet~\cite{li2018lightennet}&\footnotesize RetinexNet~\cite{Chen2018Retinex}&\footnotesize KinD~\cite{zhang2019kindling}&\footnotesize CSDNet&\footnotesize Ground Truth\\
	\end{tabular}
	\caption{Visual comparison on some example images from 15 testing images in the LOL dataset. PSNR/SSIM scores are reported below each image.}
	\label{fig:VC0}
\end{figure*}

\begin{figure*}[t]
	\centering
	\begin{tabular}{c@{\extracolsep{0.3em}}c@{\extracolsep{0.3em}}c@{\extracolsep{0.3em}}c@{\extracolsep{0.3em}}c@{\extracolsep{0.3em}}c} 
		\includegraphics[width=0.158\textwidth]{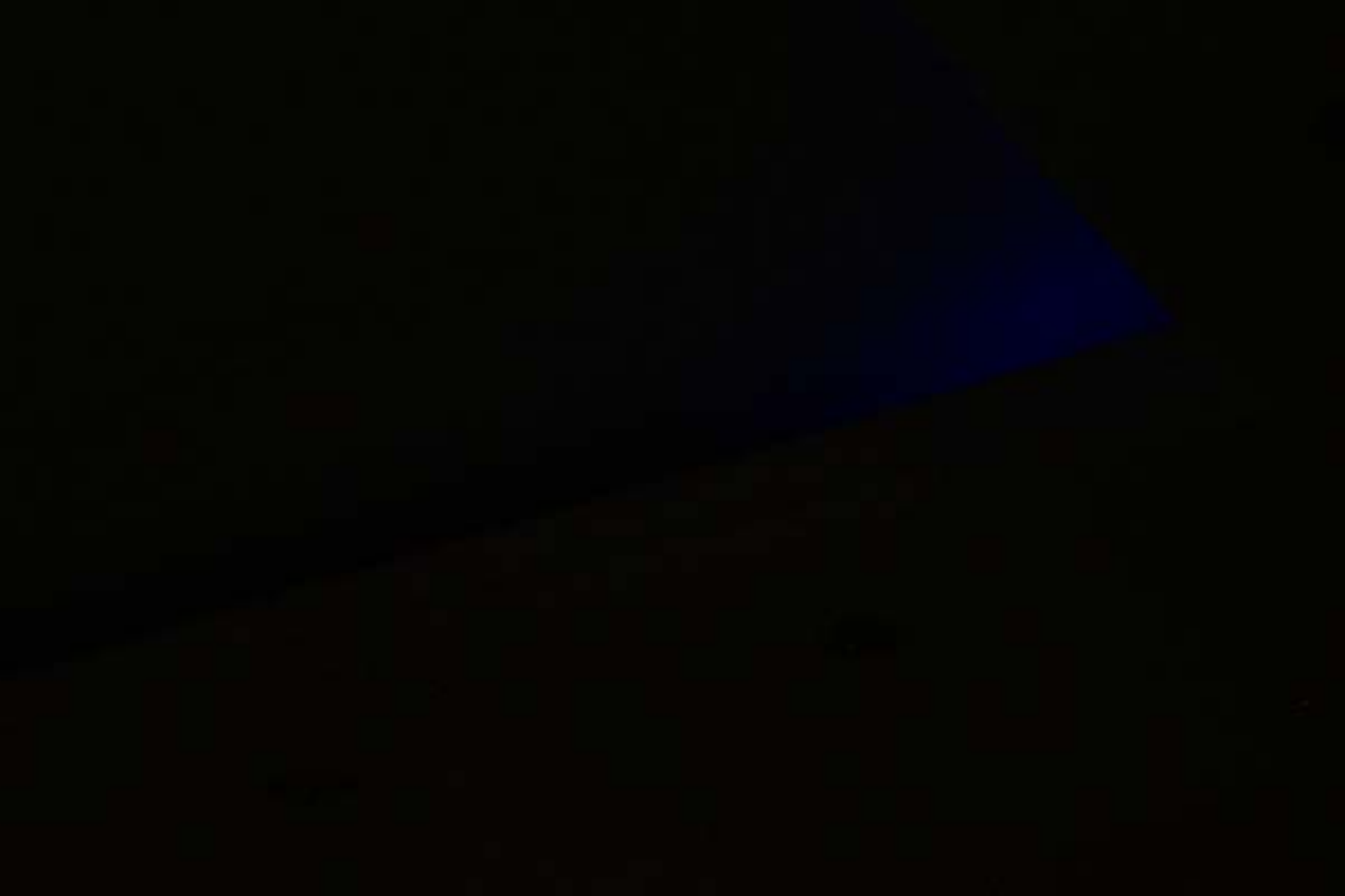}&
		\includegraphics[width=0.158\textwidth]{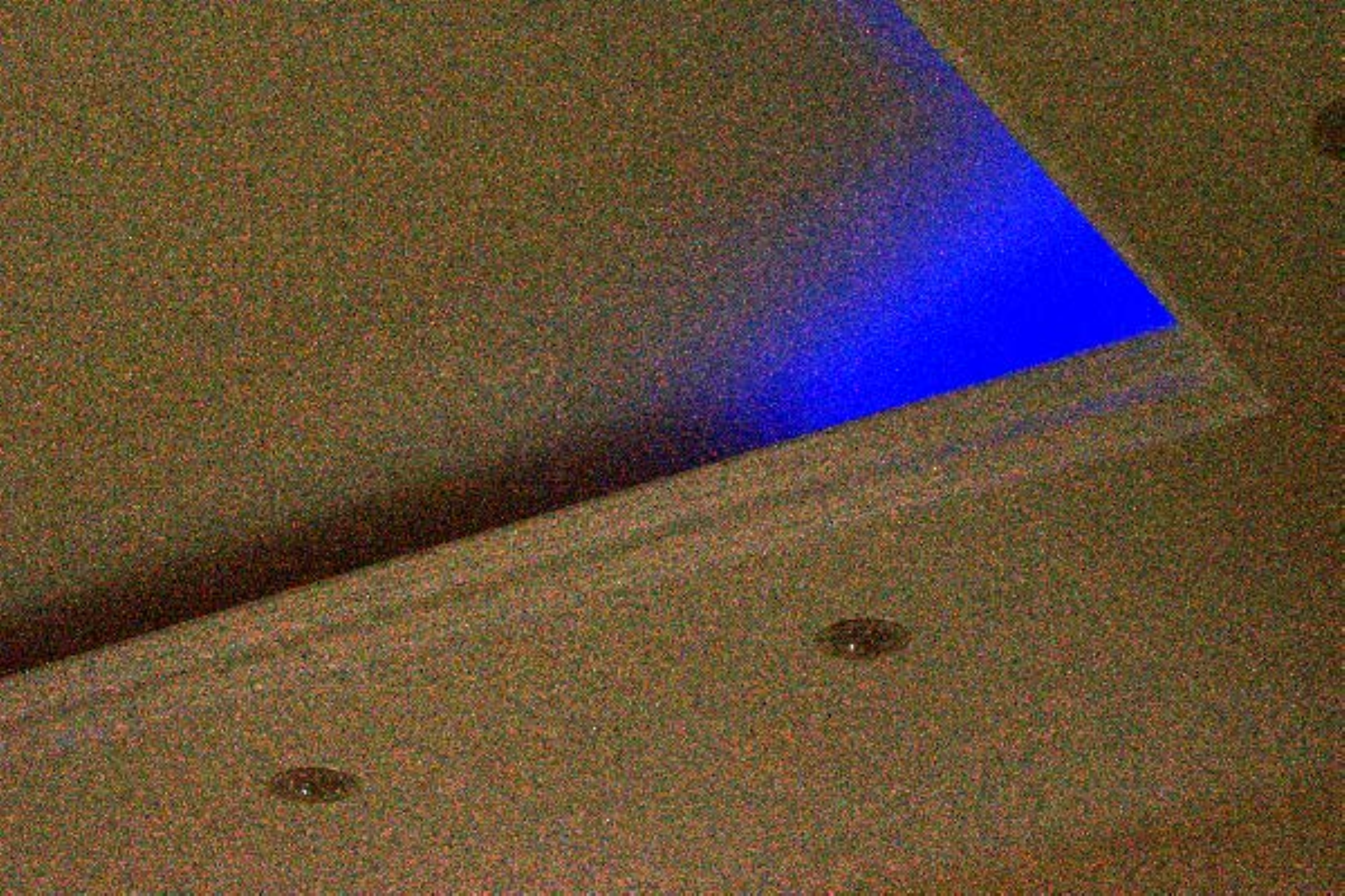}&
		\includegraphics[width=0.158\textwidth]{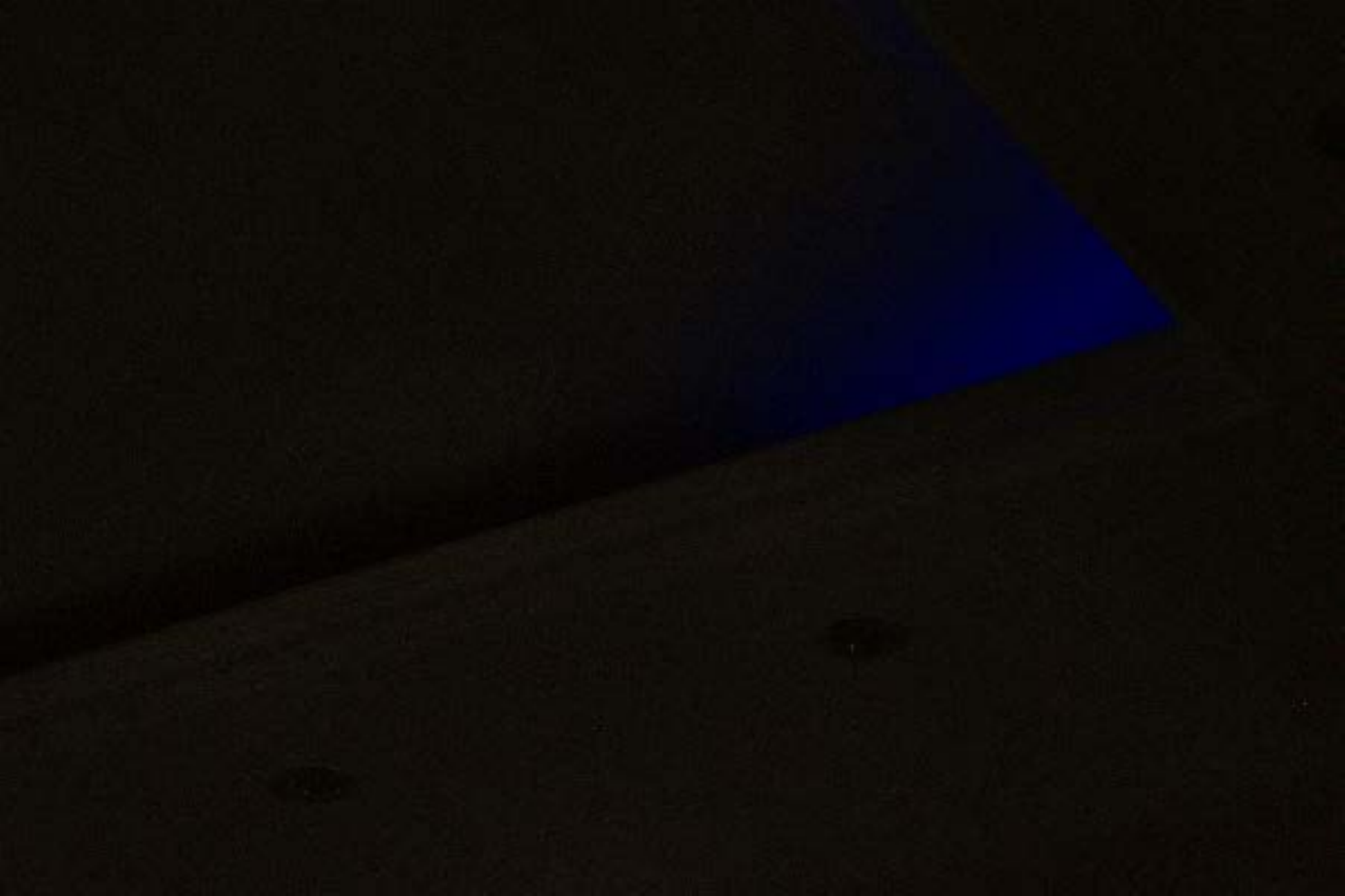}&
		\includegraphics[width=0.158\textwidth]{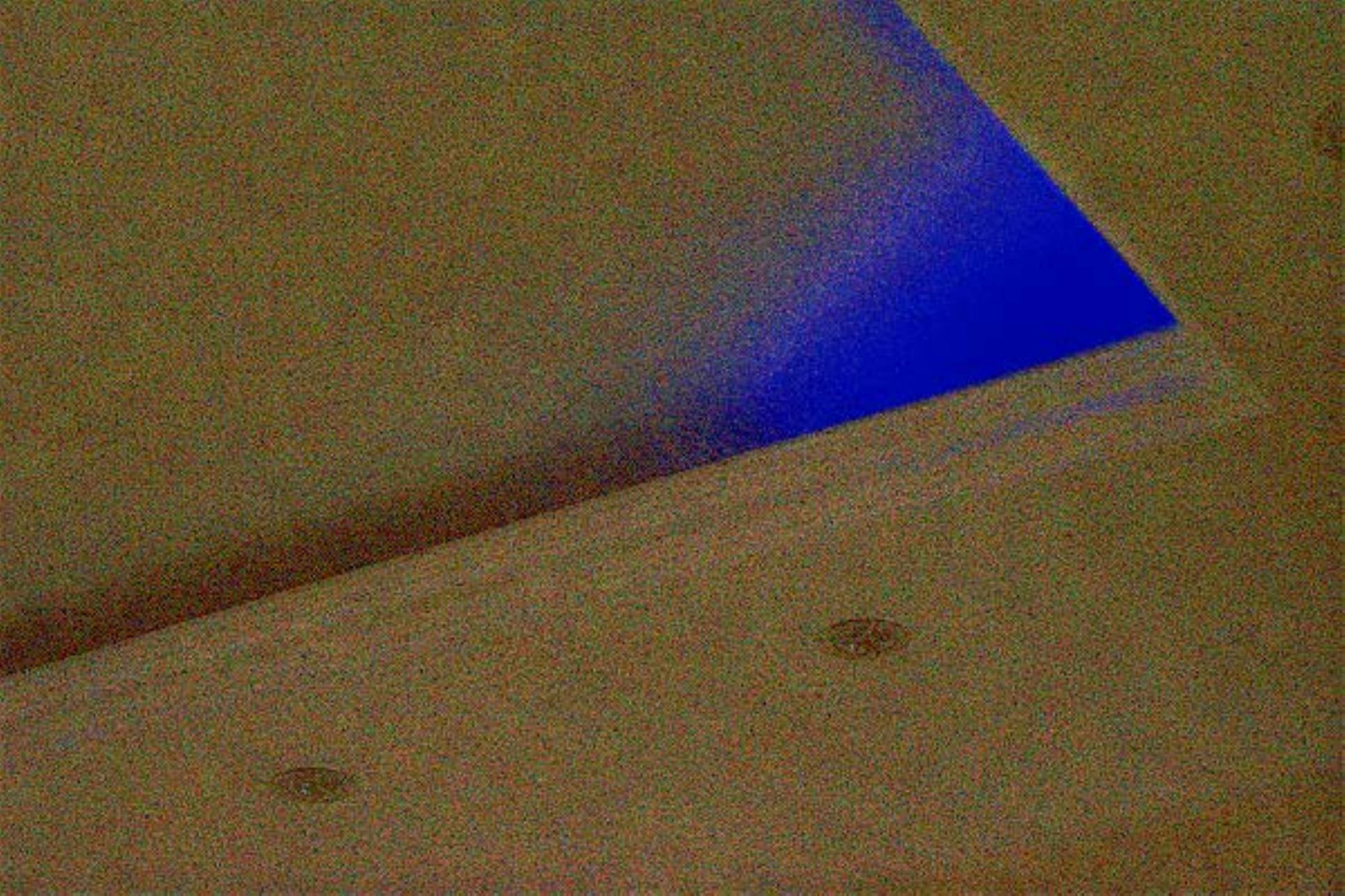}&
		\includegraphics[width=0.158\textwidth]{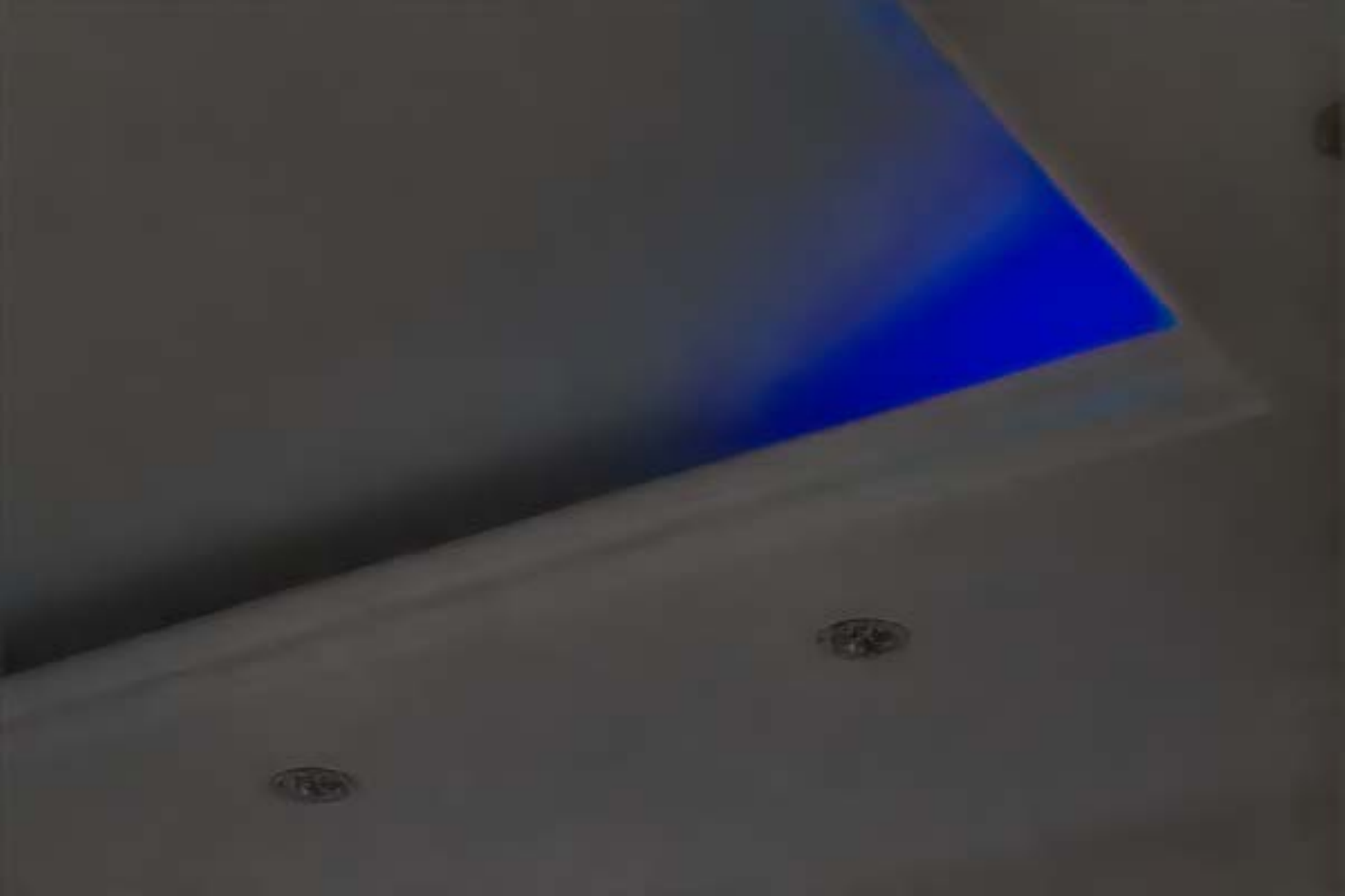}&
		\includegraphics[width=0.158\textwidth]{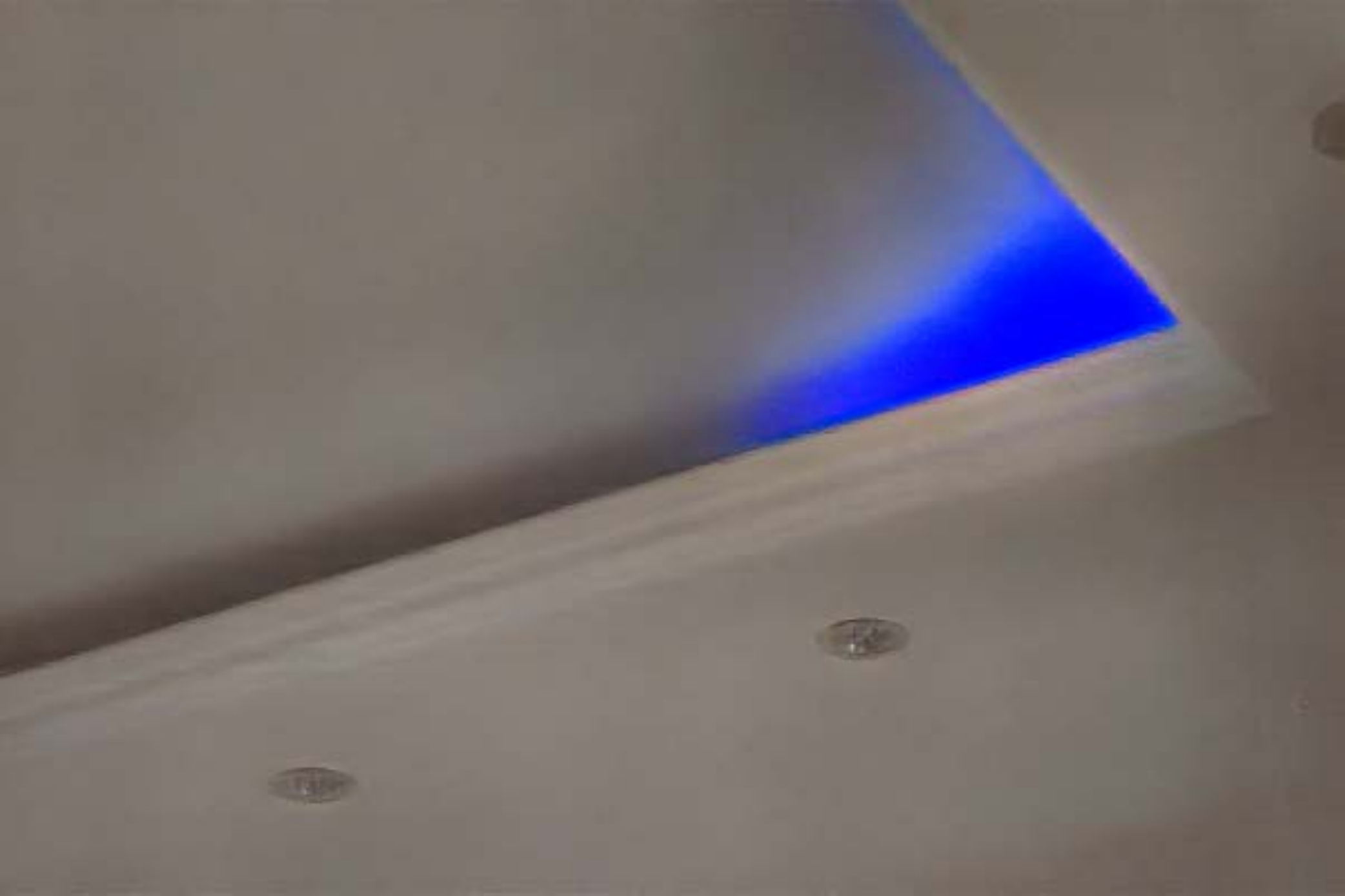}\\	
		\includegraphics[width=0.158\textwidth]{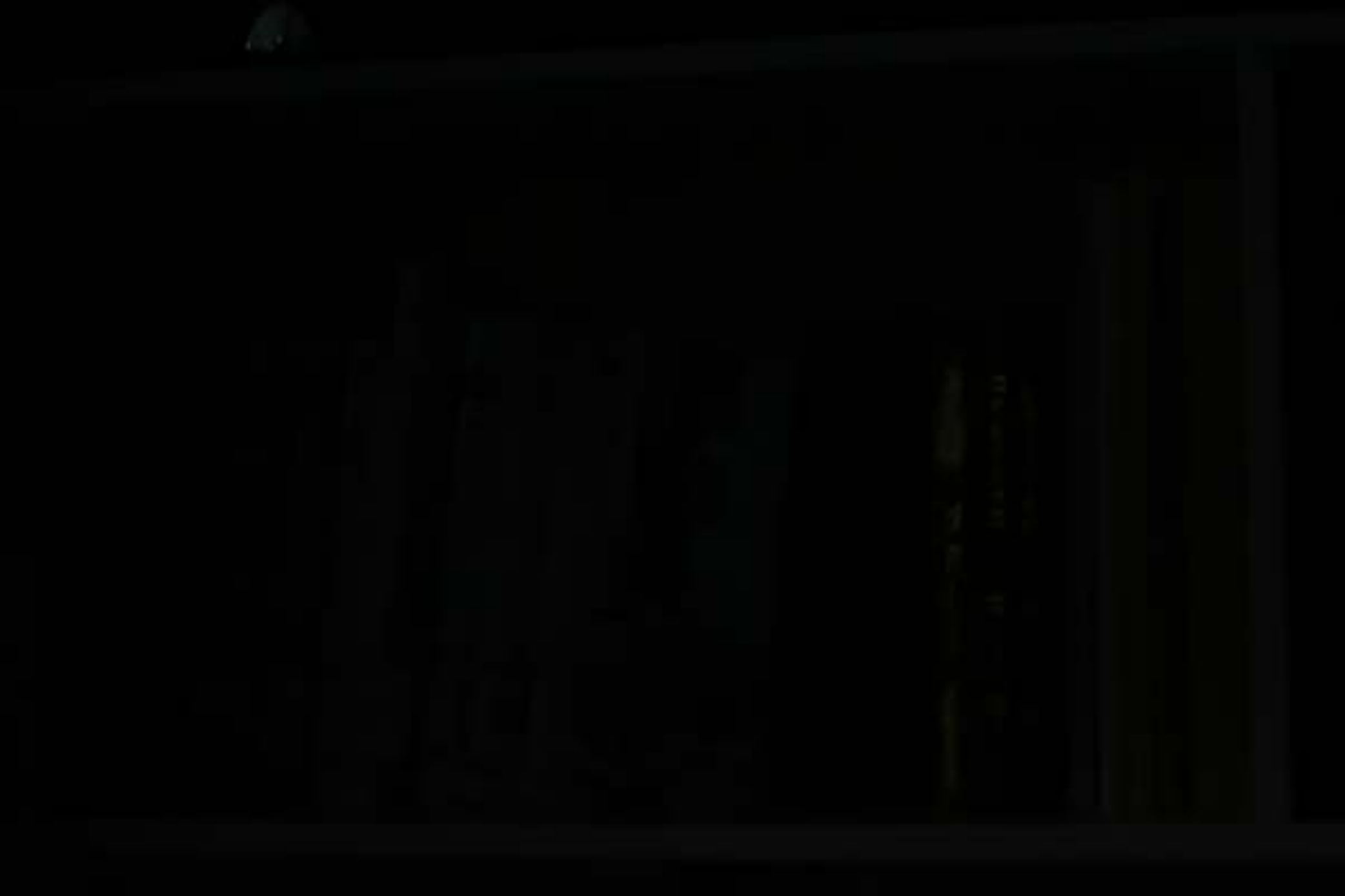}&
		\includegraphics[width=0.158\textwidth]{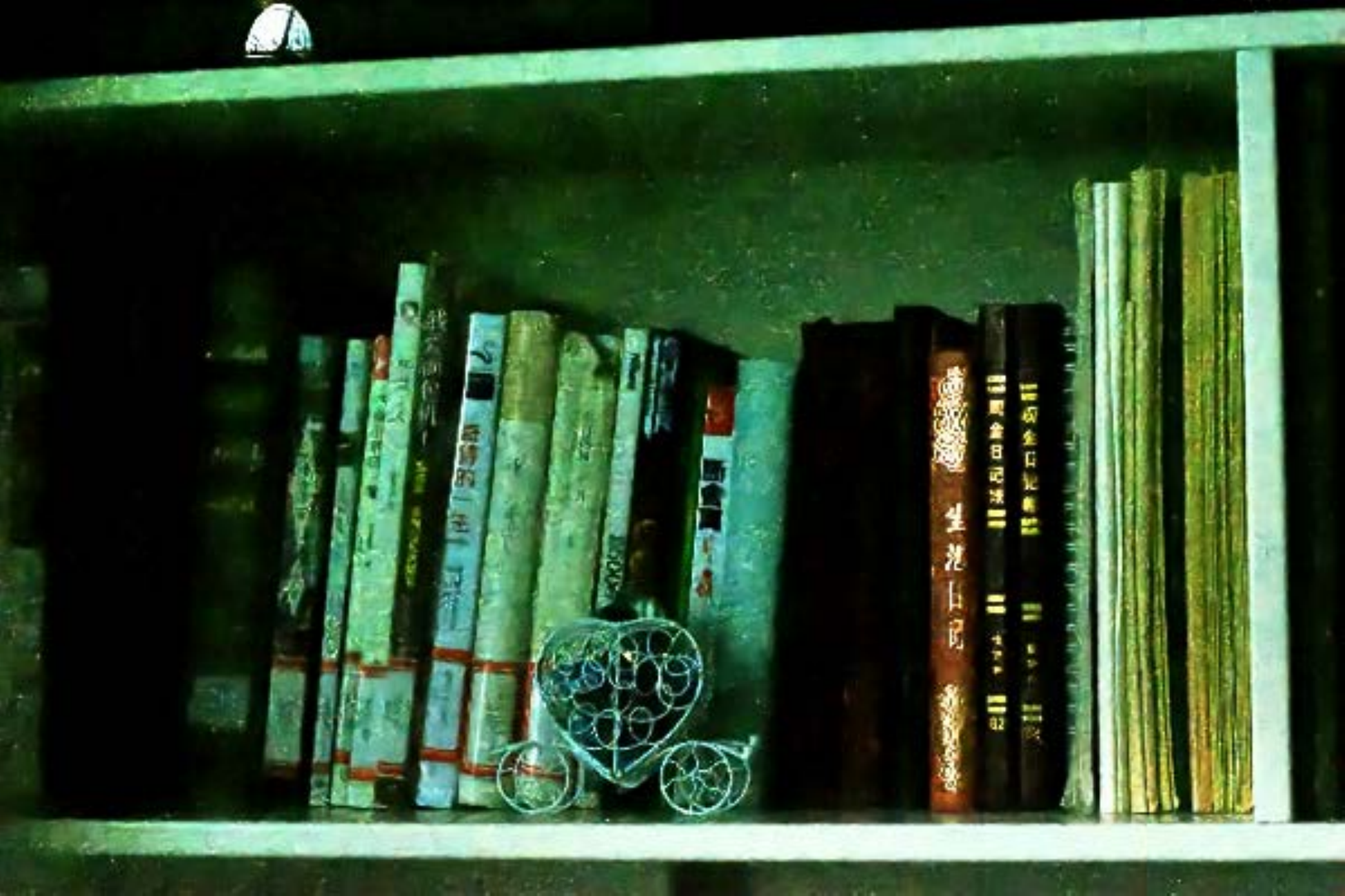}&
		\includegraphics[width=0.158\textwidth]{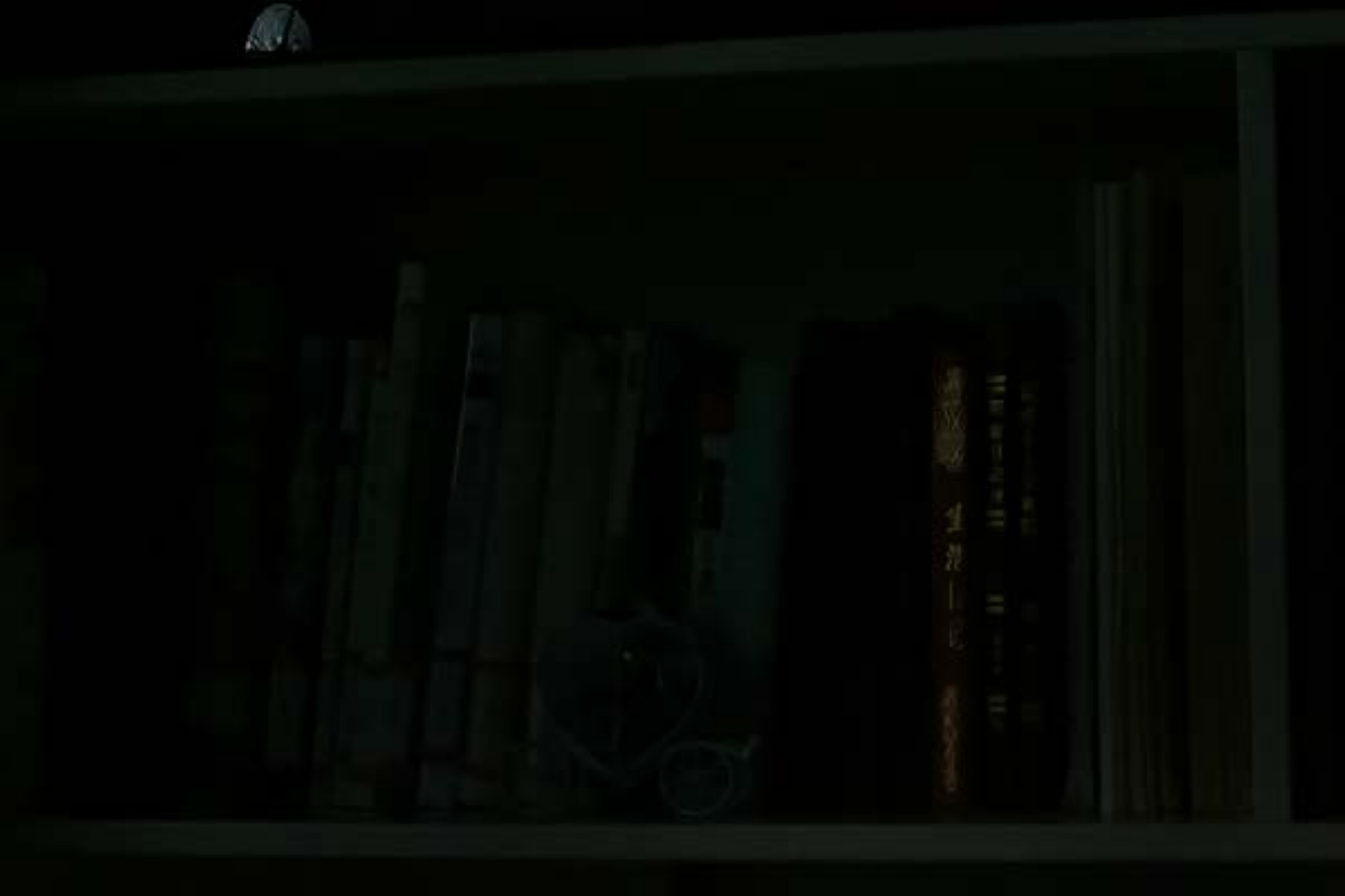}&
		\includegraphics[width=0.158\textwidth]{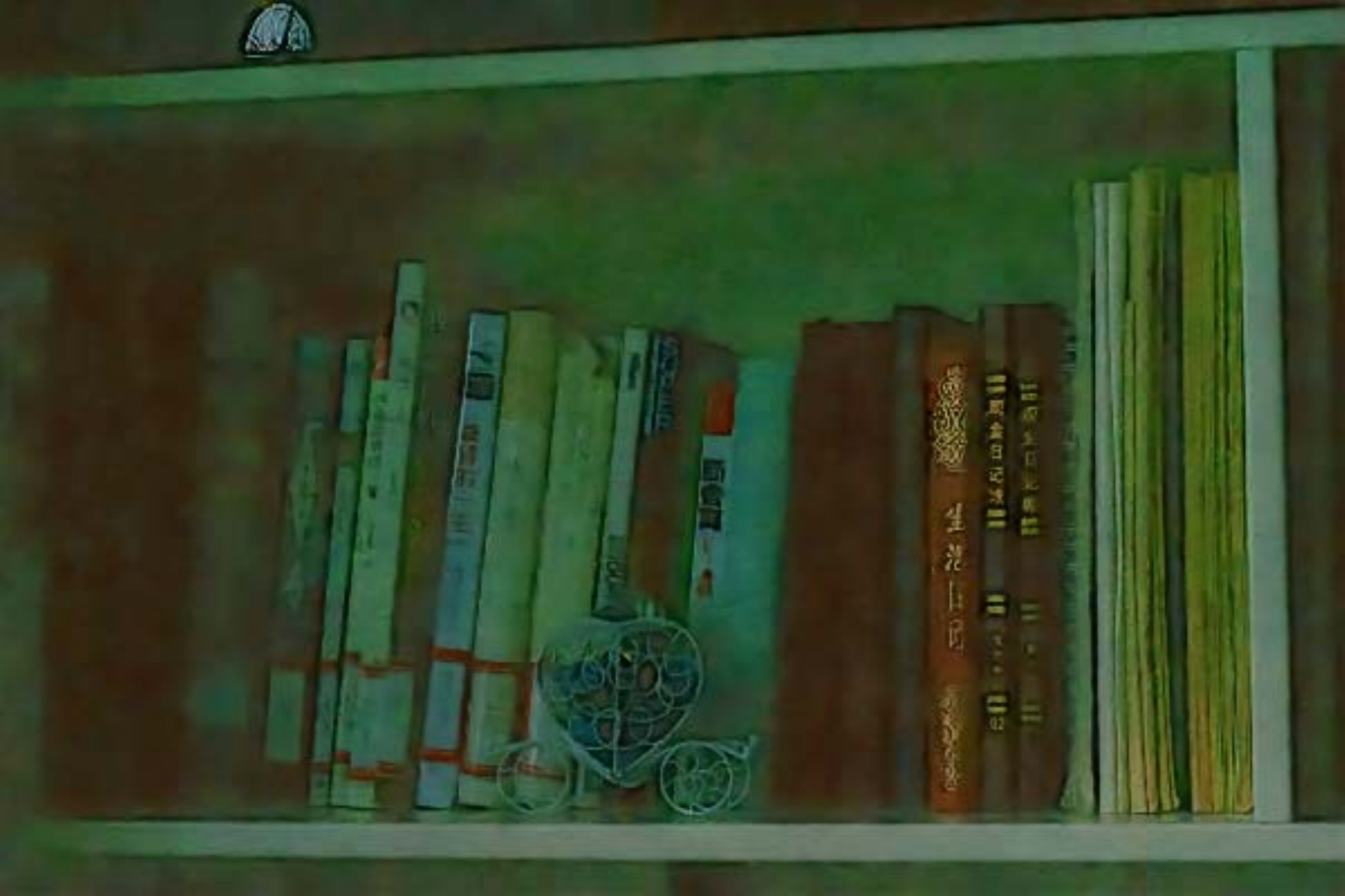}&
		\includegraphics[width=0.158\textwidth]{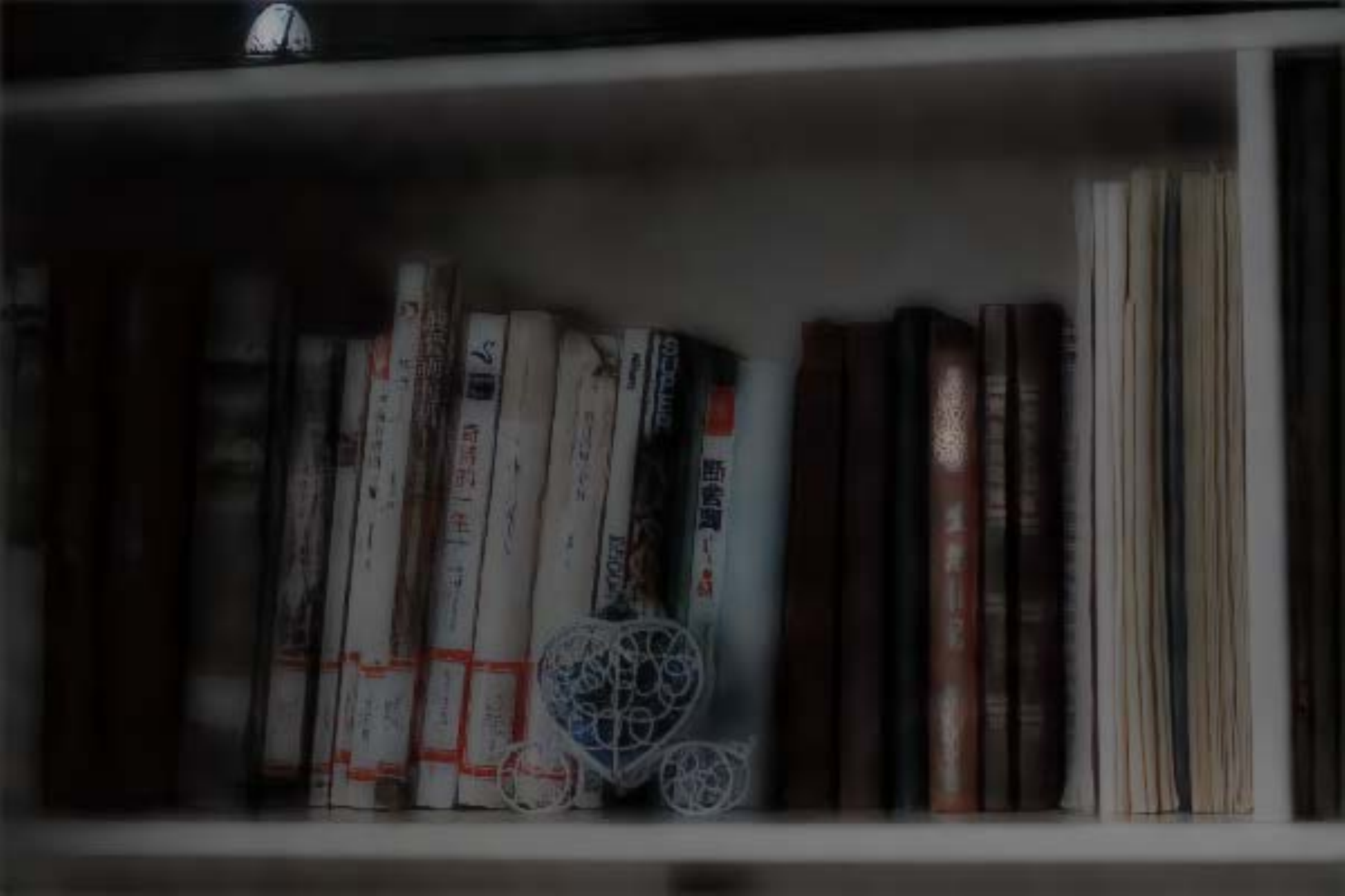}&
		\includegraphics[width=0.158\textwidth]{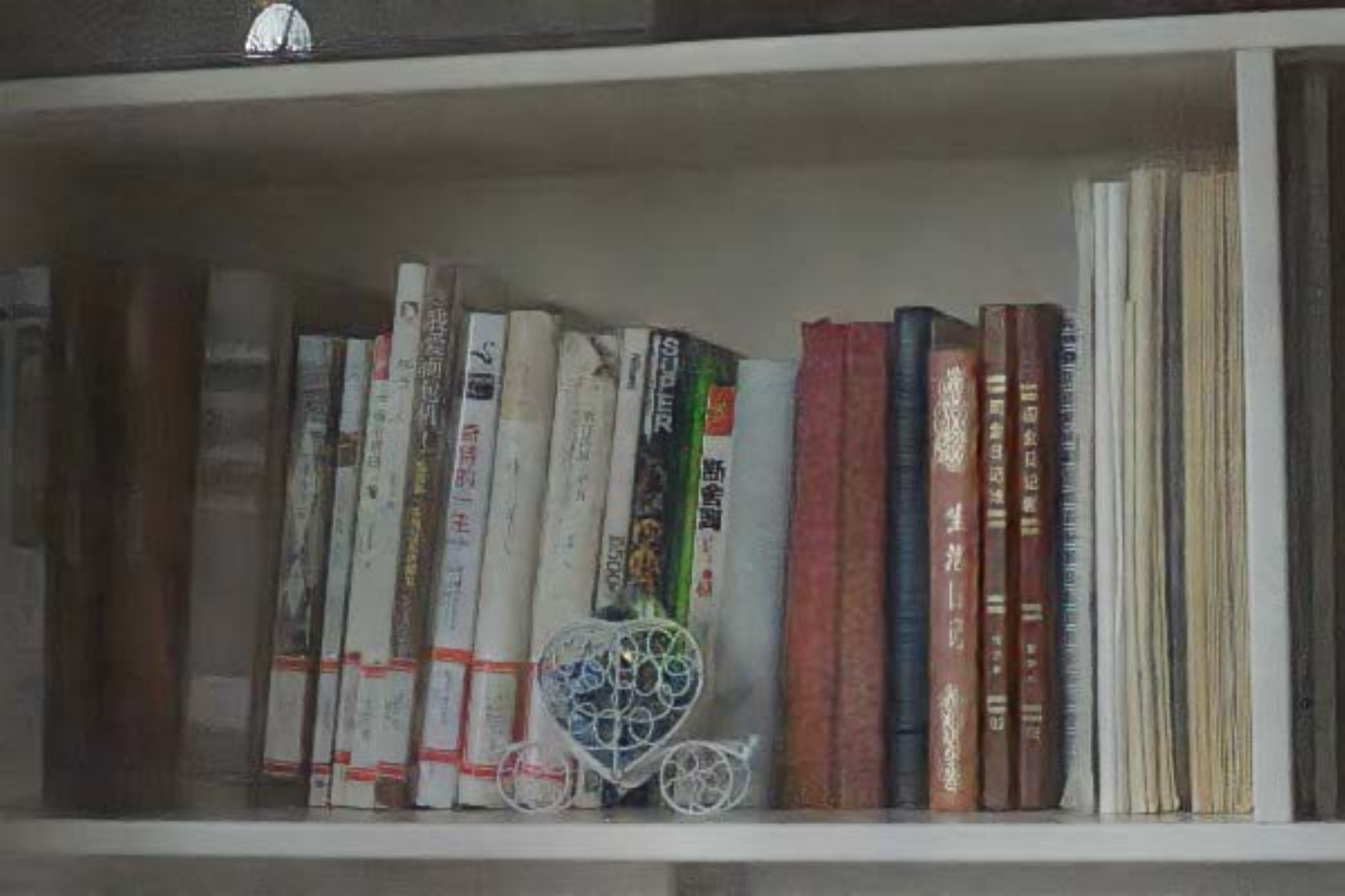}\\
		\footnotesize Input&\footnotesize LIME~\cite{guo2017lime}&\footnotesize LightenNet~\cite{li2018lightennet}&\footnotesize RetinexNet~\cite{Chen2018Retinex}&\footnotesize KinD~\cite{zhang2019kindling}&\footnotesize CSDNet\\
	\end{tabular}
	\caption{Visual comparison on two challenging images in the LOL dataset. \textbf{Top Row:} without considering the denoising operator. \textbf{Bottom Row:} all compared methods (except KinD, which considers the denoising procedure) are followed with the real noises removal network CBDNet~\cite{guo2019toward}.}
	\label{fig:LOLCBD}
\end{figure*}

\textbf{Global-Local Discriminator.}$\;$ 
We adopt the global-local discriminator presented in~\cite{jiang2019enlightengan} when training our CSDGAN. The network architecture of the discriminator is a standard PatchGAN~\cite{isola2017image}, using 3 convolutional layers with a stride of 2 to downsample the input to 1/8 of the original scale.
Additionally, about the basic generator, we make a bit change for CSDGAN compared with CSDNet, we add the residual connection between the input and output of the RENet. 

{
\textbf{Discussion.}$\;$ 
In the past few decades, there exist many GAN-based works for image processing tasks, e.g.,~\cite{ledig2017photo,zhang2020deblurring,shao2020domain}.
Actually, the adversarial learning mechanism of GAN is the core structure for them to improve the performance towards the specific task. That is, their cornerstone is the generator-discriminator framework and makes some task-oriented developments for them.  
Different from these existing works, we concentrate on constructing an effective architecture (can be viewed as the generator in GAN) towards low-light image enhancement. The introduction of discriminator just can be viewed as a training strategy for vanishing limitations of paired data to evaluate our performance under unpaired supervision. In this way, we are able to fully evaluate the effectiveness of our architecture from paired and unpaired supervision views. 
}

\section{Experimental Results}\label{sec:experiment}
In this section, we first introduced implementation details including compared methods, benchmarks description and evaluated metrics. Then we conducted plenty of experiments to evaluate our proposed CSDNet and CSDGAN, respectively. Next, we executed extensive analyses about our method. Finally, we provided two lightweight versions of CSDNet to ensure practical values.
All the following experiments are conducted on a PC with Intel Core i7-8700 CPU at 3.70GHz, 32 GB RAM and an NVIDIA GeForce GTX 1070 8GB GPU.

\begin{table*}[t]
	\centering
	\caption{Quantitative comparison on NASA, LIME, NPE, MEF datasets in terms of NIQE among existing state-of-the-art unpaired methods. For NIQE, the lower score is better.}
	\begin{tabular}{cccccccccc}
		\toprule
		Benchmark&HE~\cite{cheng2004simple}&MSRCR~\cite{jobson1997multiscale}&SRIE~\cite{fu2015probabilistic}&WVM~\cite{fu2016weighted}&LIME~\cite{guo2017lime}&JIEP~\cite{cai2017joint}&RRM~\cite{li2018structure}&ElightenGAN~\cite{jiang2019enlightengan}&CSDGAN\\
		\midrule 
		NASA&3.6778&3.6643&3.7990&3.9745&3.8638&3.7175& 4.7524&3.6154&\textbf{3.5917}\\
		LIME&4.2805&3.7644&3.9634&3.8003&3.9948&{{3.7158}}&4.6424&3.6604&\textbf{{3.6058}}\\
		NPE&3.1005&3.0384&3.0223&\textbf{{2.9445}}&3.1022&2.9697&3.9203&3.0218&{2.9578}\\
		MEF&3.7441&3.3092&3.4489&3.4687&3.7661&3.4265&5.0612&3.2206&\textbf{3.1032}\\
		\midrule
		Ave. &3.7007&3.4441&3.5584&3.5470&3.6812&3.4574&4.5941&3.3796&\textbf{3.3146}\\
		\bottomrule
	\end{tabular}
	\label{tab:rescomp}
\end{table*}

\begin{figure*}[t]
	\centering
	\begin{tabular}{c@{\extracolsep{0.25em}}c@{\extracolsep{0.25em}}c@{\extracolsep{0.25em}}c@{\extracolsep{0.25em}}c@{\extracolsep{0.25em}}c@{\extracolsep{0.25em}}c} 
		\includegraphics[width=0.136\textwidth]{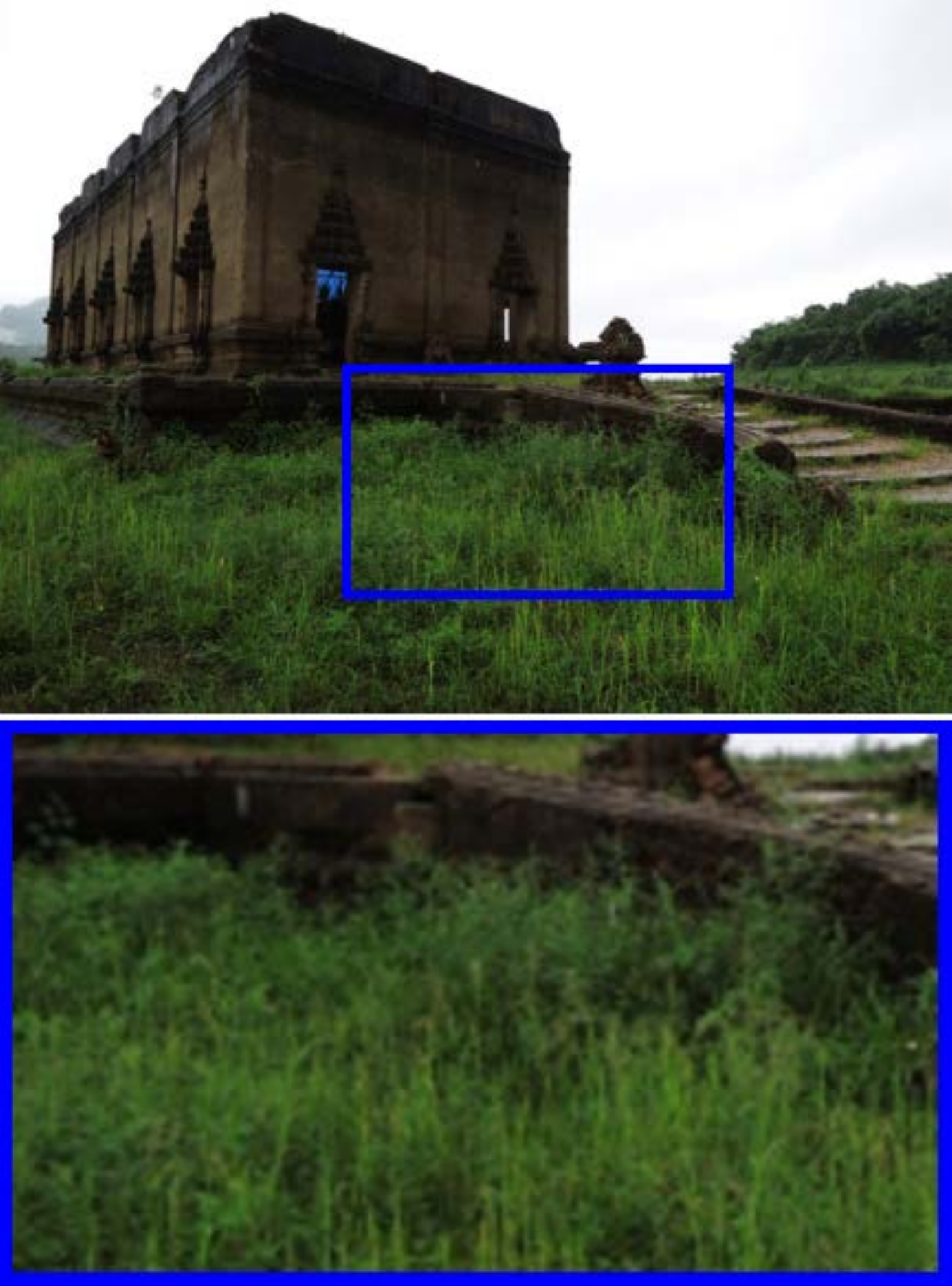}&
		\includegraphics[width=0.136\textwidth]{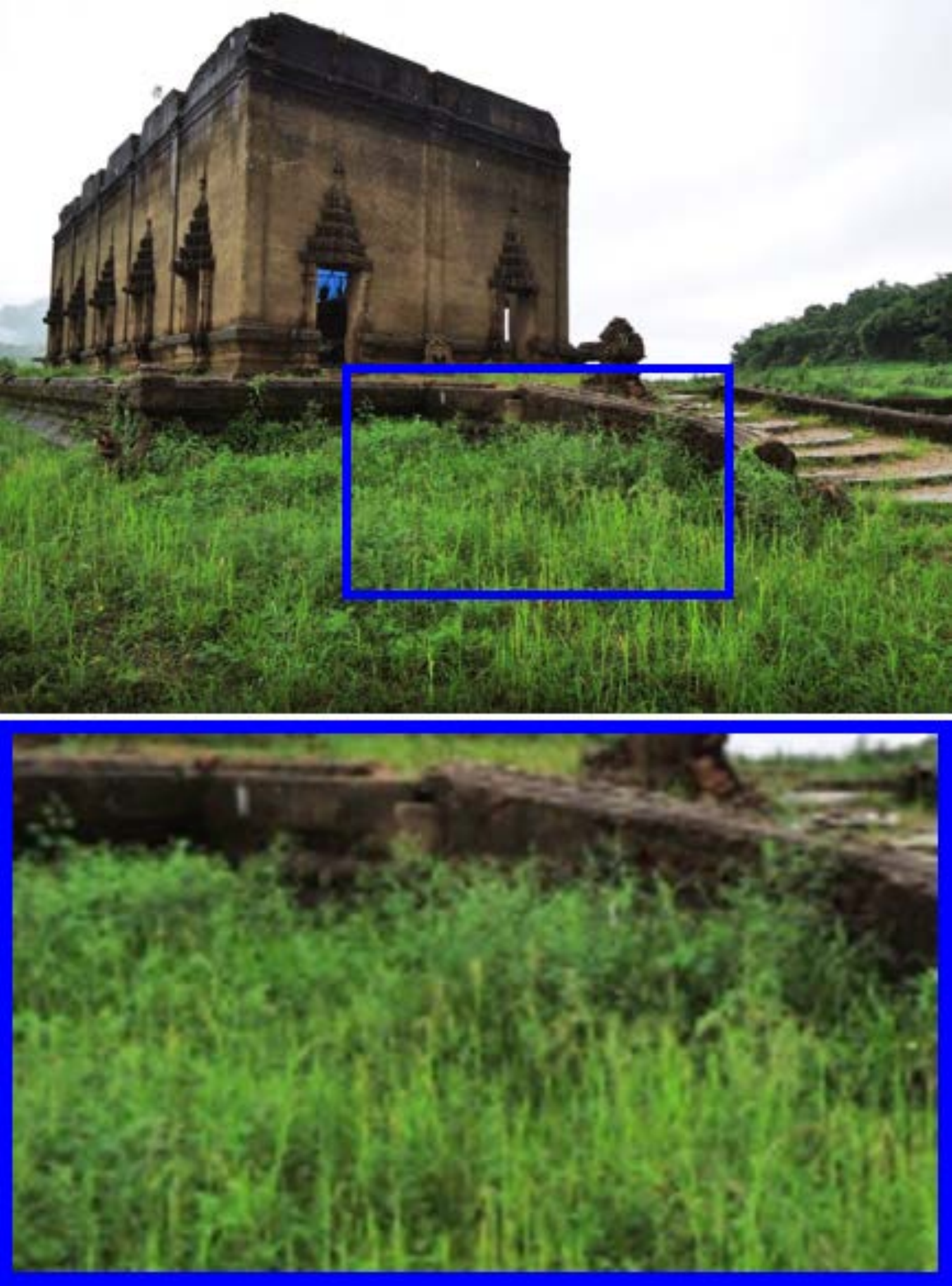}&
		\includegraphics[width=0.136\textwidth]{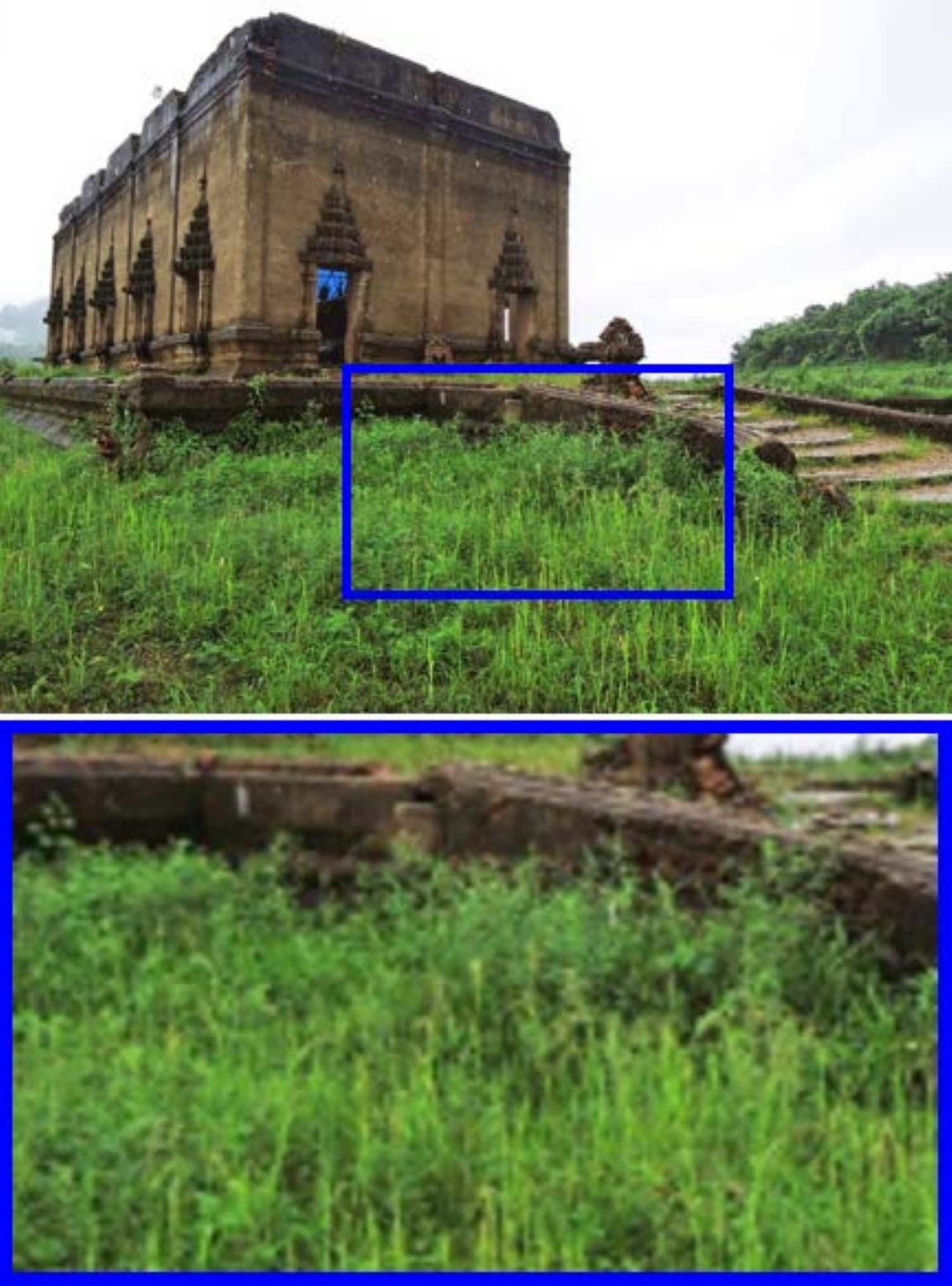}&
		\includegraphics[width=0.136\textwidth]{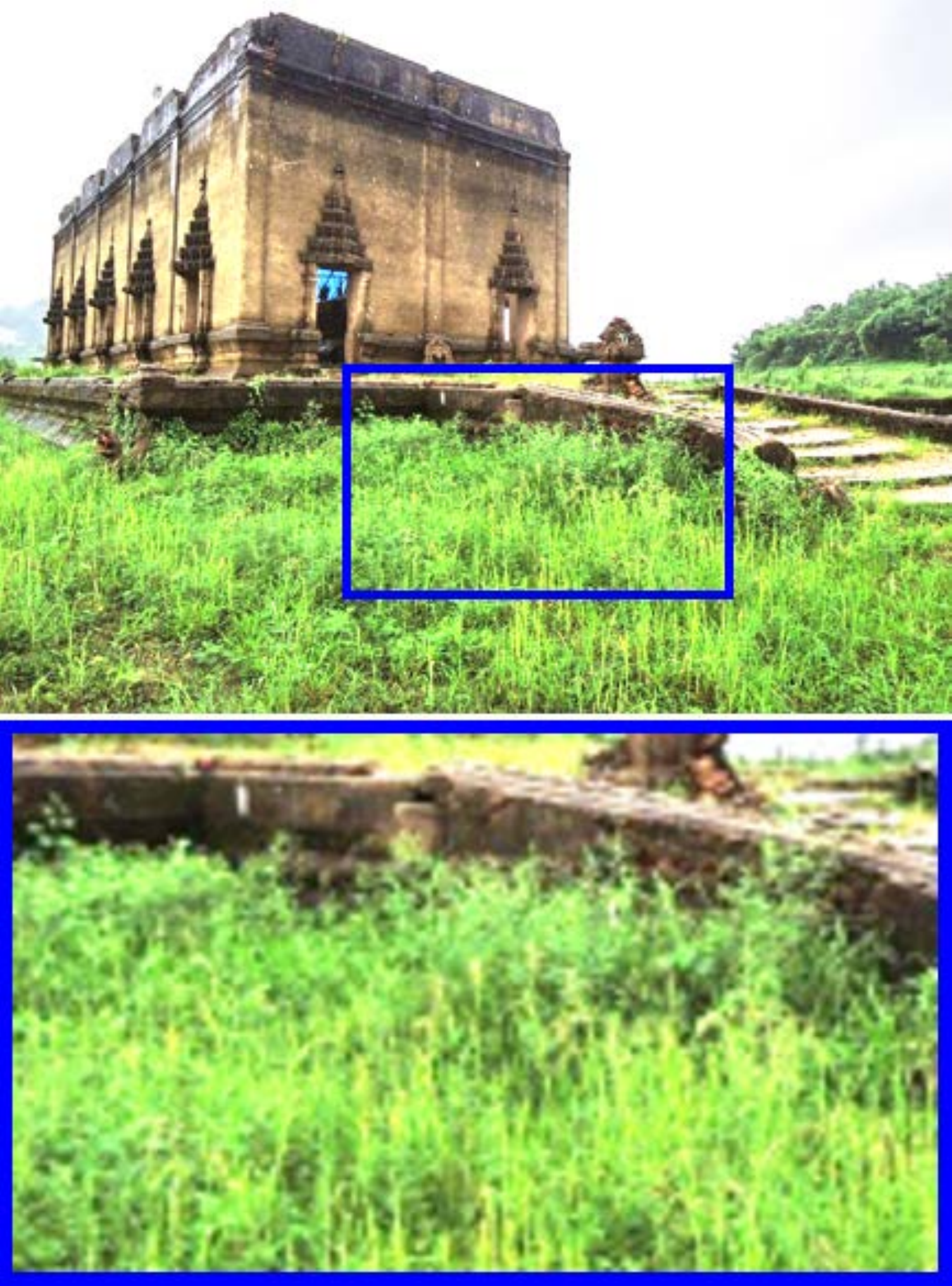}&
		\includegraphics[width=0.136\textwidth]{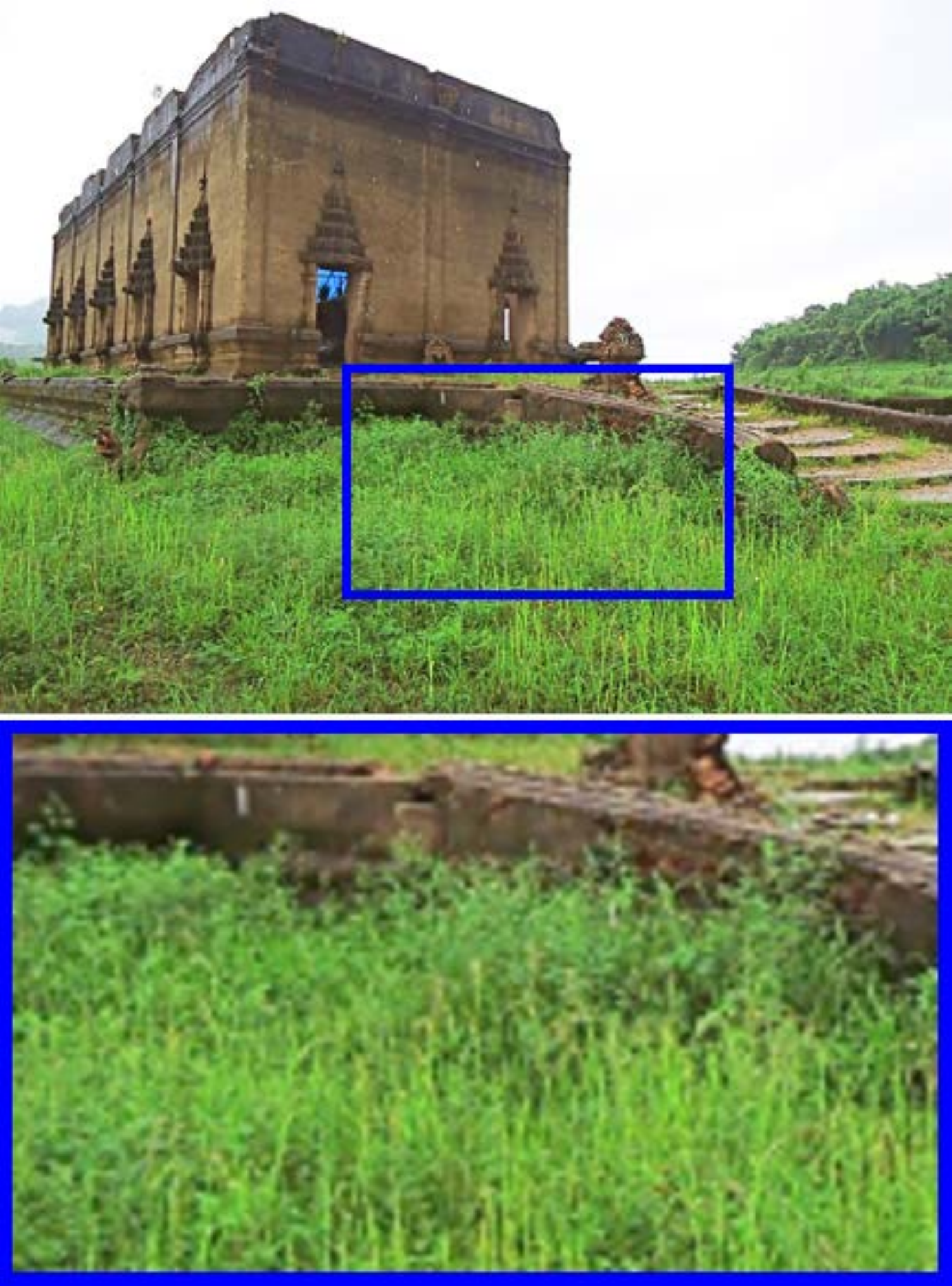}&
		\includegraphics[width=0.136\textwidth]{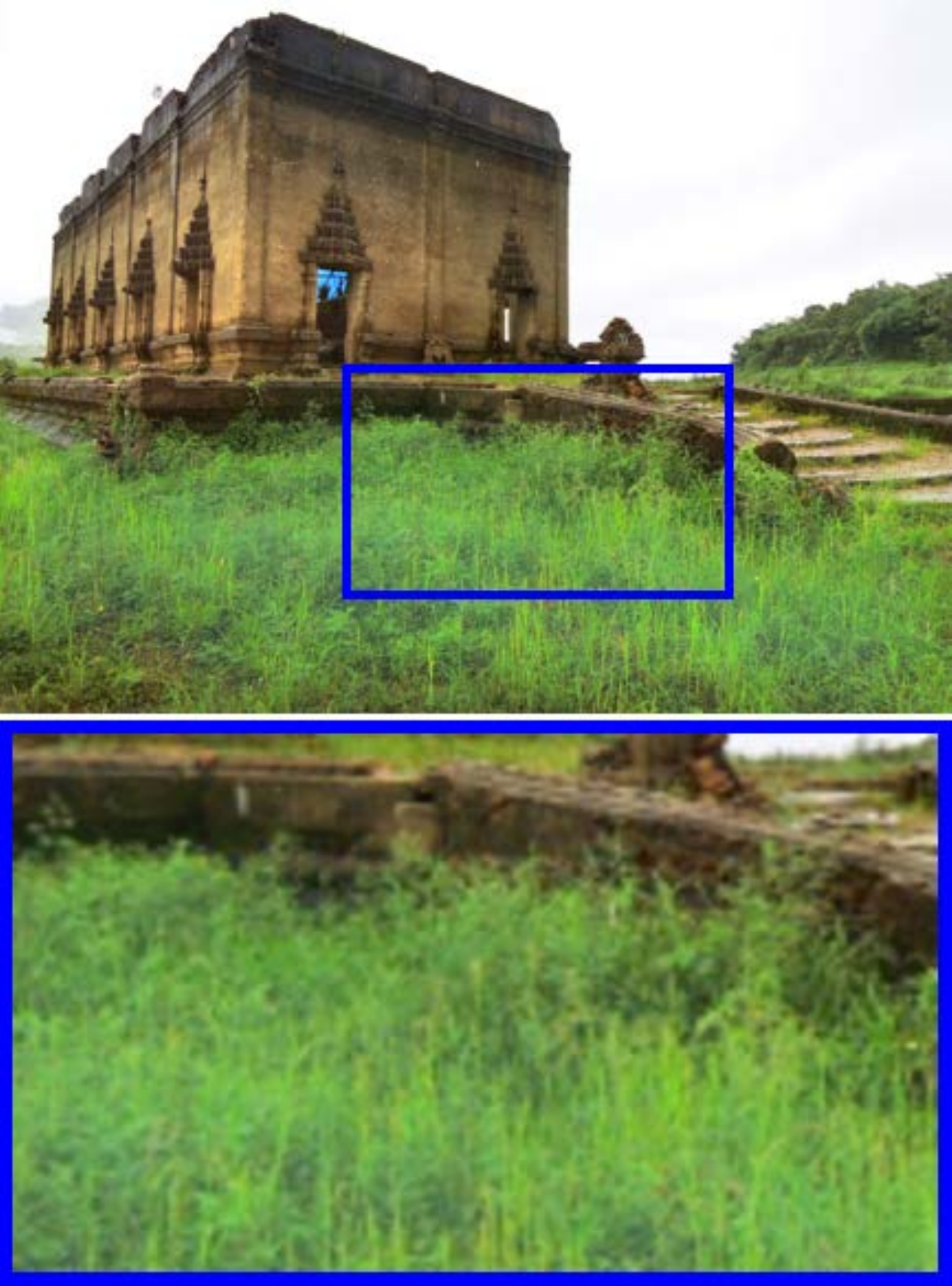}&
		\includegraphics[width=0.136\textwidth]{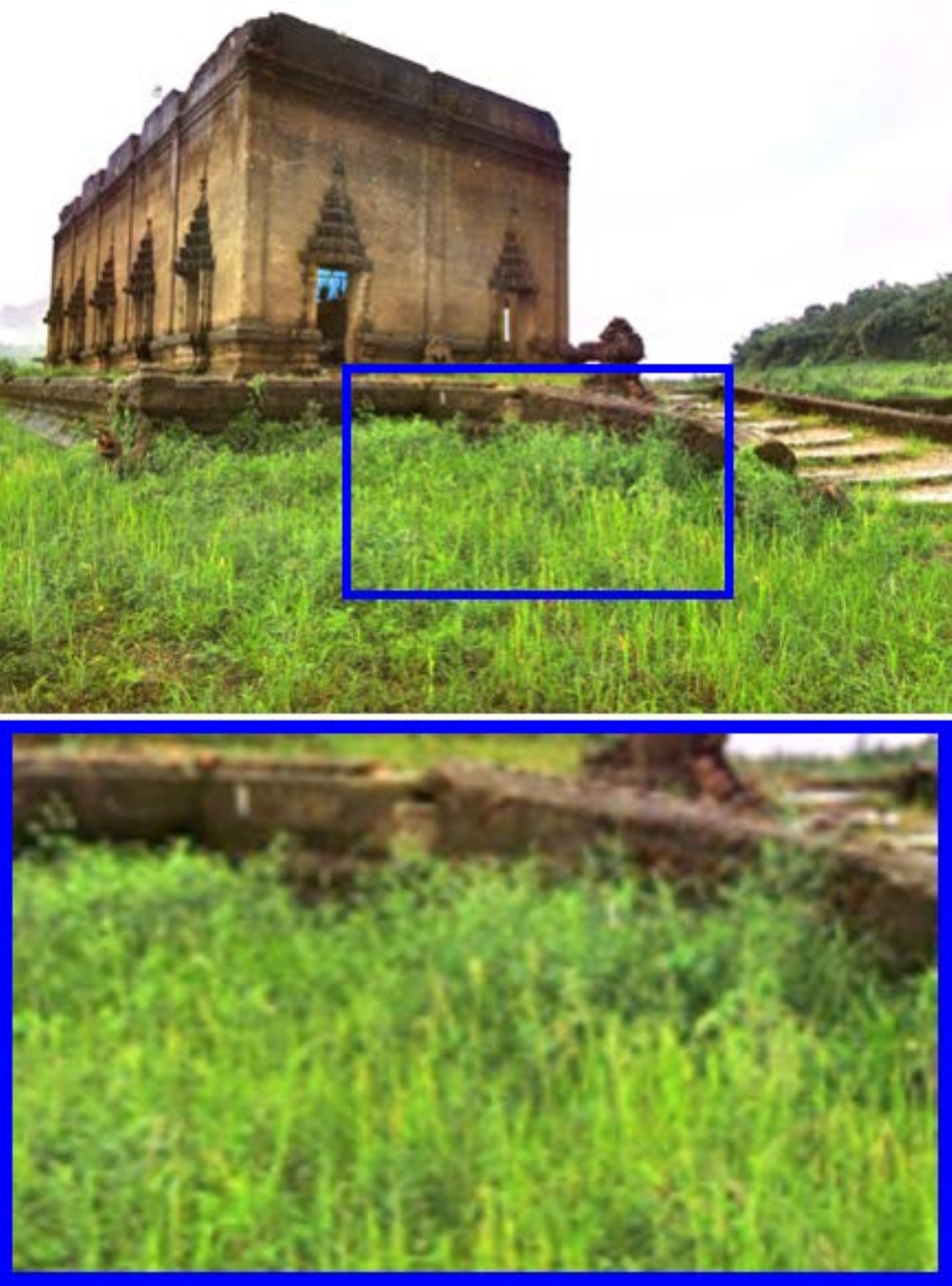}\\
		\includegraphics[width=0.136\textwidth]{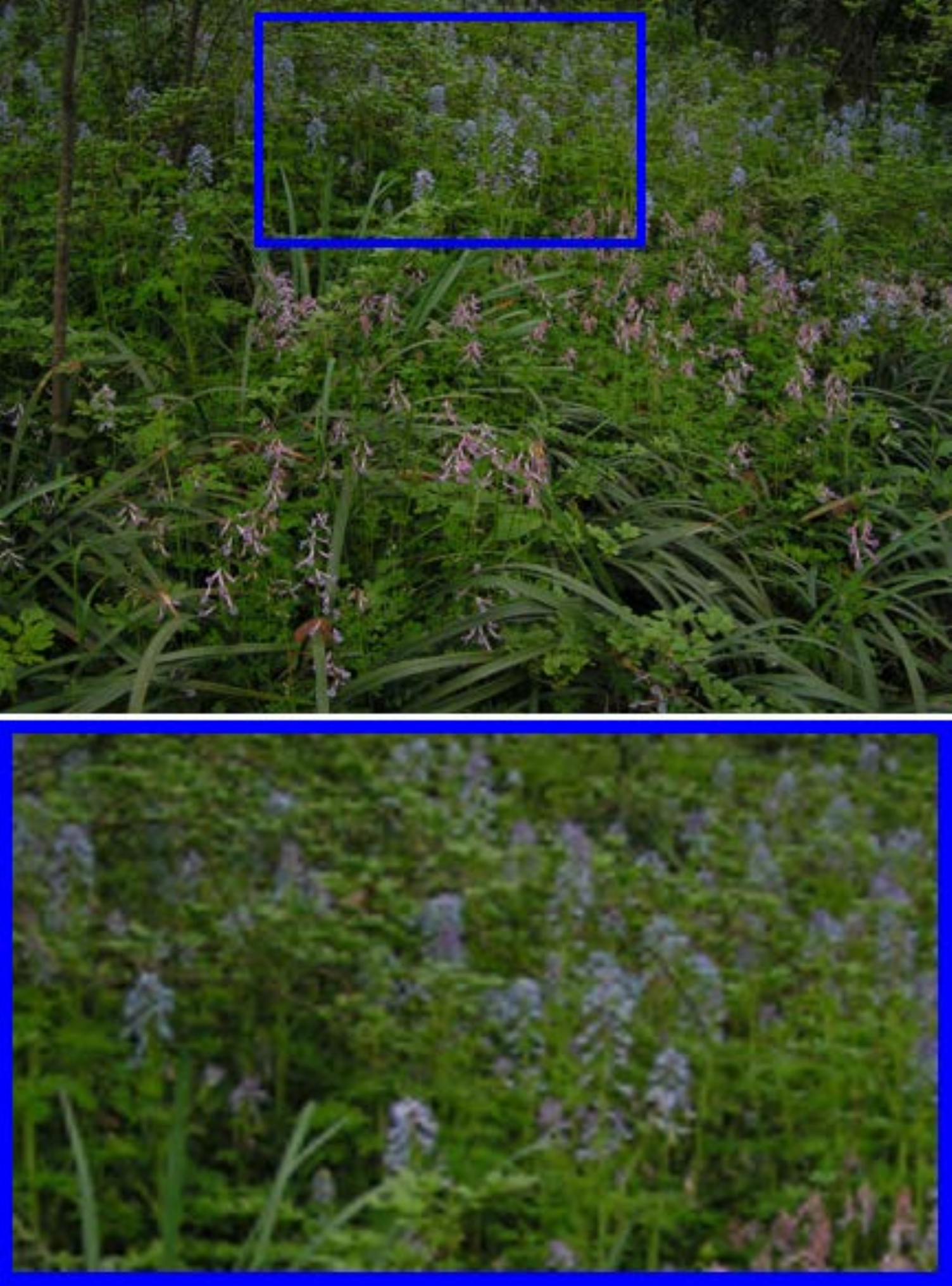}&
		\includegraphics[width=0.136\textwidth]{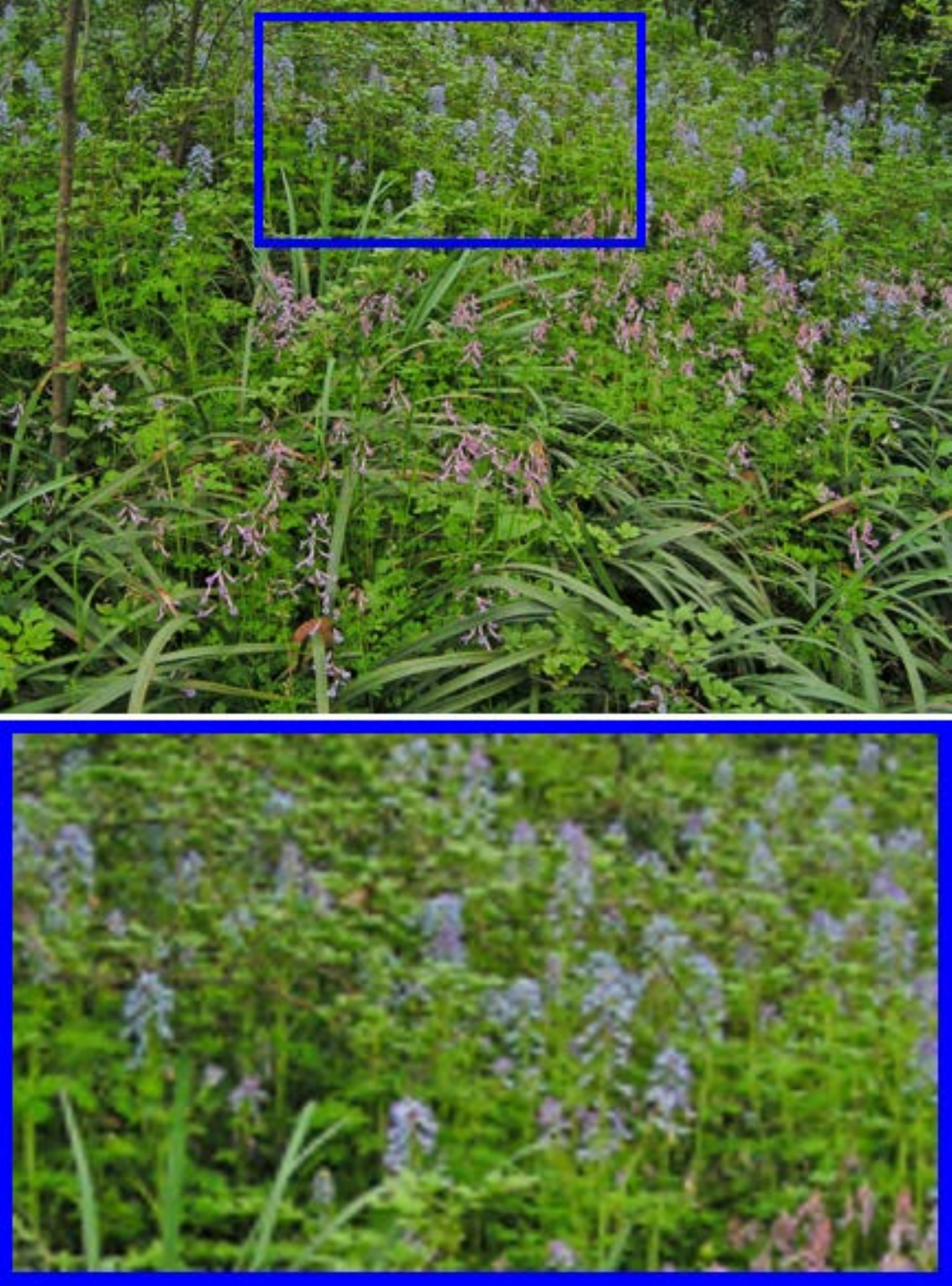}&
		\includegraphics[width=0.136\textwidth]{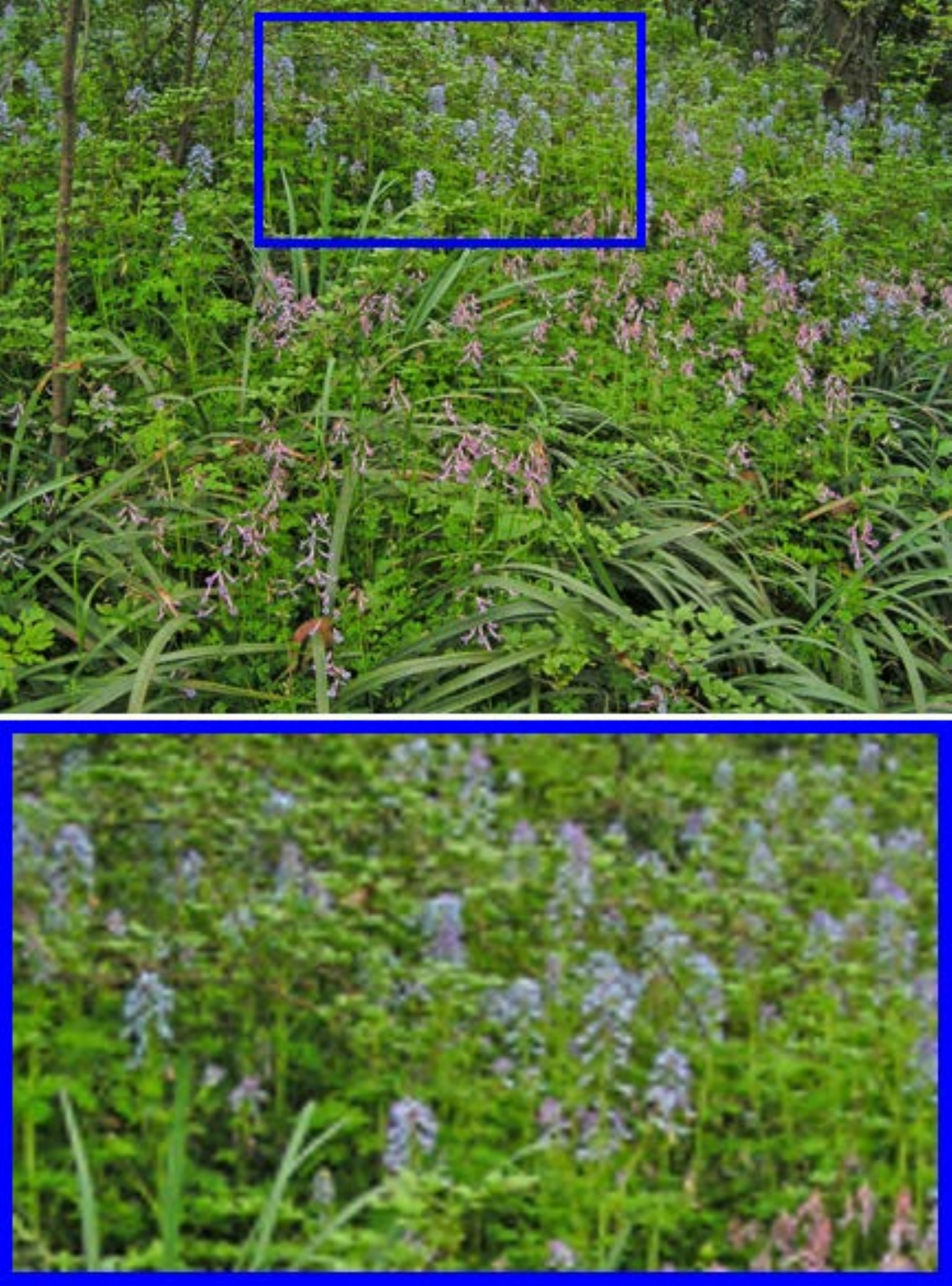}&
		\includegraphics[width=0.136\textwidth]{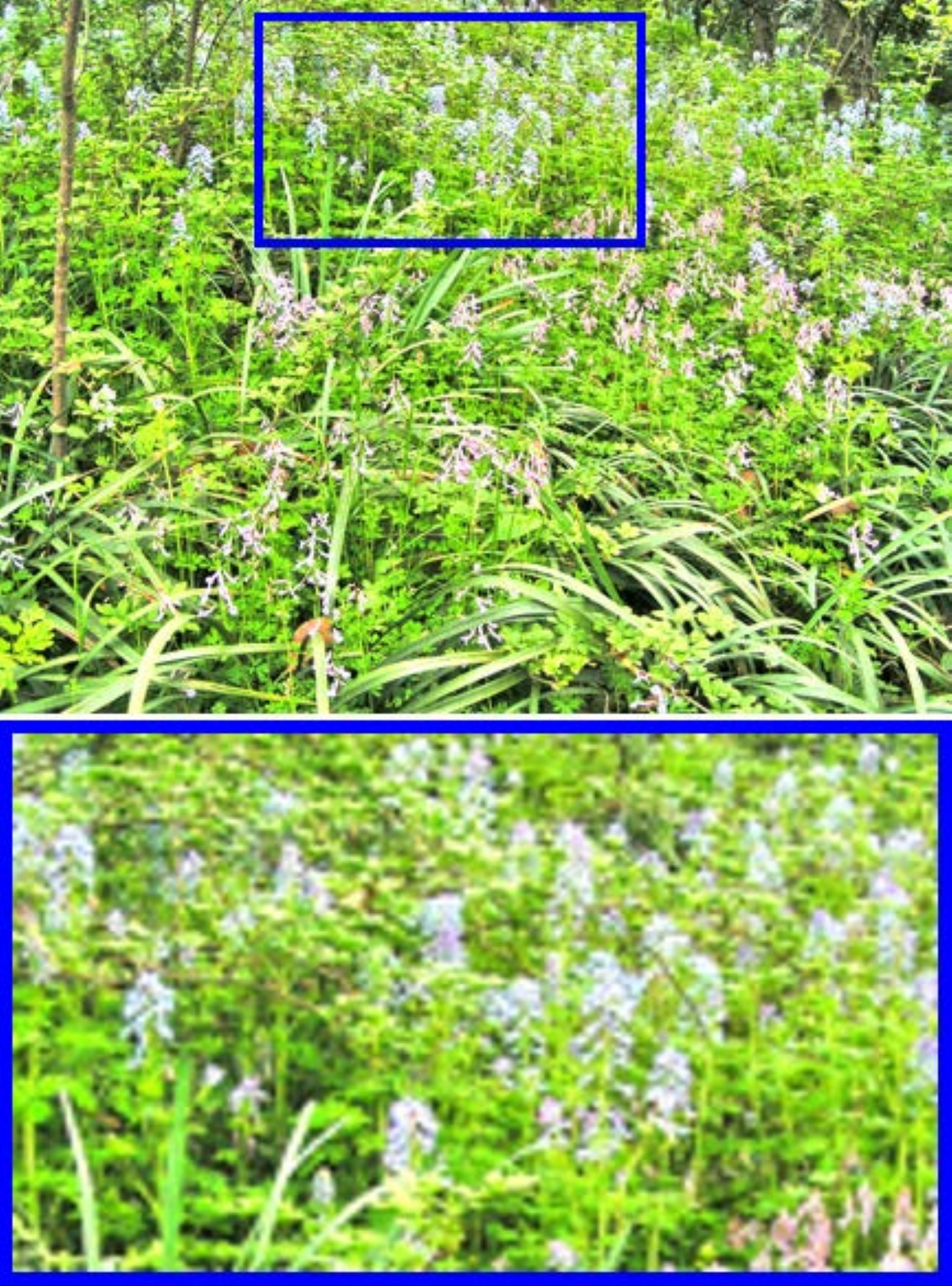}&
		\includegraphics[width=0.136\textwidth]{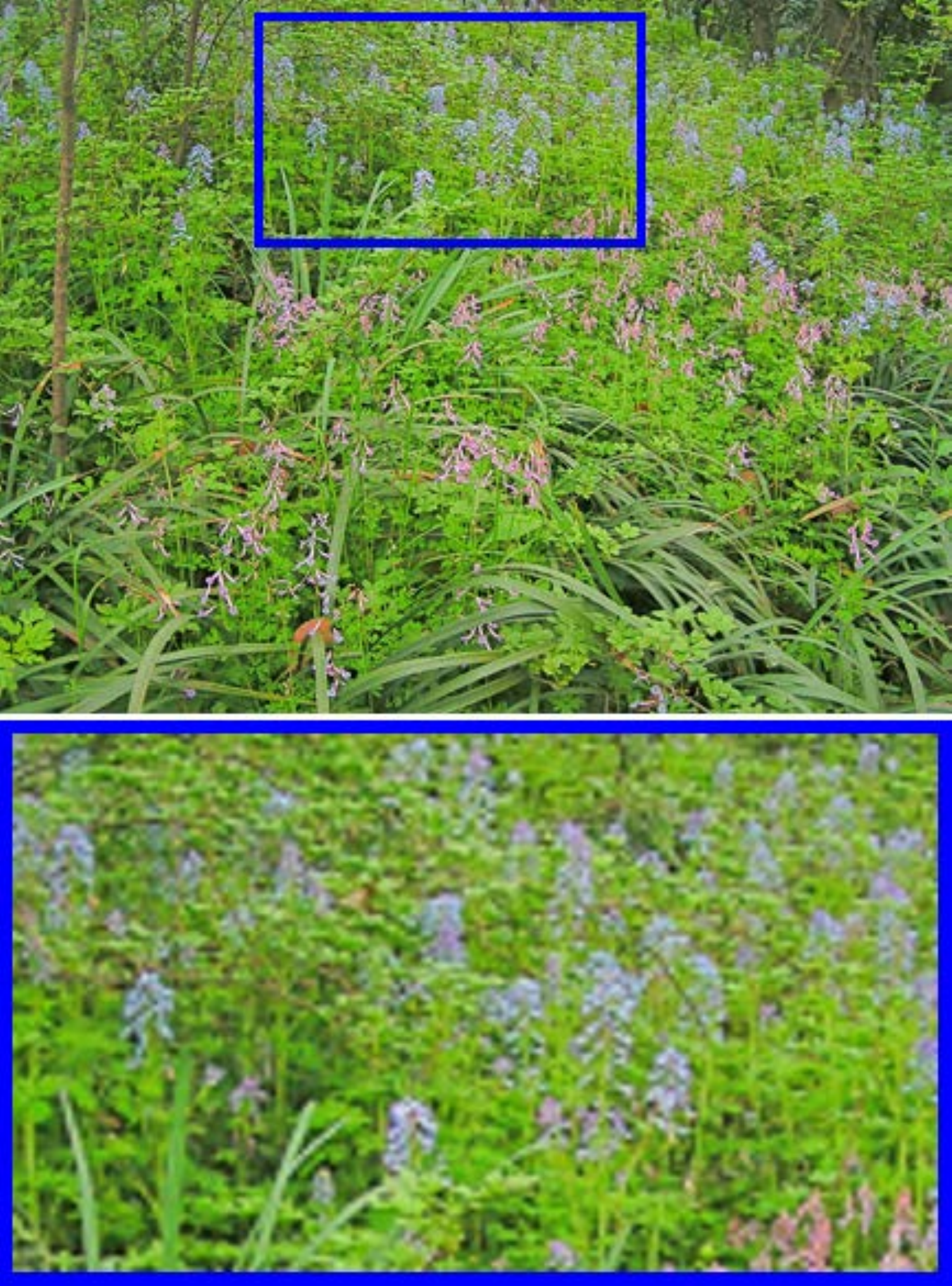}&
		\includegraphics[width=0.136\textwidth]{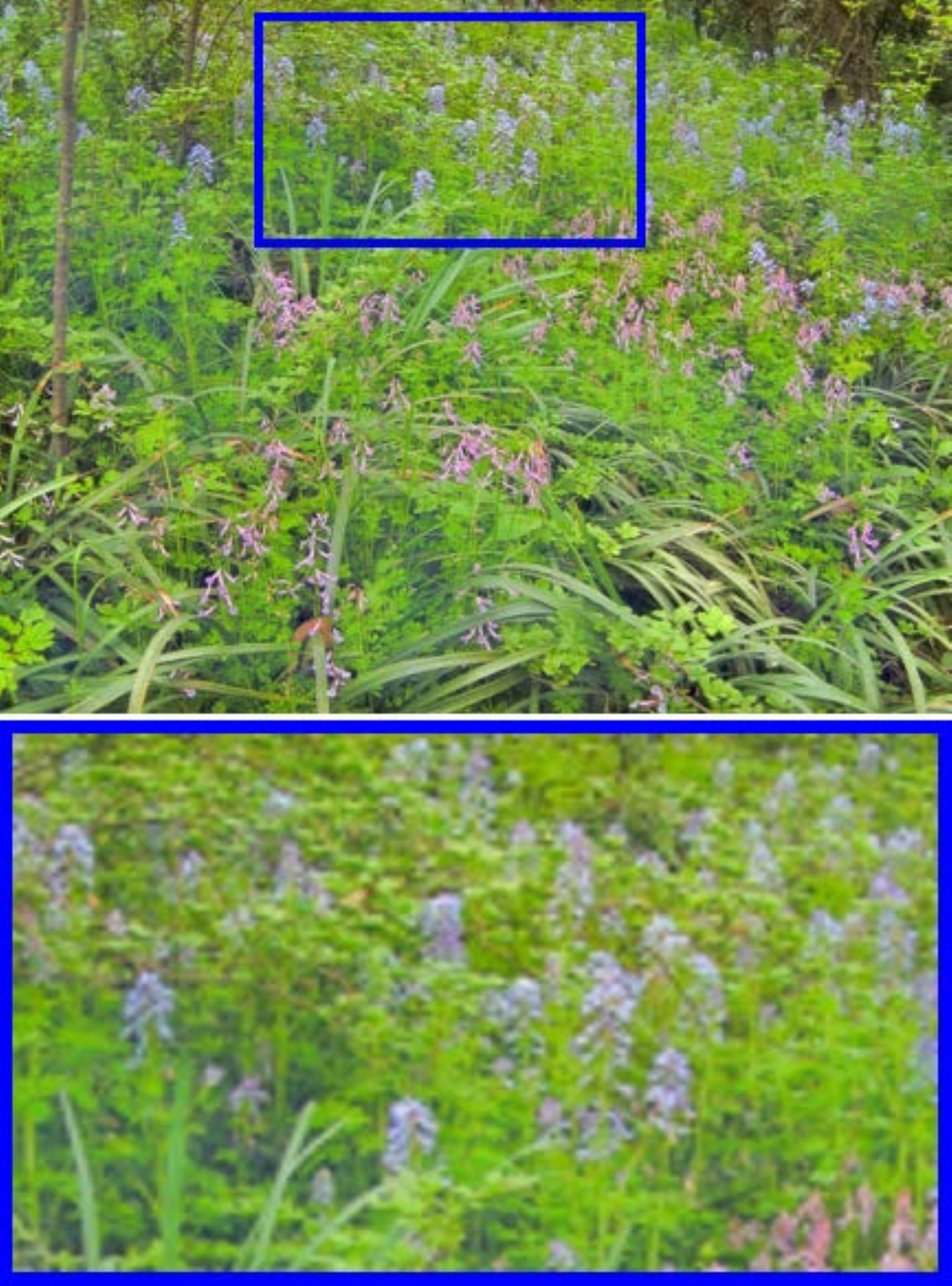}&
		\includegraphics[width=0.136\textwidth]{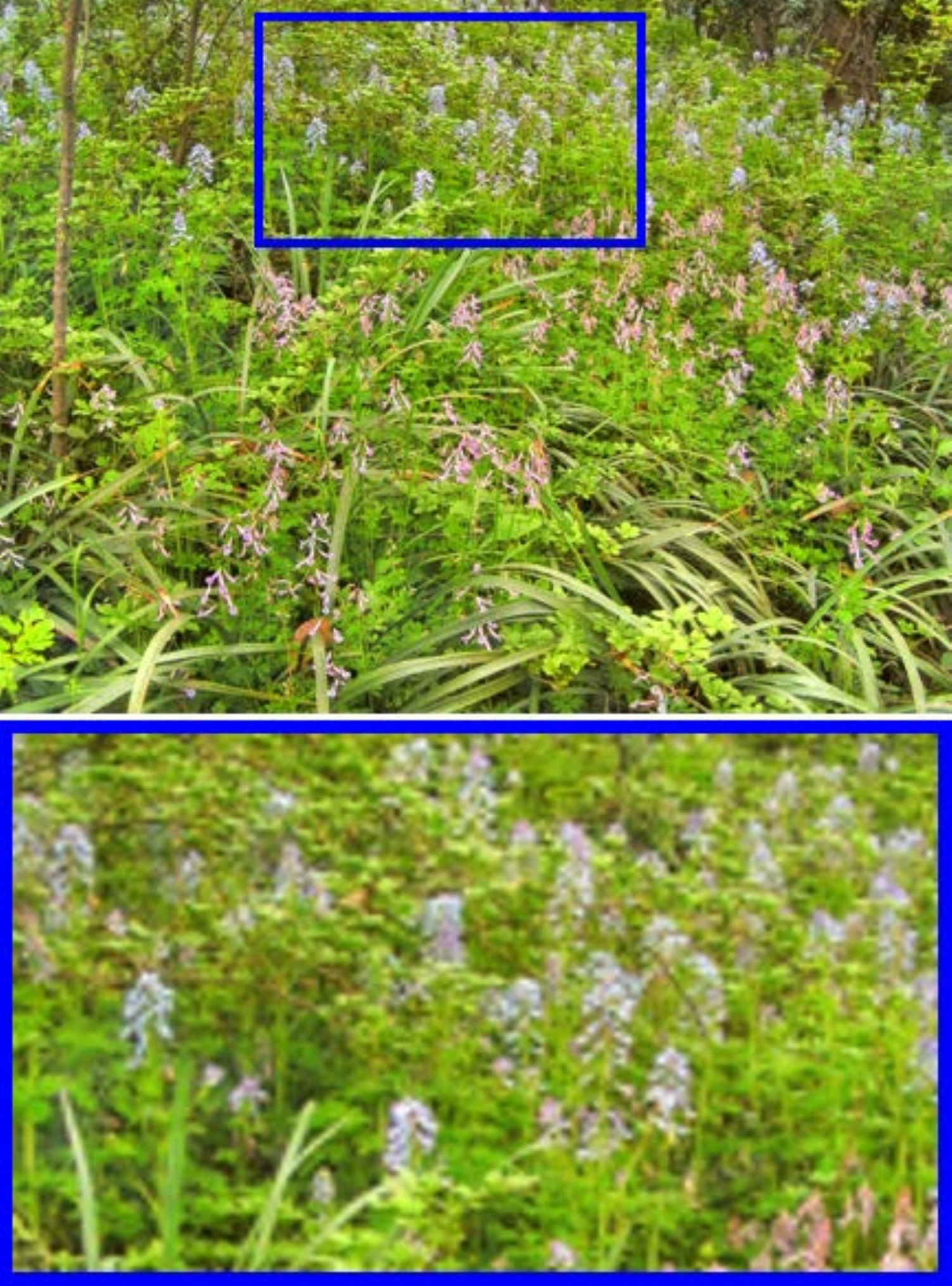}\\
		\footnotesize Input&\footnotesize WVM~\cite{fu2016weighted}&\footnotesize JIEP~\cite{cai2017joint}&\footnotesize LIME~\cite{guo2019toward}&\footnotesize RRM~\cite{li2018structure}&\footnotesize EnlightenGAN~\cite{jiang2019enlightengan}&\footnotesize CSDGAN\\
	\end{tabular}
	\caption{Visual comparison on two example images in NPE dataset.}
	\label{fig:VC1}
\end{figure*}

\subsection{Implementation Details}
\textbf{Compared Methods.}$\;$We compared our proposed method with many state-of-the-art techniques which are four representative model-based methods (i.e., WVM~\cite{fu2016weighted}, LIME~\cite{guo2017lime}, JIEP~\cite{cai2017joint}, RRM~\cite{li2018structure}), three paired end-to-end deep learning methods (i.e, WhiteBox~\cite{hu2018exposure}, LightenNet~\cite{li2018lightennet}, RetinexNet~\cite{Chen2018Retinex}, DeepUPE~\cite{wang2019underexposed}, KinD~\cite{jiang2019enlightengan}), and an unpaired end-to-end deep learning method (i.e., EnlightenGAN~\cite{jiang2019enlightengan}). For the sake of fairness, most of methods were compared according to the same training mechanisms (paired or unpaired).

\textbf{Benchmarks Description and Evaluated Metrics.}$\;$We made a series of comprehensive evaluations in terms of different training strategies on multiple commonly-used benchmarks, as is shown in Table~\ref{tab:datasets}. We achieved the paired supervision training in MIT-Adobe FiveK~\cite{fivek} and LOL~\cite{Chen2018Retinex} datasets, respectively. MIT-Adobe FiveK dataset~\cite{fivek} includes 5000 raw images, and each image includes five retouched reference images generated by different experts, where images retouched by expert-C are defined the labels. Notice that considering the labels of MIT-Adobe FiveK are generated by expert-retouched, we only presented the visual comparison in the MIT-Adobe FiveK dataset. Different from MIT-Adobe FiveK dataset, LOL dataset generates the normal exposure images by changing the exposure time in the real-world. The hardest part is that noises exist in low-light images. Considering the truthfulness of the LOL dataset, we not only tested the visual effects, but also evaluated numerical performance in terms of PSNR and SSIM.

\begin{figure}[t]
	\centering
	\begin{tabular}{c@{\extracolsep{0.2em}}c@{\extracolsep{0.2em}}c}
		\footnotesize Input&\footnotesize EnlightenGAN~\cite{jiang2019enlightengan}&\footnotesize CSDGAN\\
		\includegraphics[width=0.15\textwidth]{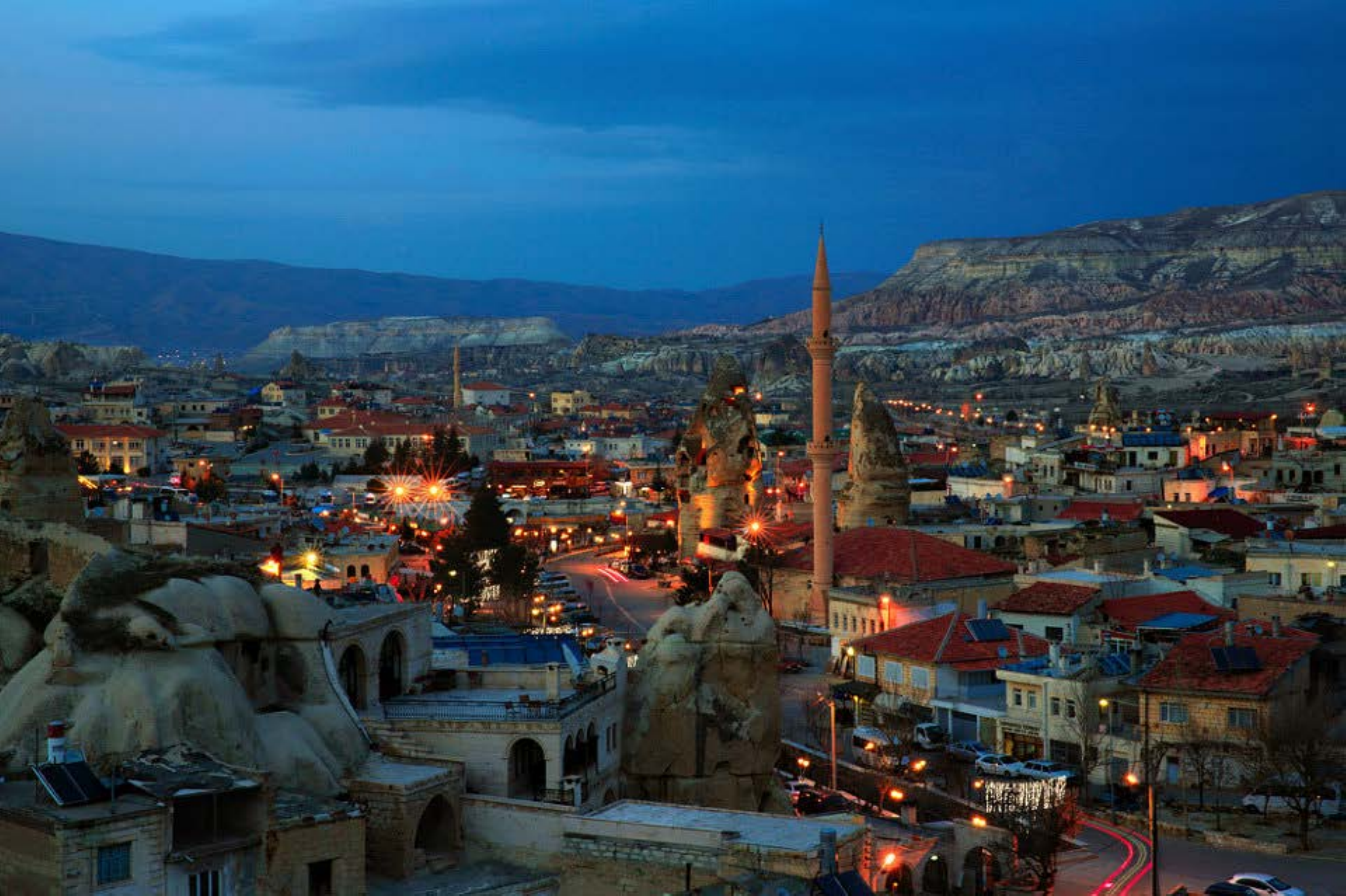}&
		\includegraphics[width=0.15\textwidth]{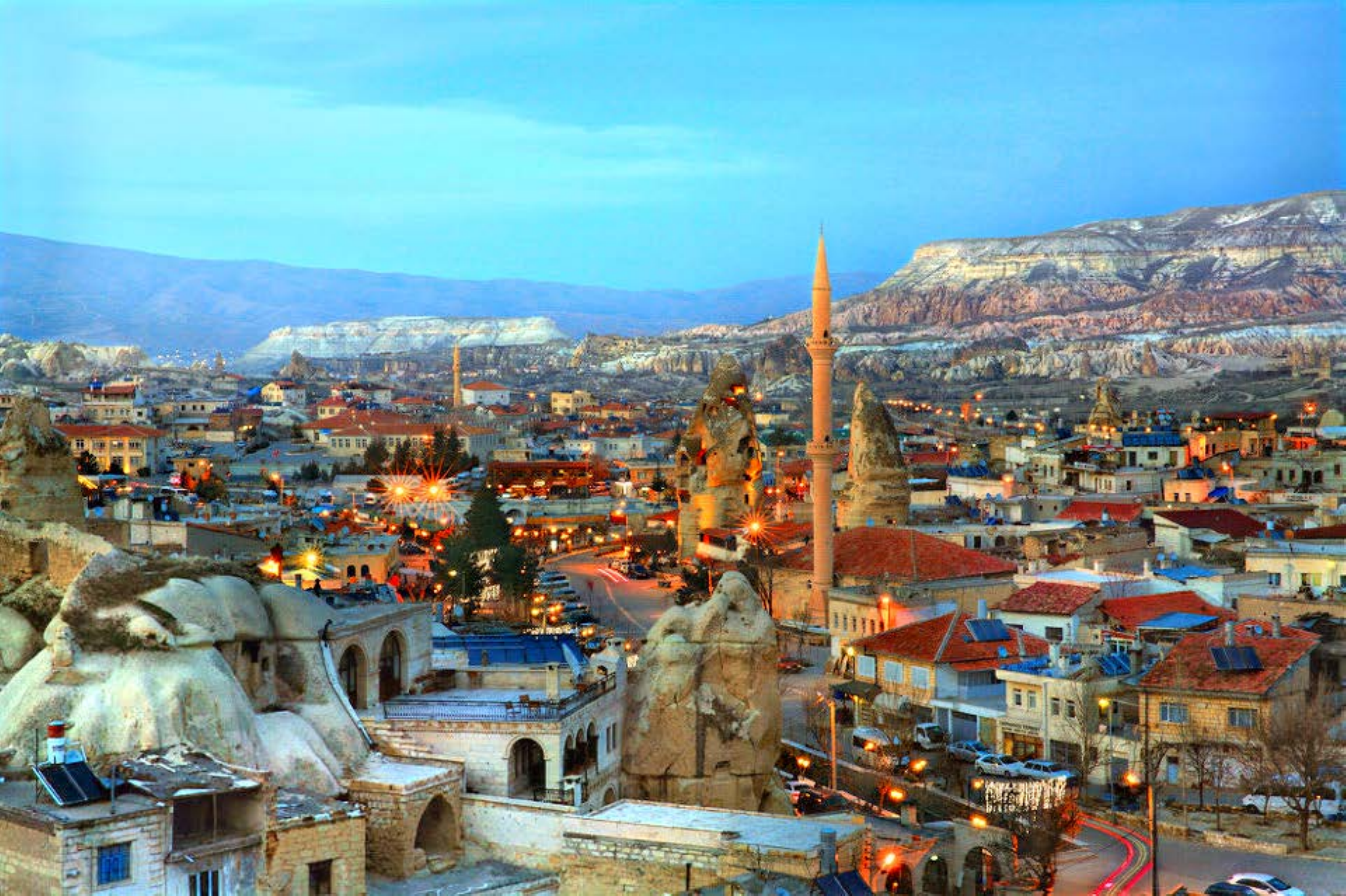}&
		\includegraphics[width=0.15\textwidth]{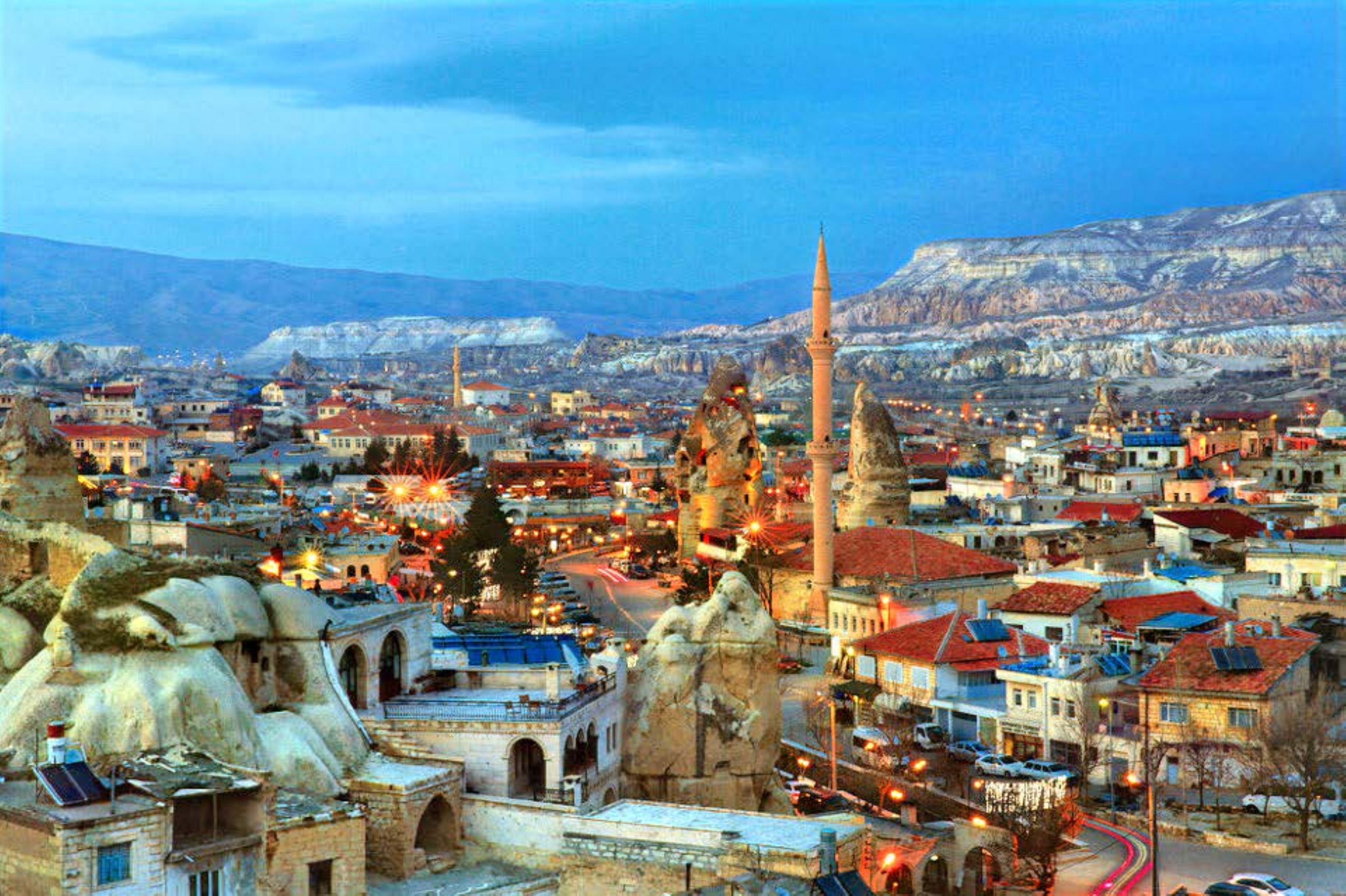}\\	
		\includegraphics[width=0.15\textwidth]{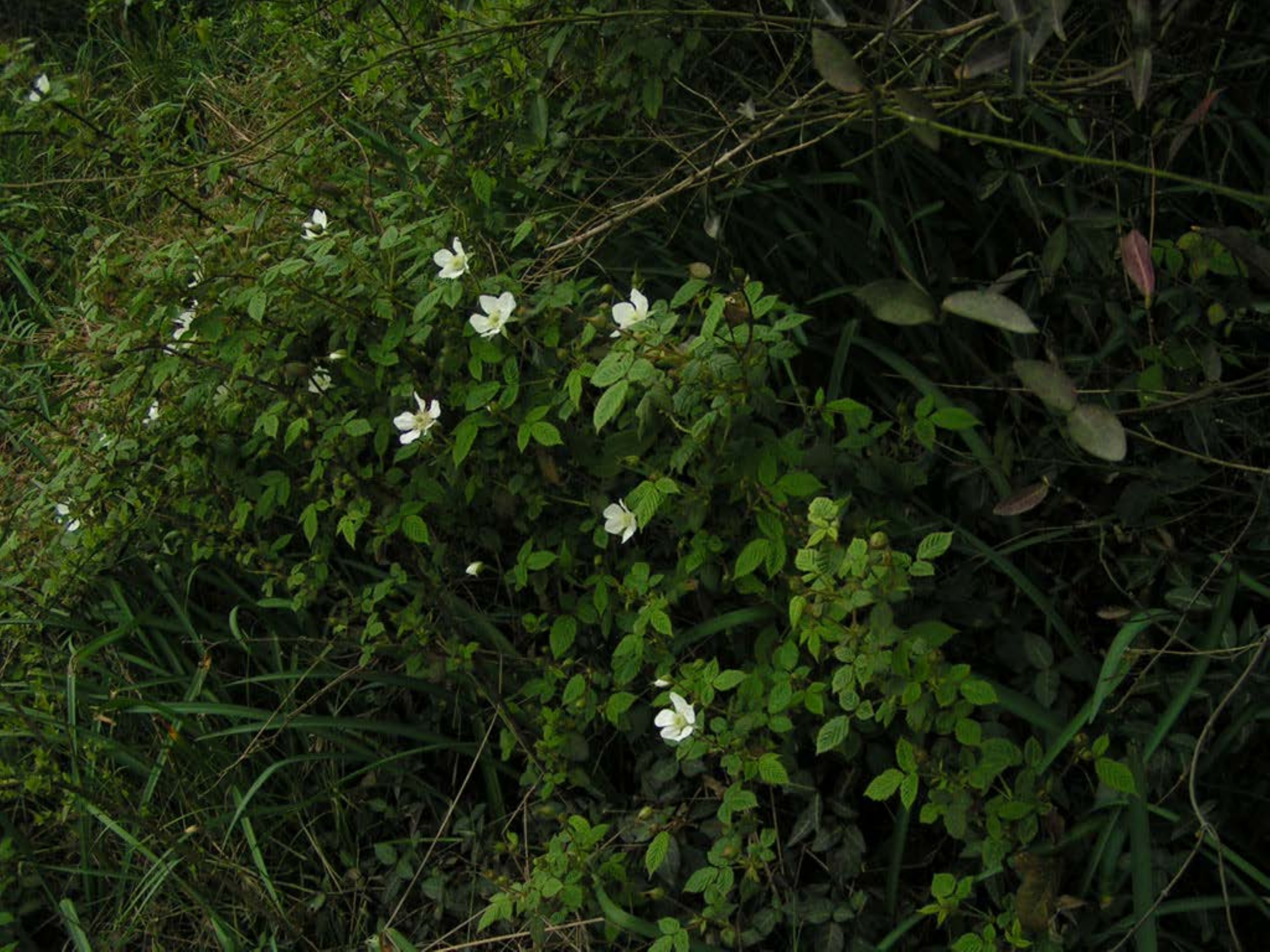}&
		\includegraphics[width=0.15\textwidth]{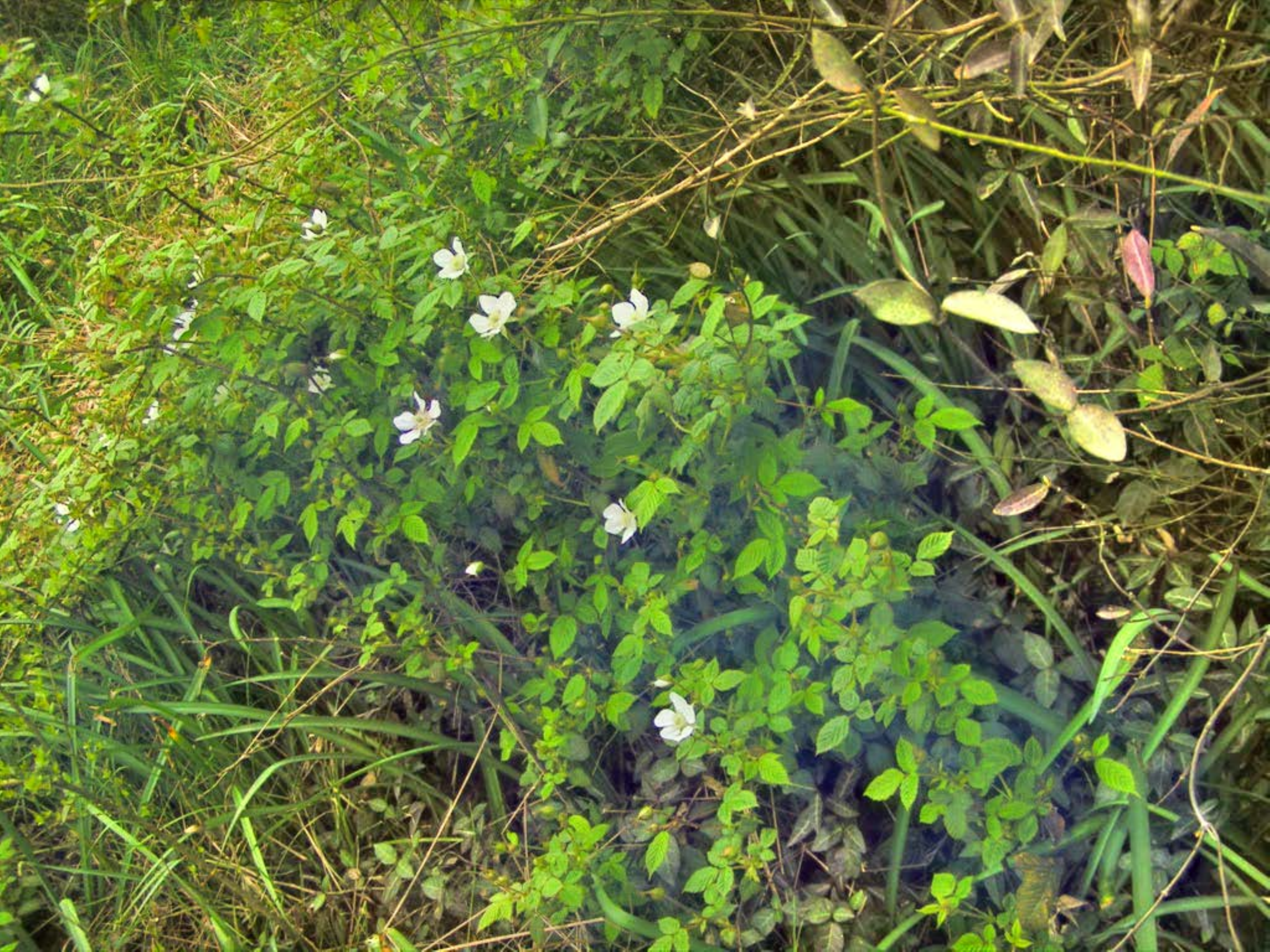}&
		\includegraphics[width=0.15\textwidth]{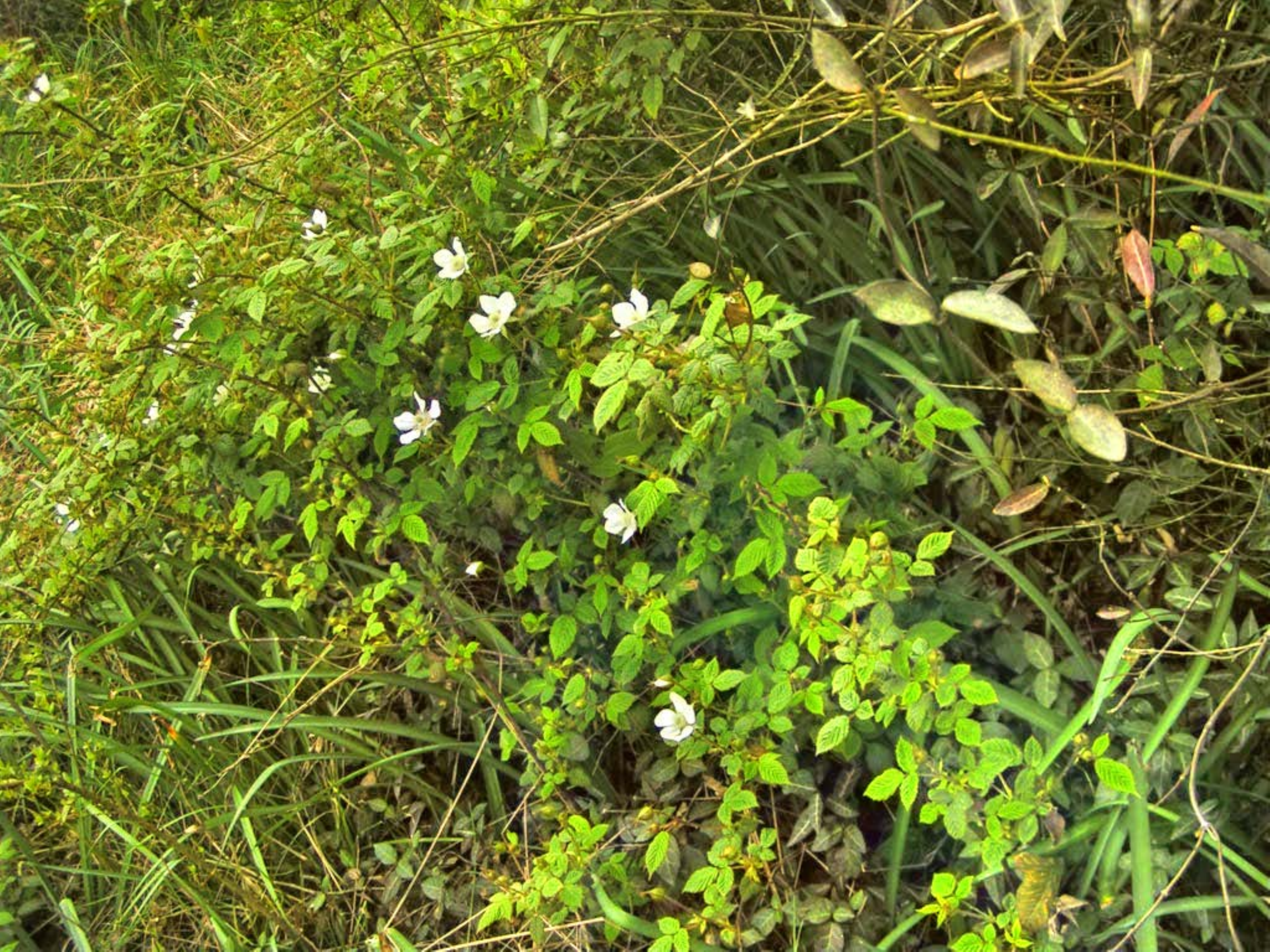}\\	
	\end{tabular}
	\caption{Visual comparison among two networks trained using the same training strategy in the same unpaired dataset.}
	\label{fig:CompGAN}
\end{figure}

\begin{figure}[t]
	\centering
	\begin{tabular}{c@{\extracolsep{0.2em}}c@{\extracolsep{0.2em}}c@{\extracolsep{0.2em}}c}	
		\footnotesize Input&\footnotesize RetinexNet~\cite{Chen2018Retinex}&\footnotesize DeepUPE~\cite{wang2019underexposed}&\footnotesize CSDGAN\\
		\includegraphics[width=0.113\textwidth]{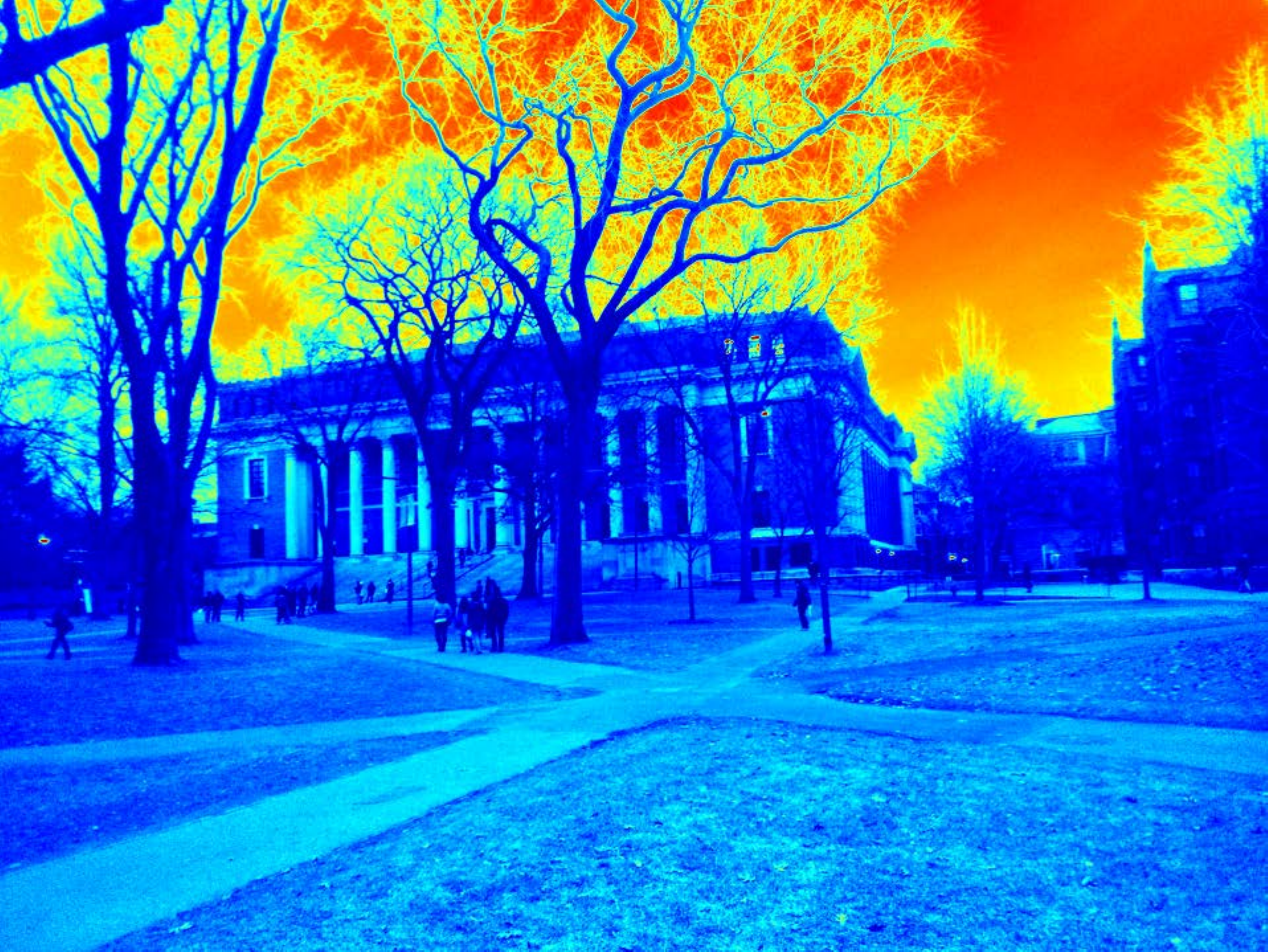}&
		\includegraphics[width=0.113\textwidth]{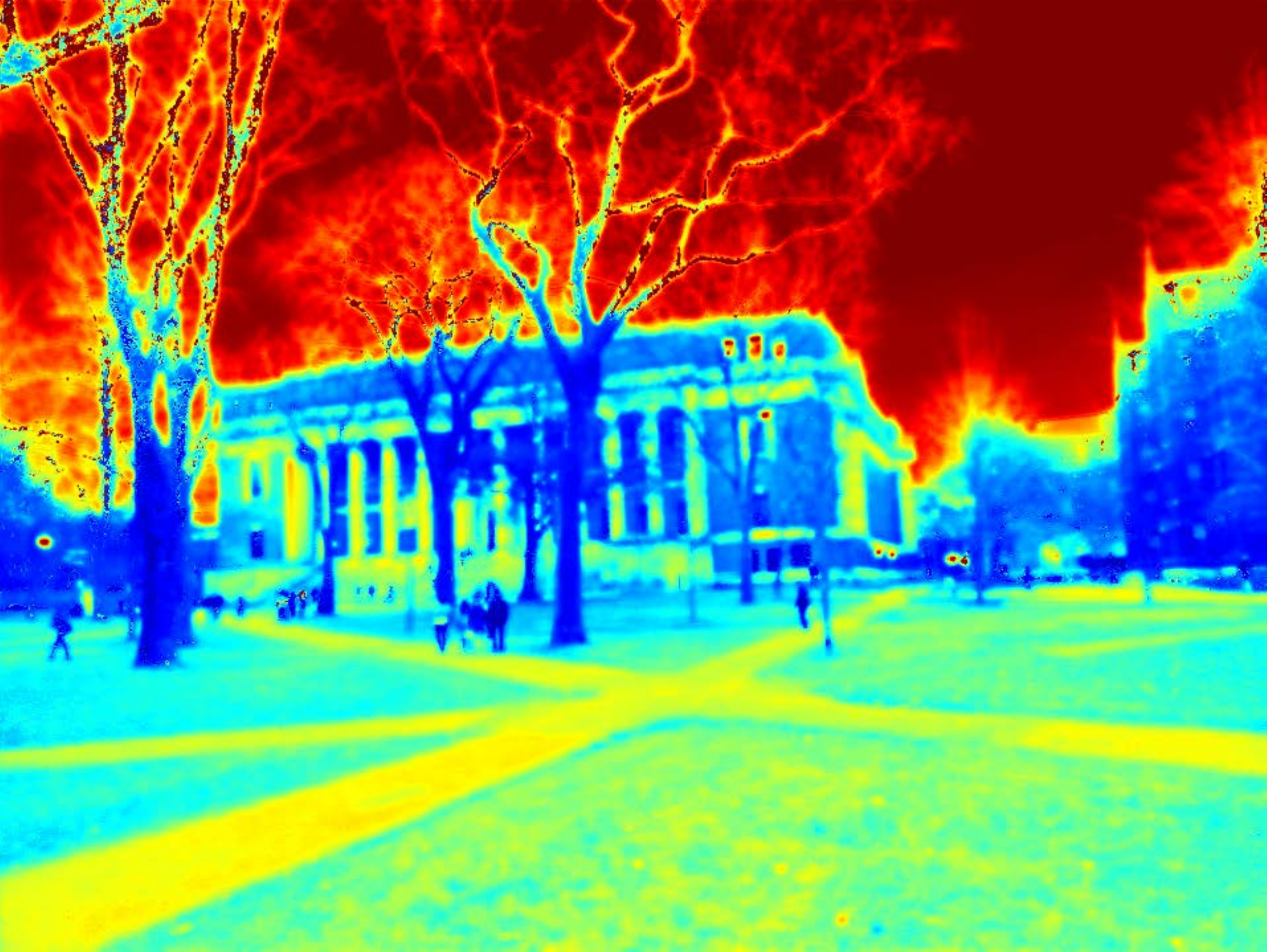}&
		\includegraphics[width=0.113\textwidth]{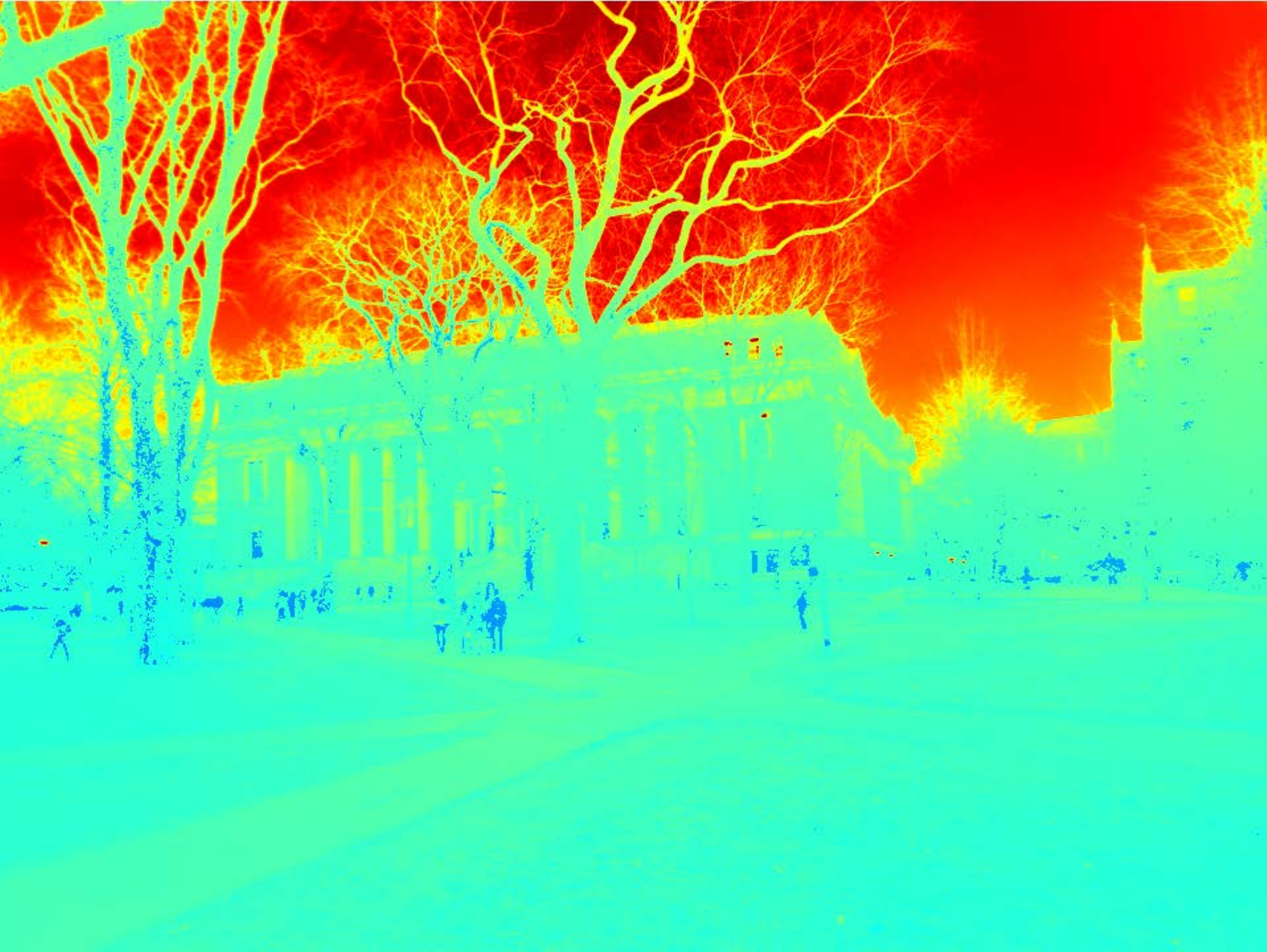}&
		\includegraphics[width=0.113\textwidth]{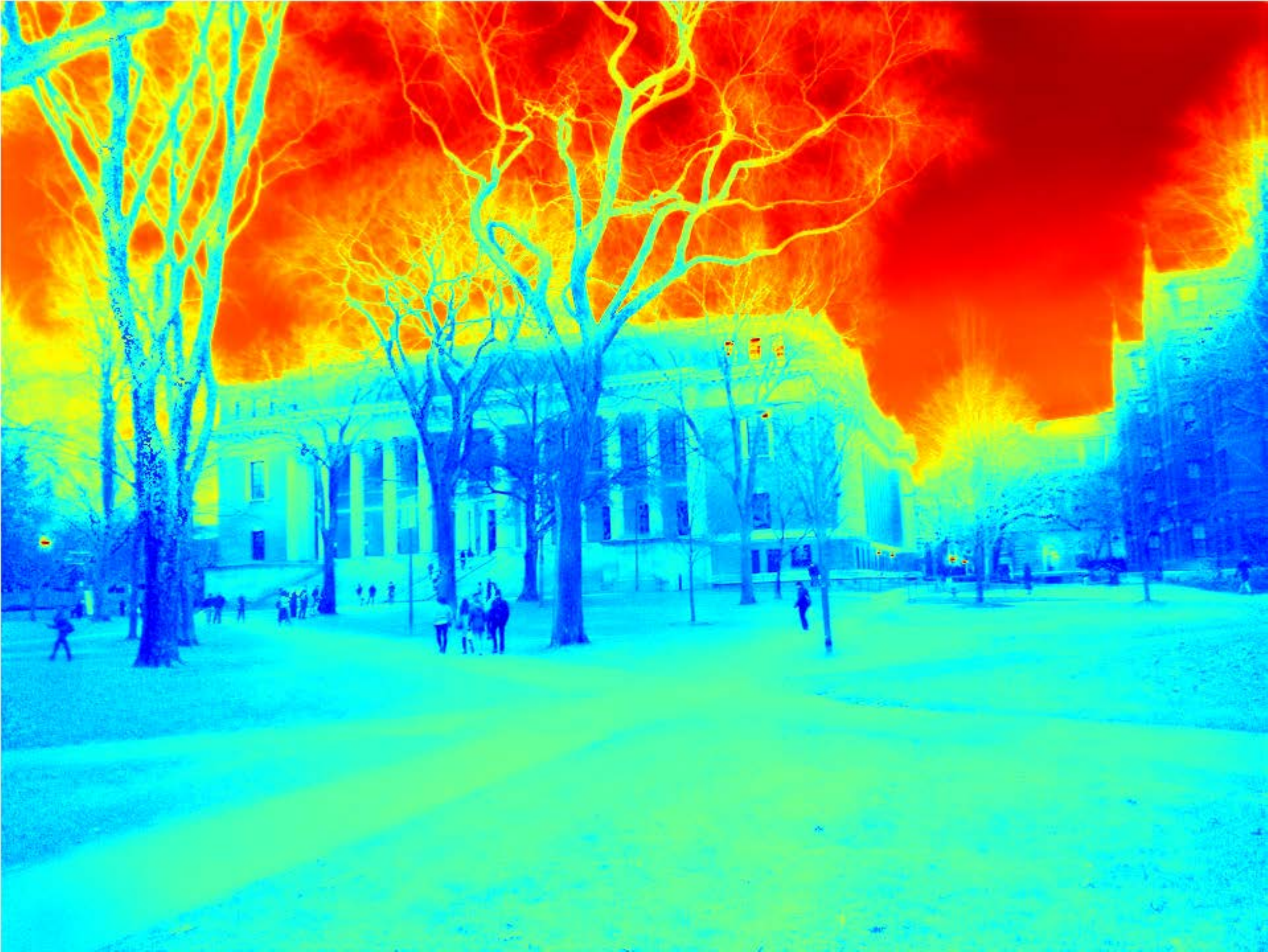}\\	
		\multicolumn{4}{c}{\footnotesize The estimated illumination}\\ 	
		\includegraphics[width=0.113\textwidth]{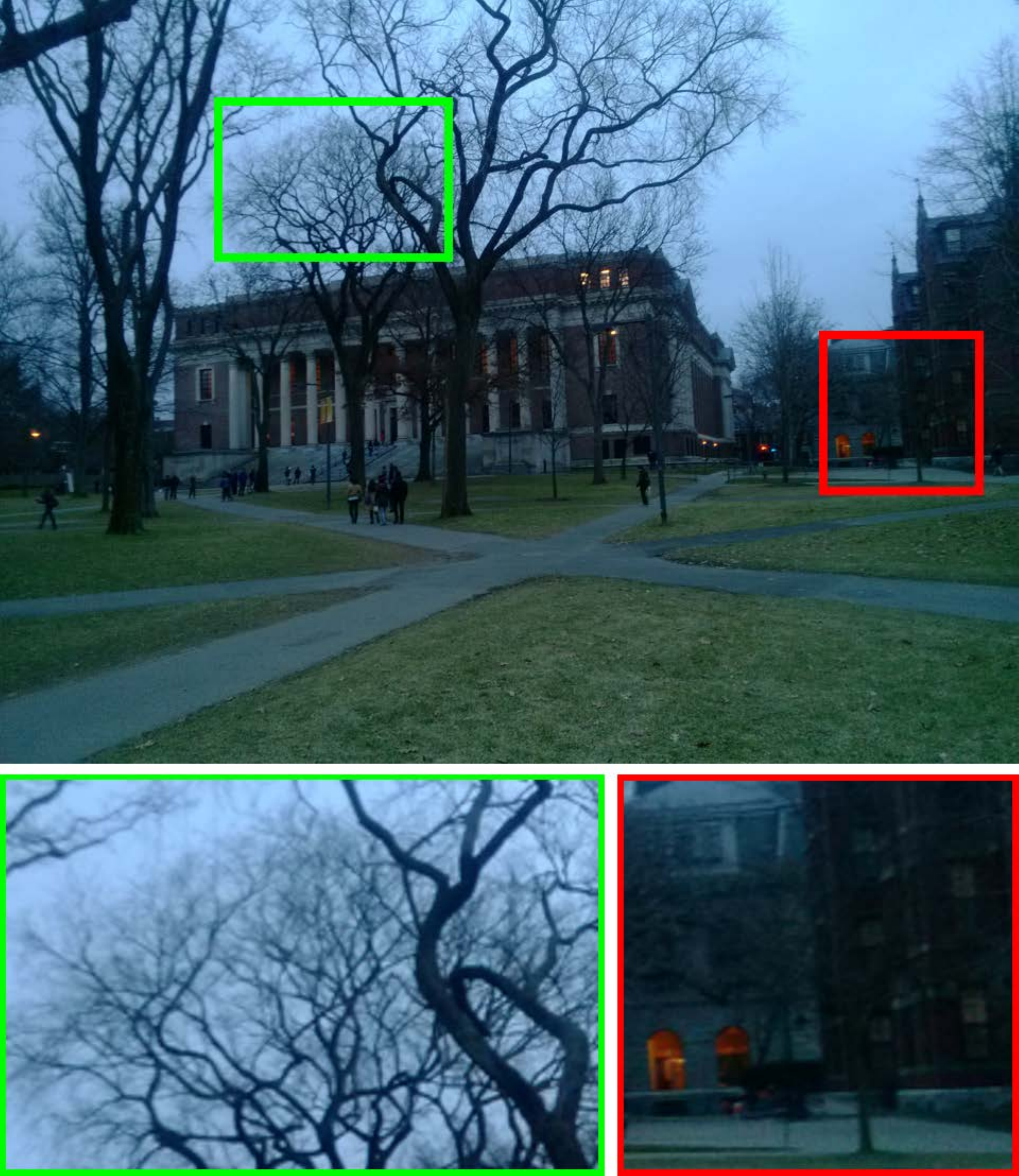}&
		\includegraphics[width=0.113\textwidth]{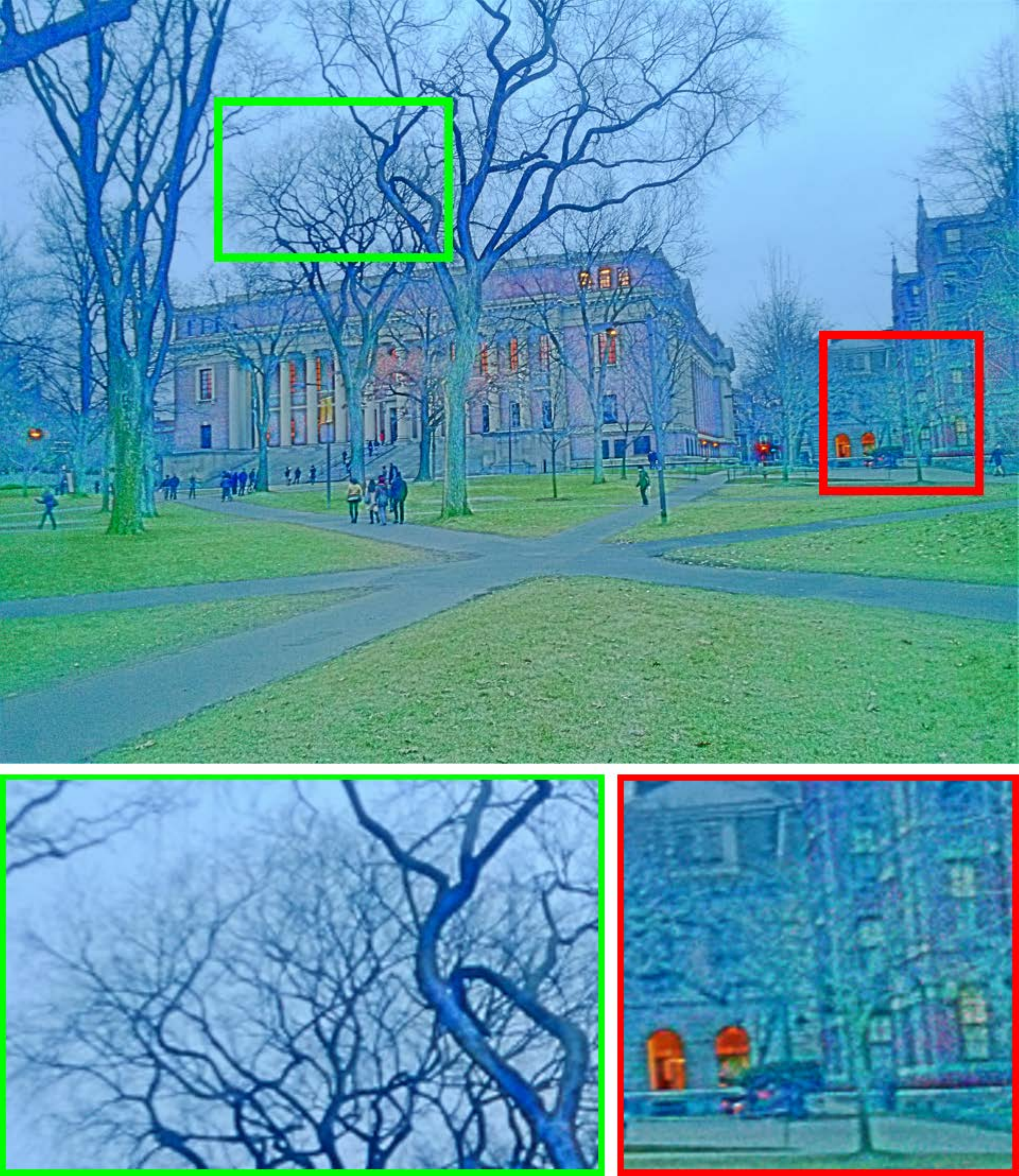}&
		\includegraphics[width=0.113\textwidth]{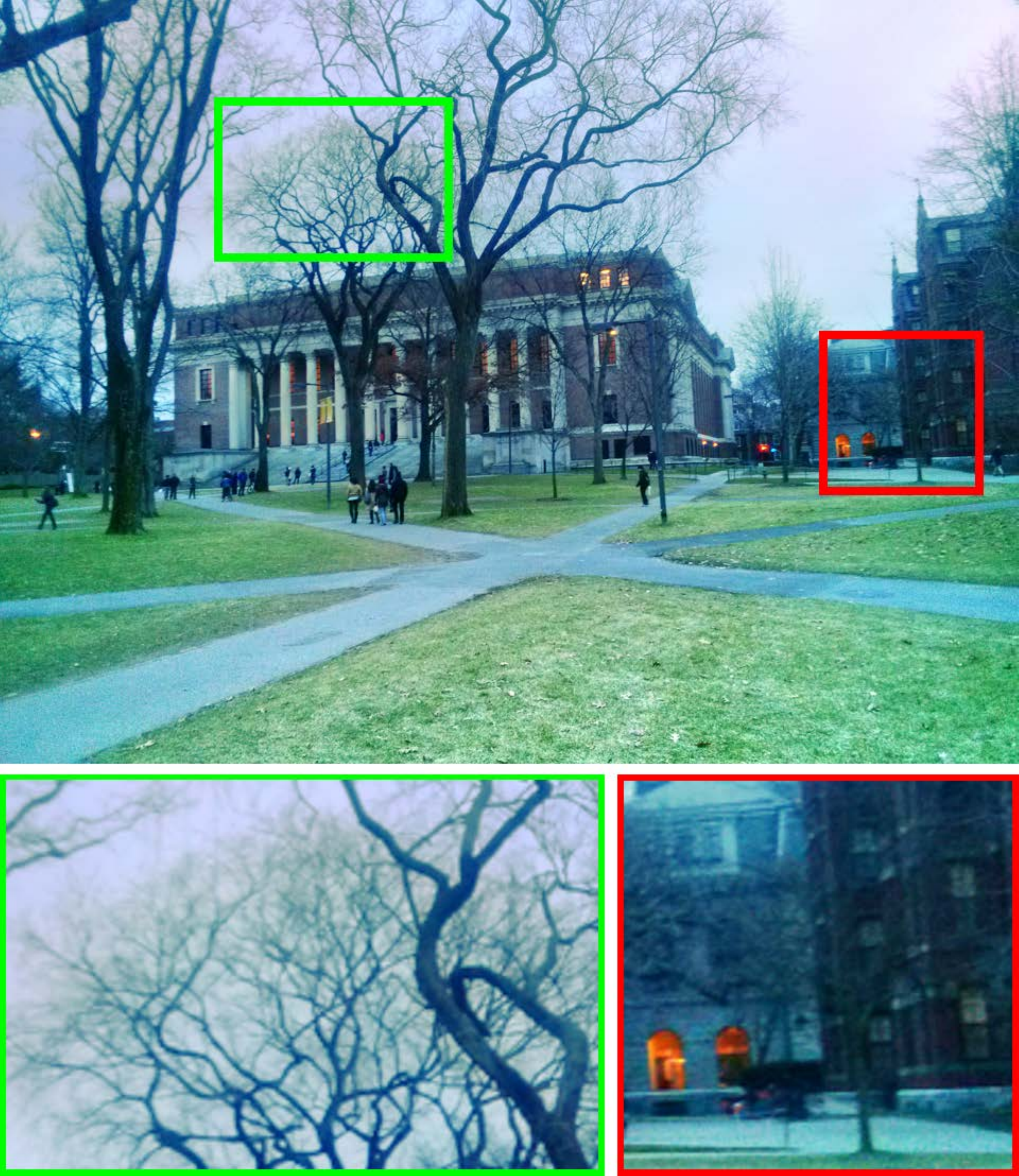}&
		\includegraphics[width=0.113\textwidth]{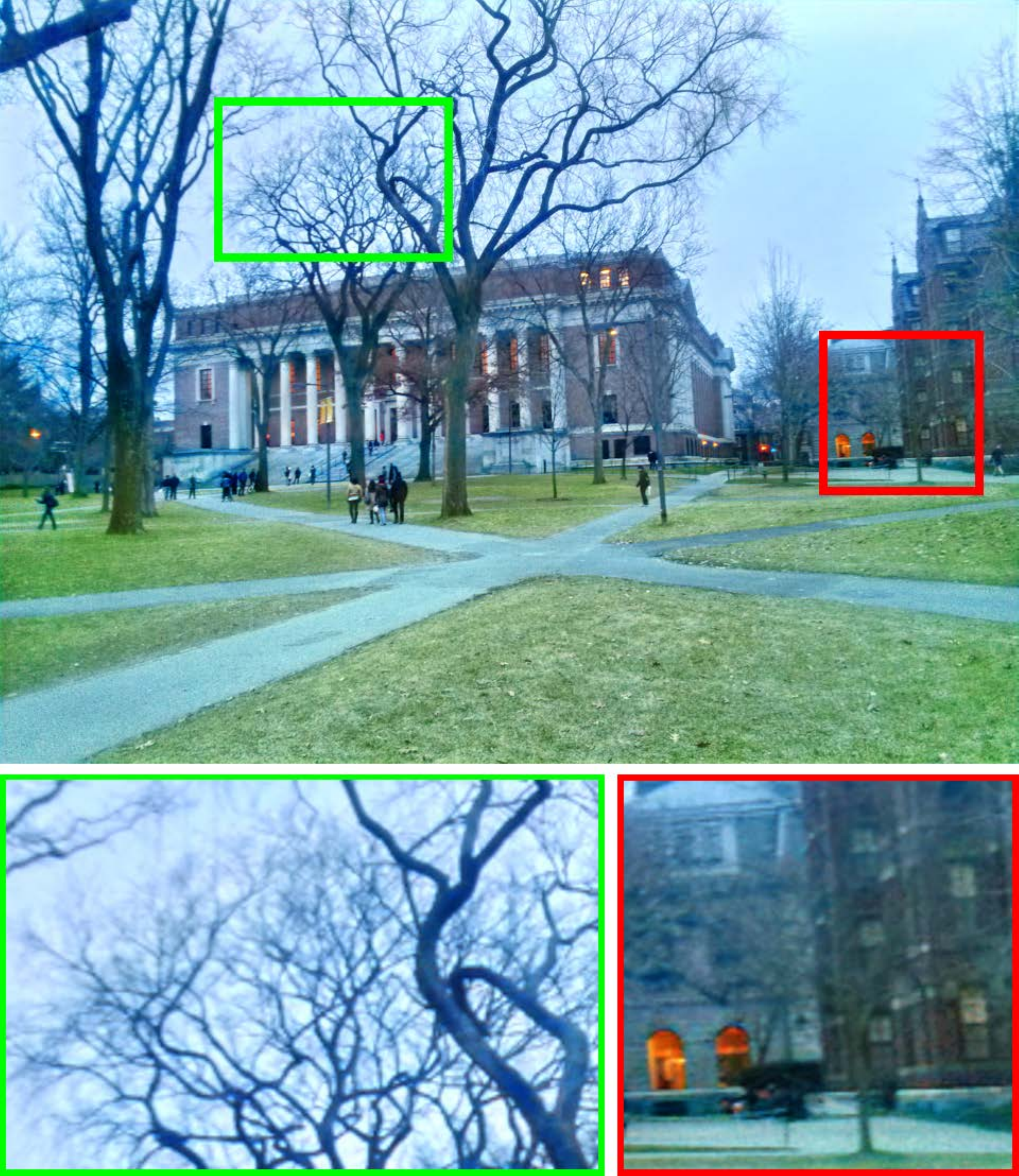}\\		
		\multicolumn{4}{c}{\footnotesize The enhanced output}\\ 	
	\end{tabular}
	\caption{Visual comparison of different components on an unknown real-world testing image among two paired end-to-end deep learning methods (i.e., Retinex and DeepUPE) and our CSDGAN.} 
	\label{fig:CompUPE}
\end{figure}

For testing in unpaired supervision, we trained CSDGAN in a newly collected unpaired dataset presented in EnlightenGAN~\cite{jiang2019enlightengan}, we then tested our trained model in four unpaired datasets that are commonly-used in the LLIE task to present the evaluation, including NPE~\cite{wang2013naturalness} consists of 130 low-light images in different natural scenarios, NASA~\footnote{https://dragon.larc.nasa.gov/retinex/pao/news/} includes 23 low-light images in the indoor and outdoor scenes, MEF~\cite{ma2015perceptual} contains 17 indoor and outdoor low-light images, and LIME~\cite{guo2017lime} which is composed of 10 nighttime low-light images.
We adopted the Natural Image Quality Evaluator (NIQE)\footnote{NIQE~\cite{mittal2012making} is a well-known no-reference image quality assessment for evaluating the naturalness of a single image. } as our evaluated standard. As for NIQE, the lower is the better.
Finally, we considered a more challenging example from ExDark dataset~\cite{loh2019getting}.

\begin{figure*}[t]
	\centering
	\begin{tabular}{c@{\extracolsep{0.3em}}c@{\extracolsep{0.3em}}c} 
		\includegraphics[width=0.32\textwidth]{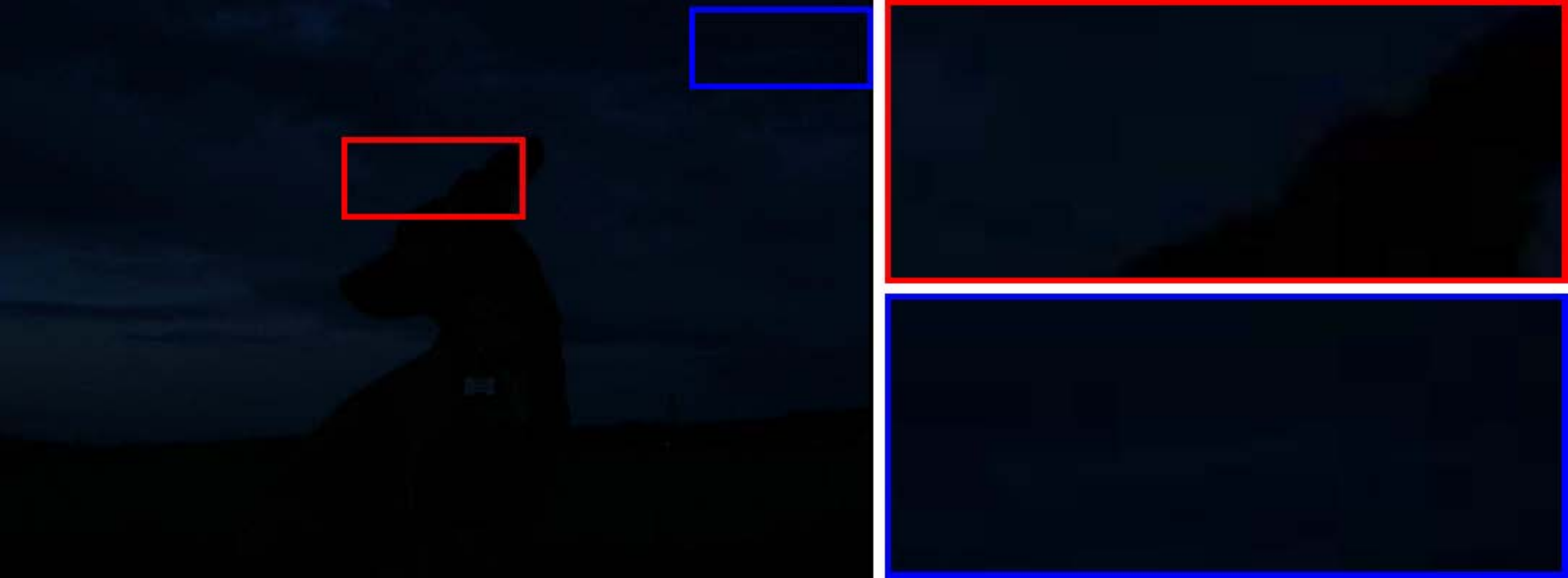}&
		\includegraphics[width=0.32\textwidth]{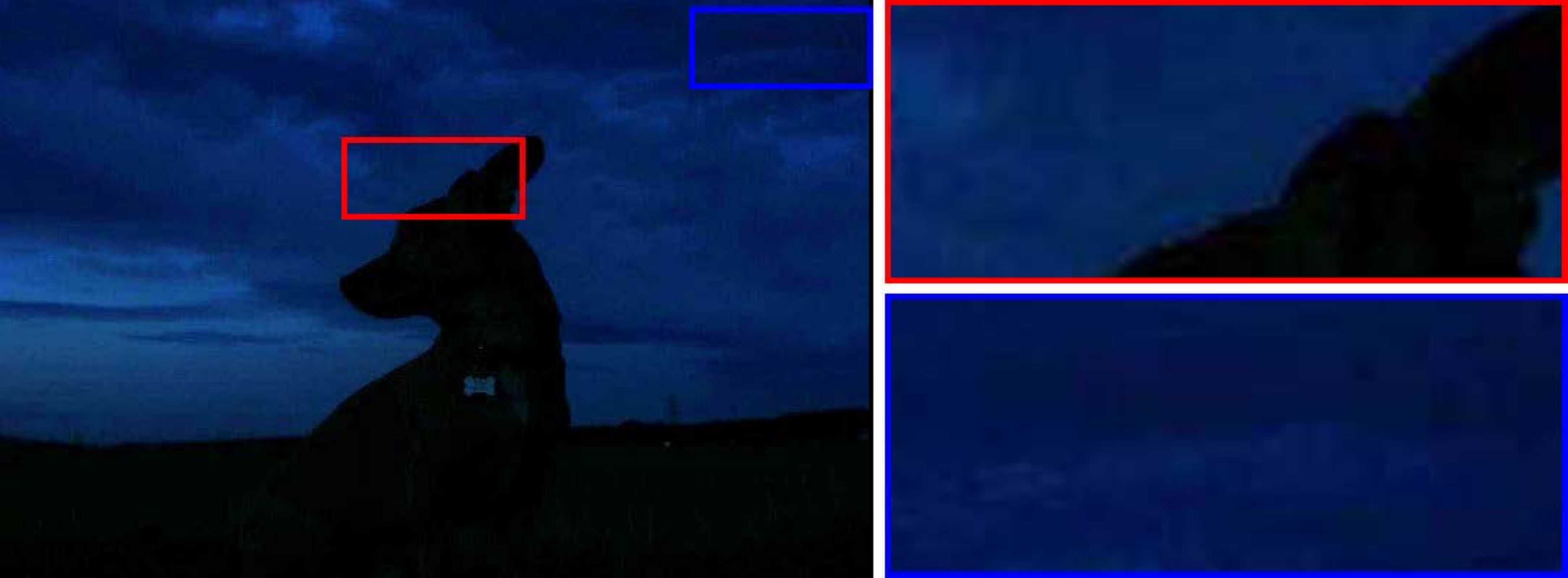}&
		\includegraphics[width=0.32\textwidth]{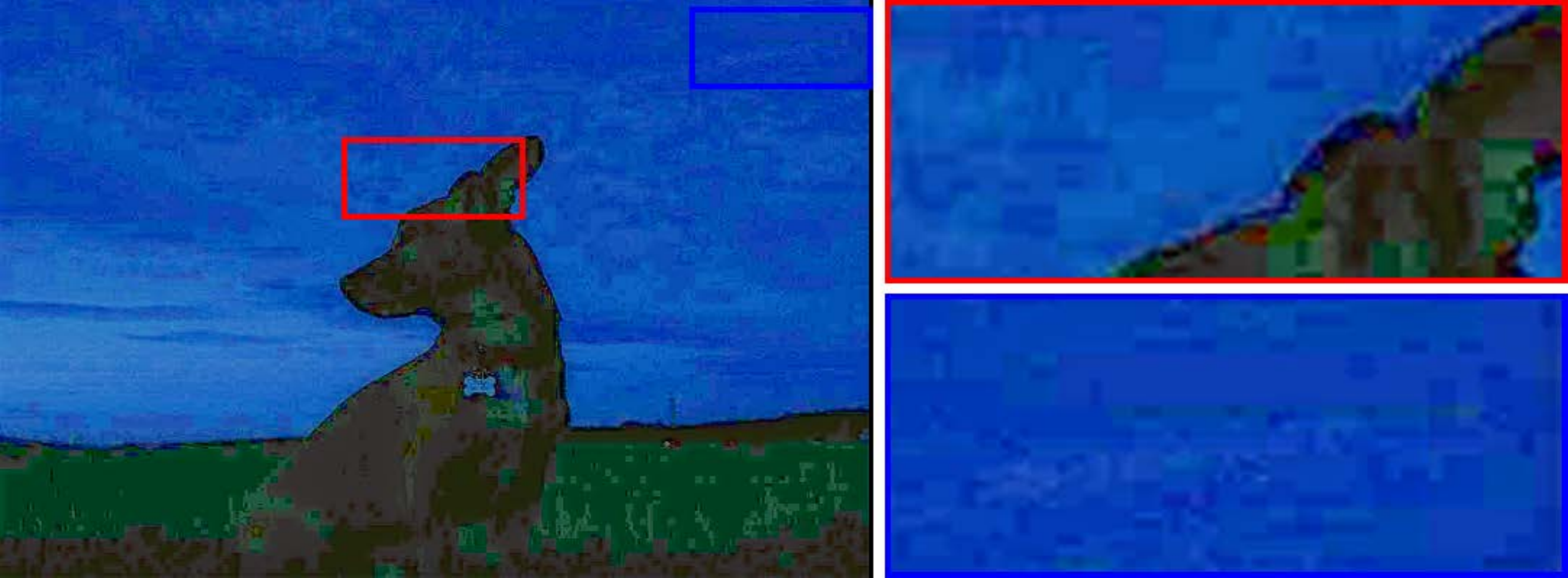}\\
		\footnotesize Input&\footnotesize DeepUPE~\cite{wang2019underexposed}&\footnotesize RetinexNet~\cite{Chen2018Retinex}\\
		\includegraphics[width=0.32\textwidth]{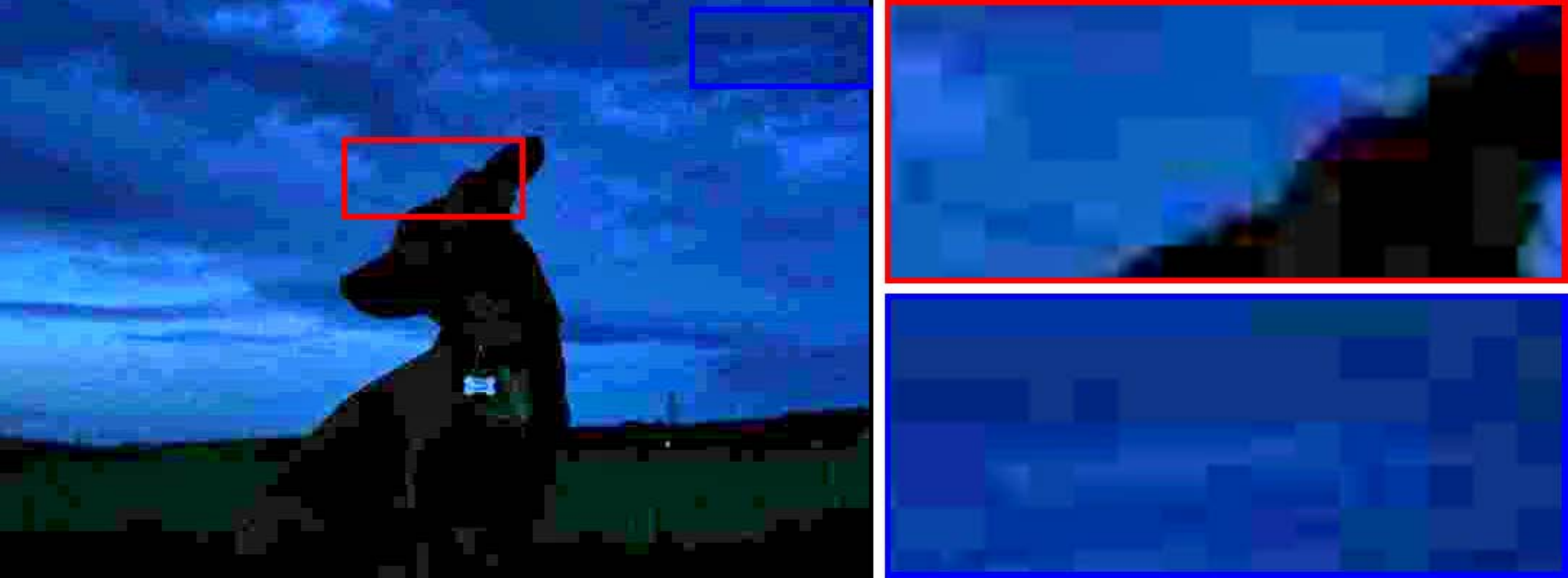}&
		\includegraphics[width=0.32\textwidth]{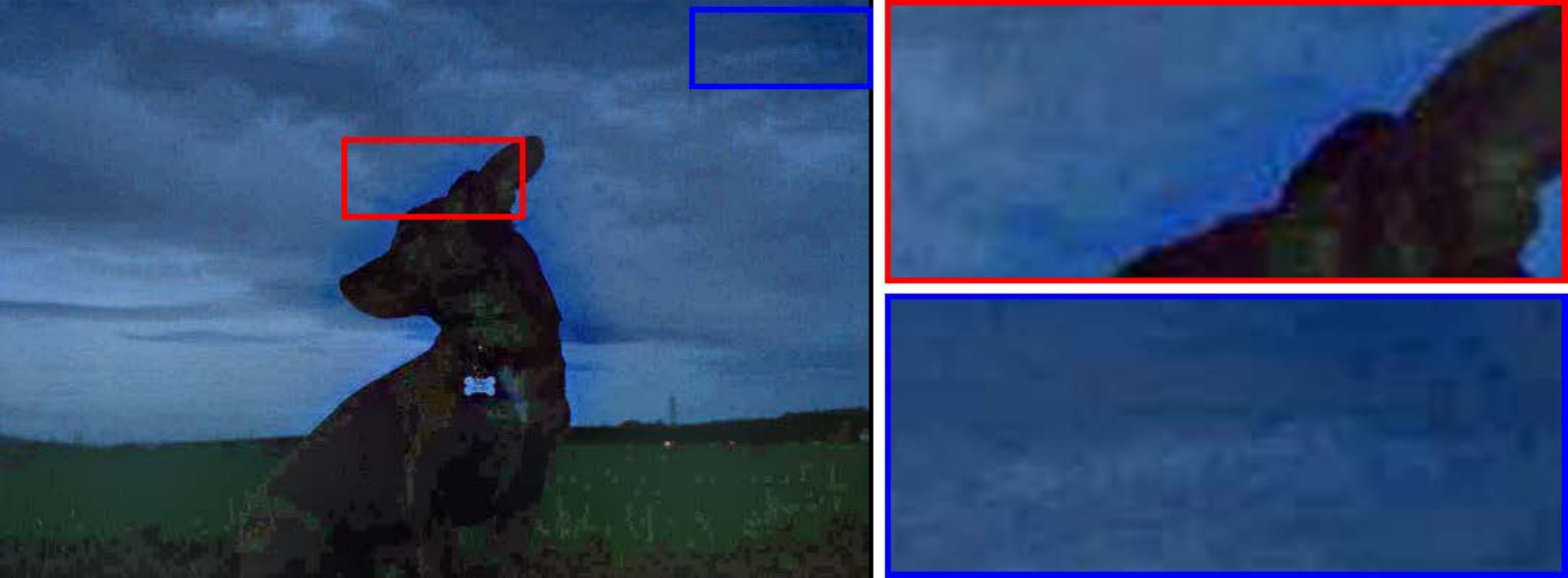}&
		\includegraphics[width=0.32\textwidth]{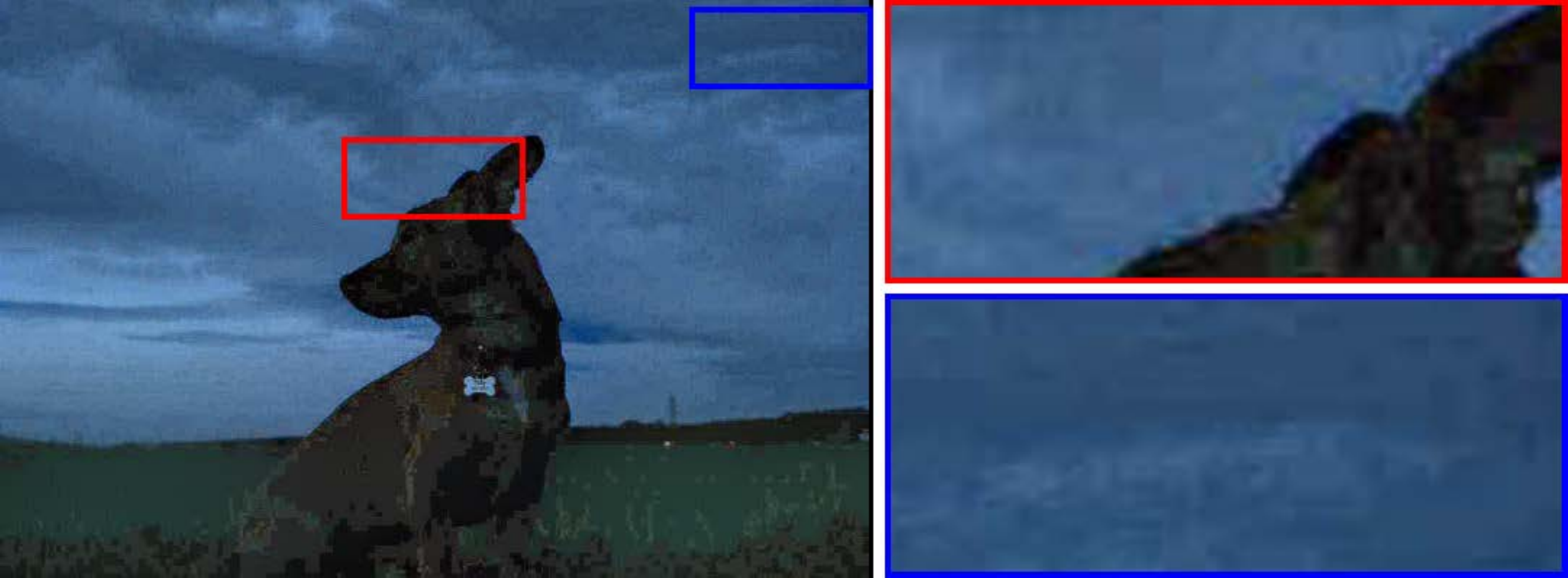}\\
		\footnotesize LIME~\cite{guo2017lime}&\footnotesize EnlightenGAN~\cite{jiang2019enlightengan}&\footnotesize CSDGAN \\
		\includegraphics[width=0.32\textwidth]{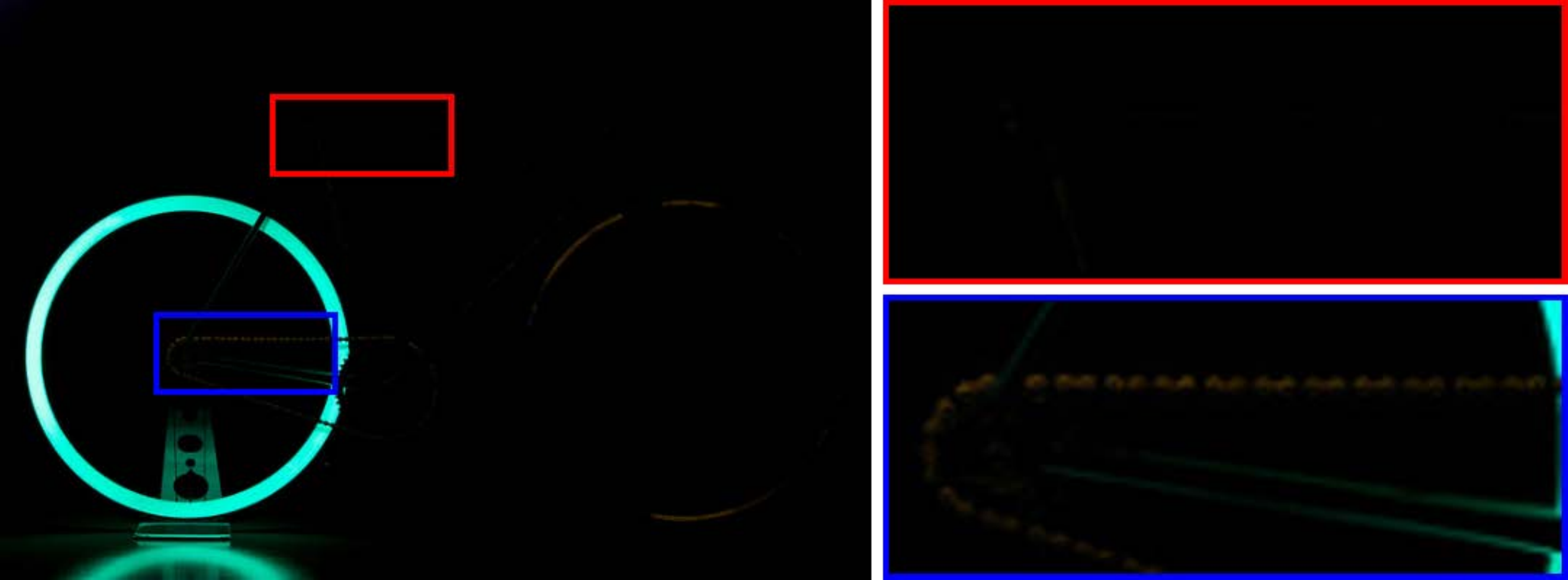}&
		\includegraphics[width=0.32\textwidth]{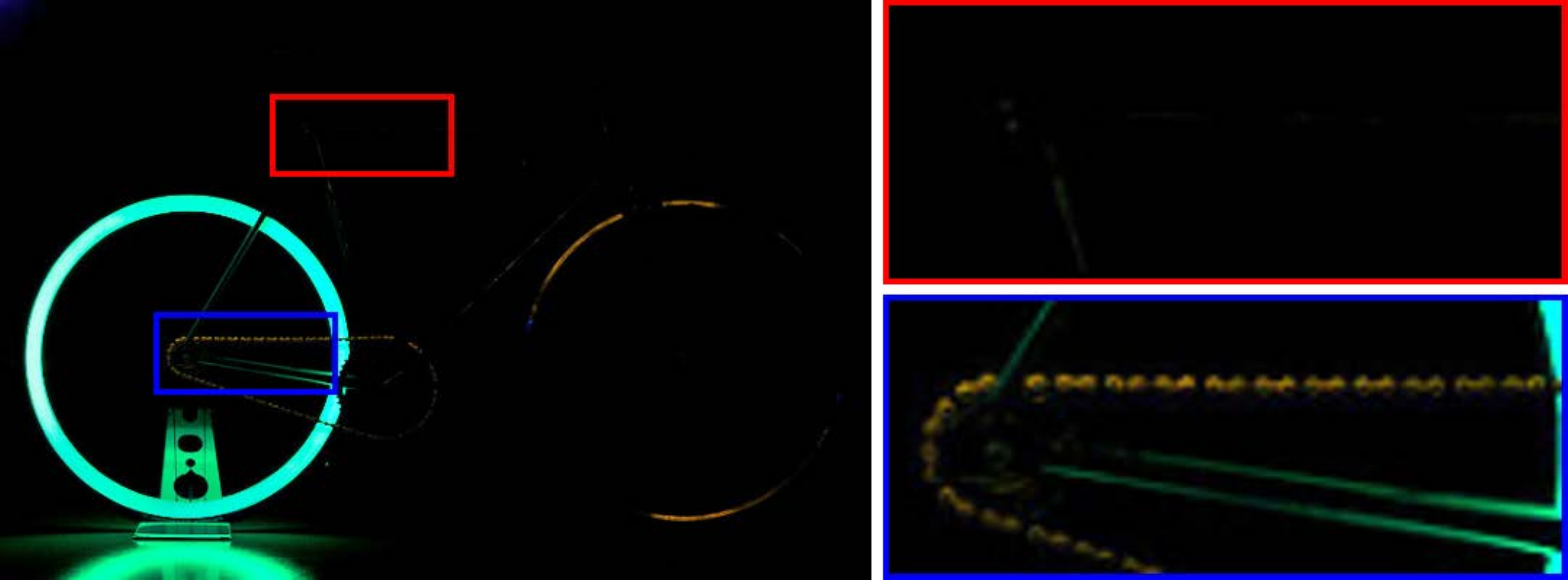}&
		\includegraphics[width=0.32\textwidth]{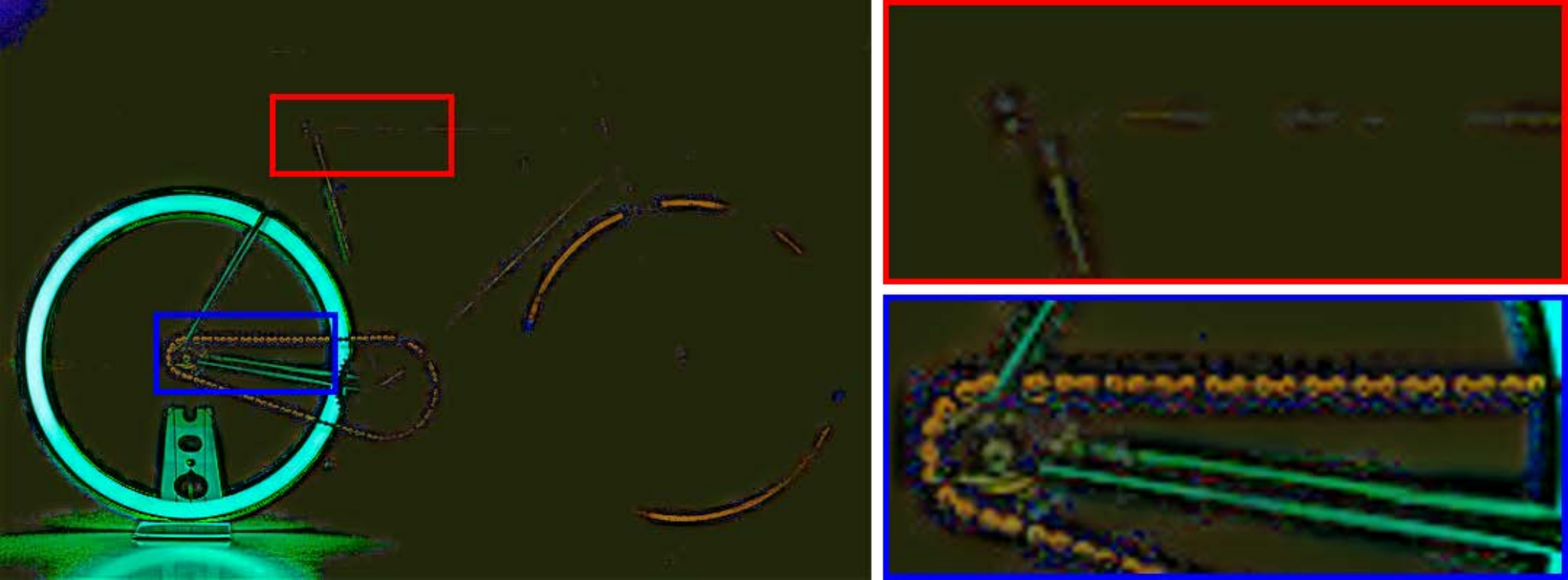}\\
		\footnotesize Input&\footnotesize DeepUPE~\cite{wang2019underexposed}&\footnotesize RetinexNet~\cite{Chen2018Retinex}\\
		\includegraphics[width=0.32\textwidth]{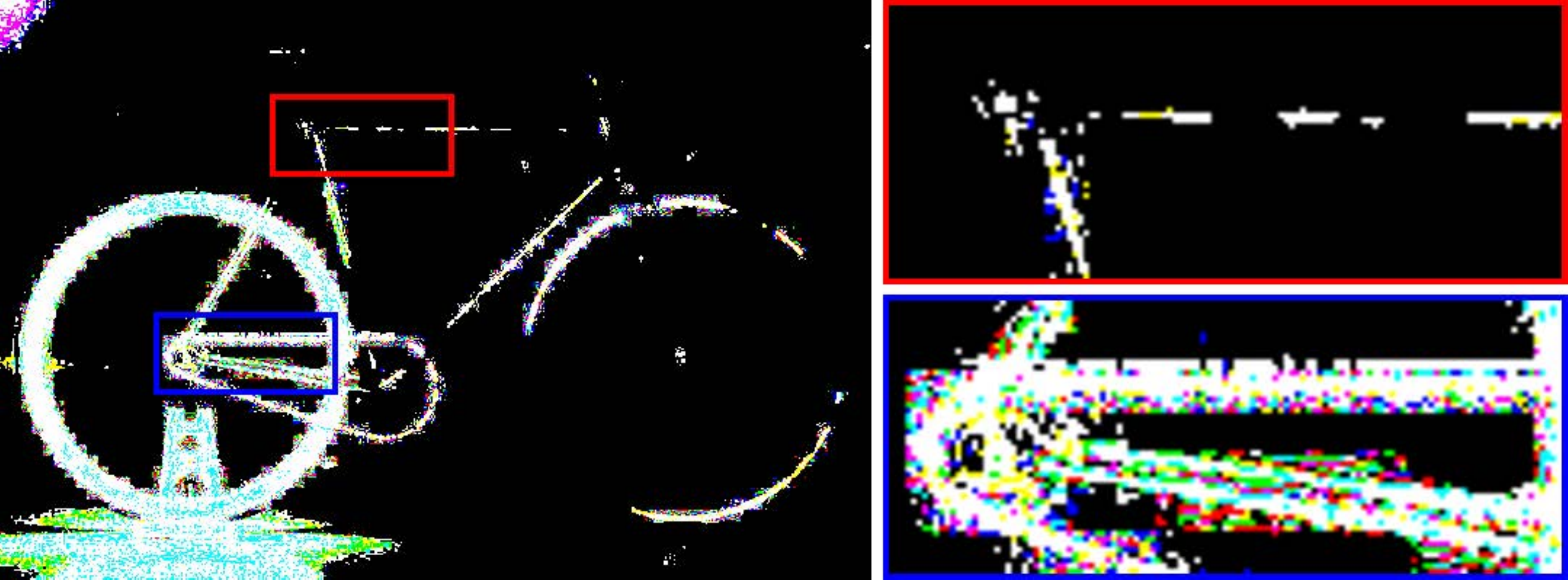}&
		\includegraphics[width=0.32\textwidth]{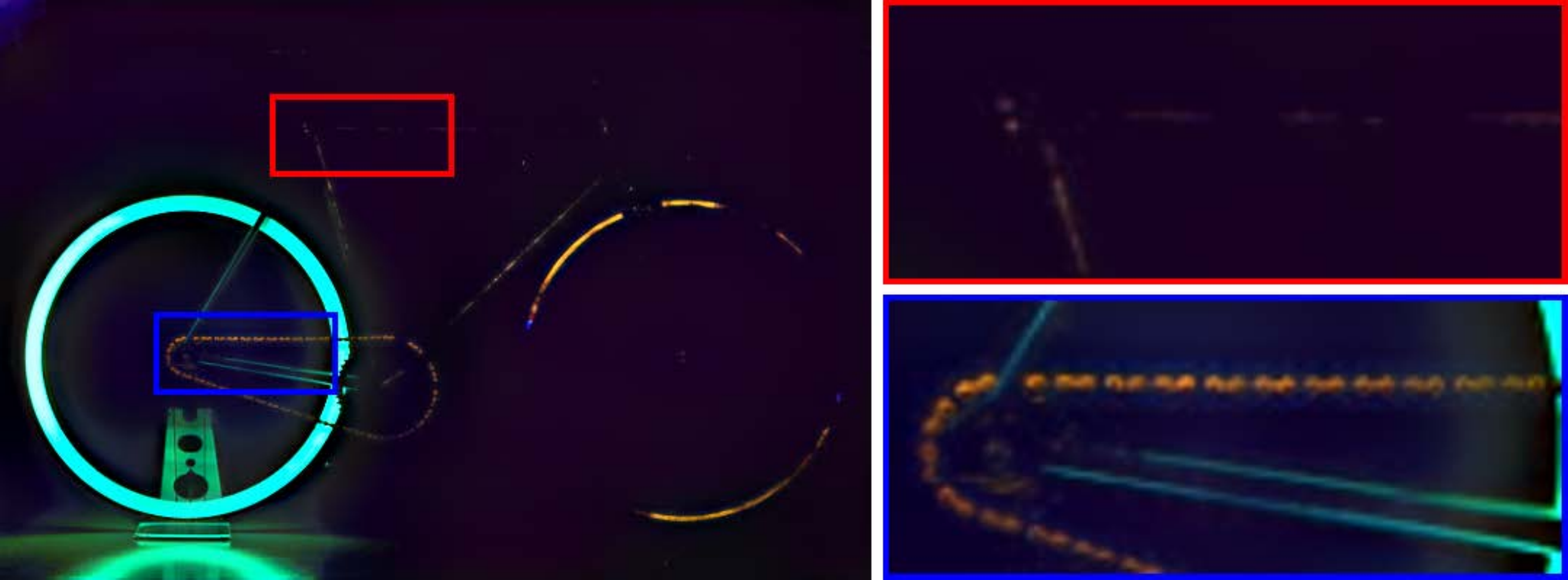}&
		\includegraphics[width=0.32\textwidth]{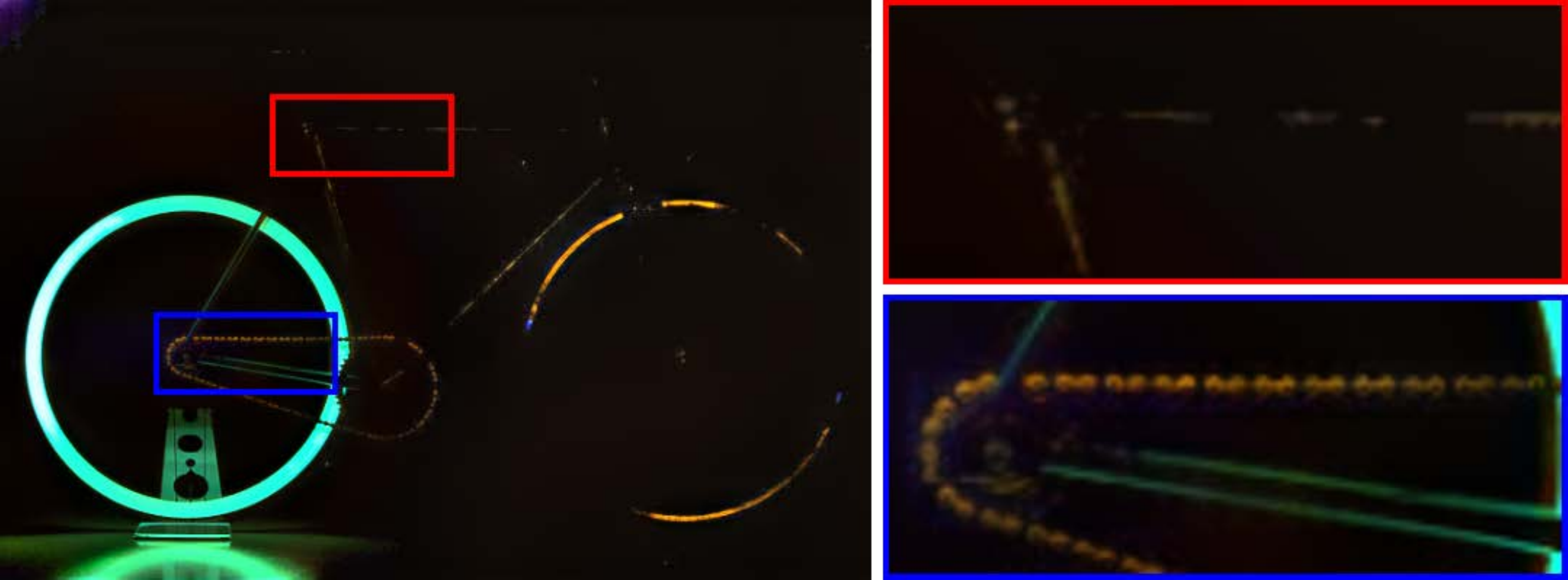}\\
		\footnotesize LIME~\cite{guo2017lime}&\footnotesize EnlightenGAN~\cite{jiang2019enlightengan}&\footnotesize CSDGAN \\
	\end{tabular}
	\caption{Visual comparison on two challenging examples from ExDark dataset~\cite{loh2019getting}. }
	\label{fig:MoreComp}
\end{figure*}

\begin{figure}[t]
	\centering
	\begin{tabular}{c@{\extracolsep{0.2em}}c@{\extracolsep{0.2em}}c@{\extracolsep{0.2em}}c}
		\footnotesize Input&\footnotesize CSDNet$_a$&\footnotesize CSDNet$_b$&\footnotesize CSDNet$_c$\\
		\includegraphics[width=0.113\textwidth]{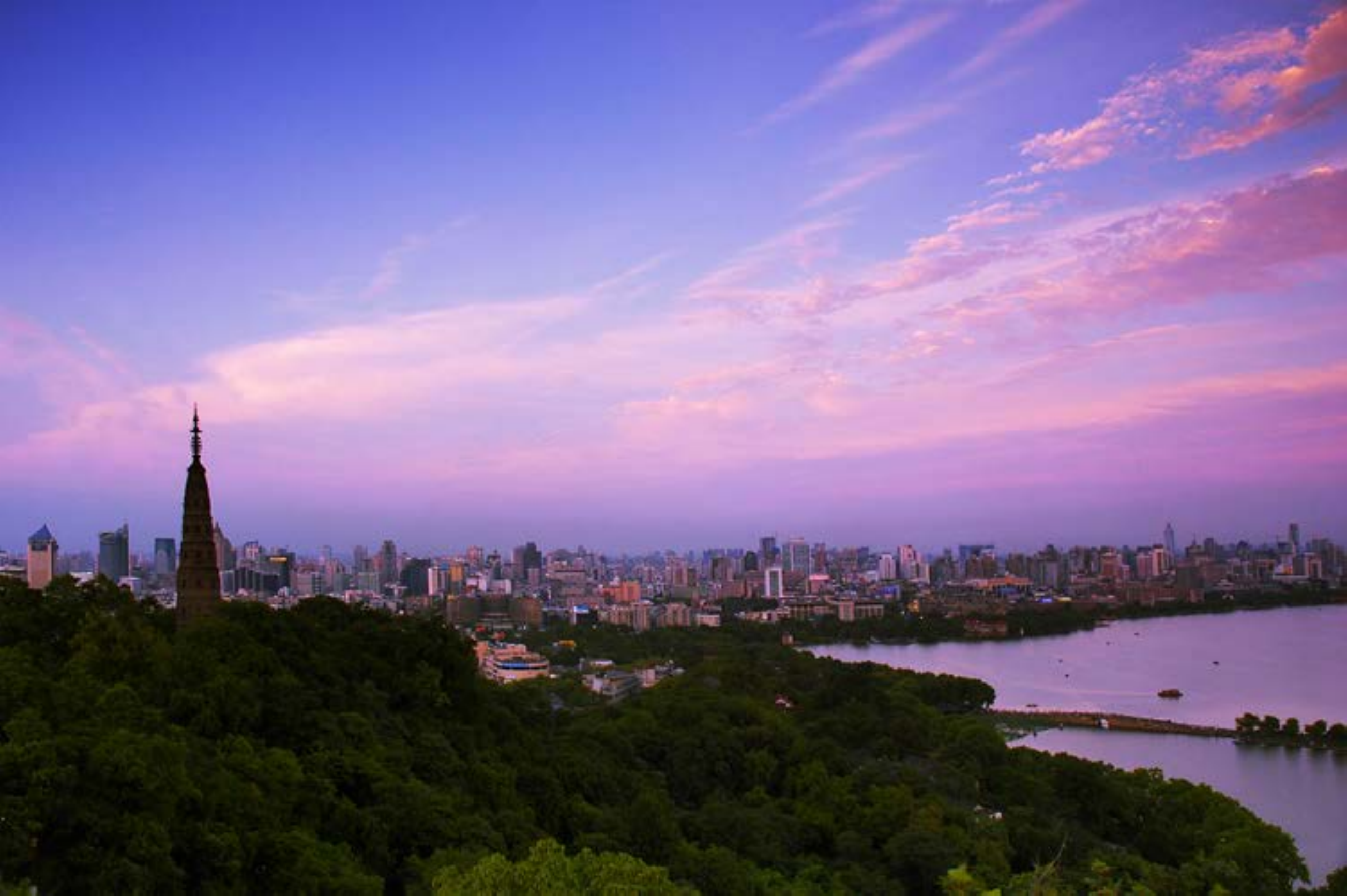}&
		\includegraphics[width=0.113\textwidth]{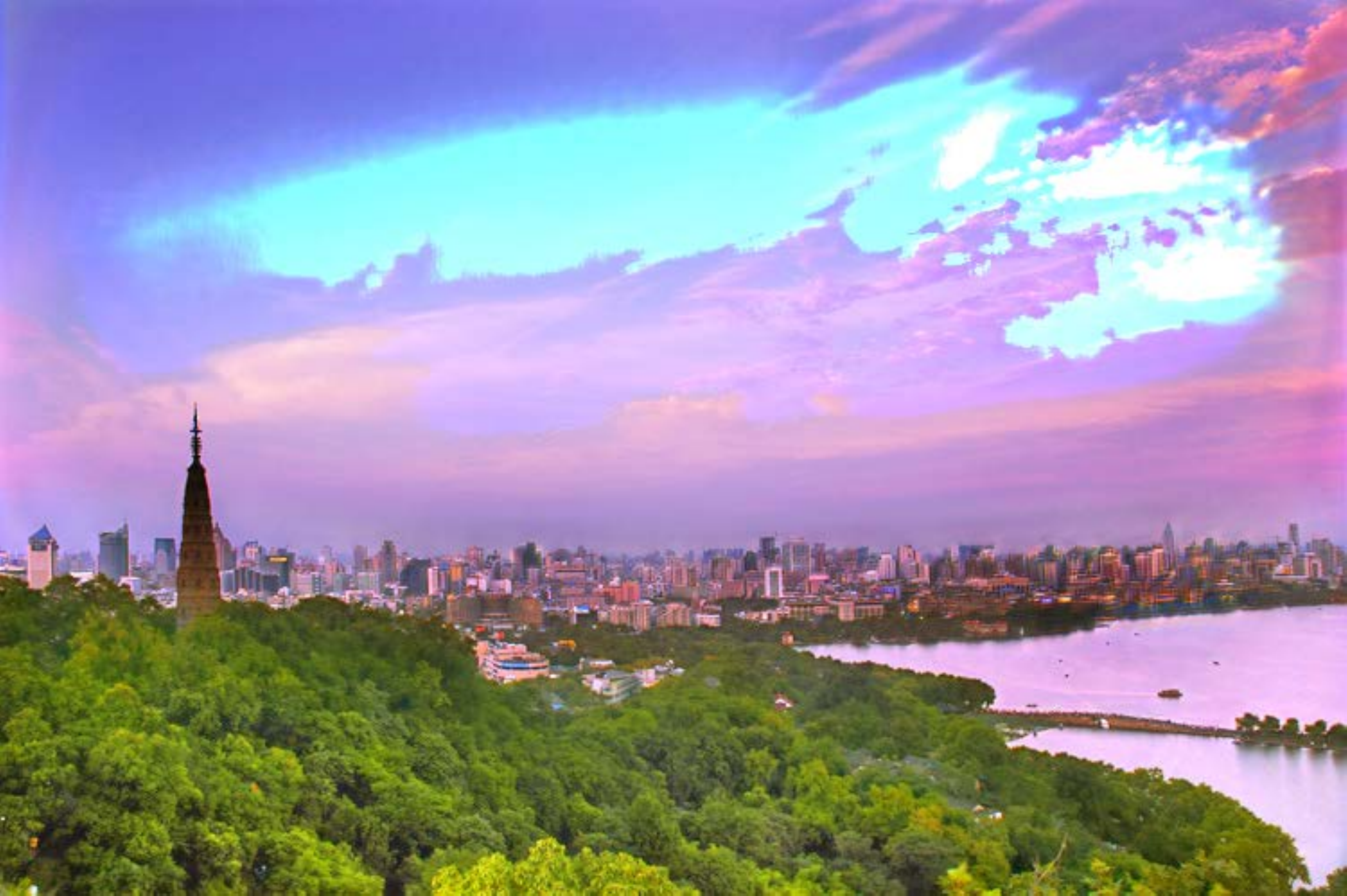}&
		\includegraphics[width=0.113\textwidth]{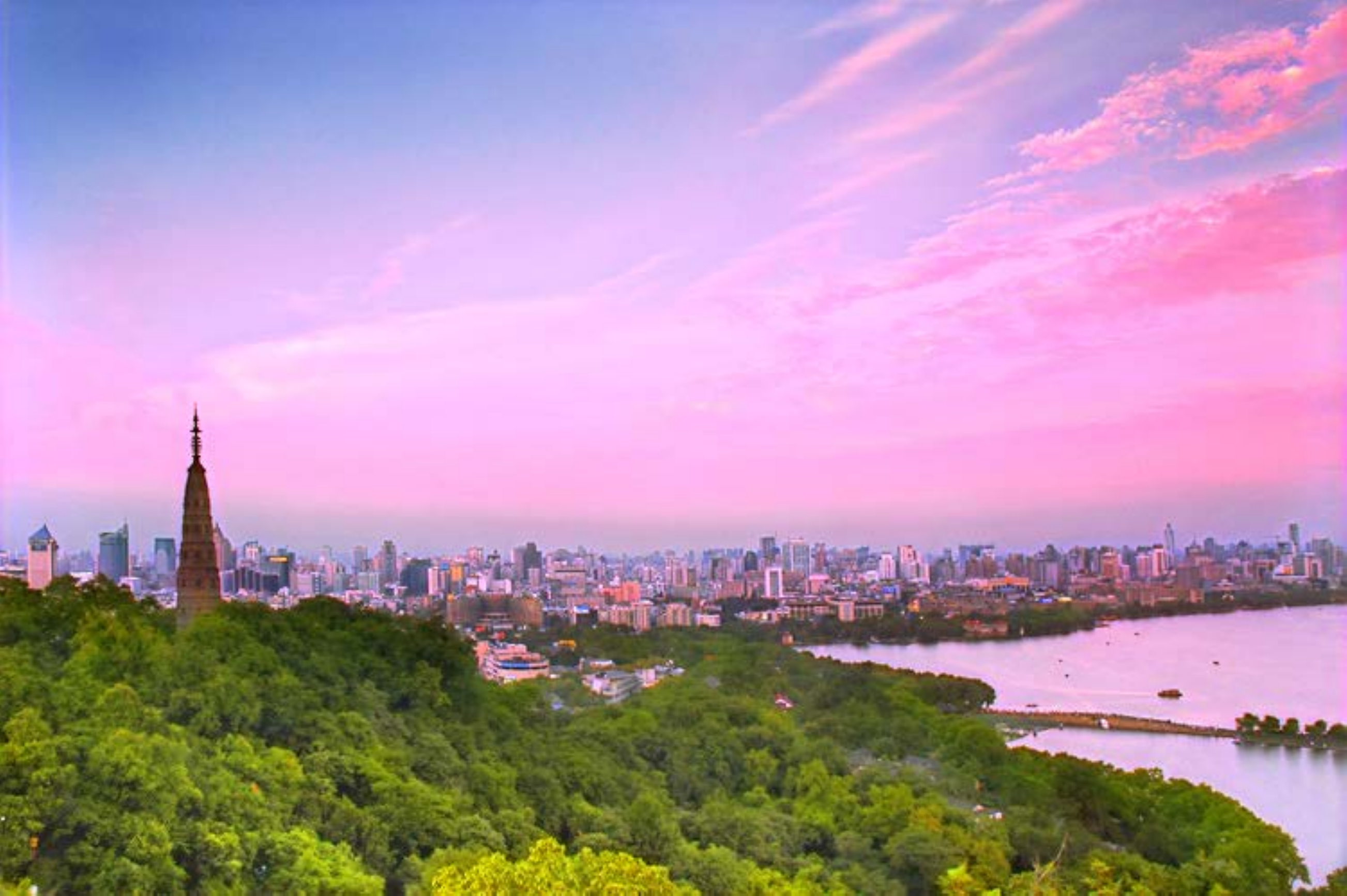}&
		\includegraphics[width=0.113\textwidth]{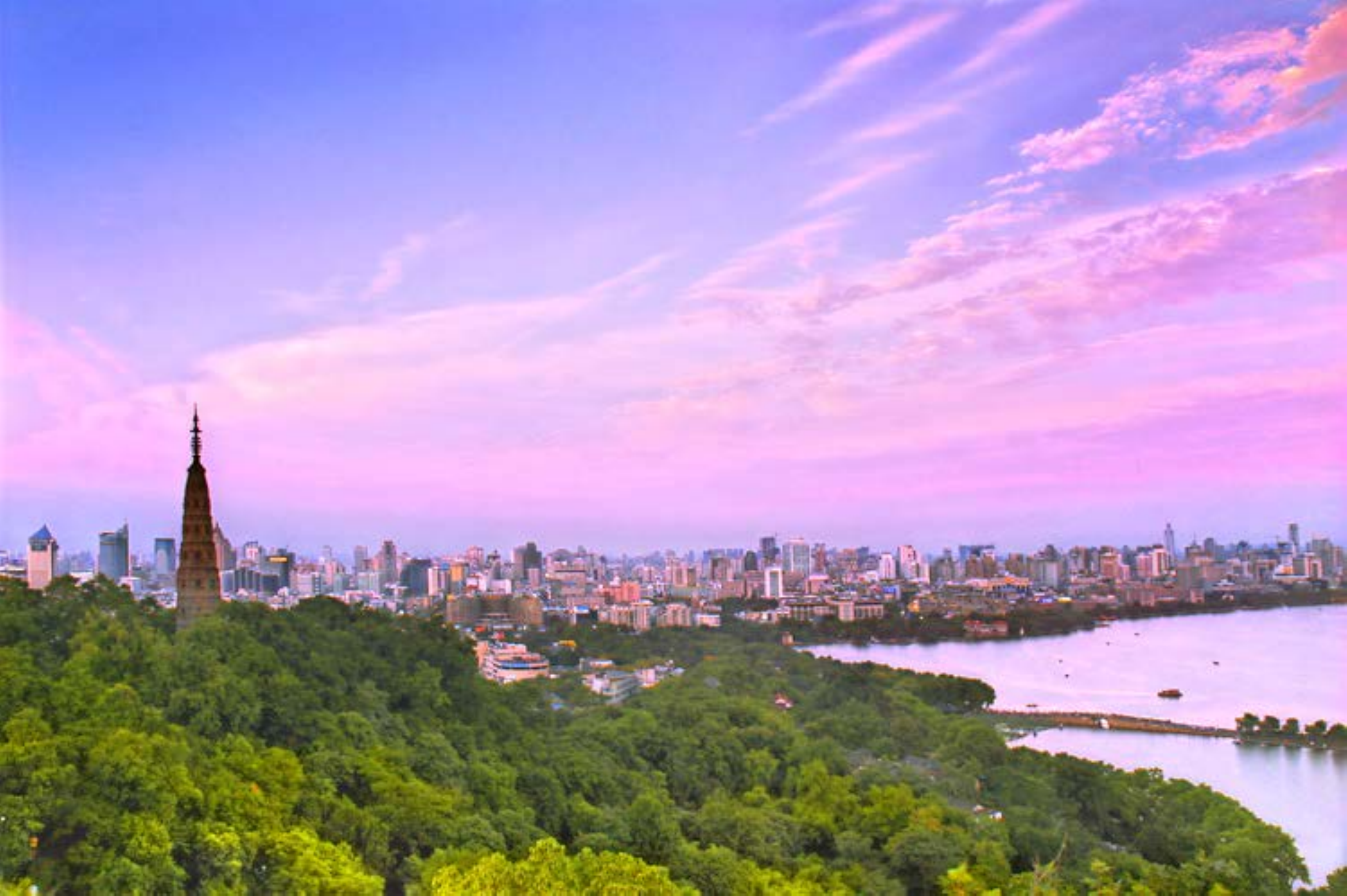}\\		
		\multicolumn{4}{c}{\footnotesize Example A}\\
		\includegraphics[width=0.113\textwidth]{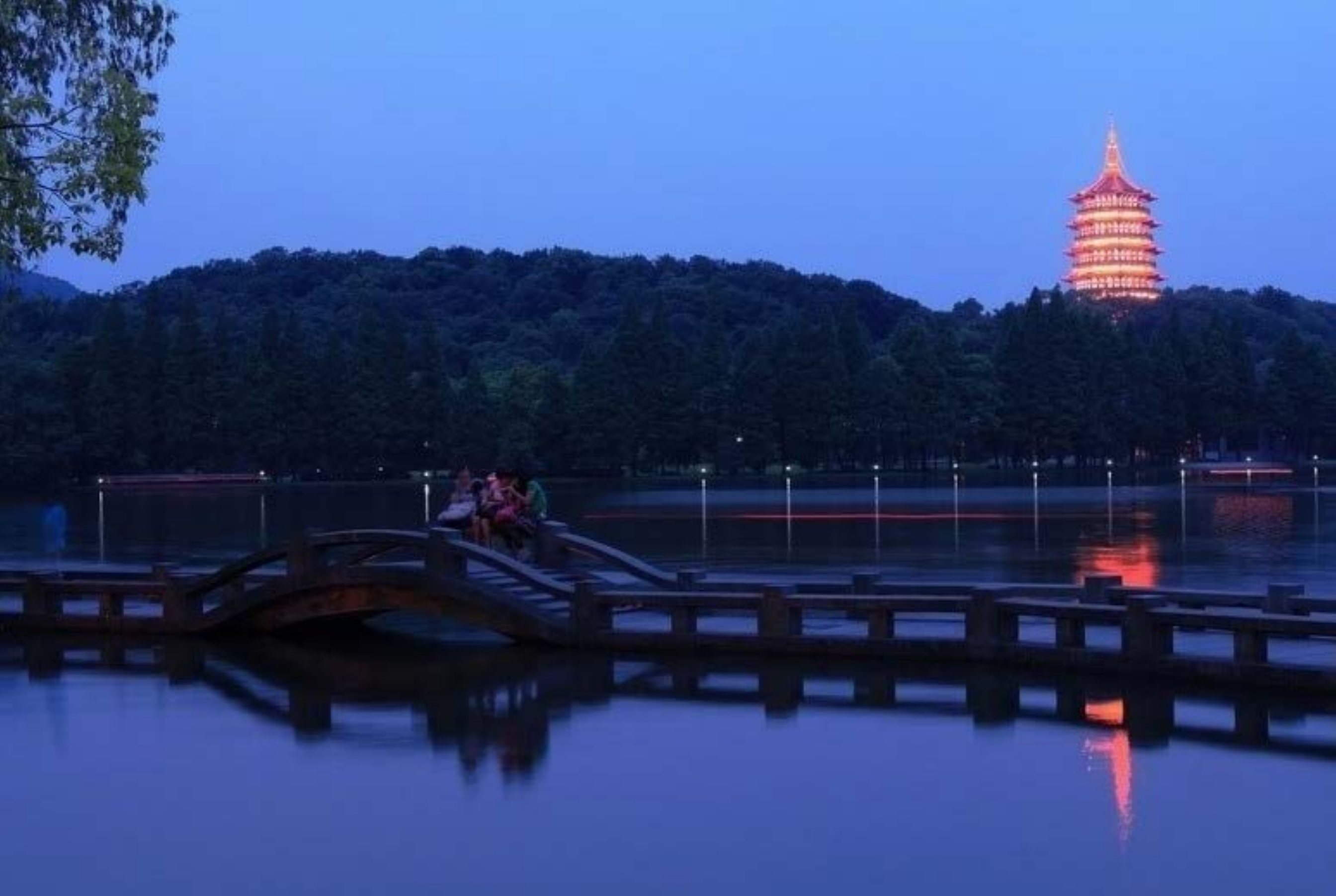}&
		\includegraphics[width=0.113\textwidth]{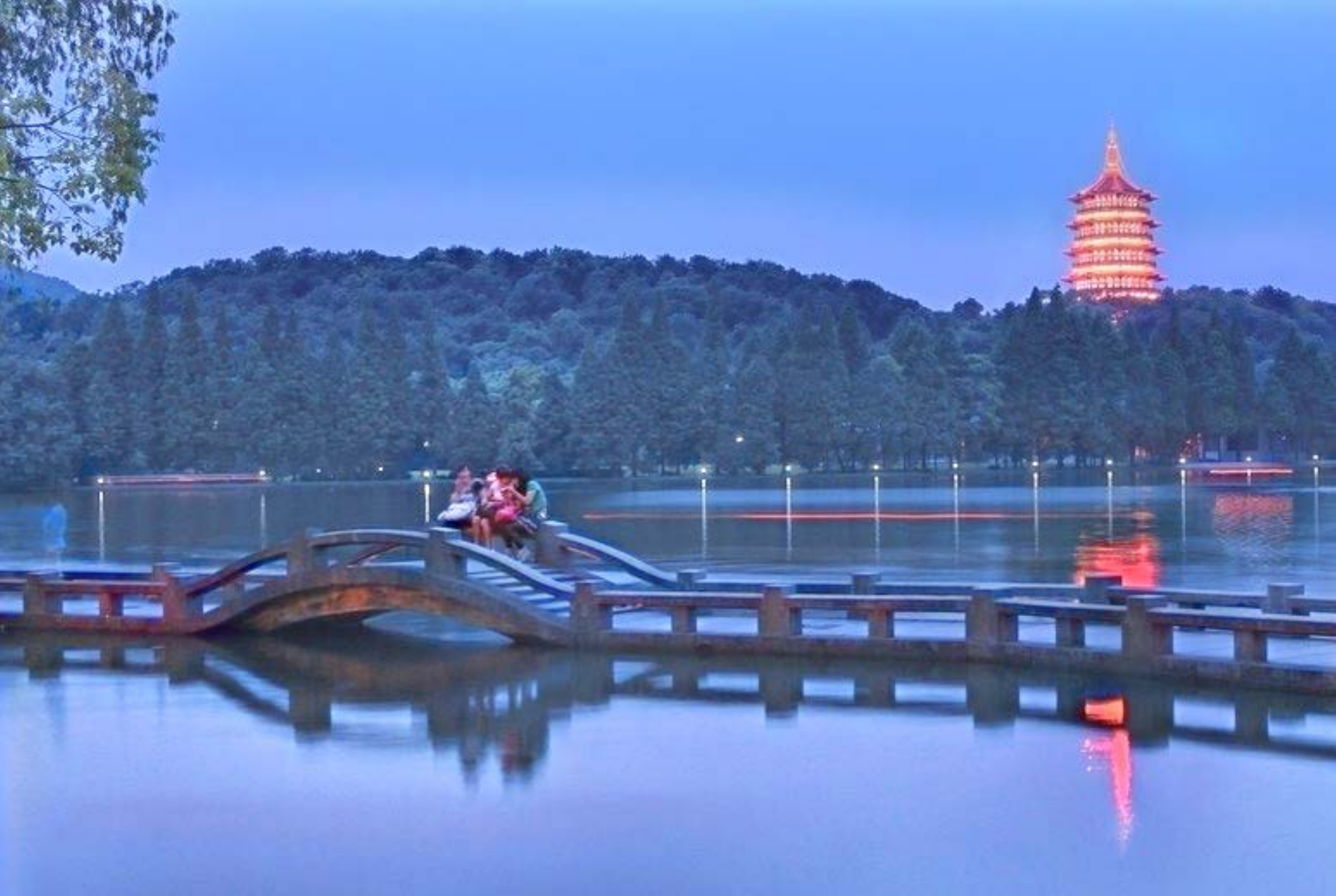}&
		\includegraphics[width=0.113\textwidth]{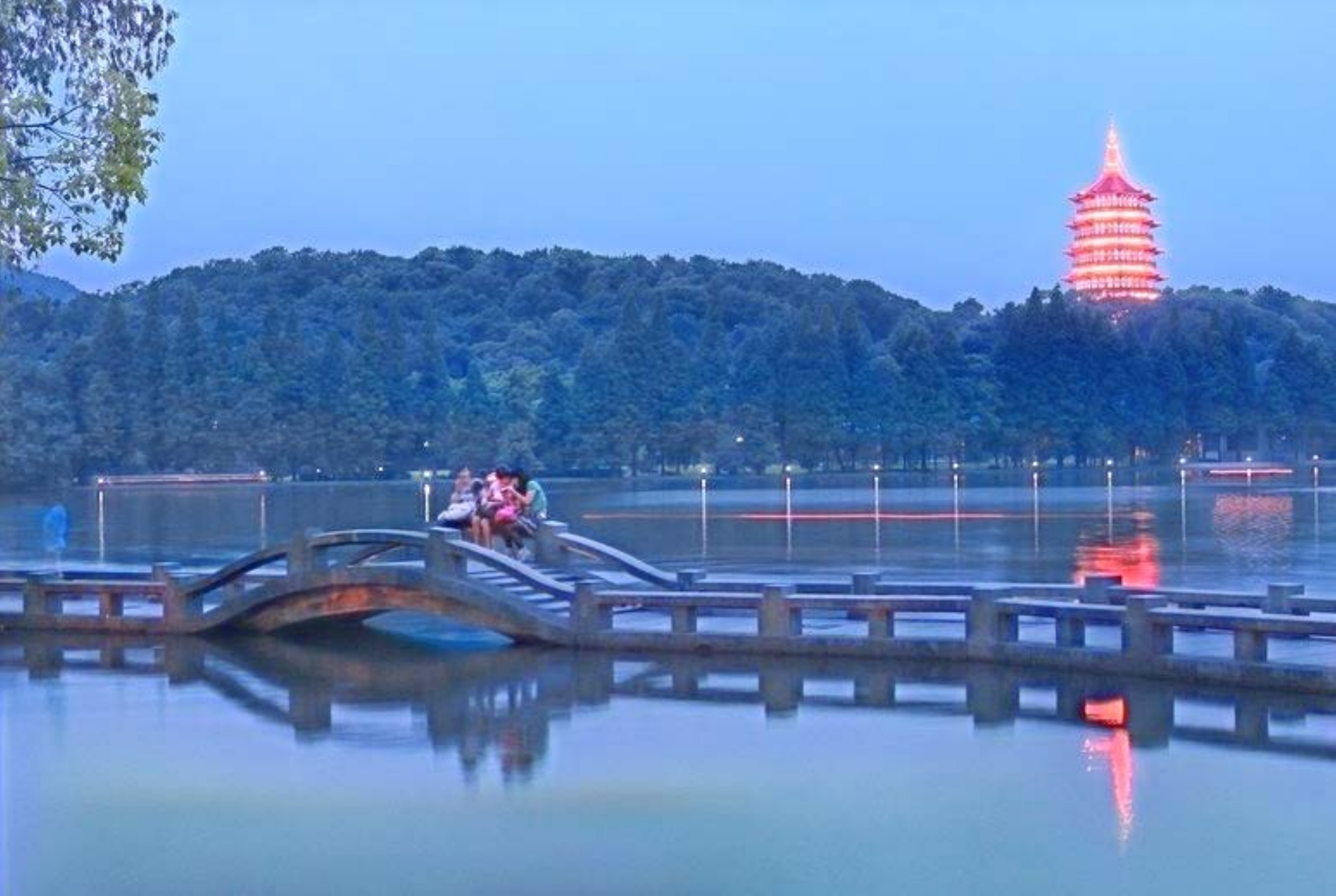}&
		\includegraphics[width=0.113\textwidth]{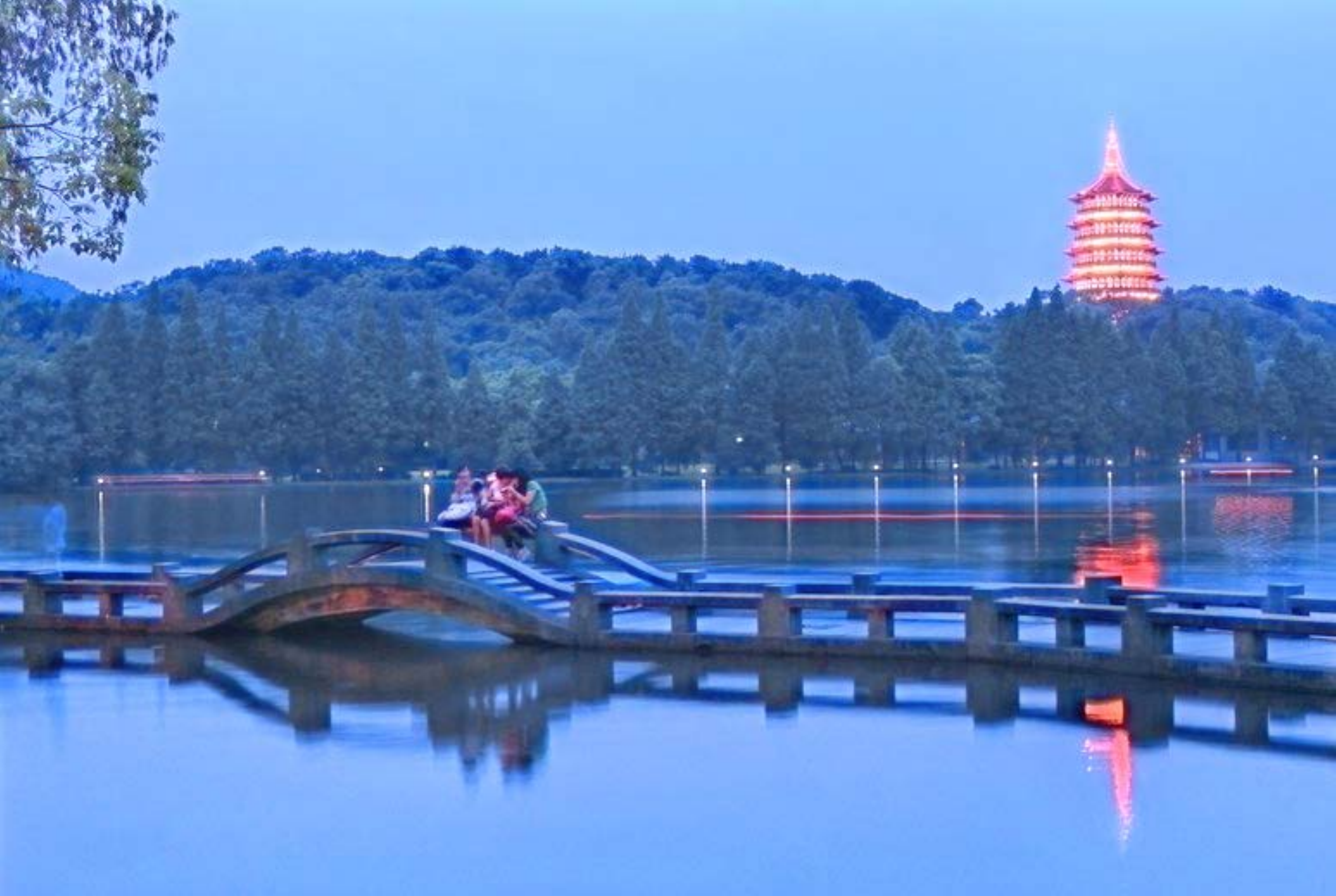}\\				
		\multicolumn{4}{c}{\footnotesize Example B}\\
	\end{tabular}
	\caption{Visual comparison among three versions of CSDNet.}
	\label{fig:RCViscomp}
\end{figure}

\begin{figure}[t]
	\centering
	\begin{tabular}{c@{\extracolsep{0.15em}}c@{\extracolsep{0.3em}}c@{\extracolsep{0.15em}}c}
		\multicolumn{2}{c}{\footnotesize Example A}&\multicolumn{2}{c}{\footnotesize Example B}\\
		\includegraphics[width=0.113\textwidth]{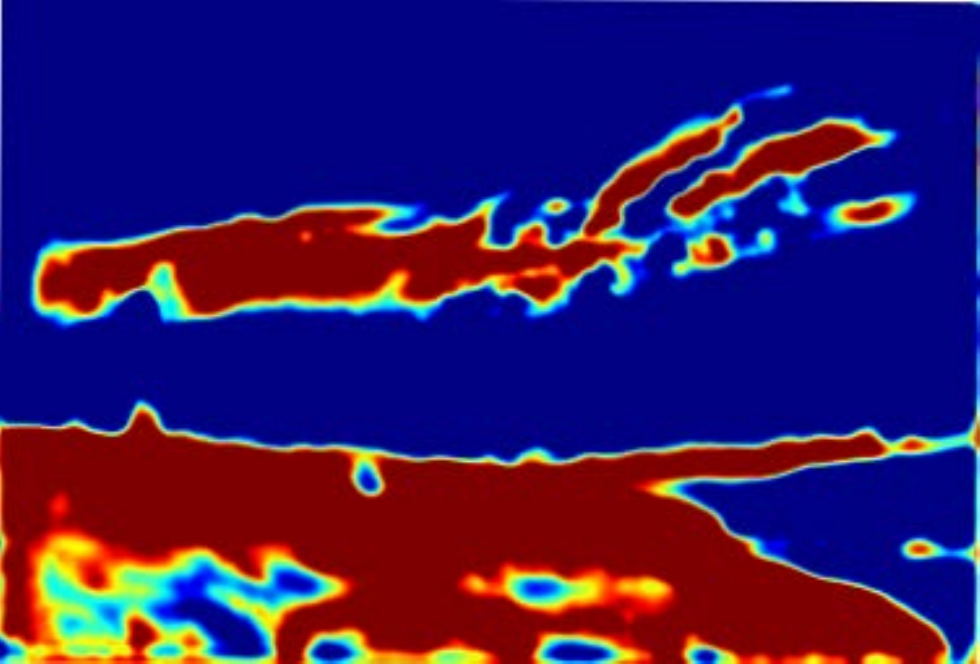}&
		\includegraphics[width=0.113\textwidth]{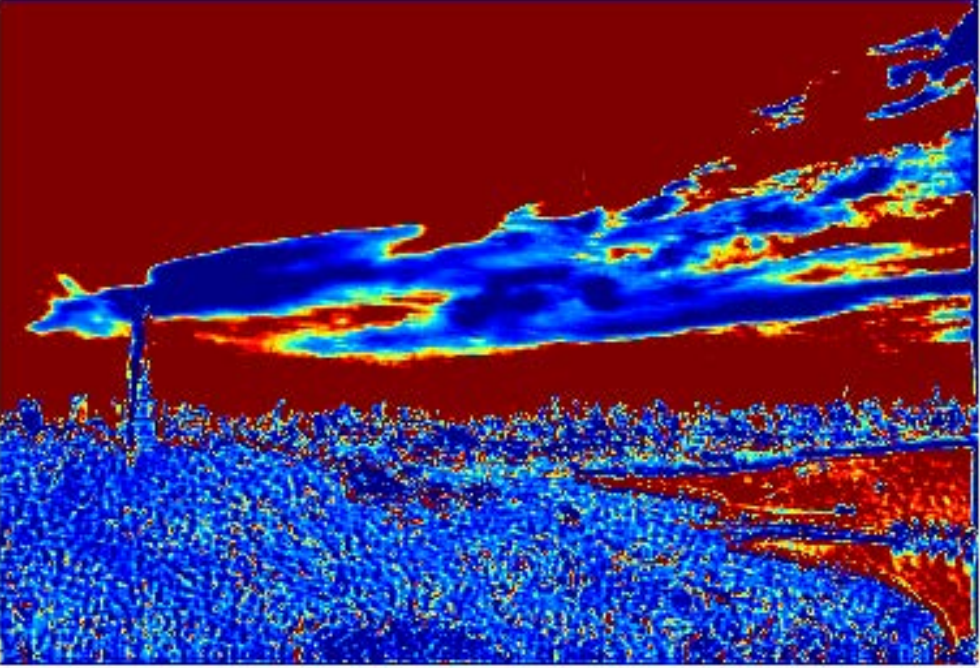}&
		\includegraphics[width=0.113\textwidth]{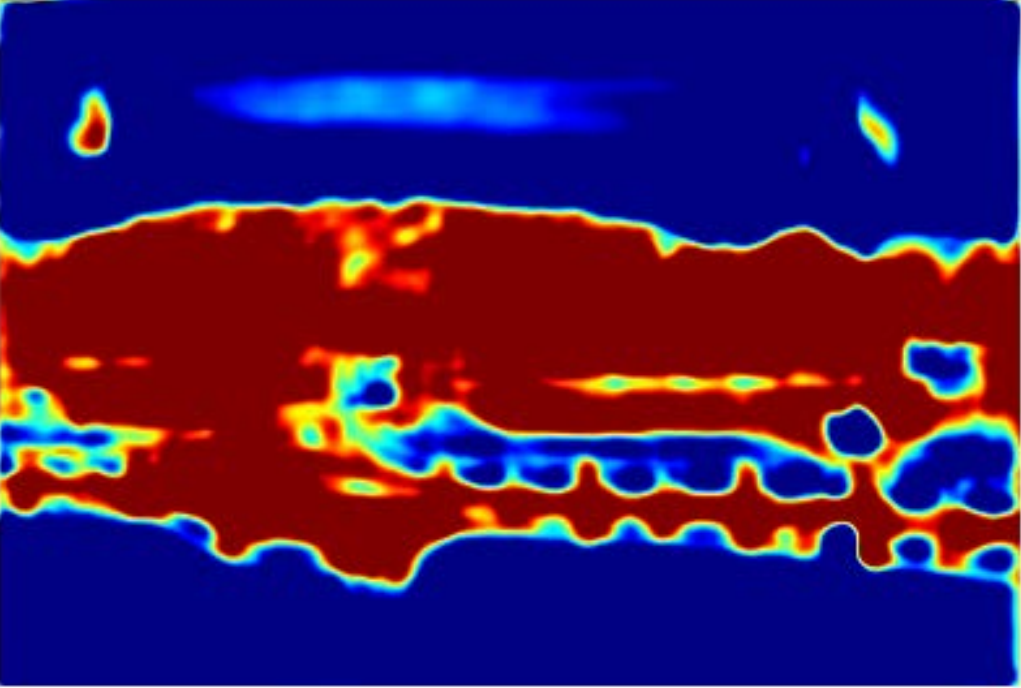}&
		\includegraphics[width=0.113\textwidth]{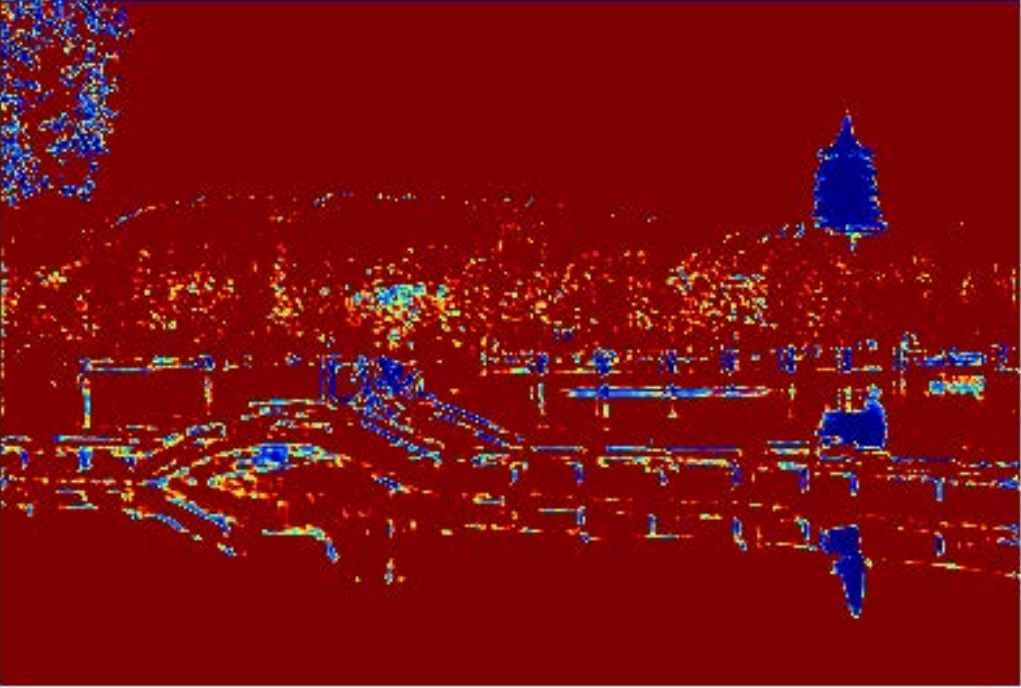}\\
		\multicolumn{4}{c}{\footnotesize Feature maps before the decomposition}\\
		\includegraphics[width=0.113\textwidth]{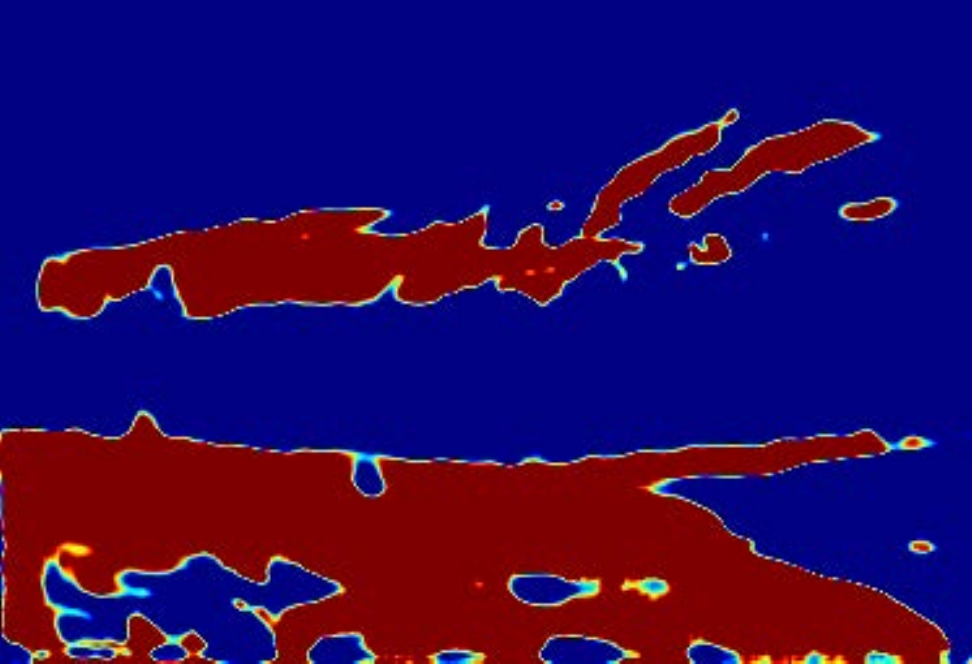}&
		\includegraphics[width=0.113\textwidth]{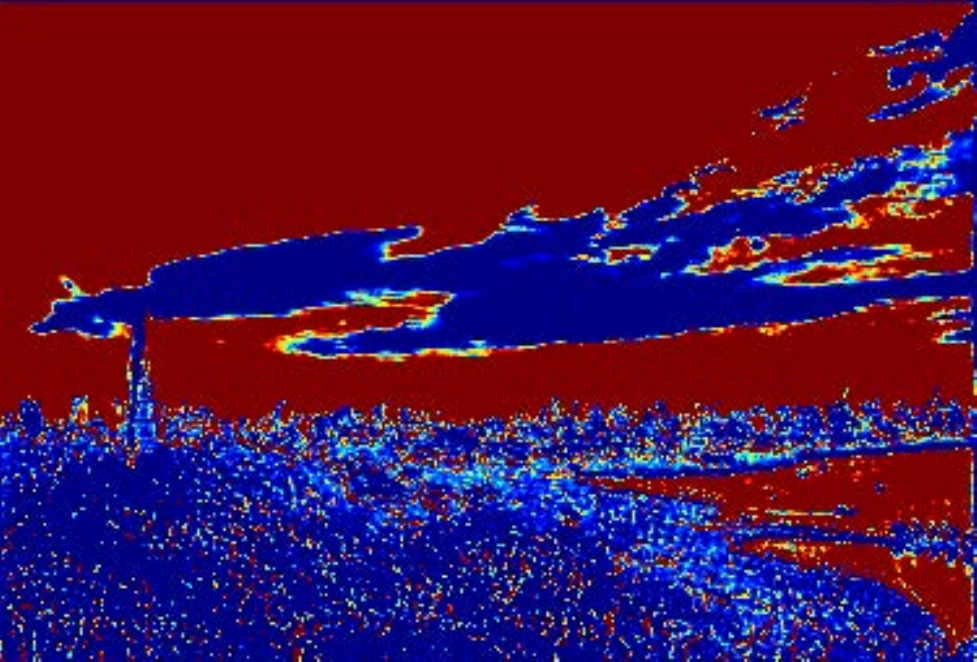}&
		\includegraphics[width=0.113\textwidth]{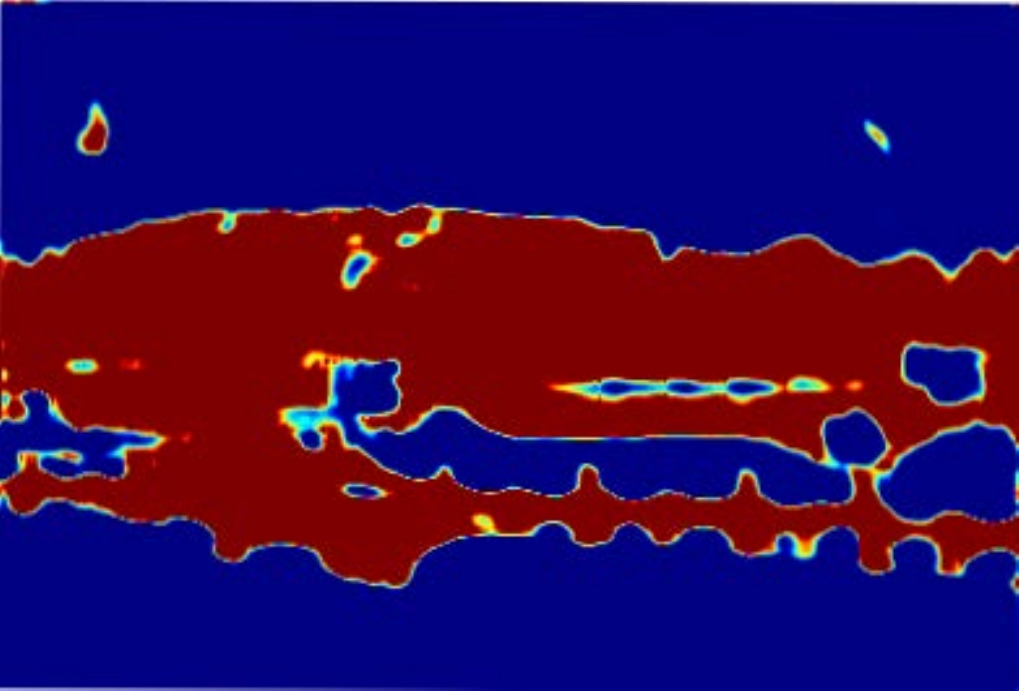}&
		\includegraphics[width=0.113\textwidth]{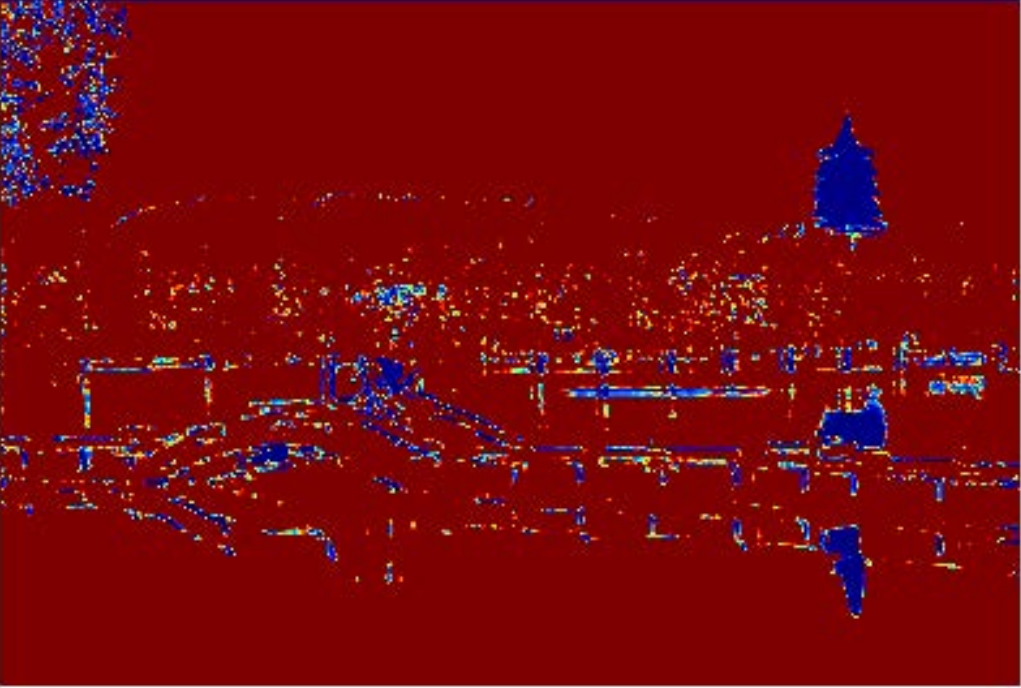}\\
		\multicolumn{4}{c}{\footnotesize Feature maps after the decomposition}\\		
	\end{tabular}
	\caption{Visual comparison of feature maps when executing the context-sensitive decomposition connection in CSDNet$_c$. After the third upsample layer of the decoder, the feature maps are generated in the 32-th channel.}
	\label{fig:RCFeature}
\end{figure}

\subsection{Performance Evaluation of CSDNet}
\textbf{MIT-Adobe FiveK Dataset.}$\;$ Similar to DeepUPE~\cite{wang2019underexposed}, we used MIT-Adobe FiveK as our training dataset. We only need to adopt 500 pairs for training, while DeepUPE utilized 4500 pairs to achieve that. We compared our algorithm with the most representative model-based (LIME~\cite{guo2017lime}), and two end-to-end deep learning (WhiteBox~\cite{hu2018exposure} and DeepUPE~\cite{wang2019underexposed}). 
In Fig.~\ref{fig:MITComparison}, it is obvious that LIME tended to whiten the enhanced results. WhiteBox brought about color distortion and non-uniform exposed performance. DeepUPE weakened the details and textures, resulting in unclear enough.
By comparison, our CSDNet realized the best performance with saturated color presentation and comfortable visual presentation. 
Fig.~\ref{fig:CompDeepUPE} demonstrated visual comparison among DeepUPE and our CSDNet, it indicates that our method estimates the illumination more precisely, account for our designed context-sensitive decomposition on spatial scales and illumination guidance.

\begin{table*}[t]
	\centering
	{
		\caption{Quantitative comparison among different cases. The specific architectures of $\mathtt{Arc}_{a}$ to $\mathtt{Arc}_{f}$ are plotted in fig.~\ref{fig:Arc}. $\mathtt{IG}$ denotes the illumination guidance. }
		\begin{tabular}{c@{\extracolsep{1.1em}}c@{\extracolsep{1.1em}}c@{\extracolsep{1.1em}}c@{\extracolsep{1.1em}}c@{\extracolsep{1.1em}}c@{\extracolsep{1.1em}}c@{\extracolsep{1.1em}}c@{\extracolsep{1.1em}}c@{\extracolsep{1.1em}}c@{\extracolsep{1.1em}}c@{\extracolsep{1.1em}}c@{\extracolsep{1.1em}}c}
			\toprule
			Methods&$S_1$&$S_2$&$S_3$&$S_4$&$S_5$&$S_6$&$S_7$&$S_8$&$S_9$&$S_{10}$&$S_{11}$&Ours\\
			\midrule
			$\mathtt{Arc}_a$&$\surd$&$\surd$&$\times$&$\times$&$\times$&$\times$&$\times$&$\times$&$\times$&$\times$&$\times$&$\times$\\
			$\mathtt{Arc}_b$&$\times$&$\times$&$\surd$&$\surd$&$\times$&$\times$&$\times$&$\times$&$\times$&$\times$&$\times$&$\times$\\
			$\mathtt{Arc}_c$&$\times$&$\times$&$\times$&$\times$&$\surd$&$\surd$&$\times$&$\times$&$\times$&$\times$&$\times$&$\times$\\
			$\mathtt{Arc}_d$&$\times$&$\times$&$\times$&$\times$&$\times$&$\times$&$\surd$&$\surd$&$\times$&$\times$&$\times$&$\times$\\
			$\mathtt{Arc}_e$&$\times$&$\times$&$\times$&$\times$&$\times$&$\times$&$\times$&$\times$&$\surd$&$\surd$&$\times$&$\times$\\
			$\mathtt{Arc}_f$&$\times$&$\times$&$\times$&$\times$&$\times$&$\times$&$\times$&$\times$&$\times$&$\times$&$\surd$&$\surd$\\
			$\mathtt{IG}$&$\times$&$\surd$&$\times$&$\surd$&$\times$&$\surd$&$\times$&$\surd$&$\times$&$\surd$&$\times$&$\surd$\\
			\midrule	
			PSNR&15.0230&17.1924&17.5082&17.7691&17.3863&17.4006&17.7416&19.3865&17.1623&17.2573&19.5499&\textbf{21.6370}\\
			SSIM&0.7227&0.7943&0.7620&0.7960&0.7467&0.8092&0.7809&\textbf{0.8579}&0.7176&0.7504& 0.7888&{0.8526}\\
			\bottomrule
		\end{tabular}
		\label{tab:AblationStudy}
	}
\end{table*}

\begin{table*}[t]	
	\centering
	{
		\caption{Quantitative comparison among state-of-the-art networks and our different versions on the MIT-Adobe FiveK dataset. }	\begin{tabular}{c@{\extracolsep{0.5em}}c@{\extracolsep{0.5em}}c@{\extracolsep{0.5em}}c@{\extracolsep{1em}}c@{\extracolsep{1em}}c@{\extracolsep{1em}}c@{\extracolsep{1em}}c@{\extracolsep{1em}}c}
			\toprule
			\multirow{2}{*}{}&RetinexNet~\cite{Chen2018Retinex}&WhiteBox~\cite{hu2018exposure}&EnlightenGAN~\cite{jiang2019enlightengan}&KinD~\cite{zhang2019kindling}&DeepUPE~\cite{wang2019underexposed}&{CSDNet}&{LiteCSDNet}&{SLiteCSDNet}\\
			\midrule
			PSNR$\uparrow$ &12.8852&17.0580&15.3547&15.5382&16.3446&18.4761&17.0638&17.8976\\
			SSIM$\uparrow$ &0.6654&0.7822&0.7783&0.7772&0.7905&0.8501&0.8289&0.8270\\
			\midrule
			Numbers of Parameters$\downarrow$&2.1103M&25.6901M&8.6455M&8.0343M&2.9963M&{17.2948}M&{0.0602}M&\textbf{0.0301}M\\
			Reduction Rate &(98.57\%)&(99.88\%)&(99.65\%)&(99.63\%)&(99.00\%)&(99.83\%)&(50.00\%)&(---)\\
			\midrule
			Time$\downarrow$ &0.0302s&1.4561s&0.0080s&0.0451s&0.0102s&0.0150s&0.0080s&\textbf{0.0050}s\\
			\bottomrule
		\end{tabular}
		\label{tab:Parameters}
	}
\end{table*}

\textbf{LOL Dataset.}$\;$Following the allocation strategy in the work~\cite{Chen2018Retinex}, we divided this dataset into 485 pairs for training, and 15 pairs for testing. 
Fig.~\ref{fig:VC0} and Table~\ref{tab:LOLTab} report qualitative and quantitative comparisons, respectively. It can be easily seen that noises are sensible in LIME and RetinexNet, insufficient exposure appearance performs in RRM and LightenNet. Our method is outstanding among all compared methods both in numerical scores and visual effects. We then considered two more challenging examples which are also from the LOL dataset, but more difficult to recover with extreme darkness and visible noises. we added a real noises removal network (CBDNet~\cite{guo2019toward}) in all compared approaches (except RRM, which considers the denoising model) to highlight our strengths. As is shown in Fig.~\ref{fig:LOLCBD}, our CSDNet still obtains the best performance. 

\subsection{Performance Evaluation of CSDGAN}
Here, we executed the unpaired supervision manner by utilizing the training pairs presented in EnlightenGAN~\cite{jiang2019enlightengan}, which includes 914 low-light and 1061 normal light images. We further evaluated our trained model in four standard datasets (NPE, NASA, MEF, and LIME) and a few examples from a challenging Exdark dataset.

\textbf{Four Standard Datasets.}$\;$
As is shown in Table~\ref{tab:rescomp}, our CSDGAN obtained competitive performance in all datasets.
From the reported average scores, CSDGAN reached the lowest NIQE, which reflected our CSDGAN generated the most naturalness enhanced results on the statistical level.
Fig.~\ref{fig:VC1} demonstrated that model-based methods are hard to handle the exposure level, as the under-exposure occurs in JIEP and RRM, over-exposure happened in LIME. The result of EnlightenGAN lost the prominent structural expression, especially in the zoomed-in region. In contrast, CSDGAN performed the best visual expression with clear structural and color presentation.

\begin{figure}[t]
	\centering
	{
		\begin{tabular}{c} 
			\includegraphics[width=0.47\textwidth]{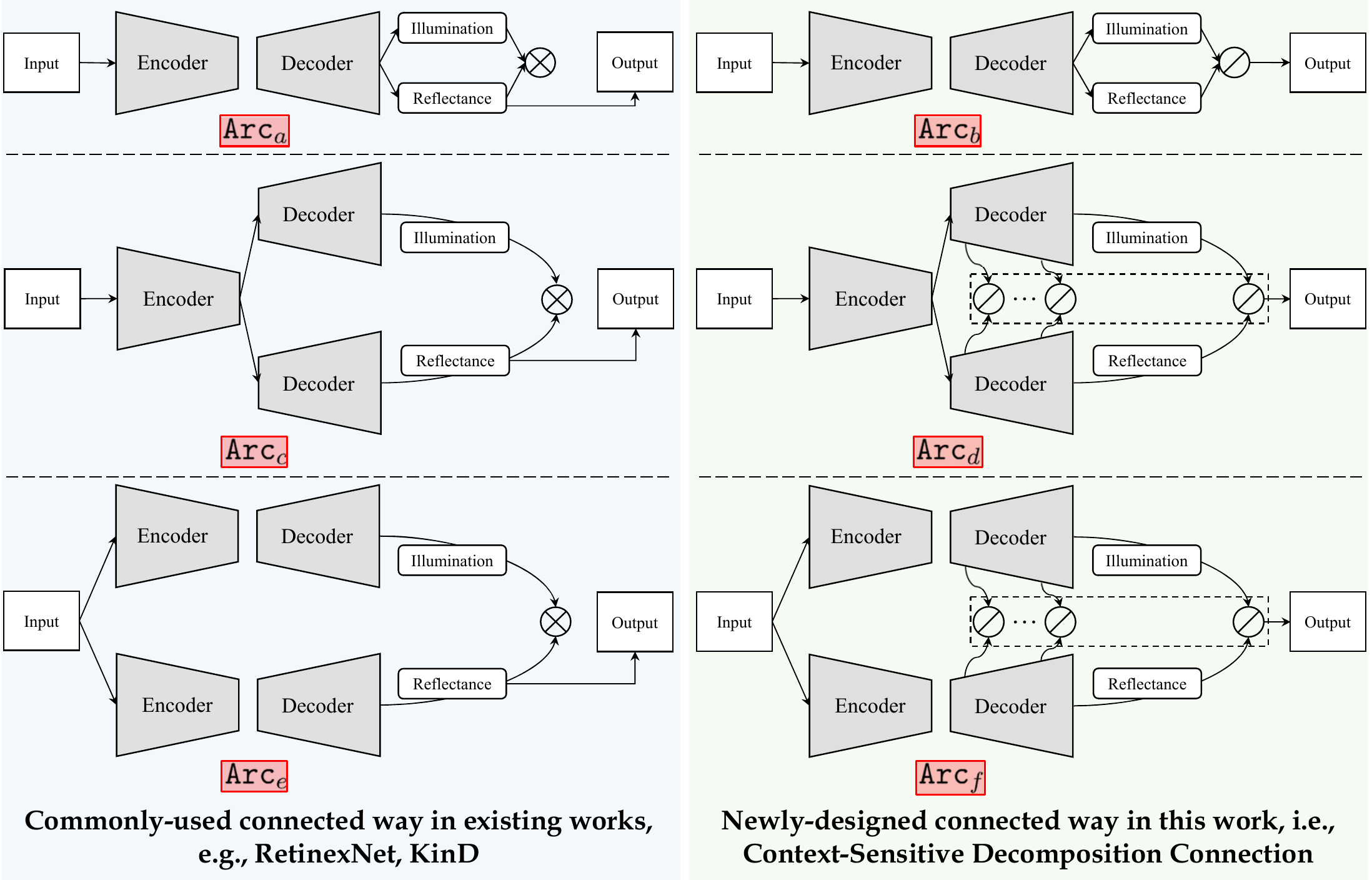}\\
		\end{tabular}
		\caption{The architectures under different connected manners. In which, the commonly-used way in existing works (e.g., RetinexNet~\cite{Chen2018Retinex}, KinD~\cite{zhang2019kindling}) is to connect the input and reconstructed output that is obtained by multiplying the illumination and reflectance. Its final desired output is the reflectance. Different from it, our designed context-sensitive decomposition connection is to remove the generated illumination by using the generated reflectance to obtain the final result. Here we consider three different U-Net based frameworks in the first to third rows. }
	\label{fig:Arc}
}
\end{figure}

\begin{figure*}[t]
	\centering
	{
		\begin{tabular}{c@{\extracolsep{0.25em}}c@{\extracolsep{0.25em}}c@{\extracolsep{0.25em}}c@{\extracolsep{0.25em}}c@{\extracolsep{0.25em}}c} 
			\includegraphics[width=0.158\textwidth]{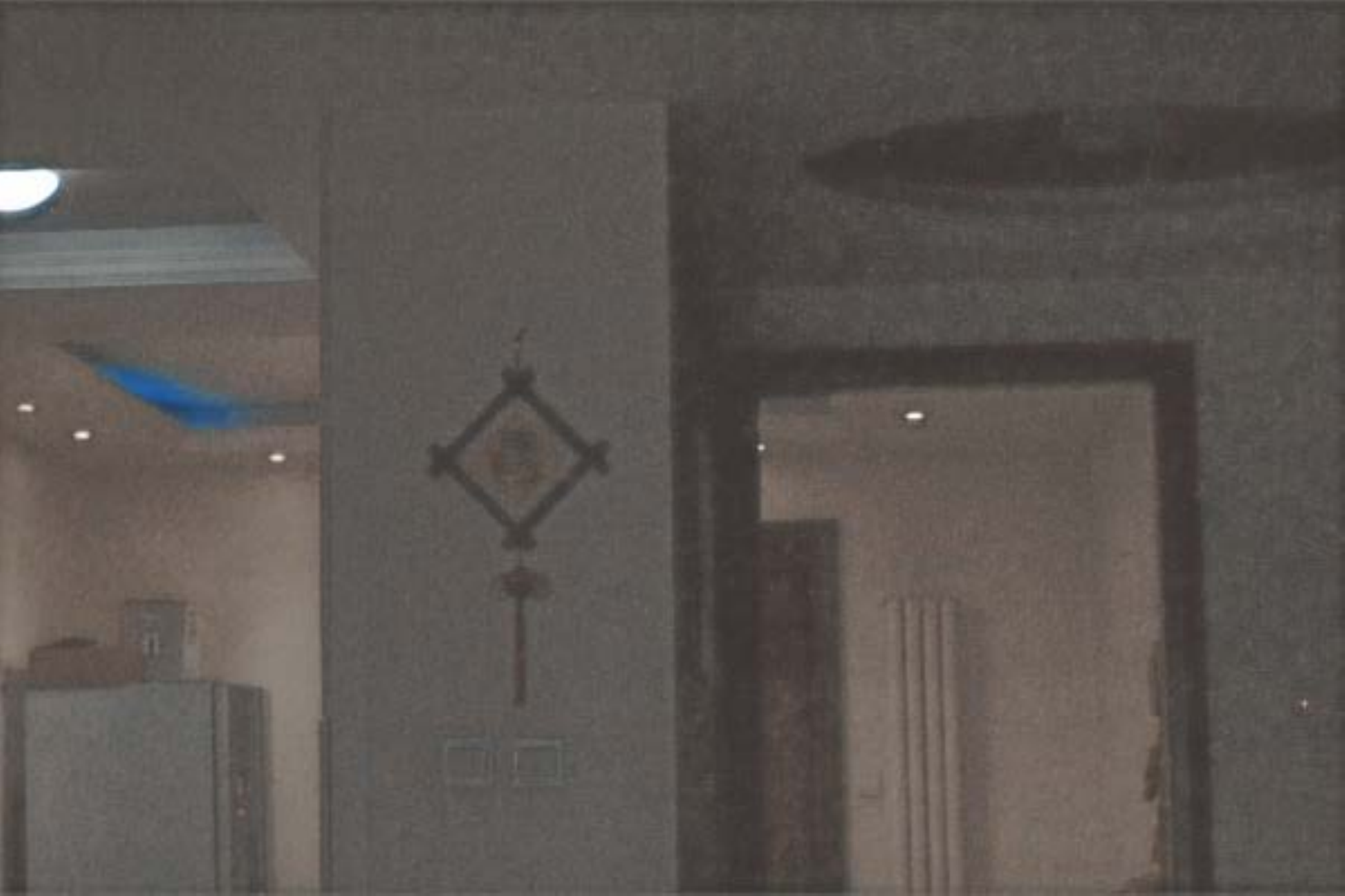}&
			\includegraphics[width=0.158\textwidth]{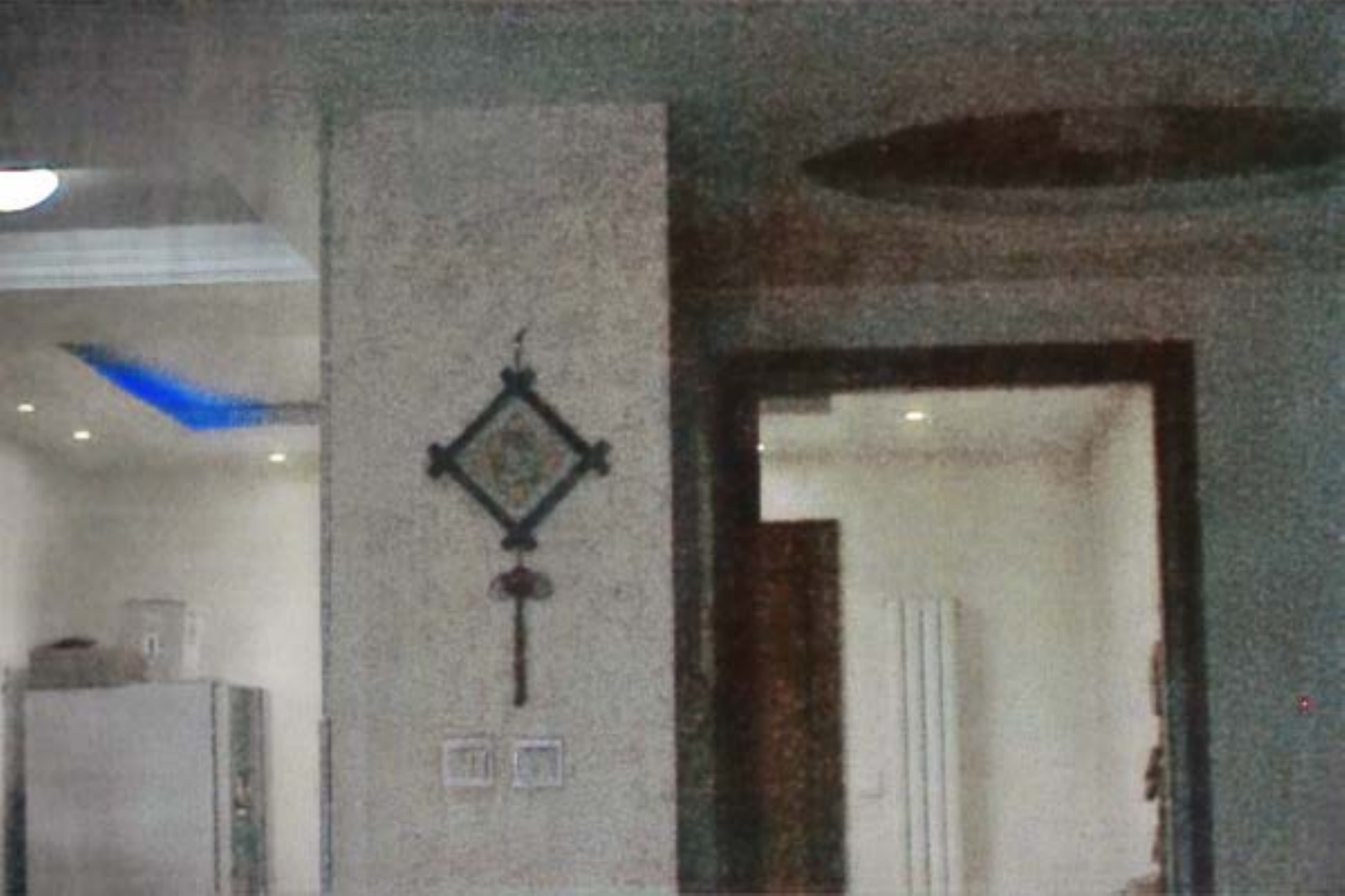}&
			\includegraphics[width=0.158\textwidth]{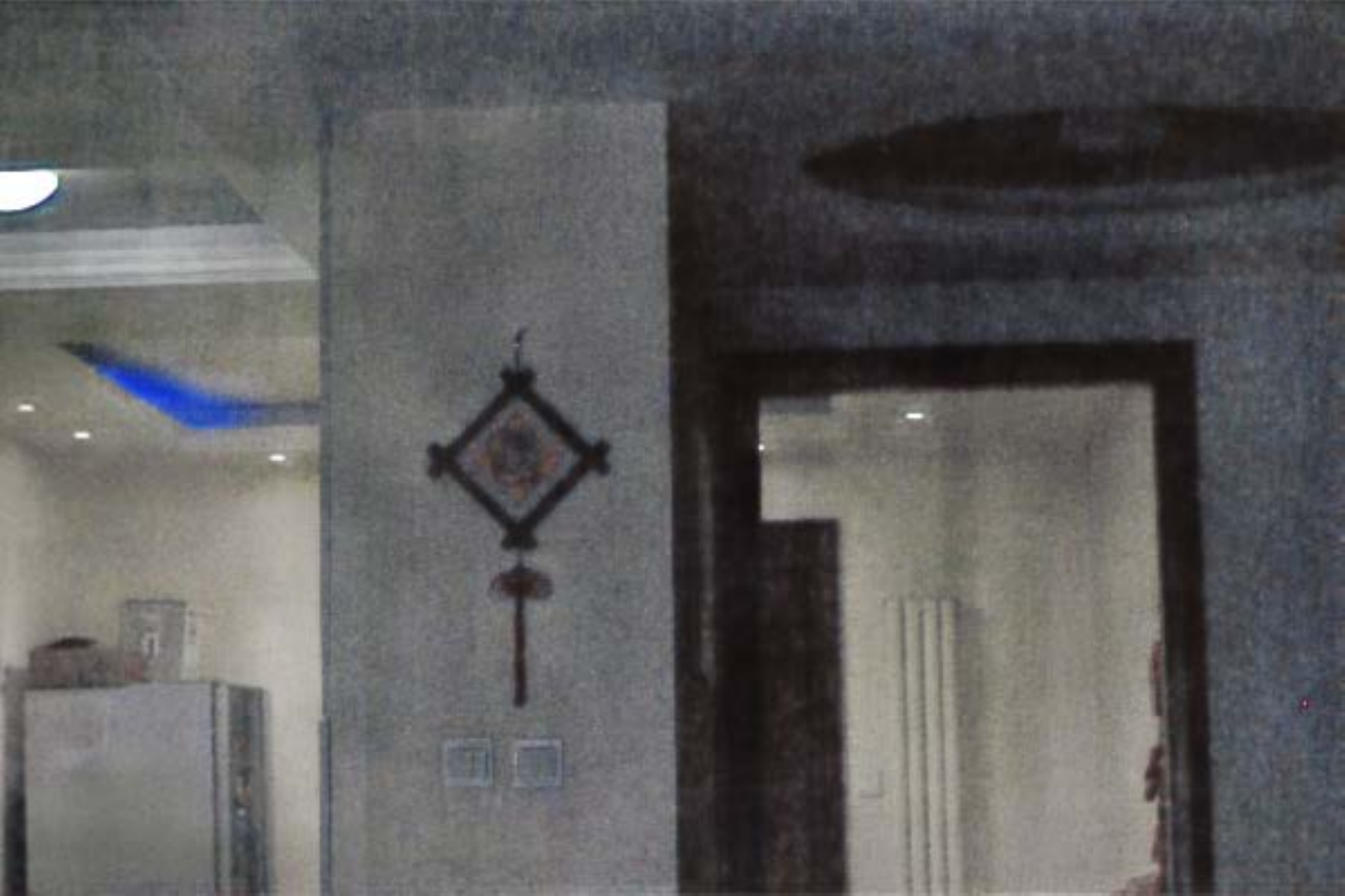}&
			\includegraphics[width=0.158\textwidth]{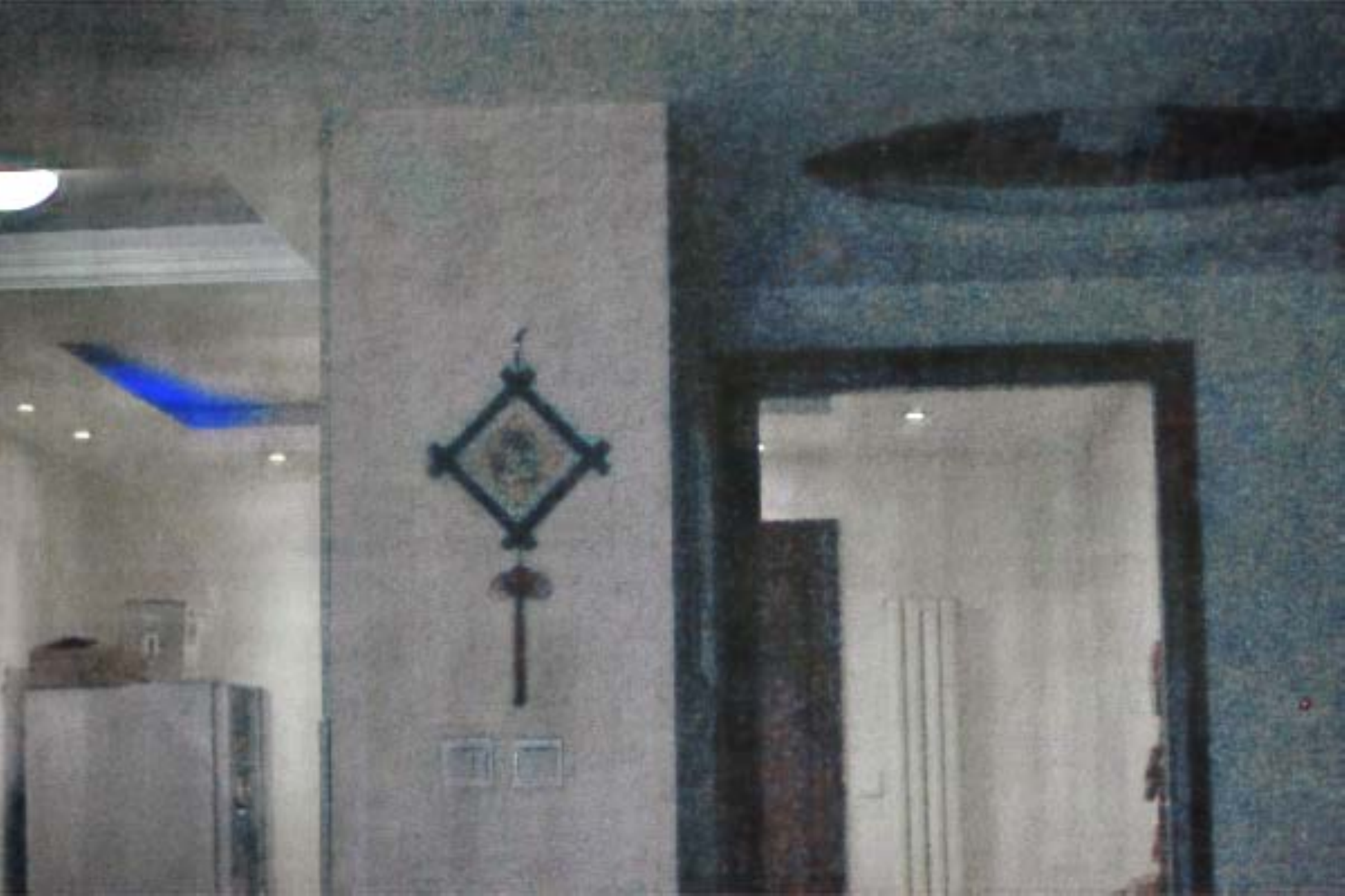}&
			\includegraphics[width=0.158\textwidth]{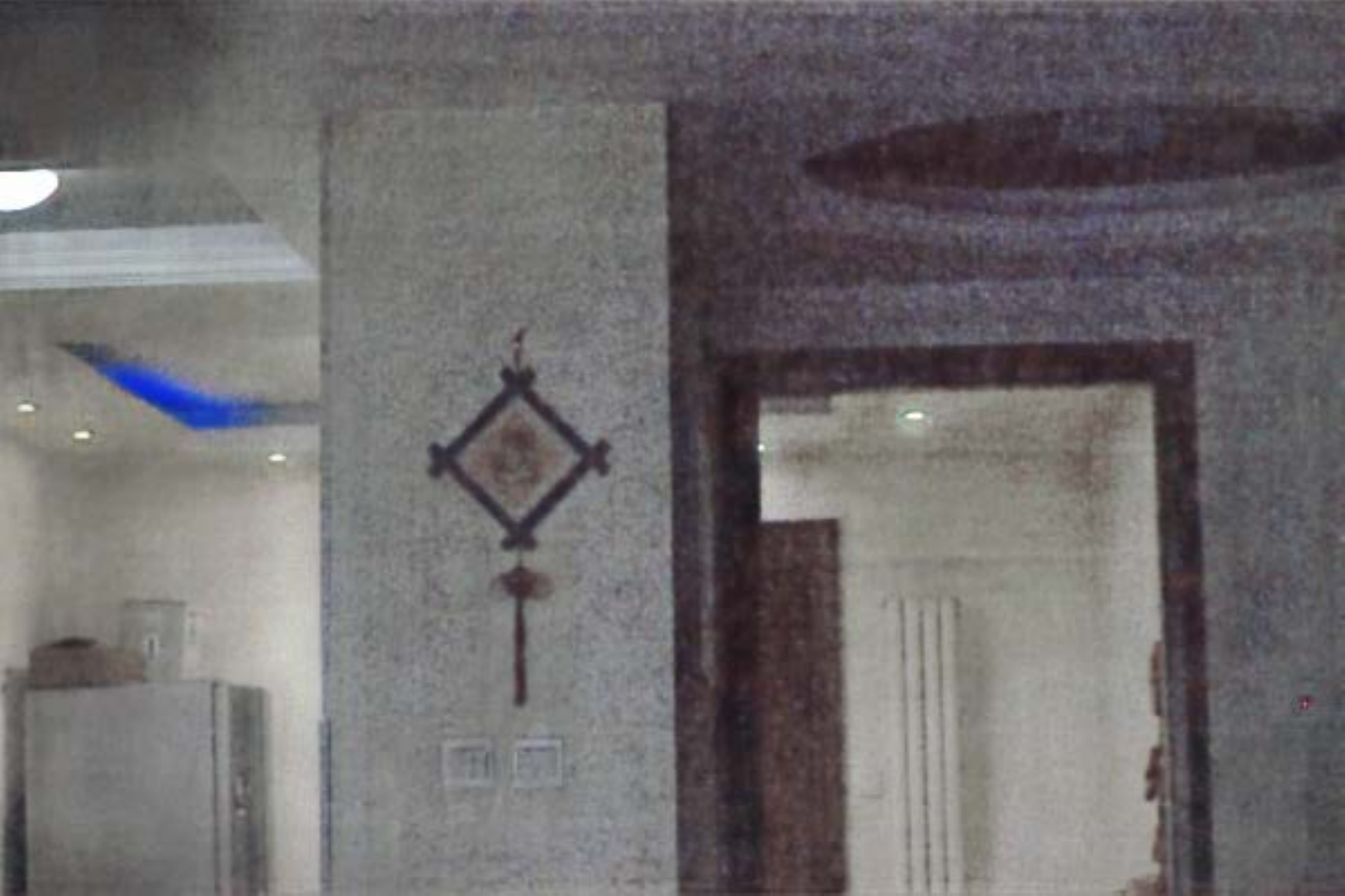}&
			\includegraphics[width=0.158\textwidth]{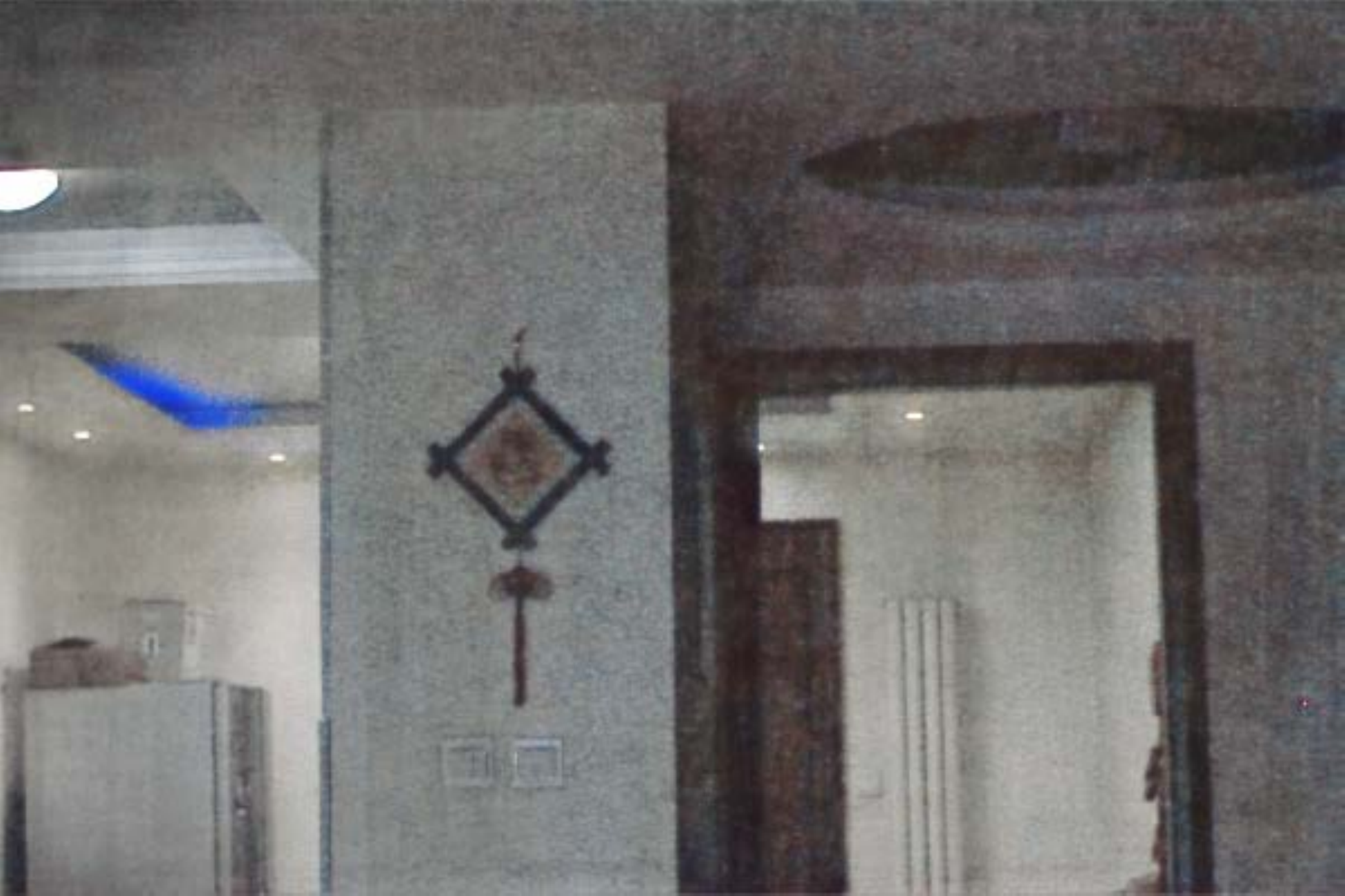}\\
			\footnotesize $S_1$&\footnotesize $S_2$&\footnotesize $S_3$&\footnotesize $S_4$&\footnotesize $S_5$&\footnotesize $S_6$\\
			\includegraphics[width=0.158\textwidth]{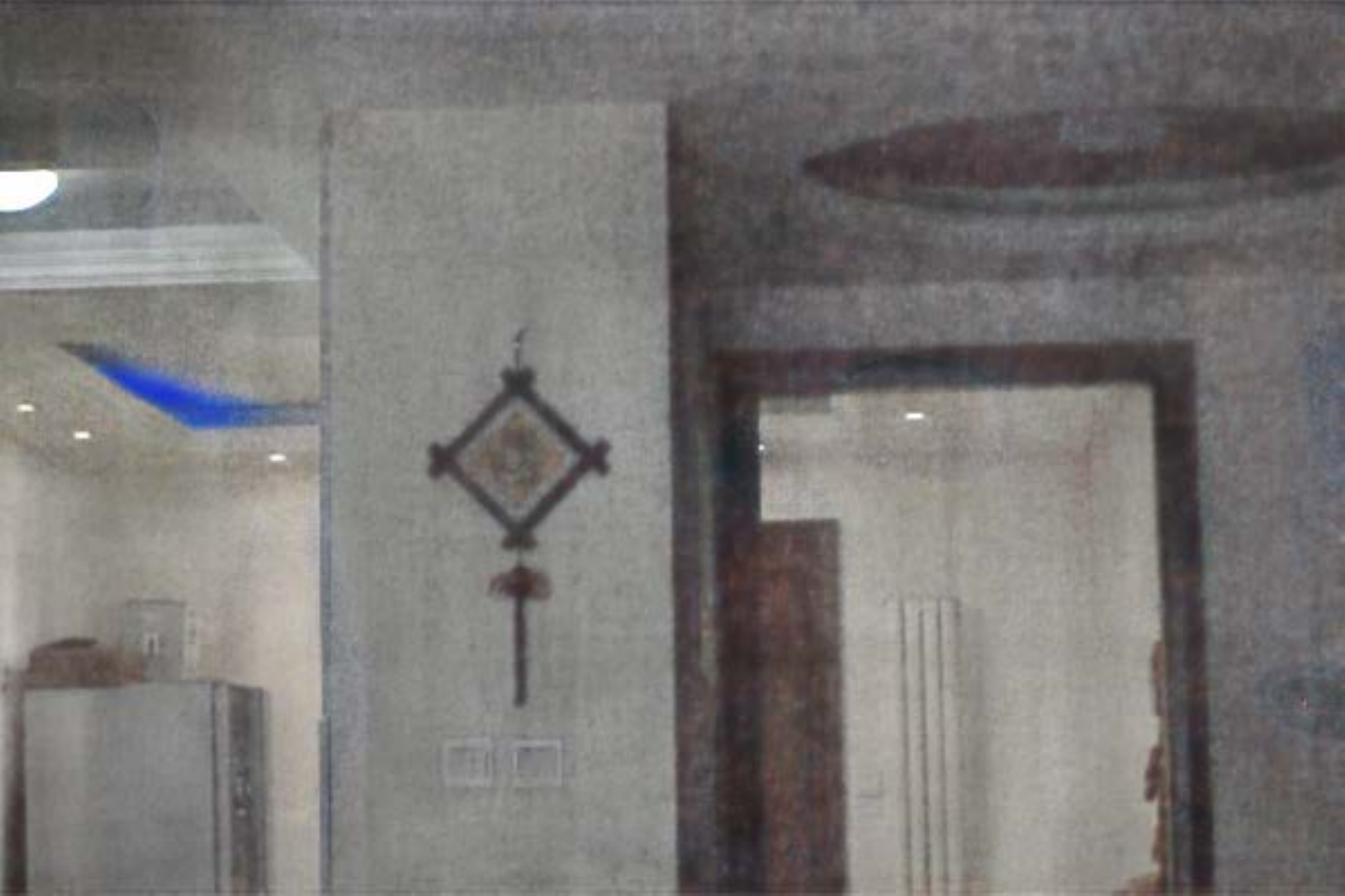}&
			\includegraphics[width=0.158\textwidth]{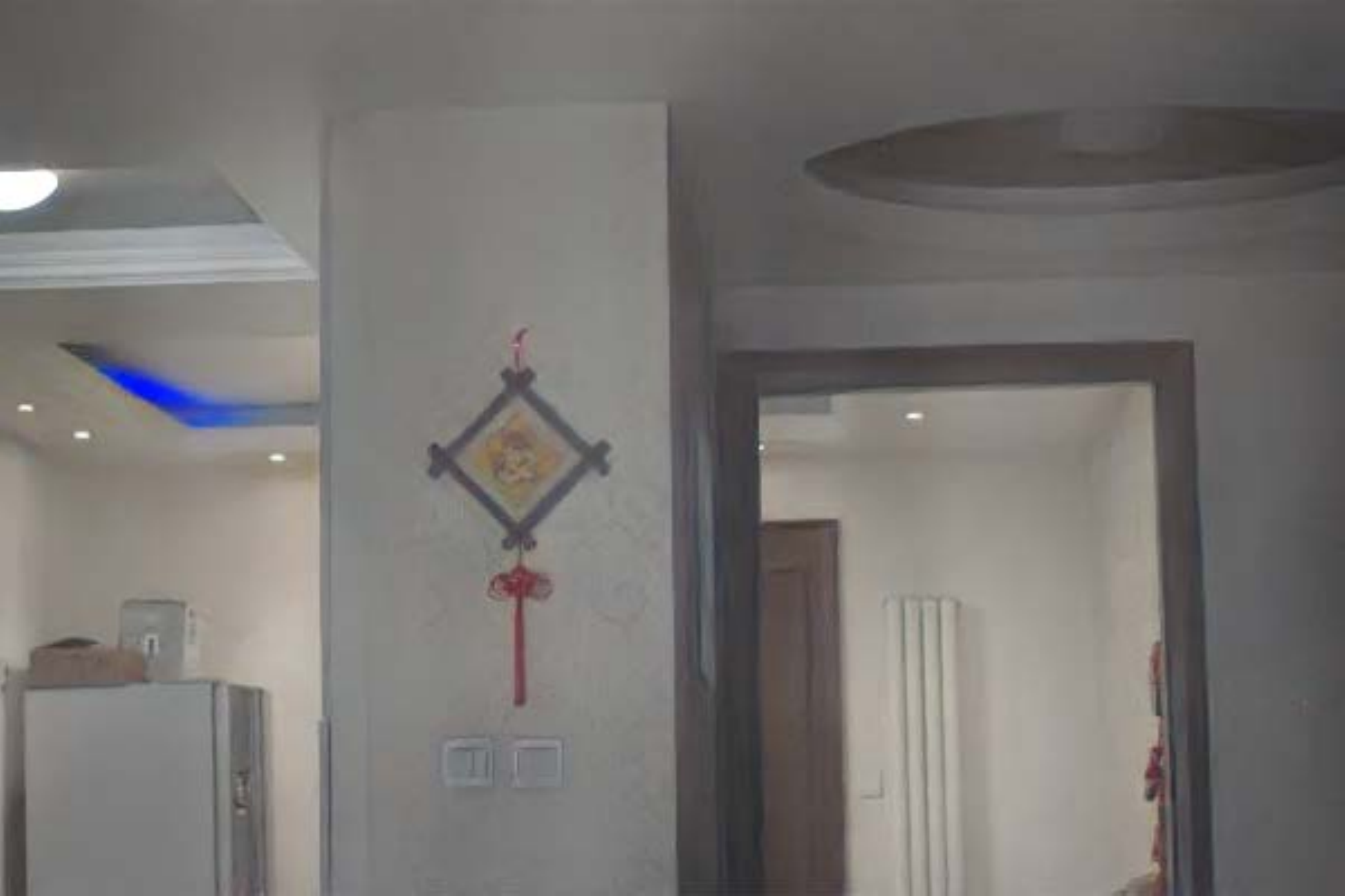}&
			\includegraphics[width=0.158\textwidth]{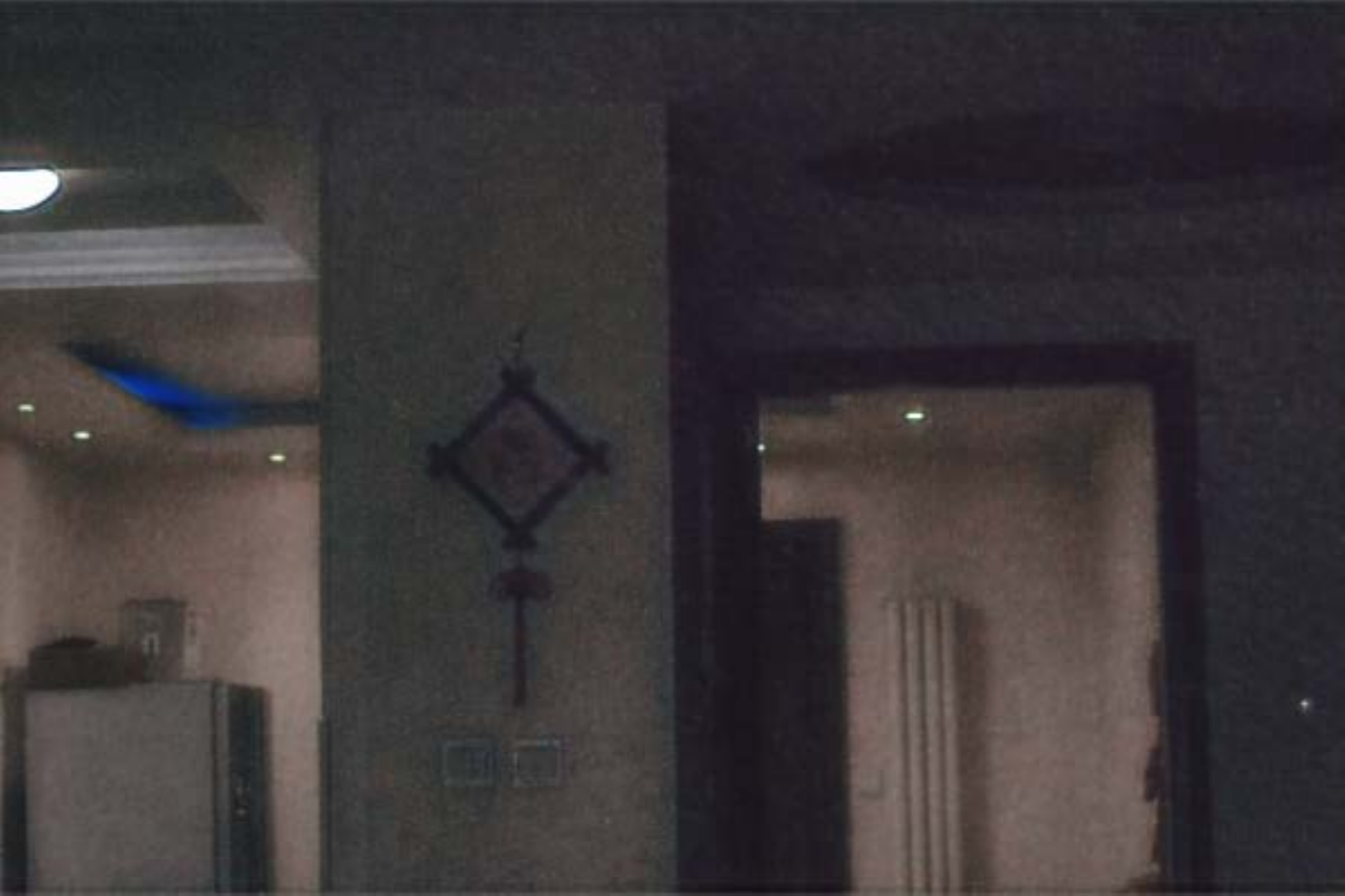}&
			\includegraphics[width=0.158\textwidth]{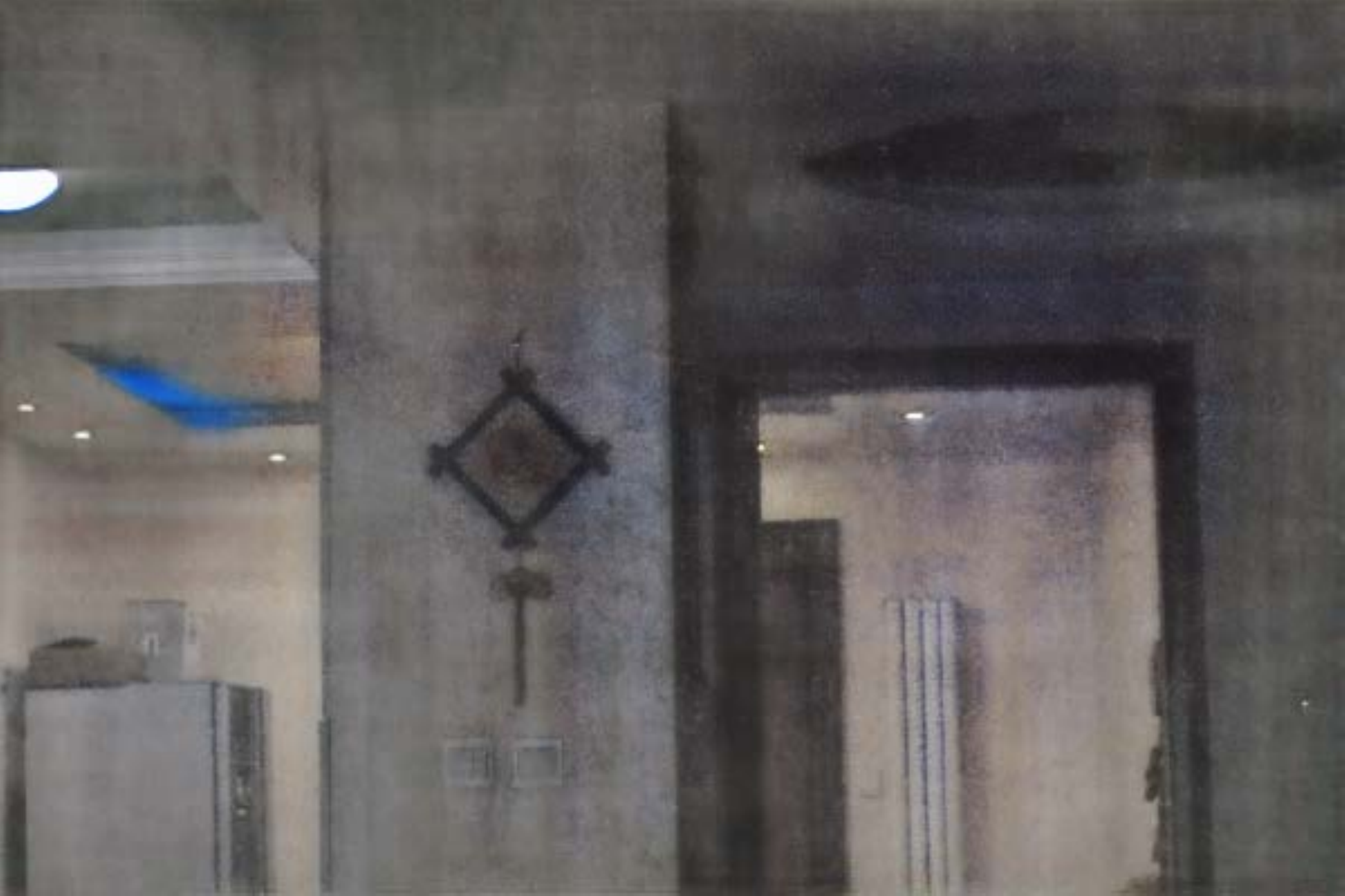}&
			\includegraphics[width=0.158\textwidth]{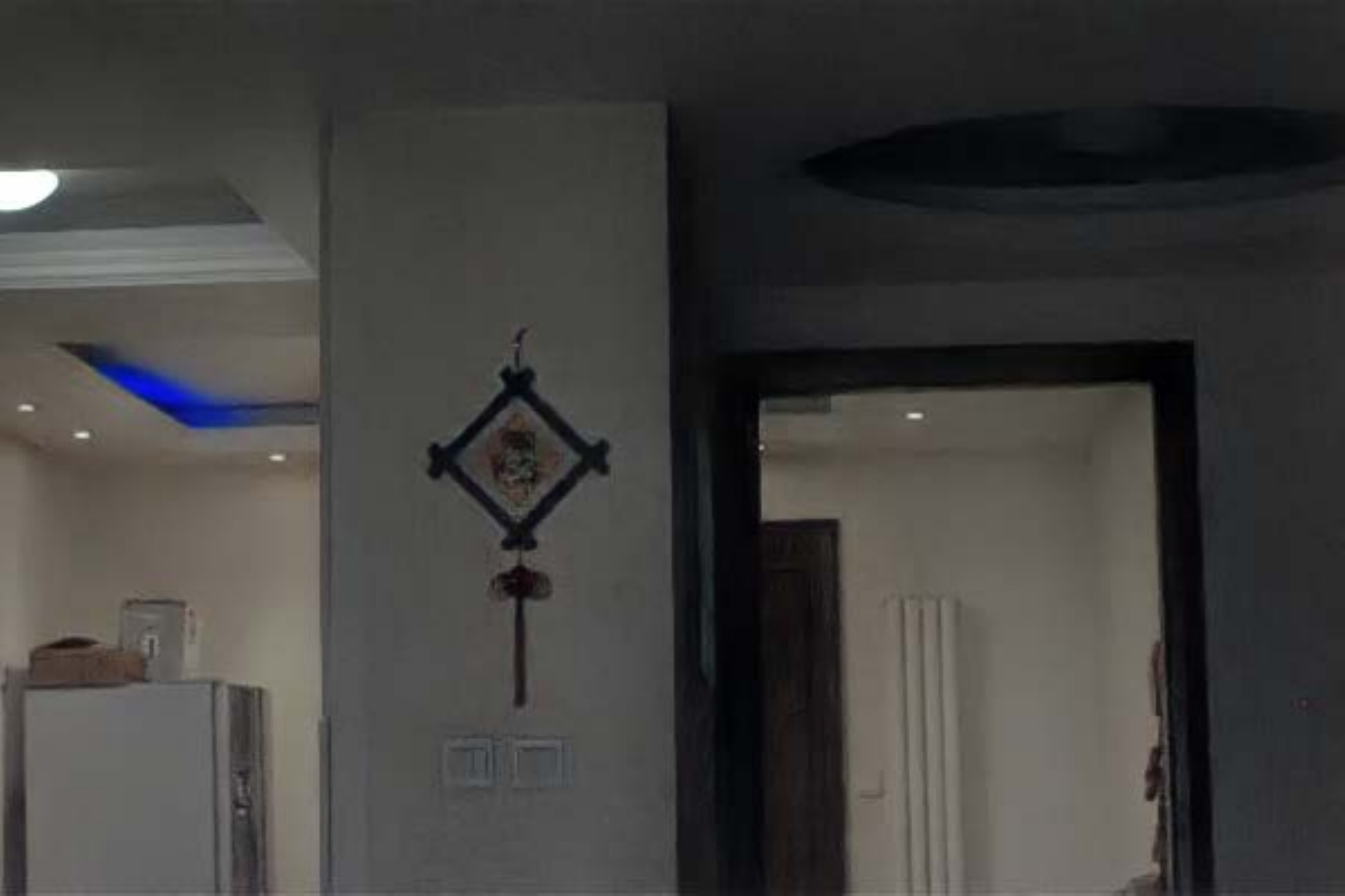}&
			\includegraphics[width=0.158\textwidth]{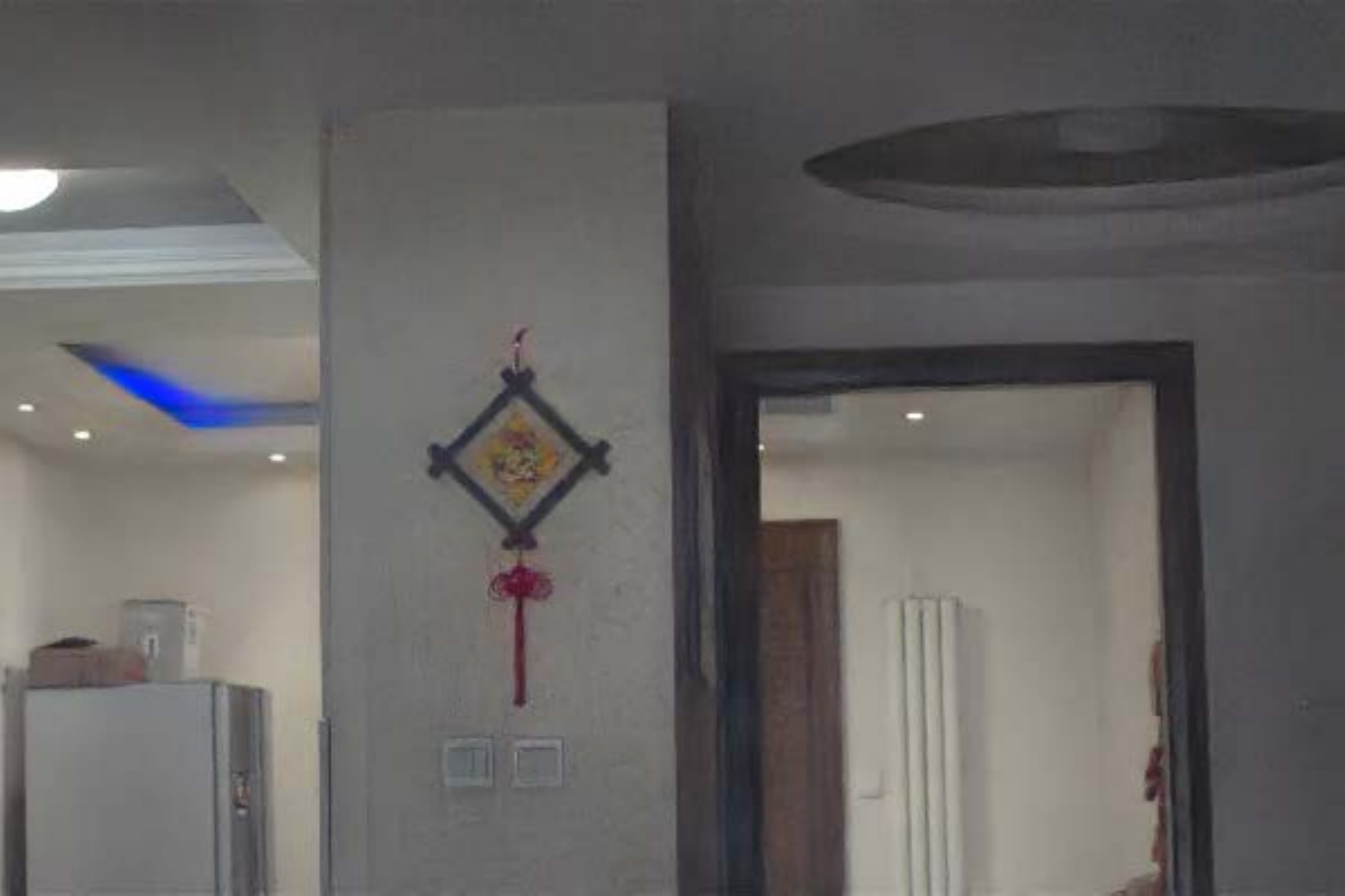}\\
			\footnotesize $S_7$&\footnotesize $S_8$&\footnotesize $S_9$&\footnotesize $S_{10}$&\footnotesize $S_{11}$&\footnotesize Ours\\
		\end{tabular}
		\caption{Visual comparison of different cases presented in Table~\ref{tab:AblationStudy}.}
		\label{fig:AblationAll}
	}
\end{figure*}

Furthermore, we conducted two groups of analytical evaluations. In Fig.~\ref{fig:CompGAN}, we can easily find that EnlightenGAN presented color distortion and unknown artifacts because of ignoring the exploitation of contextual dependencies and utilizing the physical principle. The performance of our CSDGAN was more satisfying.
Fig.~\ref{fig:CompUPE} demonstrated the enhanced result and its estimated illumination among RetinexNet, DeepUPE, and CSDGAN. It indicated that RetinexNet produced unnatural performance with redundant details. DeepUPE failed to depict the correct edge information (see the green zoomed-in region) and the appropriate exposure (see the red zoomed-in region). The result of CSDGAN kept better exposure and expression. Visual comparisons of the illumination layer further realized our superiority.

\textbf{More Challenging Cases in ExDark Dataset.}$\;$
Ulteriorly, we considered two more challenging cases in more complex real-world scenarios (e.g., low-resolution and low-quality). As is shown in Fig.~\ref{fig:MoreComp}, we can readily observe that these advanced end-to-end deep learning approaches are occasionally unstable (e.g., under-exposure, block artifacts, and noises) for these challenging images. On the contrary, our CSDGAN achieved the best exposure and visual performance, especially in the zoomed-in regions. It manifests that the higher generalization ability and practical values in terms of our CSDGAN.

\begin{figure}[t]
	\centering
	{
	\begin{tabular}{c@{\extracolsep{0.2em}}c@{\extracolsep{0.2em}}c@{\extracolsep{0.2em}}c}
		\footnotesize (a)&\footnotesize (b)&\footnotesize (c)&\footnotesize (d)\\
		\includegraphics[width=0.113\textwidth]{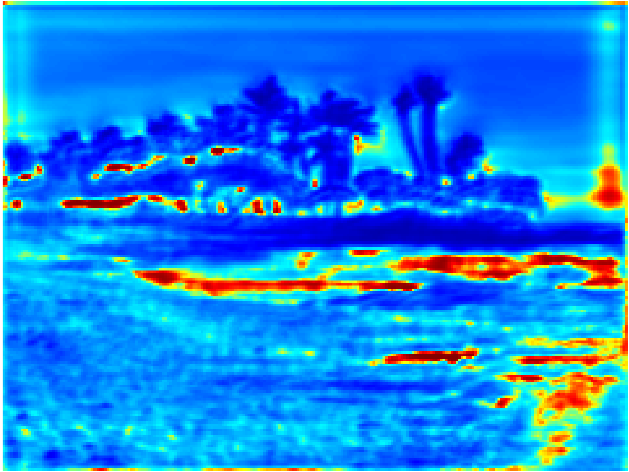}&
		\includegraphics[width=0.113\textwidth]{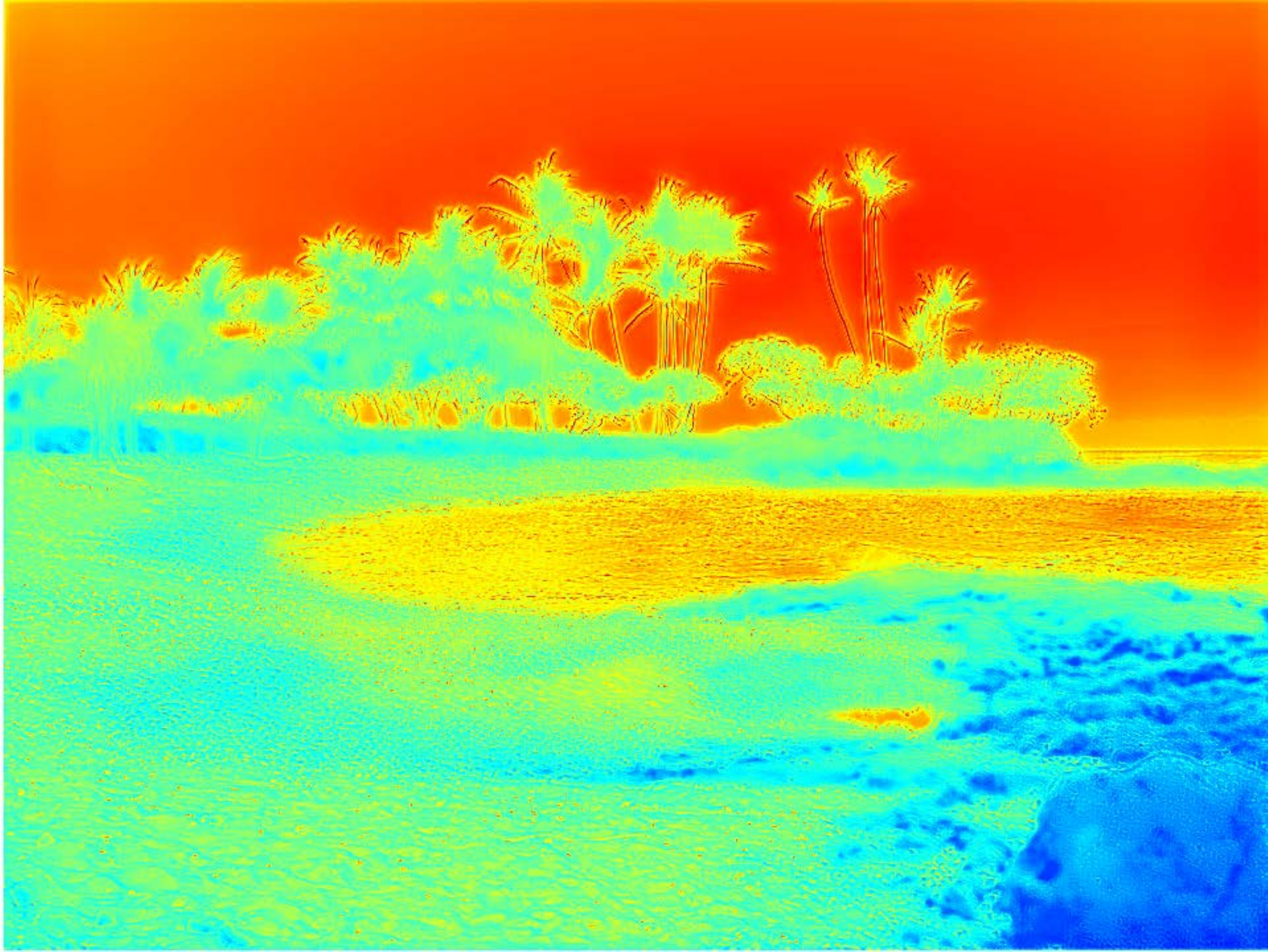}&
		\includegraphics[width=0.113\textwidth]{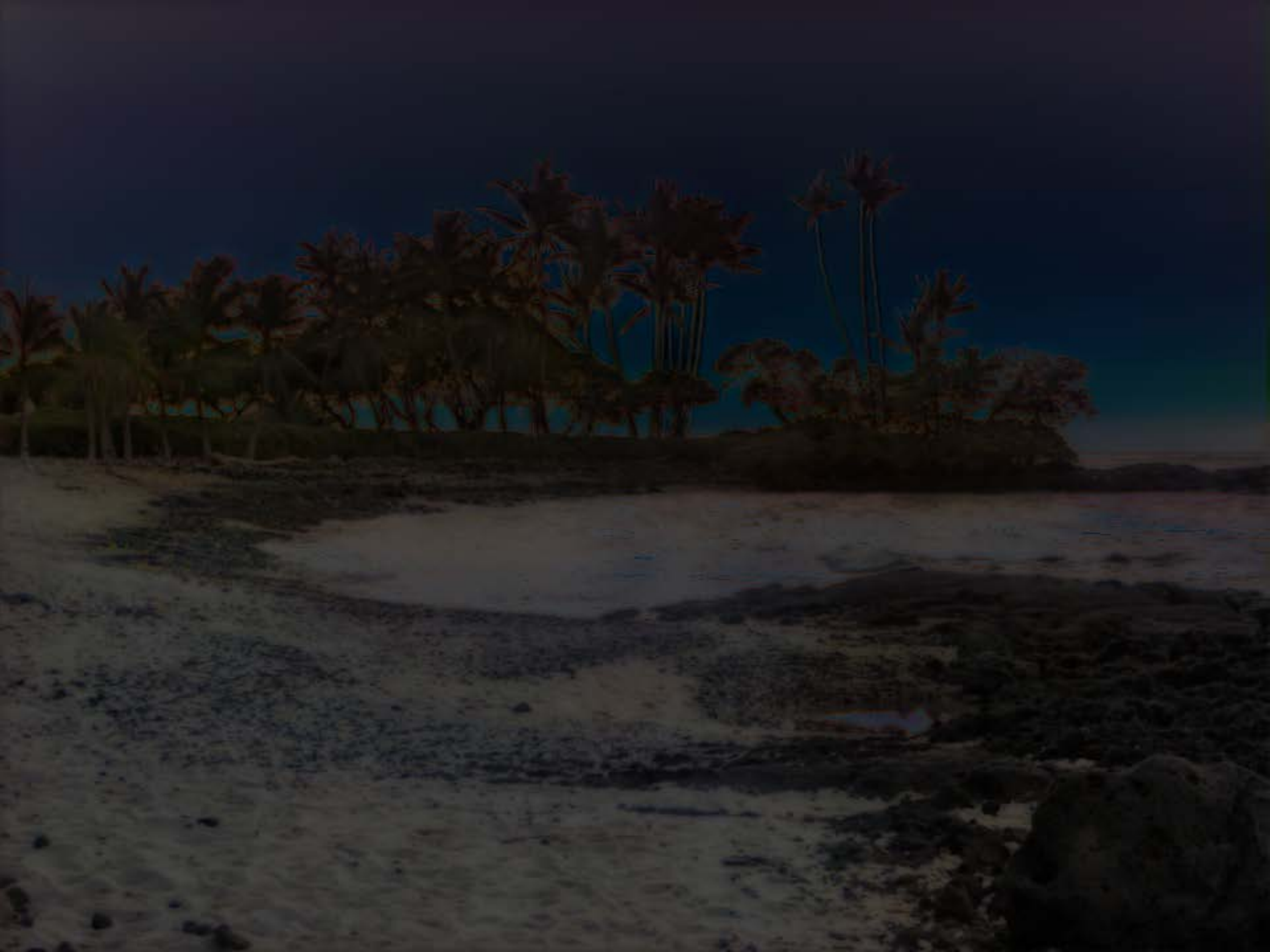}&
		\includegraphics[width=0.113\textwidth]{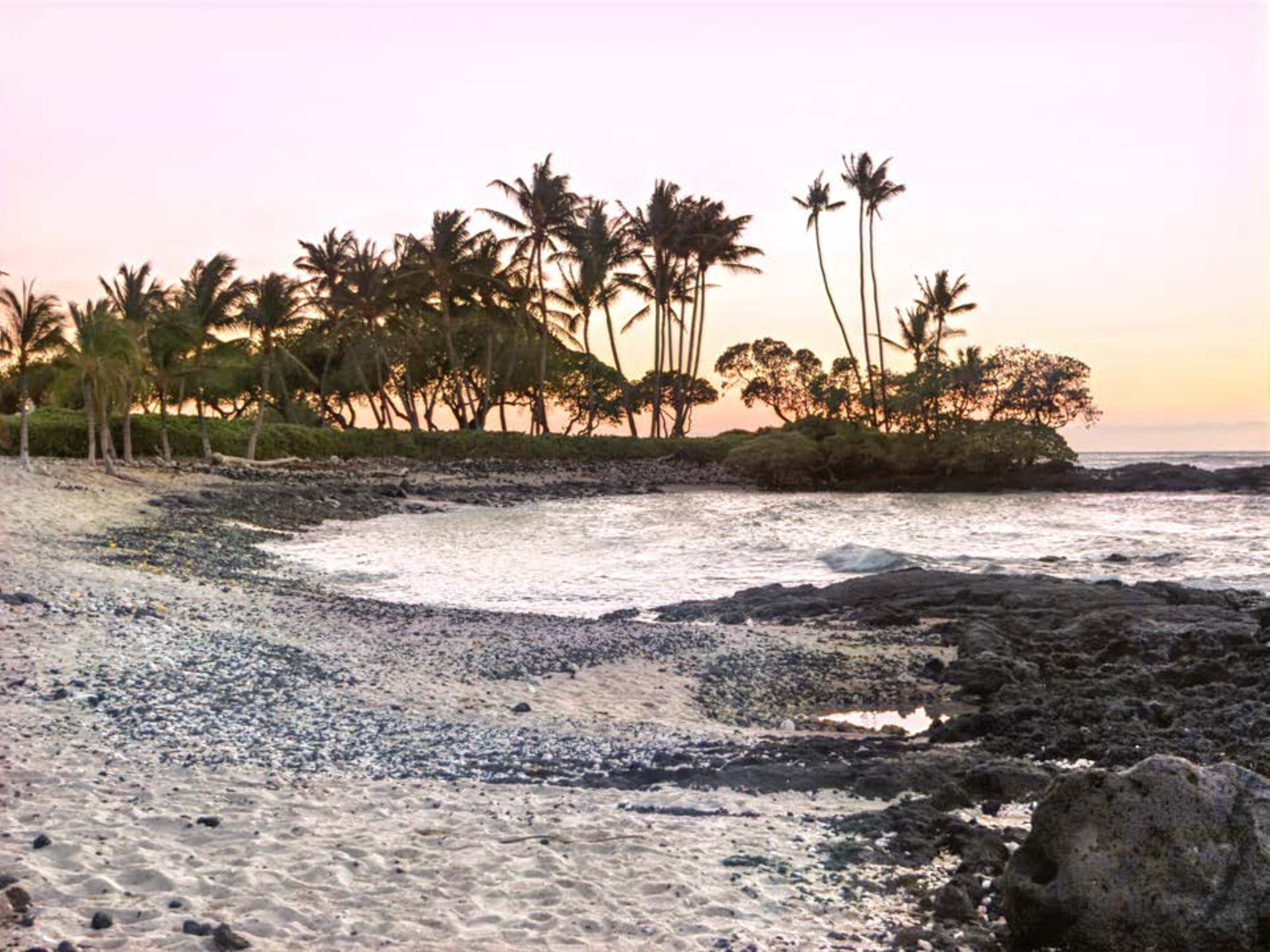}\\
		\multicolumn{4}{c}{\footnotesize w/o Illumination Guidance}\\
		\includegraphics[width=0.113\textwidth]{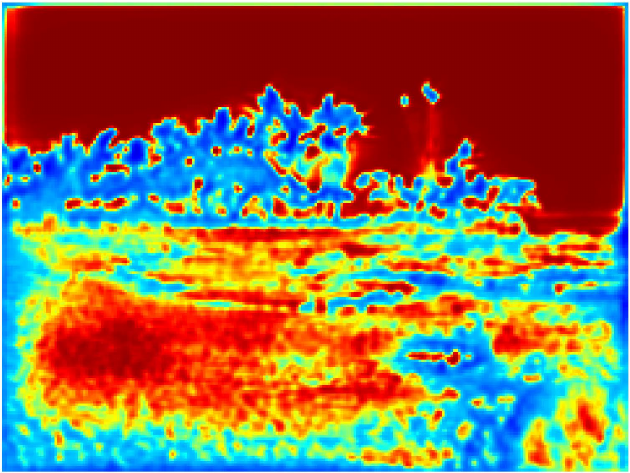}&
		\includegraphics[width=0.113\textwidth]{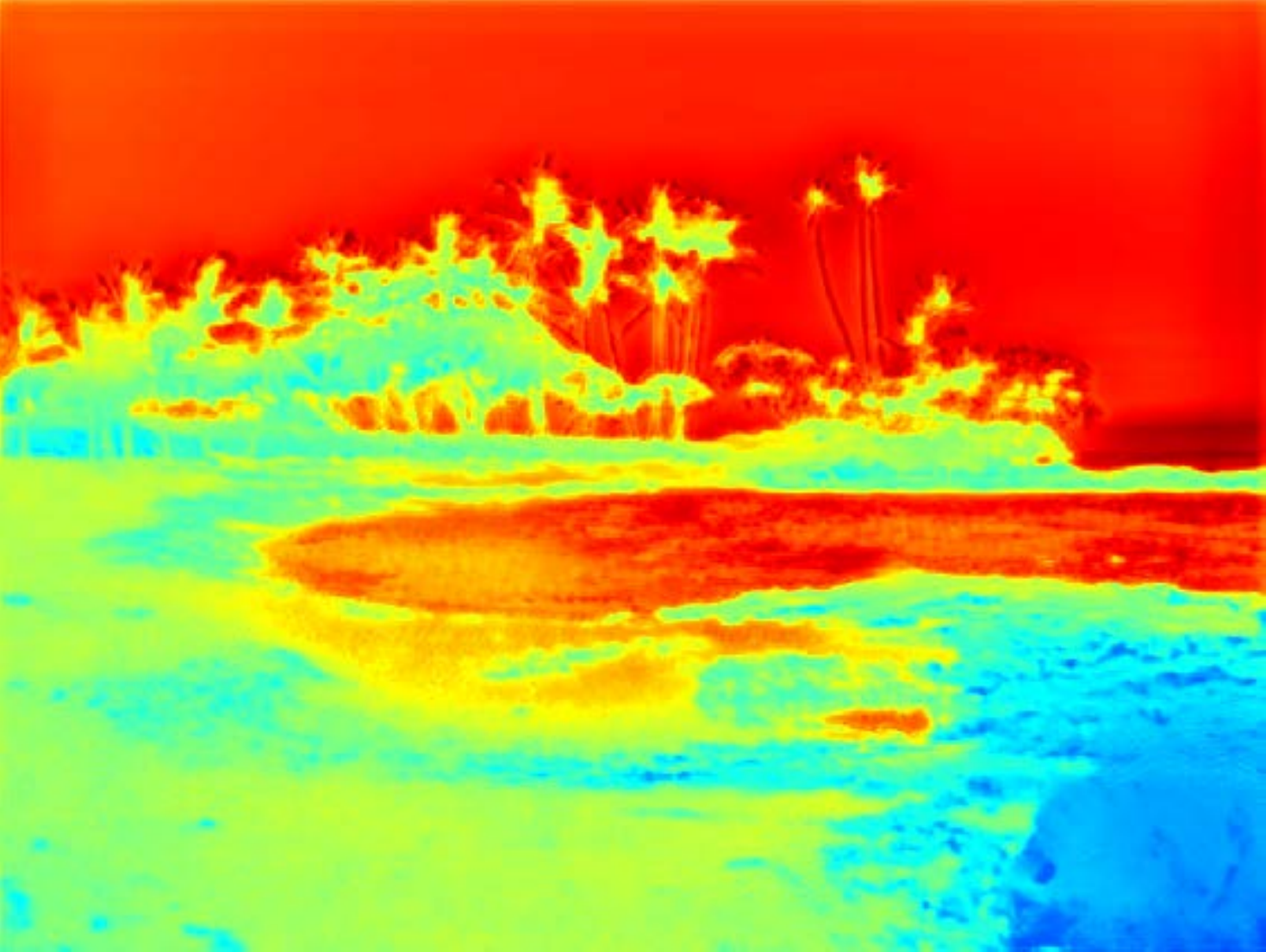}&
		\includegraphics[width=0.113\textwidth]{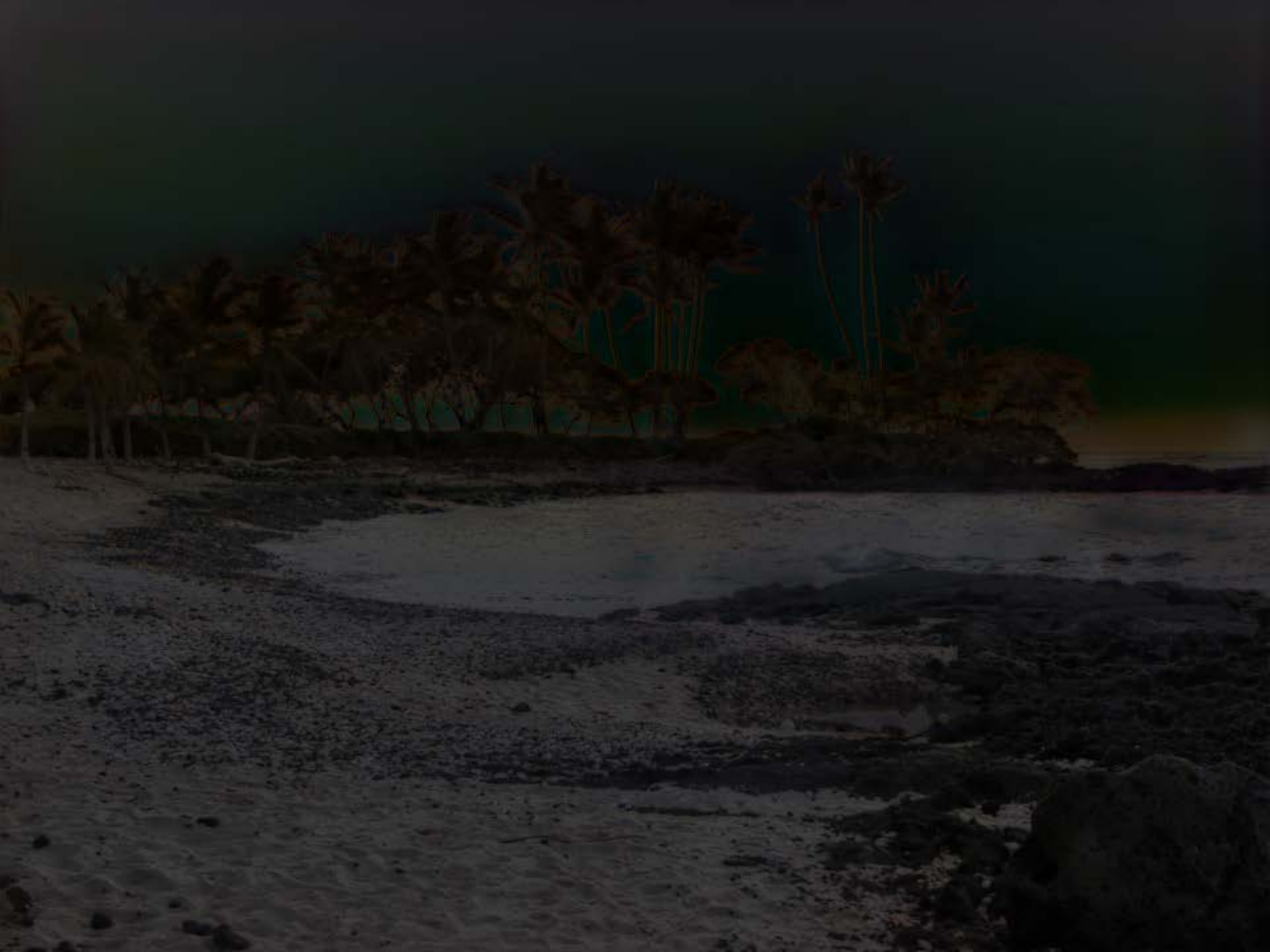}&
		\includegraphics[width=0.113\textwidth]{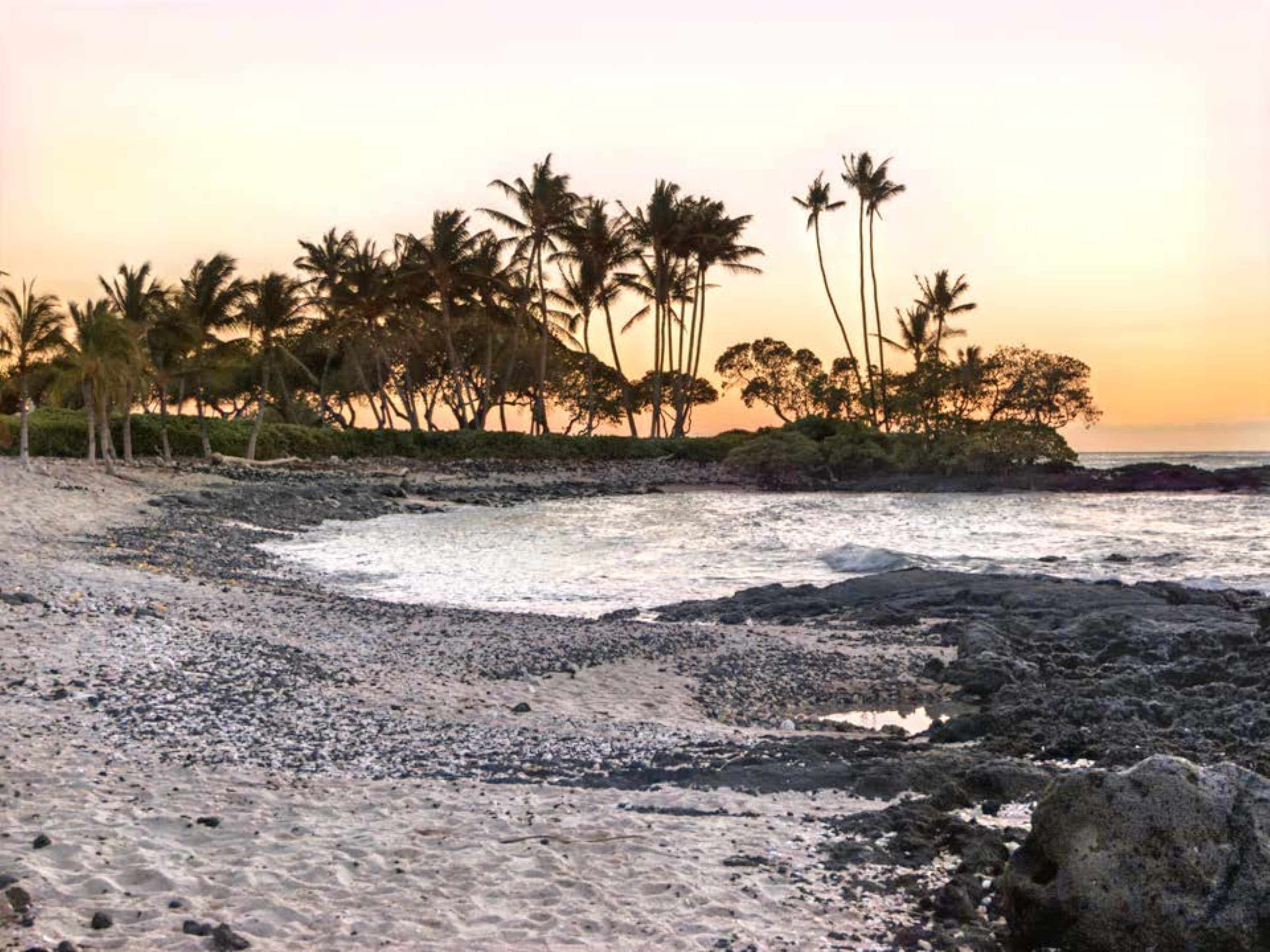}\\
		\multicolumn{4}{c}{\footnotesize w/ Illumination Guidance}\\ 
	\end{tabular}
	\caption{Ablation study on illumination guidance. (a) The feature maps generated in the 128-th channel of the third convolution block; (b) The output of IENet; (c) The output of RENet; (d) The final enhanced result. }
	\label{fig:AblationIG}
}
\end{figure}

\subsection{Effects of Context-Sensitive Decomposition Connection} \label{sec:CSDEffects}
In this part, we analyze the effects of context-sensitive decomposition connection from two perspectives. On the one hand, we defined three versions of CSDNet with different numbers of connections. On the other hand, we analyzed the ablation study in terms of this connection. 

\textbf{Three Versions of CSDNet.$\;$} As for the property of additional feature in U-Net, it brings some different options of context-sensitive decomposition. Therefore we defined three types of architectures including CSDNet$_a$ (only performing Eq.~\eqref{eq:RetinexConnection} on upsample layer in RENet), CSDNet$_b$ (only performing Eq.~\eqref{eq:RetinexConnection} on addition layer in RENet) and CSDNet$_c$ (performing Eq.~\eqref{eq:RetinexConnection} on upsample and addition layers in RENet, it is also our used strategy in the above experiments). 
We made a series of evaluations in terms of the above cases. 

As is shown in Fig.~\ref{fig:RCViscomp}, we can easily find that the best color and structural expression happen on CSDNet$_c$. This is a fully conceivable conclusion. That is to say, when the contextual dependencies are sufficiently exploited, the structural information is more completely presented and the color is more saturatedly performed. It fully indicates the procedure of context-sensitive decomposition is necessary and meaningful for low-light image enhancement. 

We also explored the change of feature maps in Fig.~\ref{fig:RCFeature}, which demonstrates the features in the 32-th channel after the third upsample layer of the decoder. Obviously, the mechanism of context-sensitive decomposition indeed strengthens the structural edges and textures, and simultaneously removes most of worthless artifacts and noises.

\begin{figure}[t]
	\centering
	{
		\begin{tabular}{c@{\extracolsep{0.3em}}c@{\extracolsep{0.3em}}c}
			\includegraphics[width=0.15\textwidth]{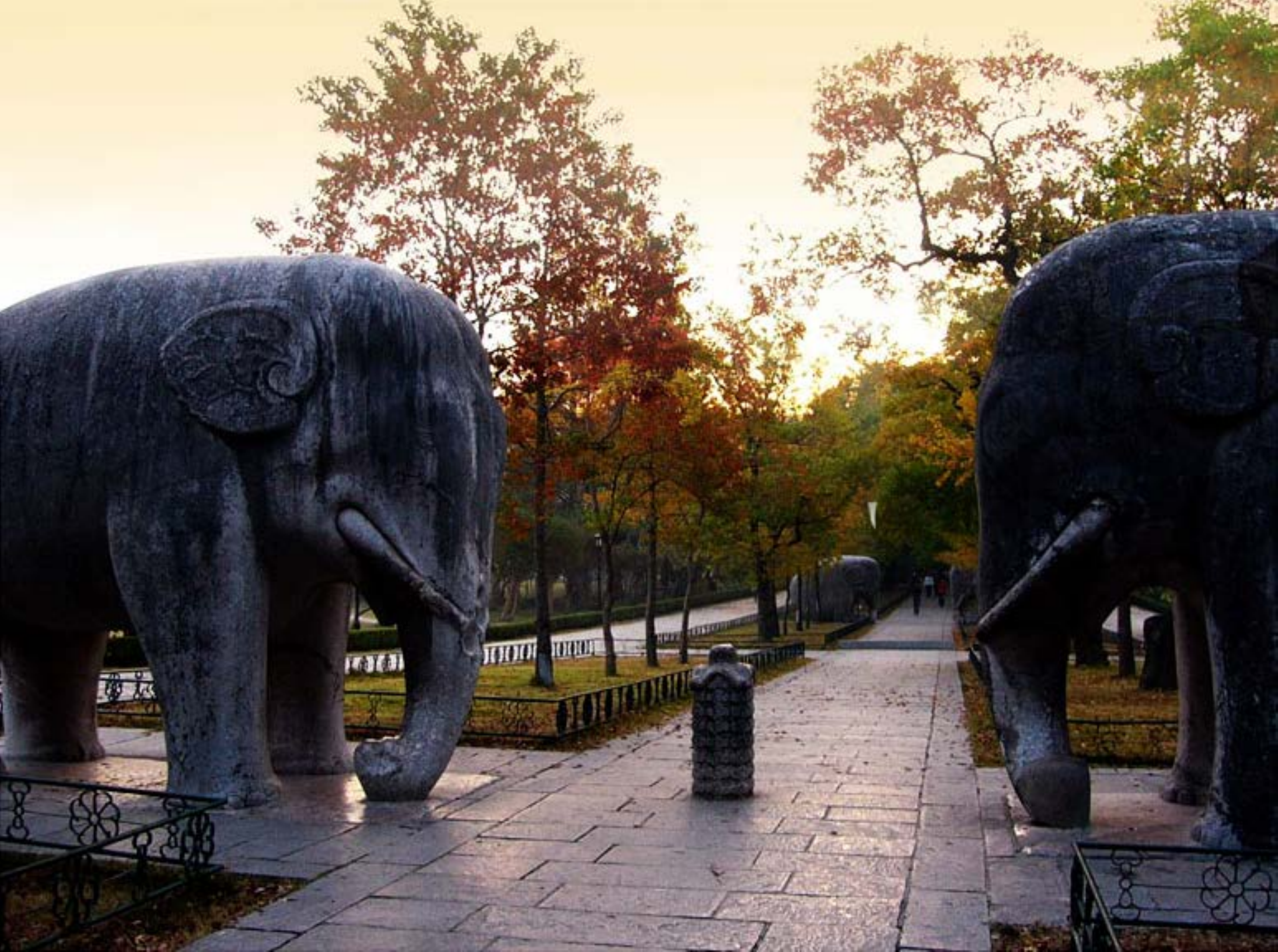}&
			\includegraphics[width=0.15\textwidth]{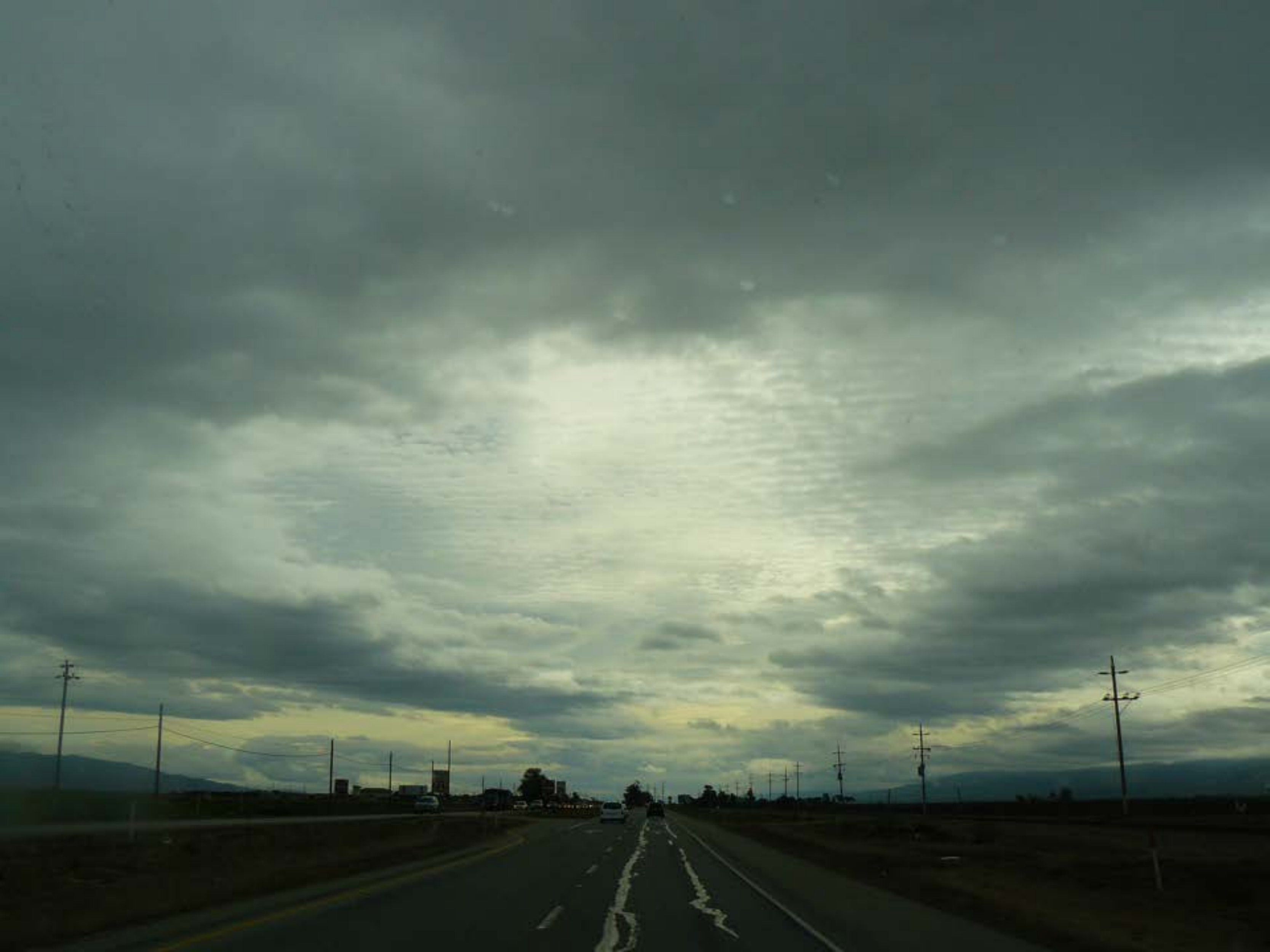}&
			\includegraphics[width=0.15\textwidth]{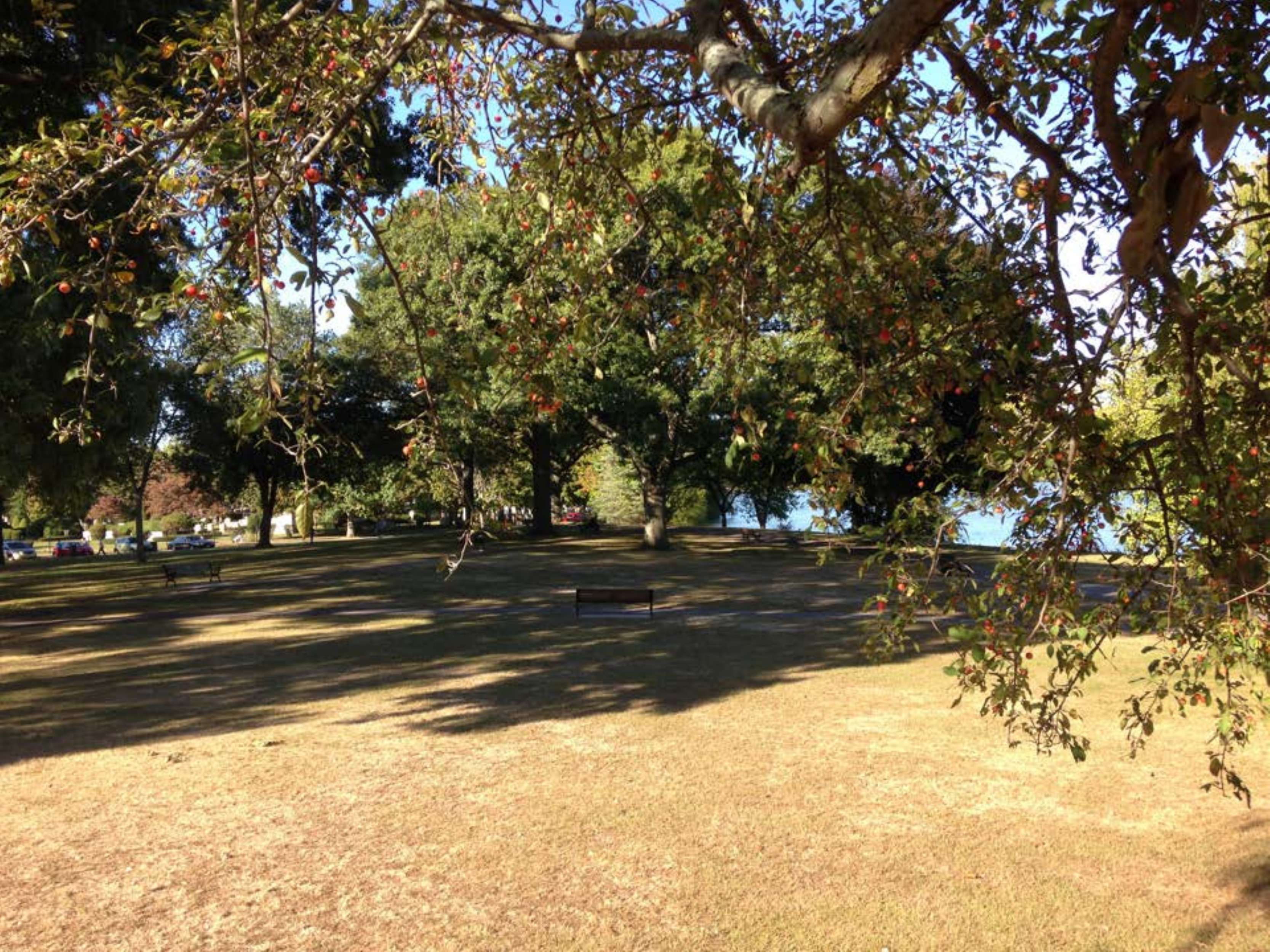}\\
			\multicolumn{3}{c}{\footnotesize Input}\\
			\includegraphics[width=0.15\textwidth]{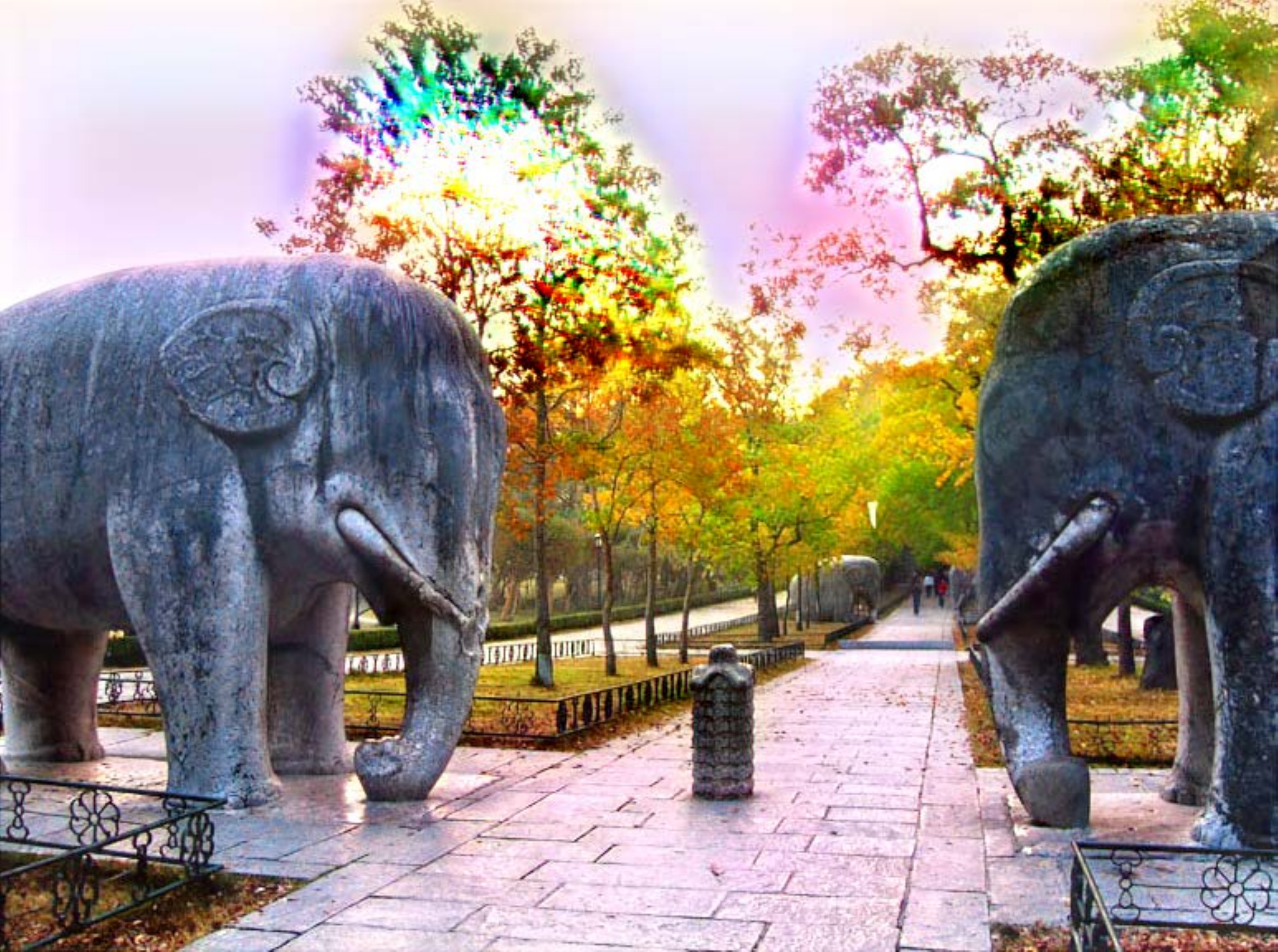}&
			\includegraphics[width=0.15\textwidth]{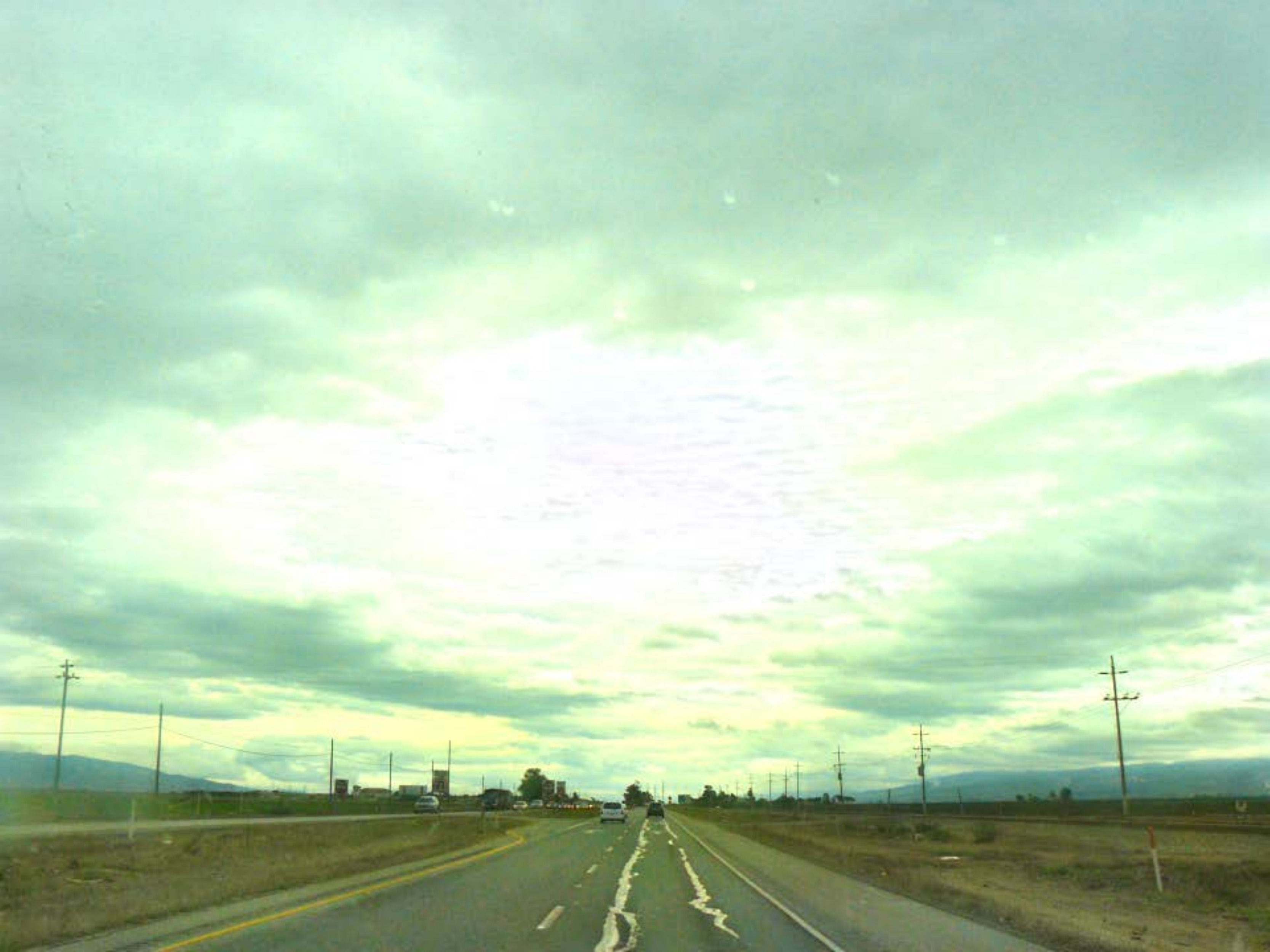}&
			\includegraphics[width=0.15\textwidth]{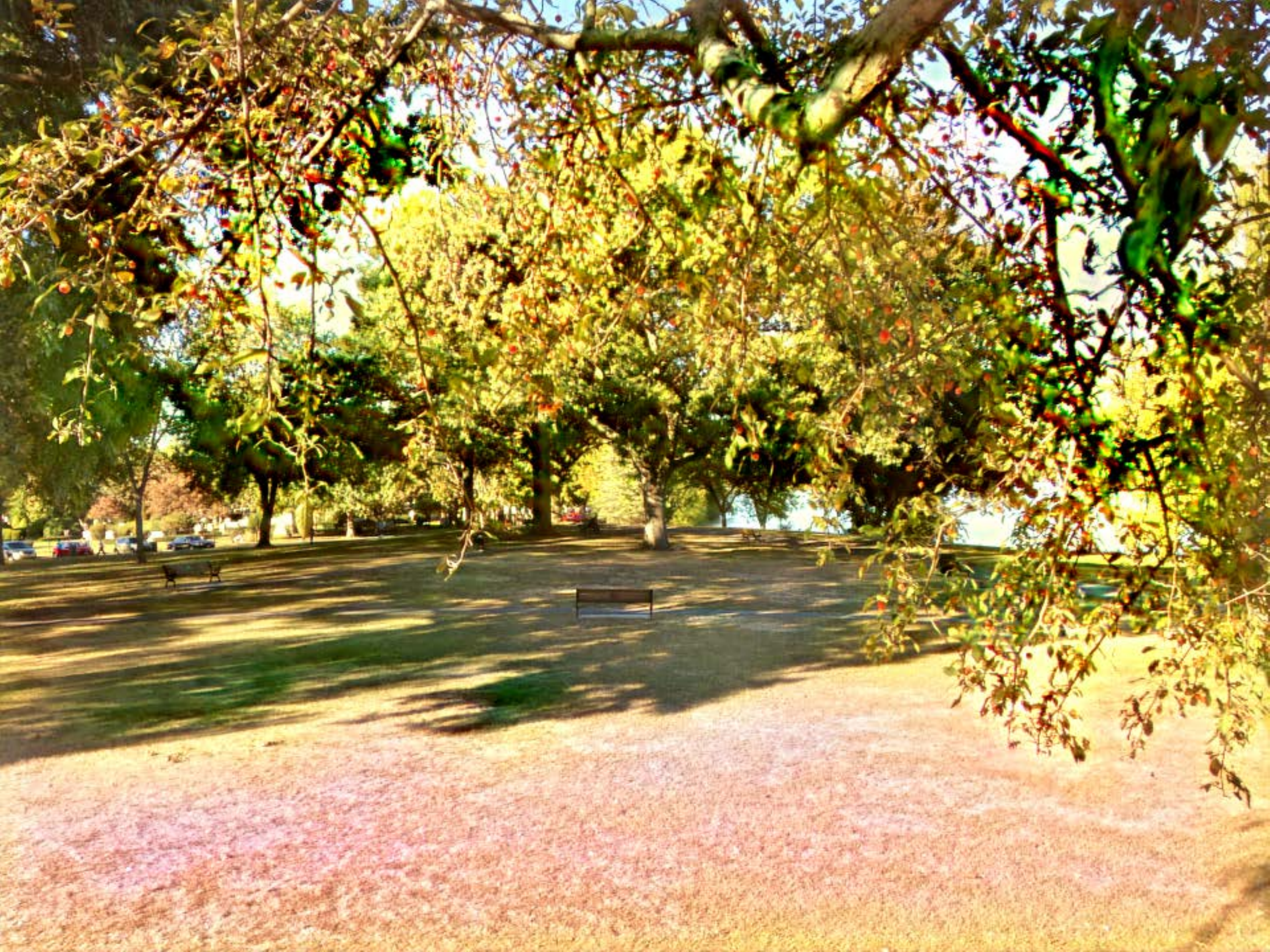}\\
			\multicolumn{3}{c}{\footnotesize w/o Residual Connection}\\
			\includegraphics[width=0.15\textwidth]{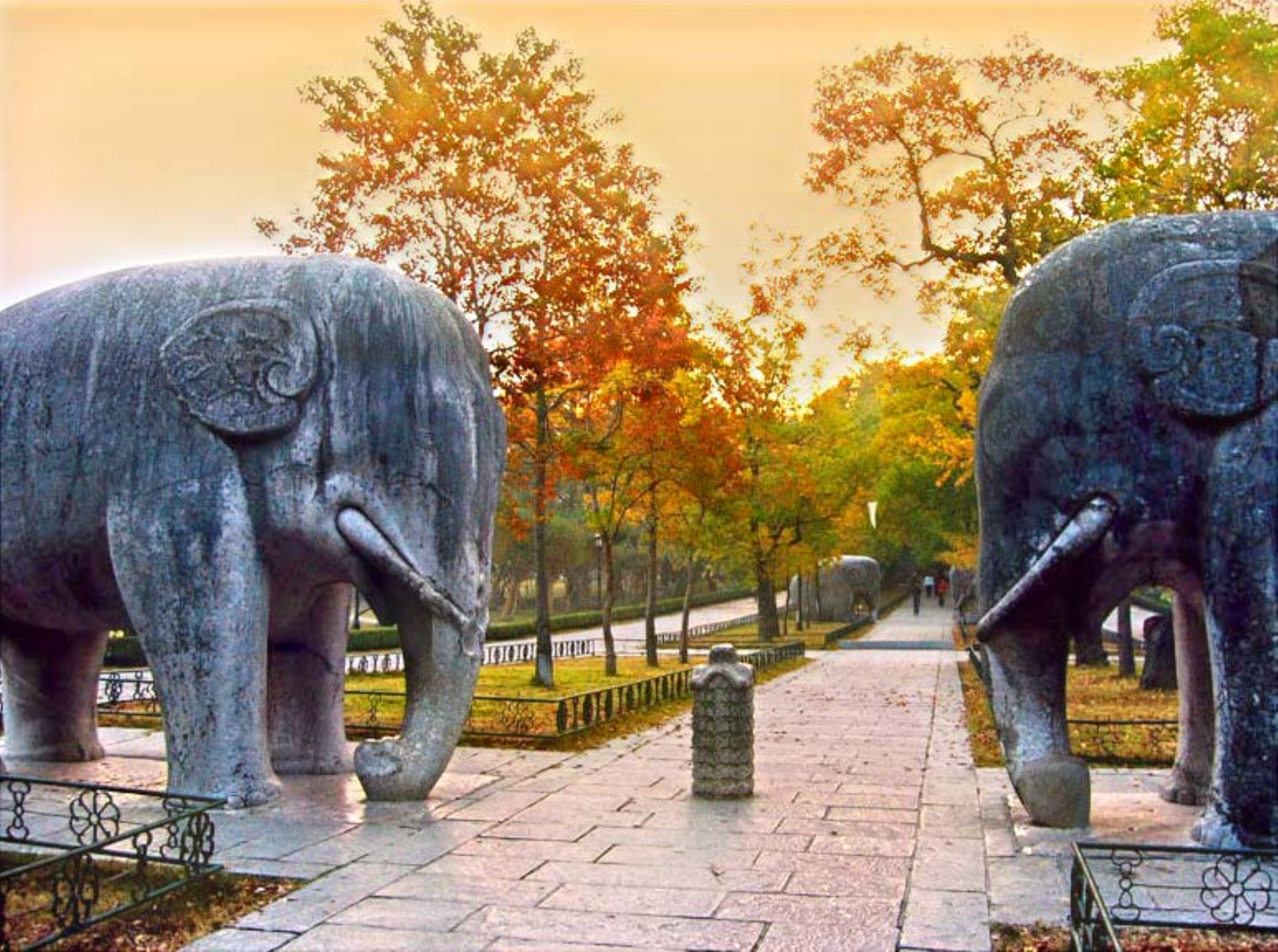}&
			\includegraphics[width=0.15\textwidth]{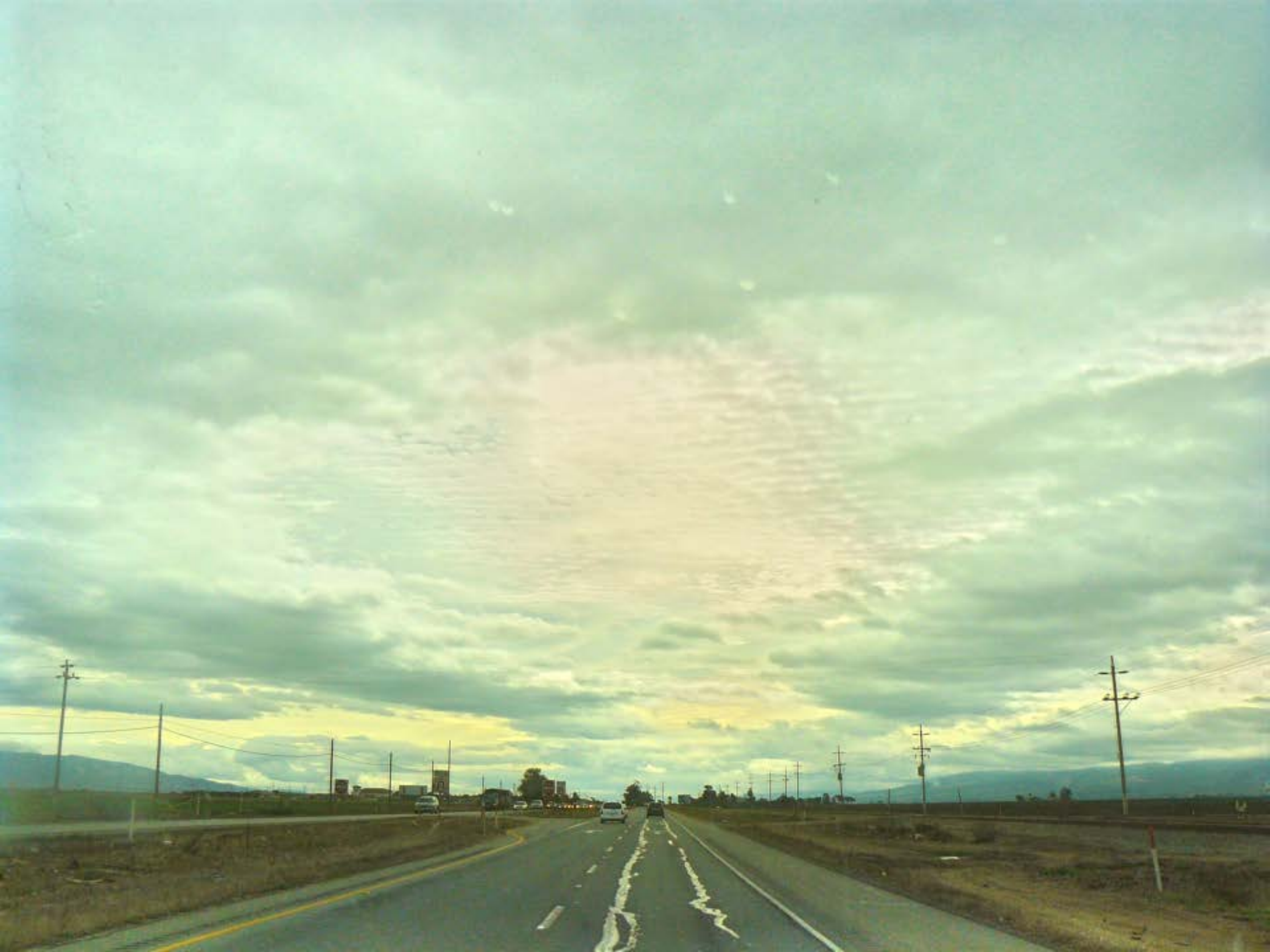}&
			\includegraphics[width=0.15\textwidth]{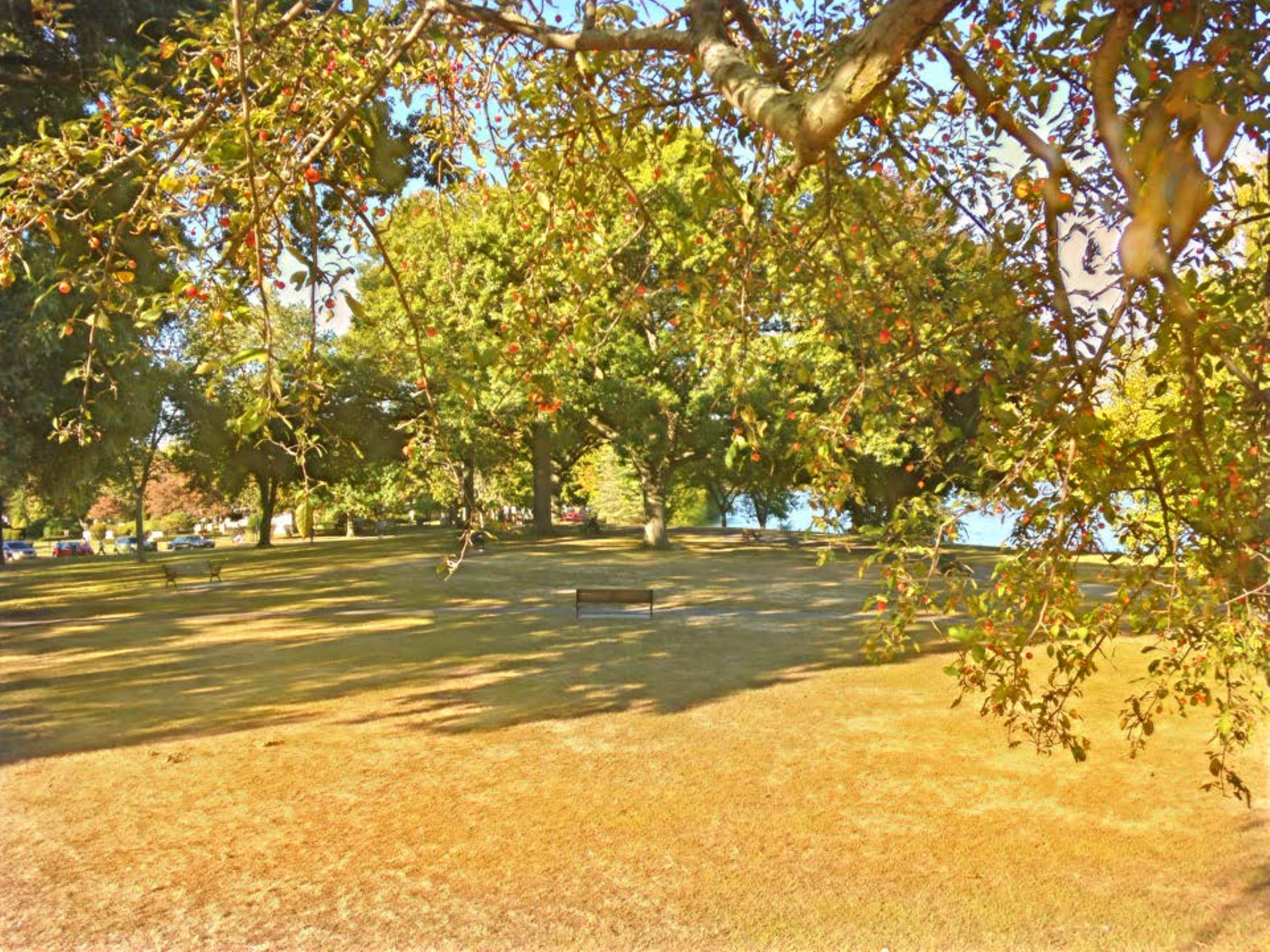}\\
			\multicolumn{3}{c}{\footnotesize w/ Residual Connection}\\
		\end{tabular}
		\caption{Effects of residual connection in CSDGAN.}
		\label{fig:GANres}
	}
\end{figure}

{
\textbf{Ablation Study.$\;$} Here we considered three different U-Net based architectures from simple to complex, and two different bridging manners for estimating the illumination and reflectance. As is shown in Fig.~\ref{fig:Arc}, inspired by the representative works in~\cite{Chen2018Retinex} and~\cite{zhang2019kindling}, $\mathtt{Arc}_a$, $\mathtt{Arc}_c$, and $\mathtt{Arc}_e$ connected the illumination and reflectance by the training loss $\|\mathbf{L}-\mathbf{R}\odot\mathbf{I}\|$. Their final enhanced outputs are the reflectance. $\mathtt{Arc}_b$, $\mathtt{Arc}_d$, and $\mathtt{Arc}_f$ connected two components by our designed context-sensitive decomposition connection. As for how to add the illumination guidance, we made some slight changes according to different architectures. To be specific, as for $\mathtt{Arc}_a$ and $\mathtt{Arc}_b$, we added the illumination guidance in the last convolution layer. As for $\mathtt{Arc}_c$ and $\mathtt{Arc}_d$, we resized the illumination guidance to the minimum scale and added to the input of illumination decoder. As for $\mathtt{Arc}_e$ and $\mathtt{Arc}_f$, we followed the manner presented in Fig.~\ref{fig:Flow}. 

Table~\ref{tab:AblationStudy} reported the numerical results on 15 LOL testing images. In terms of the connected way (commonly-used and our newly-designed), it can be easily seen that our context-sensitive decomposition connection (i.e., $S_3, S_7, S_{11}$) demonstrated the consistently superior performance against the commonly-used connection (i.e., $S_1, S_5, S_{9}$) under different architectures. 
Note that using the architecture $\mathtt{Arc}_a$ with our guidance (i.e., $S_4$) was poorer than other architectures defined by our method (i.e., $S_8$ and Ours).  It is because our intention is to carefully optimize different components using component-specific architecture. However, in $S_4$, these two components are simultaneously optimized, causing inaccurate estimations. 
Fig.~\ref{fig:AblationAll} can further verify our advantages in the newly-introduced connection. That is, our connection can effectively eliminate noises/artifacts, but the commonly-used connected way cannot realize it.  
}

{
\subsection{Ablation Study on Illumination Guidance} \label{sec:IGEffects}
In Table~\ref{tab:AblationStudy}, we can see that illumination guidance significantly improves the numerical scores under different architectures and different connected ways. As for how to add illumination guidance in different architectures, we made some slight changes according to different architectures. To be specific, as for $\mathtt{Arc}_a$ and $\mathtt{Arc}_b$, we added the illumination guidance in the last convolution layer. As for $\mathtt{Arc}_c$ and $\mathtt{Arc}_d$, we resized the illumination guidance to the minimum size and added to the input of illumination decoder. As for $\mathtt{Arc}_e$ and $\mathtt{Arc}_f$, we followed the manner presented in our framework. It can fully indicate the effectiveness of our designed illumination guidance. Visual comparisons in Fig.~\ref{fig:AblationAll} can further reflect its effects under different cases. Concretely, illumination guidance can assist in removing artifacts for the cases of $\mathtt{Arc}_a$, $\mathtt{Arc}_b$, $\mathtt{Arc}_c$, and $\mathtt{Arc}_d$ (see $S_1\rightarrow S_2$, $S_3\rightarrow S_4$, $S_5\rightarrow S_6$, and $S_7\rightarrow S_8$). As for $\mathtt{Arc}_e$ and $\mathtt{Arc}_f$, illumination guidance can significantly improve brightness (see $S_9\rightarrow S_{10}$, $S_{11}\rightarrow$ Ours). In a word, illumination guidance can provide positive support for enhanced results on different cases. 

}

{
\subsection{Effects of Residual Connection in CSDGAN}
Different from CSDNet, we added an additional residual connection between the input and output. In this part, we explored the reason for it. As is shown in Fig.~\ref{fig:GANres}, it can be easily seen that if w/o residual connection, color distortion, over-exposure, and details loss were inevitable drawbacks. Fortunately, by adding a simple residual connection, these issues were significantly ameliorated. In a word, the introduction of residual connection ensured the stability of CSDGAN. 
}

\begin{figure}[t]
	\centering
	{
		\begin{tabular}{c}
			\includegraphics[width=0.46\textwidth]{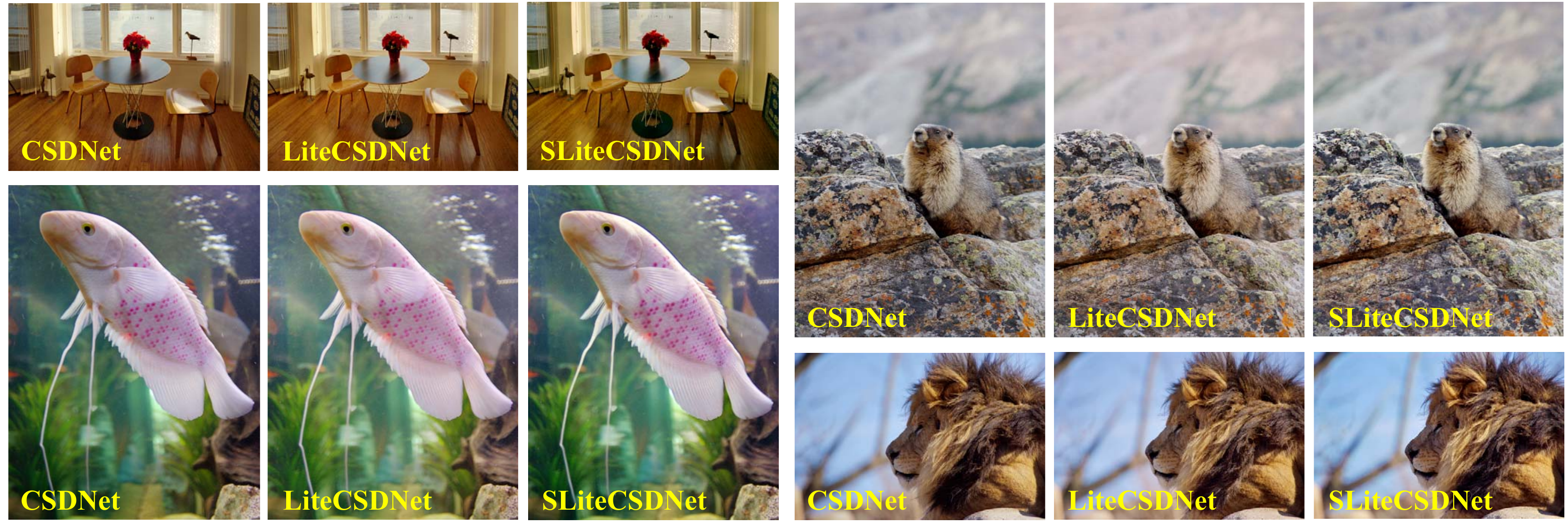}\\
			\footnotesize (a) Visual results on the MIT-Adobe FiveK dataset\\	
			\includegraphics[width=0.46\textwidth]{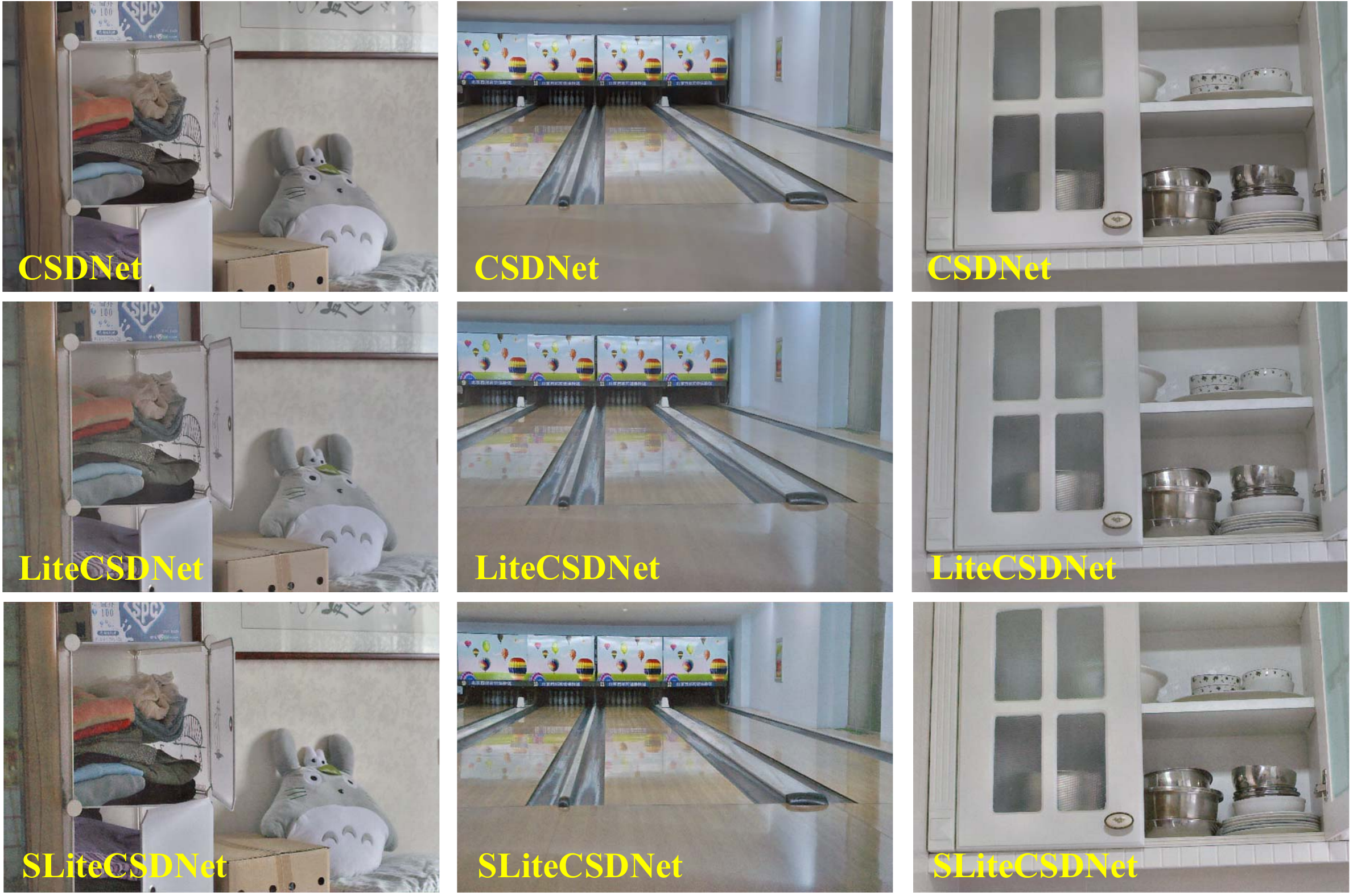}\\	
			\footnotesize (b) Visual results on the LOL dataset\\	
		\end{tabular}
		\caption{Visual comparison on the MIT-Adobe FiveK~\cite{fivek} and LOL~\cite{Chen2018Retinex} datasets among CSDNet, LiteCSDNet, and SLiteCSDNet.}
		\label{fig:LitevsCSDNet}
	}
\end{figure}

{
\subsection{LiteCSDNet, SLiteCSDNet --- Lightweight CSDNet}\label{sec: LiteCSDNet}
From the above comparisons, our proposed algorithm indeed outperformed other state-of-the-art works. In some practical applications, the computational efficiency is more critical and widely concerned, even a little accuracy loss can be acceptable. In our designed architecture, we directly adopted the original U-Net~\cite{ronneberger2015u} to construct our architecture. However, to ensure a more powerful characterization, U-Net considered too many numbers of feature channels that spent too much inference time. To improve the execution efficiency, we defined a LiteCSDNet by reducing the number of feature channels. Concretely, the channel of each block is set to 12. In this way, we avoided redesigning the architecture, but significantly improving the efficiency. 

Table~\ref{tab:Parameters} reported the number of parameters among different state-of-the-art works. Obviously, our LiteCSDNet had the least number of parameters, which can ensure fast inference. More importantly, it achieved the reduction rate of more than 97\% compared with others. We also reported the quantitative scores on 100 testing images randomly sampled from MIT-Adobe FiveK dataset. Our all versions were significantly superior to other works. 
Further, we shared an encoder for these two components to try to improve the efficiency again, named SLiteCSDNet (i.e., $\mathtt{Arc}_d$ in Fig.~\ref{fig:Arc}). Fortunately, we indeed obtained the desired improvement, that is, SLiteCSDNet just needed 0.00301M parameters which reduced by half the parameters of LiteCSDNet. In addition, to evaluate the visual difference between these versions, Fig.~\ref{fig:LitevsCSDNet} demonstrated the visual comparison of the images sampled from MIT-Adobe FiveK and LOL datasets. Fortunately, the entire visual performance of LiteCSDNet and SLiteCSDNet almost kept the same expression with CSDNet. All in all, our constructed lightweight networks really achieved the high-efficiency without losing high-accuracy.
}

\section{Conclusions and Future Work}
{
In this paper, we proposed a new context-sensitive decomposition network to tackle LLIE. By exploiting the contextual dependencies on spatial scales, color and structural information can be recovered to most extent. We also designed a spatially-varying operation to incorporate illumination guidance into our network. Both paired (CSDNet) and unpaired (CSDGAN) training strategies were adopted to fit different real-world scenarios. 
Extensive quantitative and qualitative experiments on seven datasets demonstrated our superiority against existing state-of-the-art approaches. Two lightweight versions of CSDNet (i.e, LiteCSDNet and SLiteCSDNet) were also developed by different mechanisms to achieve high-efficiency and high-accuracy in the real-world scenarios.
}

This paper provides a new perspective to exploit the contextual information in the internal of the deep network by incorporating the physical principle, to ensure the desired enhanced results. Actually, this idea can be applied for other
low-level computer vision tasks such as image denoising, image dehazing, image deraining and so on. We will verify these tasks in our future works.


%


\bibliographystyle{IEEEtran}
\bibliography{egbib}

\begin{IEEEbiography}[{\includegraphics[width=1in,height=1.25in,clip,keepaspectratio]{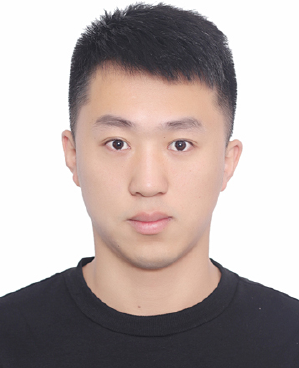}}]{Long Ma} received the B.E. degree in information and computing science from Northeast Agricultural University, Harbin, China, in 2016. received the M.S. degree in software engineering at Dalian University of Technology, Dalian, China, in 2019. He is currently pursuing the Ph.D. degree in software engineering at Dalian University of Technology, Dalian, China. His research interests include computer vision and image enhancement. 
\end{IEEEbiography}

\begin{IEEEbiography}[{\includegraphics[width=1in,height=1.25in,clip,keepaspectratio]{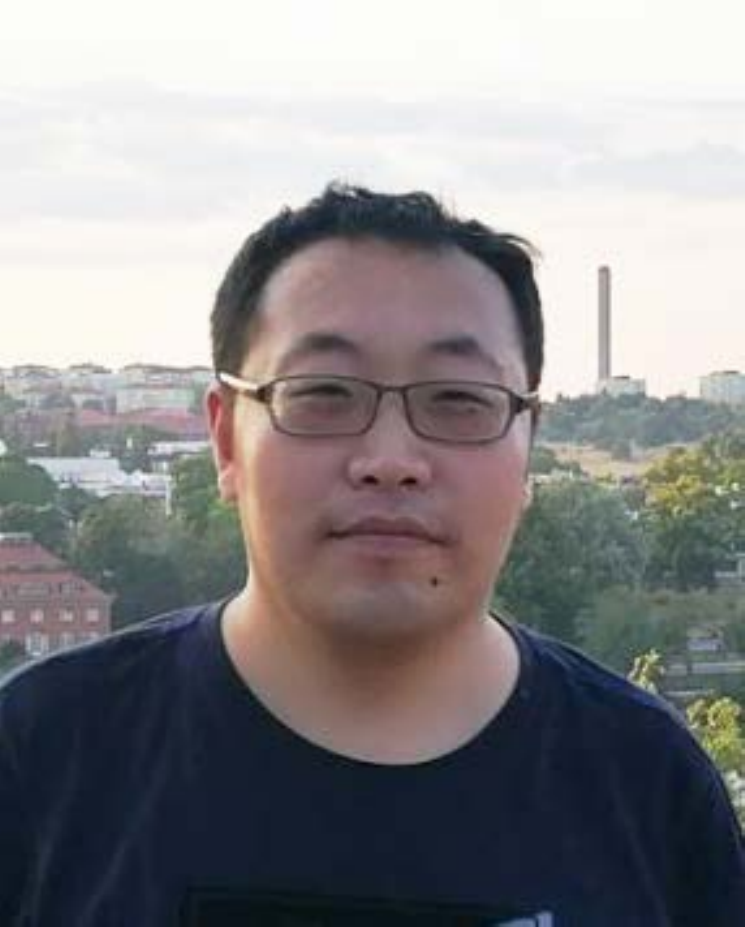}}]{Risheng Liu} received his B.Sc. (2007) and Ph.D. (2012) from Dalian University of Technology, China. From 2010 to 2012, he was doing research as joint Ph.D. in robotics institute at Carnegie Mellon University. From 2016 to 2018, He was doing research as Hong Kong Scholar at the Hong Kong Polytechnic University. He is currently a full professor with the Digital Media Department at International School of Information Science \& Engineering, Dalian University of Technology (DUT). He was awarded the ``Outstanding Youth Science Foundation" of the National Natural Science Foundation of China. He serves as editor for the Journal of Electronic Imaging (Senior Editor), The Visual Computer, and IET Image Processing. His research interests include optimization, computer vision and multimedia.
\end{IEEEbiography}

\begin{IEEEbiography}[{\includegraphics[width=1in,height=1.25in,clip,keepaspectratio]{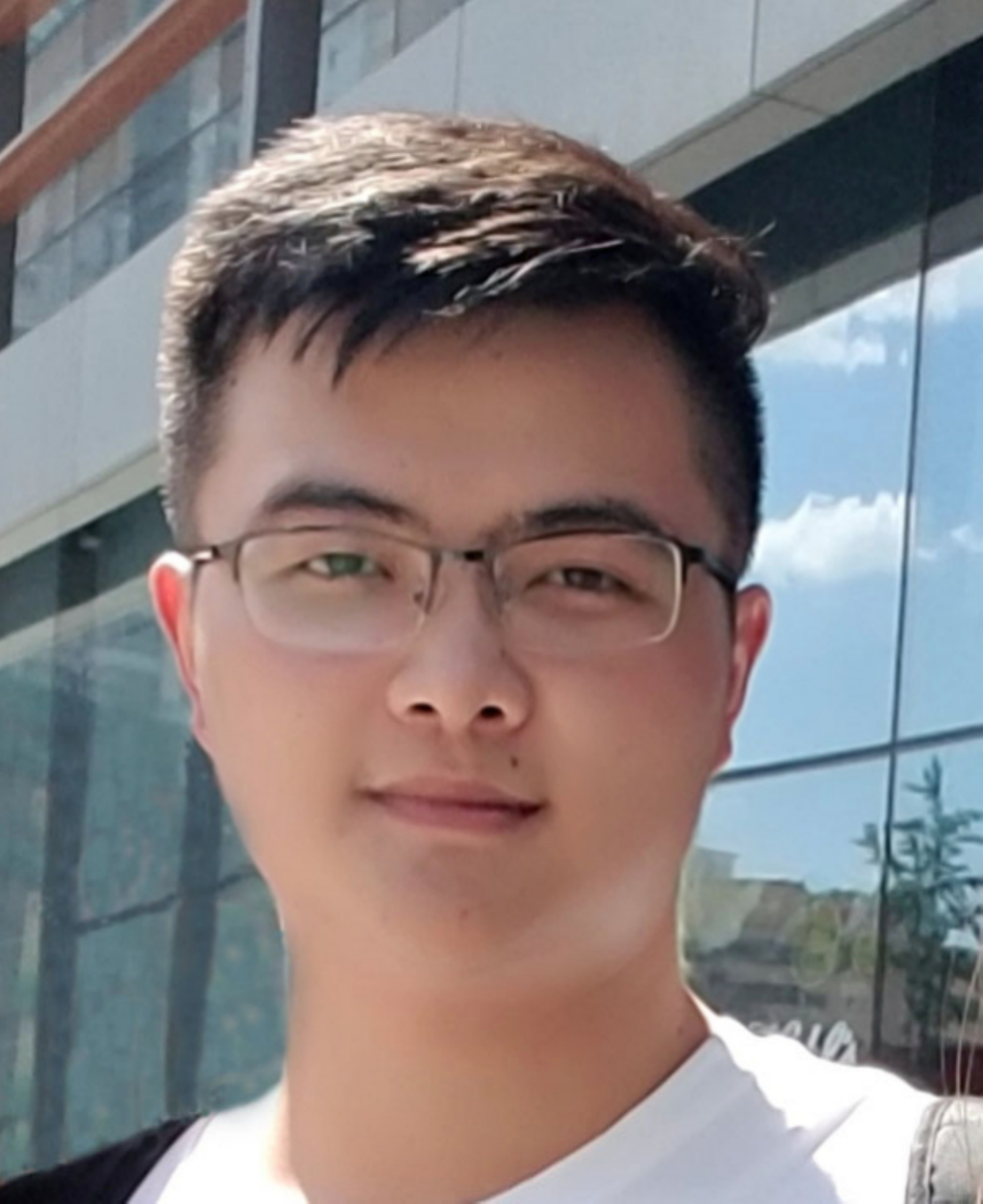}}]{Jiaao Zhang} received the B.E. degree in software engineering from Dalian University of Technology, Dalian, China, in 2019. He is currently pursuing the M.S. degree in software engineering at Dalian University of Technology, Dalian, China. His research interests include computer vision, image enhancement and deep learning. 
\end{IEEEbiography}

\begin{IEEEbiography}[{\includegraphics[width=1in,height=1.25in,clip,keepaspectratio]{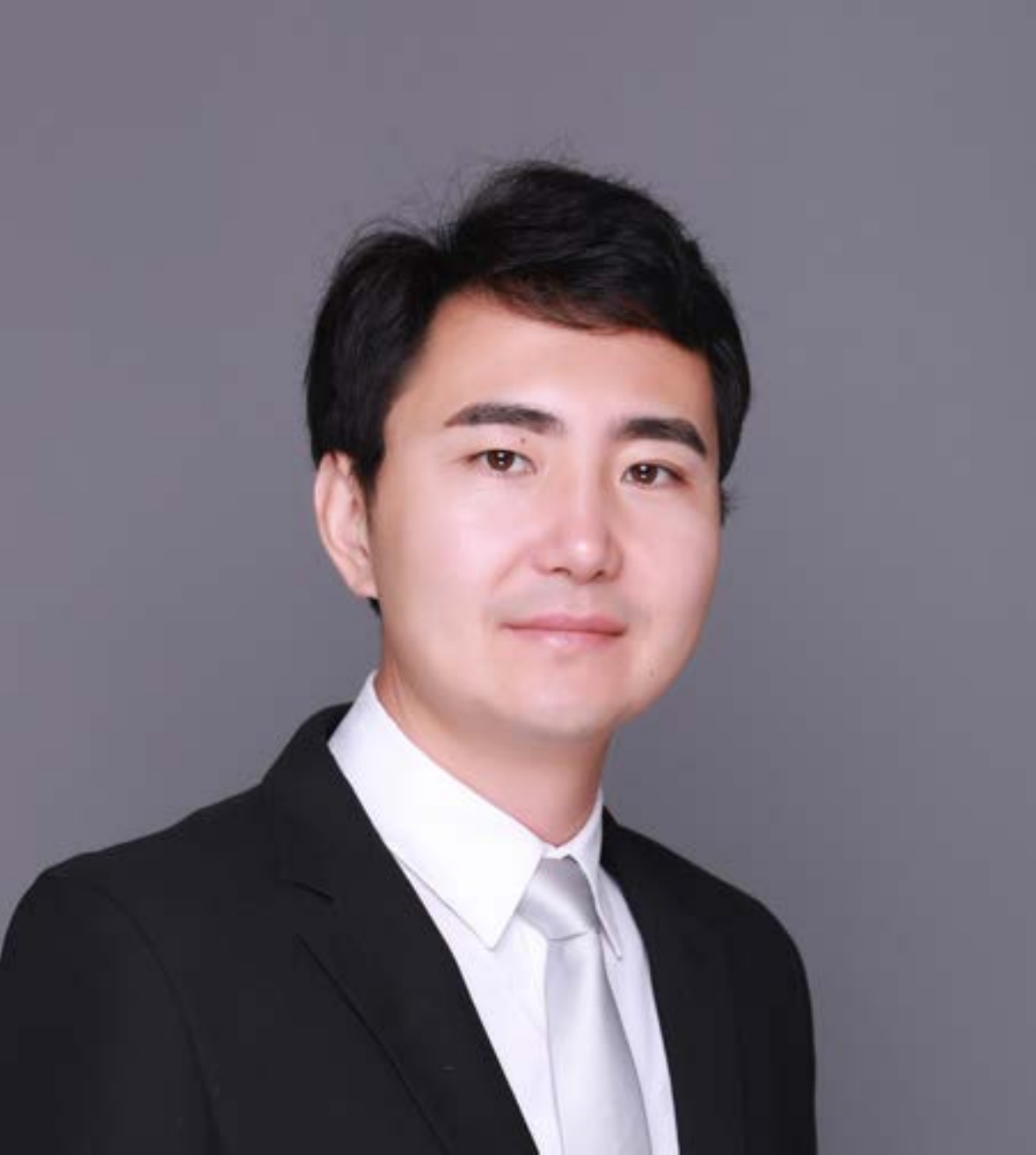}}]{Xin Fan} received the B.E. and Ph.D. degrees in information and communication engineering from Xian Jiaotong University, Xian, China, in 1998 and 2004, respectively. He was with Oklahoma State University, Stillwater, from 2006 to 2007, as a post-doctoral research
	Fellow. He joined the School of Software, Dalian University of Technology, Dalian, China, in 2009. His current research interests include computational geometry and machine learning, and their applications to lowlevel image processing and DTI-MR image analysis.
\end{IEEEbiography}

\begin{IEEEbiography}[{\includegraphics[width=1in,height=1.25in,clip,keepaspectratio]{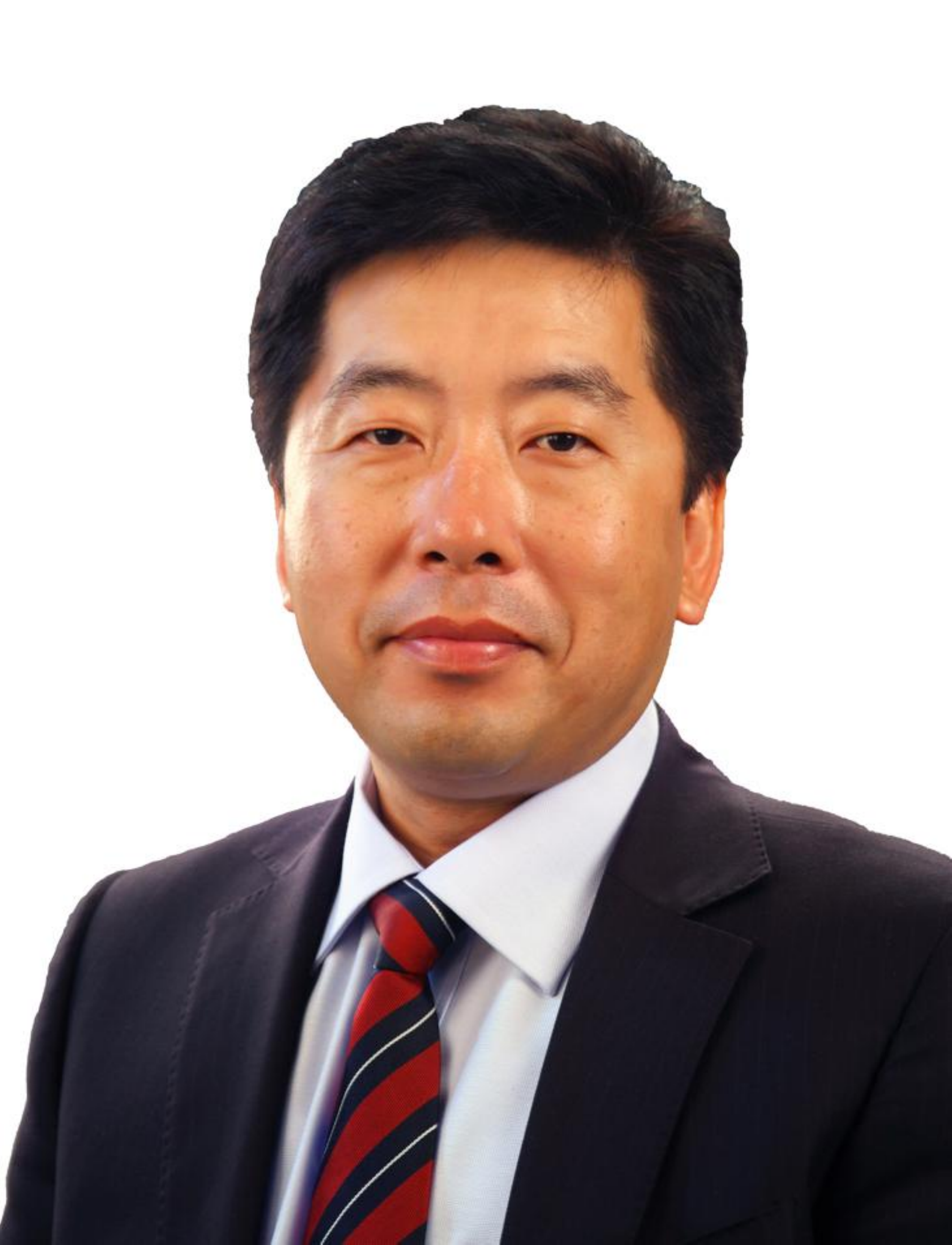}}]{Zhongxuan Luo} received the B.S. degree in Computational Mathematics from Jilin University, China, in 1985, the M.S. degree in Computational
	Mathematics from Jilin University in 1988, and the Ph.D. degree in Computational Mathematics from Dalian University of Technology, China, in 1991. He has been a full professor of the School of Mathematical Sciences at Dalian University of Technology since 1997. His research interests include computational geometry and computer vision.
\end{IEEEbiography}

\end{document}